\documentclass[a4paper,fleqn,usenatbib]{mnras}
\usepackage{mathptmx}

\usepackage[T1]{fontenc}
\usepackage{ae,aecompl}

\usepackage{pdflscape}
\usepackage{longtable}
\usepackage{rotating}

\usepackage{graphicx}	
\usepackage{amsmath}	
\usepackage{amssymb}	
\usepackage{booktabs}
\usepackage{float}
\usepackage{array}
\usepackage{wrapfig}
\usepackage{multirow}
\usepackage{tabu}
\usepackage{adjustbox}
\usepackage{tabularx}
\usepackage{booktabs,caption,fixltx2e}
\usepackage[flushleft]{threeparttable}
\usepackage{subcaption}
\usepackage{mathtools}
\usepackage{longtable} 

\usepackage[dvipsnames]{xcolor}
\hypersetup{colorlinks=true,linkcolor=Magenta,urlcolor=Plum,citecolor=DarkOrchid}

\newcommand\T{\rule{0pt}{2.6ex}}       
\newcommand\B{\rule[-1.2ex]{0pt}{0pt}} 
\newcommand{\HI}{\mbox {\sc H\thinspace{i}}}

\newcommand{\kms}{\mbox{km\thinspace s$^{-1}$}}

\newcommand{\HIPASS}{\mbox {\sc {Hipass}}}
\newcommand{\ALFALFA}{\mbox {\sc {Alfalfa}}}
\newcommand*{\ditto}{\texttt{"}}

\title[Environmental processing of galaxies in Choirs]{Environmental processing of galaxies in \HI-rich groups}

\author[D\v{z}ud\v{z}ar et al.]{
Robert D\v{z}ud\v{z}ar,$^{1}$\thanks{E-mail: rdzudzar@swin.edu.au; robertdzudzar@gmail.com}
Virginia Kilborn,$^{1,2}$
Sarah M. Sweet,$^{1,2,3}$
Gerhardt Meurer,$^{4}$
\newauthor T.H. Jarrett,$^{5}$
Dane Kleiner$^{6}$
\\
$^{1}$Centre for Astrophysics and Supercomputing, Swinburne University of Technology, P.O. Box 218, Hawthorn, VIC 3122, Australia\\
$^2$ARC Centre of Excellence for All Sky Astrophysics in 3 Dimensions (ASTRO 3D)\\
$^{3}$School of Mathematics and Physics, University of Queensland, Brisbane, QLD 4072, Australia\\
$^4$International Centre for Radio Astronomy Research, ICRAR M468, 35 Stirling Highway, Crawley, WA 6009, Australia\\
$^5$Department of Astronomy, University of Cape Town, Rondebosch, South Africa\\
$^{6}$INAF - Osservatorio Astronomico di Cagliari, Via della Scienza 5, I-09047 Selargius (CA), Italy
}

\date{Accepted XXX. Received YYY; in original form ZZZ}

\pubyear{2020}

\begin{document}
\label{firstpage}
\pagerange{\pageref{firstpage}--\pageref{lastpage}}
\maketitle

\begin{abstract}

We present and explore the resolved atomic hydrogen (\HI) content of 13 \HI-rich and late-type dominated groups denoted `Choirs'. We quantify the \HI\ content of the Choir galaxies with respect to the median of the \HI-mass fraction (f$_{\textrm{\HI}}$) of their grandparent \HIPASS\ sample. We find that the \HI\ mass fraction of the Choir galaxies is dispersed around the \HIPASS\ median in the range -1.4$\leq \Delta$f$_{\textrm{\HI}}\textrm{[dex]}\leq$0.7, from \HI-excess to \HI-deficient galaxy regime. 
The \HI-excess/\HI-deficient galaxies contain more/less than 2.5 times their expected \HI\ content with respect to the \HIPASS\ median. We show and discuss that the environmental processing in Choirs occurs via tidal stripping and galaxy mergers. Our analysis suggests that tidal stripping contributes to the loss of the \HI, while galaxy mergers contribute to the enhancement of the \HI. Exploring the mid-infrared properties of Choir galaxies we find possible environmental processing in only nine Choir galaxies, which indicates that  environmental processing is more perceptible in the \HI\ content than the mid-infrared properties. Moreover, we find that environmental processing occurs in Choir groups regardless of their global environment, whether they are in isolation or in proximity to the denser structures, such as cosmic web filaments. We explore possible scenarios of the Choirs evolution, taking into account their \HI\ content, velocity dispersion, crossing time and their global environment. We conclude that the most likely evolution for the majority of Choir groups is that they will become more compact as their members undergo multiple \HI-rich mergers. \\

\end{abstract}




\section{Introduction}

In the local Universe 50--60 percent of galaxies reside in groups \citep{Yang2007, Freeland2009, Tempel2014} however, environment is not a simple categorical property, it is rather a continuous one. 

Galaxy groups that have been studied in the literature have been classified as loose groups \citep{Driel2001, Pisano2007, Kilborn2009, Osterloo2018}, poor groups \citep{Zabludoff1998, Tovmassian2009}, compact groups \citep{Hickson1997, Coziol2000, VerdesMontenegro2001}, rich groups \citep{Wojtak2013}, small groups \citep{Barnes2001groups}, supergroups \citep{Brough2006supergroup, Wolfinger2016}, dwarf groups \citep{Stierwalt2017} and fossil groups \citep{Ponman1994, Jones2003}. The common property of these groups is that galaxies have relatively low velocity dispersion and thus tidal interactions and mergers are often observed (e.g. \citealt{Yun1994, Rots1990, Sengupta2017}). Therefore, groups are a suitable environment for galaxy environmental processing in the form of: (i) disturbing galaxy's stellar content and morphology, (ii) exchanging gas content between galaxies (removing or accreting), which can lead to change of galaxy's star formation properties. \\

\subsection{{\sc H\thinspace{i}} gas}

Hydrogen in its atomic form (\HI) is the raw material for condensing into molecular Hydrogen (H$_{2}$), the fuel for future star formation, thus it is an essential part of galaxies \citep{Bigiel2012, Catinella2018}. The \HI\ content in galaxies is often analysed using various scaling relations utilizing galaxy properties such as stellar mass, color, morphology, star formation rate, as well as tracing galaxy environment \citep{Hayens1984, Giovanelli1985, Denes2014, Catinella2018}. Large surveys such as The \HI\ Parkes All Sky Survey (\HIPASS, \citealt{Barnes2001}) and the Arecibo Legacy Fast ALFA Survey (\ALFALFA, \citealt{Giovanelli2005}) detected \HI\ in more than 5000 and 30000 galaxies respectively, and have significantly improved these scaling relations. 

Mapping the \HI\ gas within and around the galaxies can show its distribution, be used to determine galaxy kinematics, and to study galaxy interactions. The \HI\ features that reveal galaxy interactions, and are often found in galaxy groups, are in the form of tidal tails \citep{Scott2012}, bridges \citep{Jones2019}, rings \citep{Malphrus1997} and clouds \citep{Ryder2001}, and more of such systems are described in the brief overview by \citet{Koribalski2020}. The abundance of such \HI\ features points towards a complex history of galaxy evolution in the group environment. 

\subsection{{\sc H\thinspace{i}} in groups}

It is known that in galaxy clusters, the most densest of environments, a galaxy's evolution is influenced by its environment. Thus, we observe an increased fraction of early-type galaxies \citep{Dressler1980}, decreased star formation rates \citep{Lewis2002} and decreased gas content of galaxies \citep{Hayens1984} in high density environments. On the other hand, the evolution of galaxies in the field is thought to be free from environmental effects. Naturally, galaxy groups are in between these extreme environments. 

It has been shown that spiral galaxies are more \HI\ deficient in higher density environments, for instance, near the galaxy cluster centre or in groups, than spiral galaxies in the field \citep{Giovanelli1985, Chung2009, Solanes2001, VerdesMontenegro2001, Kilborn2009, Denes2014, Denes2016}. The difference in \HI\ content across different environments suggests that the higher density environment contributes to the exhaustion of the \HI\ gas. The depletion of the \HI\ content within clusters is most often connected to ram-pressure stripping \citep{GunnGott1972, Chung2009, Cortese2011, Brown2017, Stevens2017}. On the other hand, the depletion of the \HI\ content in groups is debated between ram-pressure stripping \citep{Brown2017}, tidal interactions \citep{Yun1994}, starvation \citep{Larson1980} and viscous stripping \citep{Nulsen1982}. 

The most well-studied type of groups are Hickson Compact Groups (HCG), which are dense galaxy configurations that contain four to 10 galaxies with a projected distance between them on the order of a galaxy size (and smaller) and velocity dispersion of $\sim$200 km s$^{-1}$ \citep{Hickson1997,Williams1990,VerdesMontenegro2001, Borthakur2015, Coziol2007}. From observations of the \HI\ content in Hickson Compact Groups \citep{Hickson1982, VerdesMontenegro2001, Borthakur2015} it has been found that gas removal is evident as the galaxies are often depleted of their \HI\ content. \citet{VerdesMontenegro2001} explains the evolutionary path of galaxy groups through a gradual removal of the \HI\ gas from the galaxies. In contrast to HCGs, loose groups are sparse and have a typical galaxy separation of few hundred kpc \citep{Sengupta2006, Pisano2011}. \citet{Kilborn2009} found that the \HI\ deficient galaxies within loose groups tend to lie within the 1 Mpc in projected distance from the group centre. Utilizing over 740 galaxy groups from the \ALFALFA\ survey, \citet{HessWilcots2013} found that the fraction of \HI-rich galaxies decreases in the central regions of groups as the group optical membership increases. They also found that \HI\ gas depletion starts first in the lowest \HI\ mass galaxies. 

It is common to make a distinction between the central galaxy which is most often the brightest (most massive) galaxy within a group (or a cluster) and satellite galaxies, being all other galaxies within the same group (e.g. \citealt{White1978, Skibba2011, Lacerna2014}. The difference between the central galaxy and satellites is that the central galaxy is theoretically at rest with respect to its halo, while satellite galaxies are moving in the gravitational potential of a halo and can experience environmental effects \citep{White1978, Yang2007, Brown2017}. \citet{Dzudzar2019} gives the example of a group with two centrals, both at rest within the halo. Recent studies of the \HI\ content of central and satellite galaxies have shown significance of the group mass regime. \citet{Janowiecki2017} have shown that central galaxies within small groups are $\sim$0.3 dex more \HI-rich than the central galaxies in isolation. \citet{Brown2017}, utilizing the stacking technique, found that satellite galaxies in high mass haloes have a lower \HI\ mass fraction with respect to those in low mass haloes.

An increasing number of observations of the \HI\ gas content show evidence and importance of galaxy environmental processing in groups \citep{VerdesMontenegro2001, Kilborn2005, Pisano2011, HessWilcots2013, Catinella2013, Brown2017, Dzudzar2019b}. How much environmental processing is important, and to what extent it has progressed in the \HI-rich and late-type dominated groups is still an open question. 

\subsection{From SINGG to Choirs}

The Survey of Ionization in Neutral Gas Galaxies (SINGG) imaged a sample of 468 \HI\ detections from the \HIPASS\ survey in the R-band and H$\alpha$ narrow-band \citep{Meurer2006}. Inspecting SINGG fields\footnote{Field-of-view$\sim$15\arcmin} \citet{Sweet2013} discovered 15 fields that have 4+ emission line galaxies within the 15\arcmin\ beam of the Parkes \HI\ detections, catalogued them and denoted them `Choirs'. Such selection overcomes optical-selection biases, however it introduces selection effects on the distance and \HI\ mass. Due to the field-of-view of the SINGG fields, selection of the Choir groups is prone towards \HI-rich systems at distance $\geq$30 Mpc \citep{Sweet2013}.

Galaxies in the Choir groups have similar properties as the single galaxies in SINGG based on their R-band radius, R-band surface brightness, H$\alpha$ equivalent width and specific star formation rate \citep{Sweet2013}. The mean group projected size is around two times larger than that of Hickson Compact Groups and around 10 times smaller than \citet{Garcia1993} groups. Based on the total group \HI\ content, \citet{Sweet2013} found that Choir groups are in an early stage of a group assembly: they have an average star formation efficiency, and they are not significantly \HI-deficient.

\citet{Sweet2014} probed the gas-phase metallicity relation of galaxies within \HI-rich Choir groups. They found two sub-populations of galaxies: metal-poor dwarfs and metal-rich giants in the metallicity-luminosity plane. \citet{Sweet2014} also found that the metallicity of the dwarf galaxies depends on the group membership and group \HI\ gas content: i) dwarf galaxies in Choir groups have lower metallicity than those in isolation; ii) at low luminosity, metallicity has a higher dispersion which indicates difference in \HI\ content and environment.

Choir groups were also used to find a population of galaxies that are formed from the tidal tail material of the interacting galaxies - tidal dwarf galaxies. \citet{Sweet2014} found three strong tidal dwarf galaxy candidates: J0205-55:S7, J0400-52:S8 and J0400-52:S9. Moreover, \citet{Sweet2016} observed 22 star-forming dwarf galaxies in four Choir groups with DEIMOS multi-object spectroscopy and analysed the galaxy kinematics, finding two new tidal dwarf galaxy candidates: J1051-17:g11 and J1403-06:g1. The rotation curves of the observed dwarf galaxies are disturbed due to recent interactions \citep{Sweet2016}. Moreover, they found that as much as half of the sample may be affected by the tidal interactions with neighbouring galaxies.

In \citet{Dzudzar2019} we analysed the integrated \HI\ properties of 27 galaxies within nine Choir groups and compared them to a sample of isolated galaxies. We explored the \HI\ mass fraction, specific star formation rate, star formation efficiency, \HI\ deficiency, \HI\ mass-size relation and stability parameter of the group galaxies with respect to the isolated galaxies. We found that the majority of central galaxies are within 2$\sigma$ scatter of the isolated galaxies, comparing \HI\ mass fraction, the specific star formation and star formation efficiency. Satellite galaxies were shown to have lower \HI\ mass fractions which is a possible indication that these galaxies experienced tidal stripping \citep{Dzudzar2019}. Furthermore, we determined that seven gas-rich Choir galaxies lie on the f$_{\textrm{atm}}$-q relation \citep{Obreschkow2016} indicating that there are no strong environmental influences on these galaxies. 

The motivation for this paper is to analyse the environmental processing in the \HI-rich Choir groups, using the resolved \HI\ content of member galaxies. For each group from our sample, we present the \HI\ column density maps, kinematic maps and the global environment for each group. We present the sample in the context of the parent SINGG and grandparent \HIPASS\ sample, and several galaxy samples from the literature. We discuss the properties of the \HI\ in galaxies and in the tidal streams. 

This paper is organised as follows: Section \ref{Sample} describes our sample of Choir groups and their \HI\ mapping, and auxiliary data used in this work. Section \ref{sec:results} presents our main results in this work: Choir galaxies and their \HI-mass fraction in context, the \HI\ distribution in Choir galaxies - highlighting \HI-deficient and \HI-excess galaxies. This section also presents the mid-infrared properties of Choir galaxies, as well as the Choir groups in context of Hickson compact groups and their evolution. Section \ref{sec:global_environment} presents the global environment around Choir groups. Section \ref{sec:conclusion} discusses our results and summarises our conclusions. Appendix \ref{sec:table_choir_properties} describes the Choir group properties, Appendix \ref{sec:note_on_groups} describes our Choir groups individually. We present \HI\ intensity maps, \HI\ kinematics, \HI\ spectra and global group environment for each Choir group in Appendix \ref{sec:appendix_images}.

Throughout this paper the adopted cosmology is H$_{0}$ = 70 km s$^{-1}$ Mpc$^{-1}$, $\Omega_{\Lambda}$ = 0.7 and $\Omega_{\text{m}}$ = 0.3. Mid-infrared star-formation follows the \citet{Cluver2017} prescription. The \textit{WISE} photometric calibration is described in \citet{Jarrett2011}. Presented \HI\ maps and spectra use velocities in the optical convention. The group distances are based on the multipole attractor model \citep{Mould2000}, as in (\citealt{Meurer2006, Sweet2013} and \citealt{Dzudzar2019}). 

\begin{figure}
    \centering
        \centering
        \includegraphics[width=1\columnwidth]{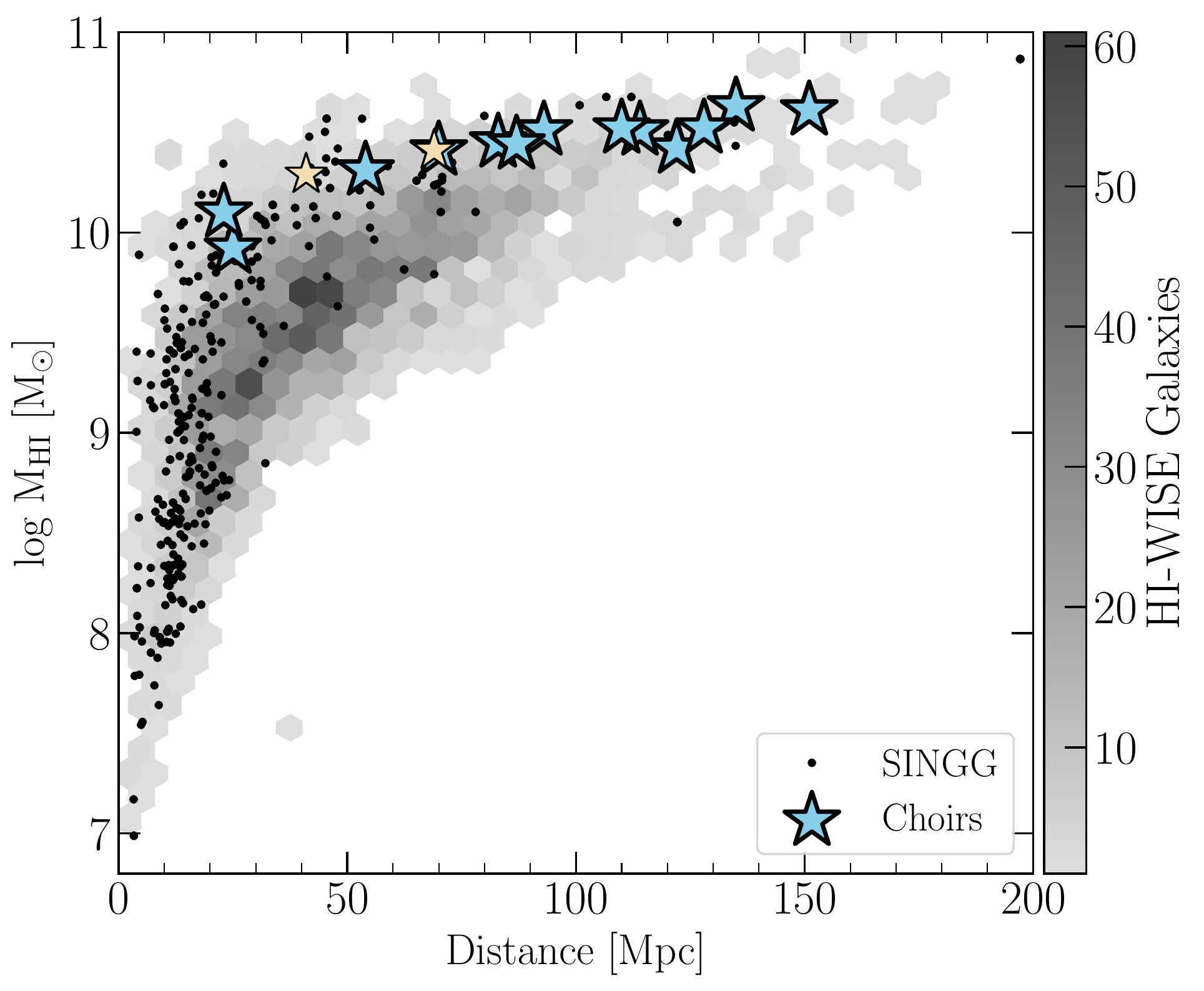} 
        \caption{The \HI\ mass versus distance. The stars show Choir groups, the blue stars are those that we have resolved \HI\ data, and the yellow stars are those for which we do not have resolved \HI\ data. The black points show galaxies from the parent SINGG sample. The hexbins in the background show the distribution of the parent \HIPASS\ sample using the galaxies from the HI-WISE catalogue of \citet{Parkash2018}. }
\label{fig:ChoirSampleDistance}
\end{figure}

\section{Sample and data}
\label{Sample}

We present \HI\ content in a total of 13 Choir groups, we have mapped 11 Choir groups and we use previously published data for three Choir groups. Eight Choir groups were observed with the Australia Compact Telescope Array (ATCA, project C2440)  using the CFB 64M-32k (34 kHz resolution) and CFB 1M-0.5k (0.5 kHz resolution) correlator configuration on the Compact Array Broadband Backend (CABB; \citealt{Wilson2011}). These groups are: HIPASS J0205-55, HIPASS J0258-74, HIPASS J0400-52, HIPASS J1051-17, HIPASS J1159-19, HIPASS J1250-20, HIPASS J1956-50 and HIPASS J2027-51. Whereas, three Choir groups were observed with the Karl G. Jansky Very Large Array (VLA, project 13A-207) in the hybrid DnC configuration, with a total bandwidth of 4 MHz (1024 channels of 3.906 kHz width). These groups are: HIPASS J1026-19, HIPASS J1059-09 and HIPASS J1408-21. 

In addition to our observations (see \citet{Dzudzar2019}, Table \ref{tab:NewObs} and Table \ref{tab:Choirs_newobs}), we use previously published data for three Choir groups: HIPASS J0209-10 \citep{Jones2019}, HIPASS J1159-19 \citep{Phookun1992} and HIPASS J2318-42a \citep{Dahlem2005} (see Table \ref{tab:Choirs_literature}). We were unable to find \HI\ data in the literature for two Choir groups: HIPASS J0443-05 and HIPASS J1403-06. We show the Choir galaxy group sample on the total \HI\ mass versus distance in Figure \ref{fig:ChoirSampleDistance}. Henceforth, we omit the HIPASS prefix for brevity. 

\subsection{\HI\ data reduction}

We processed the \HI\ data on the Green II and OzSTAR Swinburne supercomputers. ATCA and VLA data were reduced with the standard procedures in MIRIAD - Multichannel Image Reconstruction, Image Analysis and Display \citep{Sault1995} and CASA - Common Astronomy Software Applications \citep{McMulin2007} software packages, and we summarise the procedures below. In \citet{Dzudzar2019} we presented the integrated properties obtained from our \HI\ mapping. In this work we present the full maps of Choir groups including both: moment 0 - the \HI\ total intensity maps; and moment 1 - the velocity field maps. 

The ATCA data reduction was carried out as follows. Data were loaded with the task \textsc{atlod}. Raw data were flagged for radio interference using the tasks \textsc{uvflag}, \textsc{blflag} and \textsc{pgflag}. After the flagging we performed bandpass, flux and phase calibrations using the following tasks: \textsc{mfcal}, \textsc{gpcal}, \textsc{gpboot} and \textsc{mfboot}. For bandpass and flux calibration we used the standard calibration source PKS 1934-638, while phase calibrators are given in Table \ref{tab:NewObs} as well as in Table 1 in \citet{Dzudzar2019}. After calibrations we performed continuum subtraction with \textsc{uvlin} by fitting it to the line-free channels. This procedure was performed on each of the array configurations and then all visibilities from the same source were combined using \textsc{invert} to form a dirty cube (excluding antenna 6), using Briggs' robust parameter of 0.5 and a channel width of 5 km s$^{-1}$. The obtained dirty data cubes were then cleaned using the task \textsc{clean}, restored using the task \textsc{restor}. The primary beam correction was done with the task \textsc{linmos}. We determined the RMS in the data cube with the tasks \textsc{imstat} and \textsc{imhist}.

We measured the galaxy \HI\ properties using \textsc{mbspect} and obtained the moment 0 and moment 1 maps with the task \textsc{moment}, applying a 3$\sigma$ clipping. We used \textsc{kvis} from the Karma library \citep{Gooch1996} for initial visualization, while exported moment maps were visualized using python. We present the \HI-intensity maps, \HI\ velocity fields and spectra in Appendix \ref{sec:appendix_images}.

The VLA data reduction was carried out as follows. Data were cleaned from radio interference using task \textsc{flagdata}. Bandpass, flux and phase calibration was performed using the following tasks: \textsc{setju}, \textsc{gaincal}, \textsc{bandpass}, and \textsc{fluxscale}. We then split the data using task \textsc{split} and performed continuum subtraction using line-free channels with the task \textsc{uvcontsub}. We produced the \HI\ data cubes with the task \textsc{clean}, using natural weighting (for J1026-19 and J1059-09) and Briggs' robust parameter of of 0 (for J1408-21) and a channel width of 5 km s$^{-1}$. With the task \textsc{impbcor} we corrected for the primary beam, and we created moment maps with the task \textsc{immoments}. We used \textsc{viewer} for initial visualization, while exported moment maps were visualized using python and presented in Appendix \ref{sec:appendix_images}.  \\

\begin{table*}
\begin{threeparttable}
\centering
\caption{Summary of the ATCA \HI\ observations that are used in this paper in addition to those in Table 1. from \citet{Dzudzar2019}.}
\label{tab:NewObs}
\begin{tabular}{cccccc}
\toprule
Group ID    & 1.5/750/EW [h] & Phase Calibrator & f$_{C}$ [MHz] & $\theta_{\textrm{min}}$ $\times$ $\theta_{\textrm{maj}}$ [$''$] & rms  [mJy beam$^{-1}$] \\
(1)		 & (2)		& (3)			   & (4)  & (5) &  (6) \B \\ 	\hline \hline
J0205-55\tnote{a} & \dots/7.82/7.6   & PKS 0302-623 	& 1393  & 56.7 $\times$ 108.5& 1.13 \T \\
J1956-50 & 6.57/8.16/7.48    & PKS 2052-474 	& 1385  & 40.5 $\times$ 86.1& 1    \\
J1159-19 & \dots/2.35/4.14	 & PKS 1245-197 	& 1413  & 70.1 $\times$ 193.3&  2  \B  \\ 
\bottomrule
\end{tabular}         
\begin{tablenotes}
\item Columns: (1) Group name (HIPASS+ID); (2) Time on the source for the particular ATCA array configuration; (3) Phase calibrator for the source; (4) Central frequency of band in MHz. (5) Minor and major axis of the synthesized beam size, respectively; (6) Root-mean-square of the data cube.
\item[a] In \citet{Dzudzar2019} we used pre-CABB data for J0205-55, while in this work we use our observations of this group.
\end{tablenotes}
\end{threeparttable}
\end{table*}

\subsection{Auxiliary data}

We make use of imaging from the Wide-field Infrared Survey Explorer (\textit{WISE}; \citealt{Wright2010}) to obtain integrated mid-infrared photometry in the four \textit{WISE} bands: W1 (3.4$\mu$m), W2 (4.6$\mu$m), W3 (12$\mu$m) and W4 (22$\mu$m). Sources are measured using the same procedure as detailed in \citet{Jarrett2013, Jarrett2019}, and the resulting photometry used to derive stellar masses, star formation rates and mid-infrared colour properties of the Choir galaxies (see Section \ref{sec:WISE} for details).

We make use of the optical observations of Choir groups which were carried out with the Dark Energy Camera (DECam) on the CTIO Blanco 4-m telescope in the \textit{g}, \textit{r}, \textit{i} and \textit{z} bands obtained for program AAT/13A/02 (PI S. Sweet). In addition, we use optical imaging from Dark Energy Survey Data Release 1 (\textit{g}, \textit{r} and \textit{i}-band) for two Choir groups: HIPASS J0205-55 (partially covered by DES) and HIPASS J0400-52 (as used in \citealt{Dzudzar2019b}). We also make use of the Digitized Sky Survey (DSS) imaging.

\subsection{Comparison samples}

To place our sample of Choir groups in the wider context, we compare it to several galaxy samples from the literature:
    \begin{itemize}
        \item HICAT catalogue \citep{Meyer2004, Zwaan2004} from the \HI\ All Sky Survey (\HIPASS\ \citealt{Barnes2001}), which is the parent sample of the SINGG survey. We are using HICAT+\textit{WISE} catalogue (HI-WISE, $\sim$2300 galaxies) with the updated \textit{WISE} photometry from \citet{Parkash2018}. 
        \item SINGG -- Survey of Ionization in Neutral Gas Galaxies \citep{Meurer2006} comprised of 468 galaxies, and it is a parent sample of Choir groups. 
        \item AMIGA -- Analysis of the Interstellar Medium of Isolated GAlaxies \citep{VerdesMontenegro2005Amiga}. Data of the AMIGA sample is obtained combining two catalogues: \HI\ masses from \citet{Jones2018Amiga} and stellar masses from  \citet{Fernandez2013Amiga} for a total of 342 galaxies.
        \item xGASS -- The extended GALEX Arecibo SDSS Survey \citep{Catinella2018}, a gas fraction limited survey of $\sim$1200 galaxies. 
        \item HCG -- Hickson Compact Groups \citep{Hickson1982}. We use data of 72 HCGs \citep{VerdesMontenegro2001} and their \HI\ properties.
        \item \HI-deficient galaxies from \citet{Denes2014} and \citet{Murugeshan2019}.
        \item \HI-excess galaxies from \citet{Lutz2018}.
    \end{itemize}
    
\begin{figure*}
    \centering
        \centering
        \includegraphics[width=2\columnwidth]{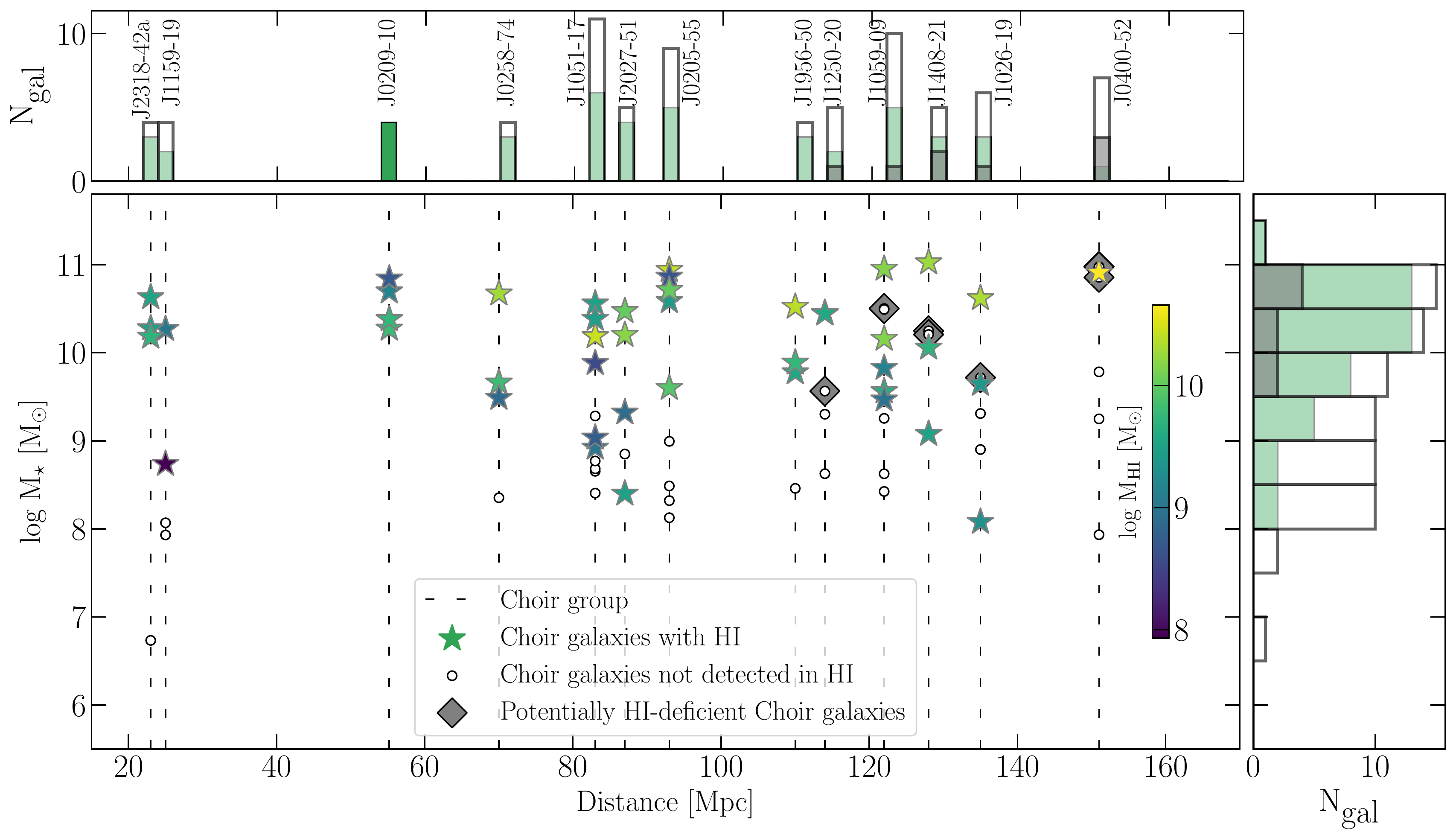} 
        \caption{Stellar mass versus distance, and their histograms, for Choir galaxies. The coloured stars show Choir galaxies detected in \HI, while the open circles show Choir galaxies without detected \HI. The circles within the grey diamonds are potentially \HI-deficient galaxies, as described in Appendix \ref{appendix:detection_limit}. The green histograms show Choir galaxies with \HI, the grey histograms show potentially \HI-deficient galaxies and the white histograms show all Choir galaxies. The vertical dashed lines connect Choir galaxies within the same group, based on the adopted group distance - their names are shown in the upper histogram. The nearest three groups are those which \HI\ data are obtained from the literature: J2318-42a \citep{Dahlem2005}, J1159-19 \citep{Phookun1992} and J0209-10 \citep{Jones2019}.
        }
\label{fig:Choir_detections}
\end{figure*}

\section{Results}
\label{sec:results}

In order to assess environmental processing of galaxies in \HI-rich groups, we analyse the \HI\ properties, star formation and environment of Choir galaxies. We show the \HI\ mass fraction of the Choir galaxies and compare them to the several other galaxy samples from the literature. We explore the mid-infrared properties of Choir galaxies. And finally, we analyse the Choir galaxies discussing their anomalous \HI\ content, specific star formation rate and tidal interactions. 

We use standard relation to derive the \HI\ mass:
\begin{equation}
\hspace{2cm} \textrm{M}_{\textrm{HI}} \hspace{0.05cm} = 2.365\times10^{5} \textrm{D}^{2} \textrm{F}_{\textrm{\HI}},
\label{eq:mass}
\end{equation}

where F$_{\textrm{\HI}}$ is the integrated flux density in Jy km s$^{-1}$, D is the distance to the galaxy in Mpc, and mass is obtained in M$\odot$. The distances are adopted from \citet{Sweet2013}. For the three Choir groups for which data are used from the literature, we re-derive their \HI\ masses in order to be consistent with our distances and show results in Table \ref{tab:Choirs_literature}. There is an excellent agreement between the literature and our derivation of the log M$_{\HI}$ for J0209-10 \citep{Jones2019} and J2318-42a \citep{Dahlem2005}, with a mean difference of $\sim$0.1 dex, while galaxies within J1159-19 have a mean difference in log M$_{\HI}$ of $\sim$0.8 dex. Such a large difference is due to the adapted distance: we use 25 Mpc (similar to the Hubble flow), while \citet{Phookun1992} used 10 Mpc. 

\begin{figure*}
    \centering
        \centering
        \includegraphics[width=2\columnwidth]{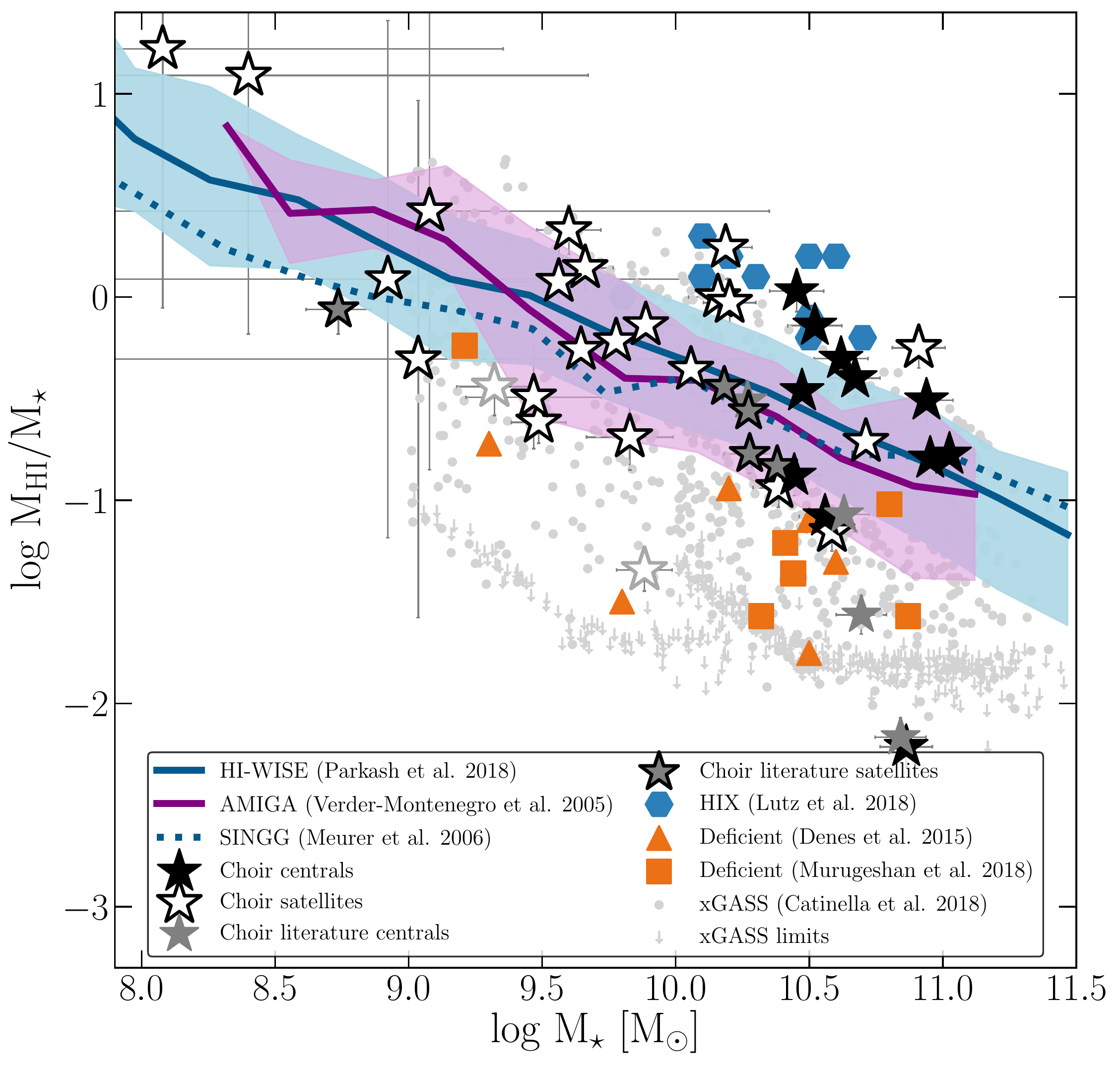} 
        \caption{\HI\ mass fraction versus stellar mass. Choirs galaxies are denoted as the star symbols, the solid black and white stars respectively correspond to central and satellite galaxies, as in \citet{Dzudzar2019}. Two stars with the grey edge are Choir galaxies that are tentative detections, below 3$\sigma$. The grey stars are Choir galaxies for which we found \HI\ fluxes in the literature (J0209-10 group from \citealt{Jones2019}, J1159-19 group from \citealt{Phookun1992}, J2318-42a group from \citealt{Dahlem2005}). The grey solid ones are centrals and the grey ones with black outlines are satellites. The orange triangles and rectangles are \HI\ deficient galaxies, respectively from \citet{Denes2016} and \citet{Murugeshan2019}. The blue hexagons are \HI-excess galaxies from \citet{Lutz2018}. The blue solid line is the running median of the \HIPASS\ parent sample (using HI-WISE data obtained from \citealt{Parkash2018}) and the shaded region shows 16th and 84th percentile. The blue dotted line is the running median of the SINGG parent sample (SINGG sample is from \citet{Meurer2006}, while the stellar masses are from \citealt{Parkash2018}). The purple solid line is the running median of isolated galaxies from the AMIGA sample \citet{VerdesMontenegro2005Amiga}, and the shaded region shows 16th and 84th percentile. The grey points and arrows in the background are galaxies and upper limits from the xGASS sample \citep{Catinella2018}.}
\label{fig:Choir_sample_context}
\end{figure*}
\subsection{\HI\ detections vs \HI\ non-detections}
\label{sec:non_detections}

Within 13 Choir groups presented in this paper, \citet{Sweet2013} catalogued 80 galaxies. We have detected \HI\ emission in 44 galaxies: nine of these are obtained from the literature and 35 with our observations, including seven new detections (hereafter, annotated with A) which are outside of the field of view in SINGG. Therefore, we obtained the \HI\ emission in 50.6\% of galaxies, and now we examine the \HI\ non-detections.

In Figure \ref{fig:Choir_detections} we show the stellar mass versus distance, for Choir galaxies. Choir galaxies detected in the \HI\ are shown with the filled coloured stars (where colour denotes their \HI\ mass), while open circles are Choirs not detected in \HI. In the mass regime M$_{\star} \leq$10$^{9}$ M$_{\odot}$, we detect four out of 23 galaxies in \HI\ emission. This is not surprising as the majority of these non-detections are small compact dwarf galaxies. In Figure \ref{fig:Choir_detections} we also show that we have \HI\  non-detections in the stellar mass range 10$^{9}$--10$^{11}$ M$\odot$, well within the range of the \HI\ detected galaxies. These galaxies have a sufficiently high stellar mass, and angular size, to be above our \HI\ detection threshold if they have normal amounts of \HI, indicating that they are \HI-deficient galaxies - they have on average eight times less \HI\ than the average galaxy of the similar stellar mass (they are marked with the grey diamonds in Figure \ref{fig:Choir_detections}, and see Appendix \ref{appendix:detection_limit} for details). These galaxies are: J1250-20:S4 [Irrs], J1059-09:S4 [SB0], J1408-21:S2 [S]; J1408-21:S4 [S0], J1026-19:S6 [S0], J0400-52:S4 [Scd], J0400-52:S5 [SB pec] and J0400-52:S6 [SB pec], morphology adopted from \citet{Sweet2013}. Choir J0400-52 has three galaxies that are \HI-deficient; as this group is infalling into a cluster \citep{Sweet2013, Dzudzar2019}, a lower/deficient \HI\ content in these galaxies is not surprising. See Table \ref{tab:deficient} for details on these \HI-deficient galaxies.

\subsection{Choir sample in context}

The \HI\ mass fraction versus stellar mass relation (M$_{\HI}$/M$_{\star}$ vs M$_{\star}$) for Choir galaxies is shown in Figure \ref{fig:Choir_sample_context}. The \HI\ mass fraction decreases with the stellar mass and has a sizeable scatter \citep{Catinella2013, Catinella2018}. It is still an open question what drives the scatter in this relation, whether it is environment or internal galaxy properties, or a combination of both. 

We show the \HI-mass fraction of Choir galaxies in the wider context, comparing it to parent SINGG \citep{Meurer2006}, grandparent HIPASS sample using HI-WISE data from the \citep{Parkash2018}, AMIGA isolated galaxy sample \citep{VerdesMontenegro2005Amiga}, xGASS sample \citep{Catinella2018}, and to \HI-deficient \citep{Denes2014, Murugeshan2019} and \HI-excess \citep{Lutz2018} galaxy samples.\\

\subsubsection{{\sc H\thinspace{i}} mass fraction of Choirs}
\label{sec:fraction}

\begin{figure}
    \centering
        \centering
        \includegraphics[width=1\columnwidth]{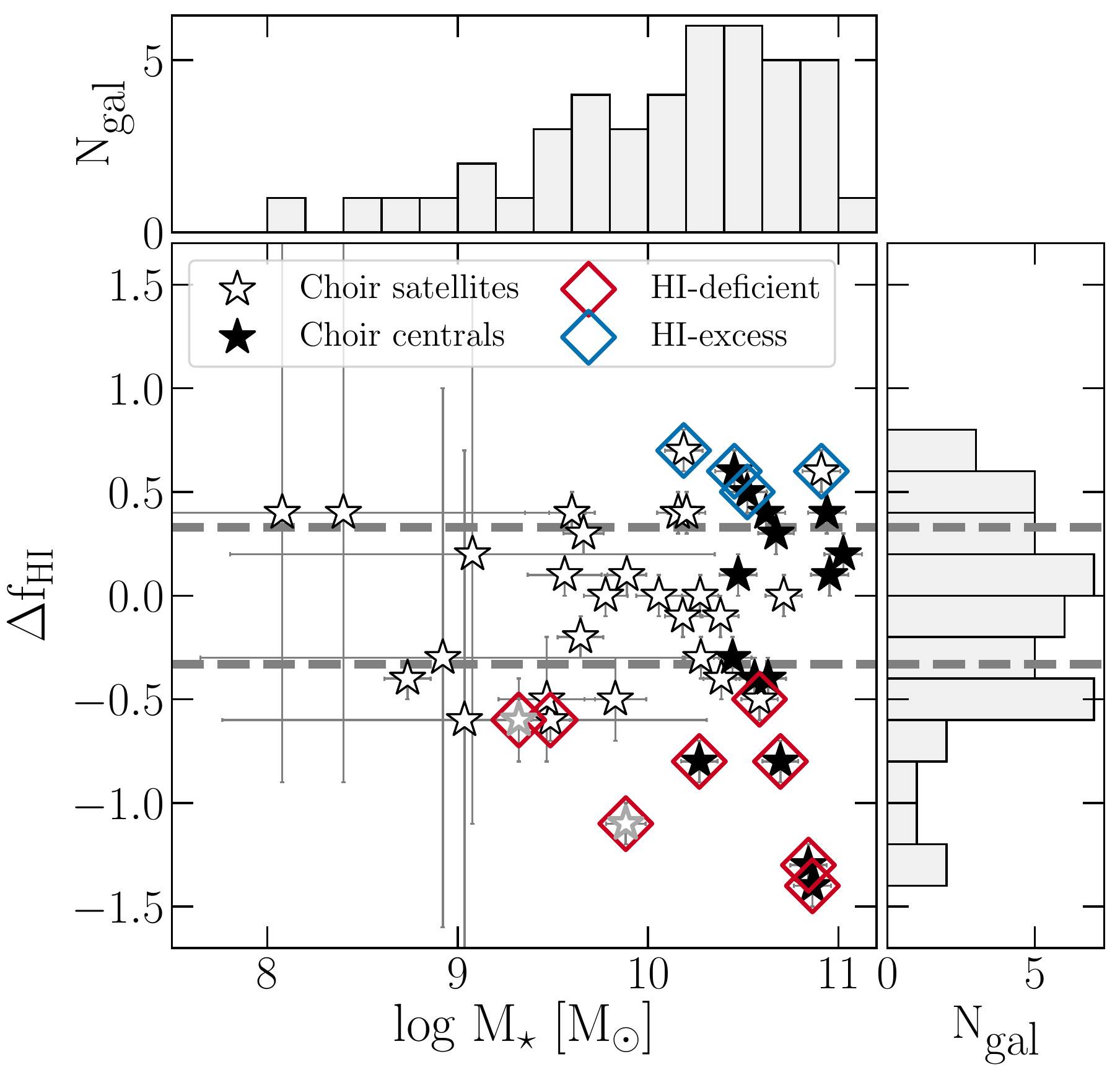} 
        \caption{Distance of the \HI\ mass fraction from the \HIPASS\ running median line, $\Delta$f$_{\textrm{\HI}}$ as a function of the stellar mass. The grey dashed horizontal lines show the mean scatter around the HIPASS running median line $\pm$0.3 dex. Choir galaxies are marked as stars, the solid black, outlined black, grey are respectively central galaxies, satellite galaxies and galaxies detected in 
        \HI\ below 3$\sigma$. The red and blue diamonds show respectively the \HI-deficient and \HI-excess galaxies. The grey histograms include all Choir galaxies.}
\label{fig:Delta_fraction}
\end{figure}

In Figure \ref{fig:Choir_sample_context} we show the comparison of the \HI\ mass fraction of our Choir sample to its parent samples \HIPASS\ and SINGG. The parent \HIPASS\ sample is shown with the running median of the \HI\ mass fraction in the stellar mass bins of 0.3 dex, the solid blue line, and its 16th and 84th percentile (the blue shaded region). The parent SINGG sample is shown with the running median, the dotted blue line. We find that the \HI\ mass fraction of the Choir galaxies is dispersed around the \HIPASS\ median, $\Delta$f$_{\textrm{\HI}}$, in the range -1.4$\leq \Delta$f$_{\textrm{\HI}}\textrm{[dex]}\leq$0.7. Using the standard deviation from the \HIPASS\ running median \HI\ mass fraction, we find that 4 galaxies are \HI-excess ($\Delta$f$_{\textrm{\HI}}>$0.4 dex), 32 galaxies have an average \HI\ content ($-$0.4 $<\Delta$f$_{\textrm{\HI}}\textrm{ [dex]}<$0.4) and 8 galaxies are \HI-deficient ($\Delta$f$_{\textrm{\HI}}<$($-$0.4 dex)). We show the distribution of the \HI\ mass fraction around the \HIPASS\ running median line, $\Delta$f$_{\textrm{\HI}}$, in Figure \ref{fig:Delta_fraction}, and their individual values in Table \ref{tab:Choirs_extremes}. In this Figure we outline the \HI-excess and \HI-deficient galaxies. Moreover, we show that the \HI average class of galaxies represents a continuity of '\HI-richness' from HI-rich to HI-poor galaxies. The existence of galaxies spanning from \HI-excess and \HI-deficient galaxies may indicate the environmental processing in Choir groups. The two central galaxies with the lowest \HI\ mass fraction are J0205-55:S3 and J0209-10:S4. J0205-55:S3 is presented in Figure \ref{fig:HI_deficient} and discussed in Section \ref{sec:extremes}. J0209-10:S4 is tidally interacting with its neighbour galaxy of similar stellar mass, J0209-10:S3 \citep{Jones2019}, which is most likely cause of its low \HI\ mass fraction. 

In Figure \ref{fig:Choir_sample_context} we also show a comparison of Choir galaxies to several samples from the literature. Overall, we show that our sample has a wider stellar mass and \HI\ mass fraction range as xGASS sample \citep{Catinella2018}. We find that a number of Choir galaxies, with high \HI\ mass fractions, are similar to the \HI-extreme galaxy sample \citep{Lutz2018}. In \citet{Dzudzar2019} we have shown that based on the specific angular momenta, gas-rich Choir galaxies are indeed comparable to the \HI-extreme galaxies. In Section \ref{sec:extremes} we analyse the most \HI-rich Choir galaxies, we show their physical sizes and the \HI\ distribution (see Figure \ref{fig:HI_excess}). From Figure \ref{fig:Choir_sample_context} it is also evident that several Choir galaxies are \HI-deficient and they are similar to samples of \citet{Denes2016} and \citet{Murugeshan2019}. We describe the most \HI-deficient Choirs in Section \ref{sec:extremes}, and present them in Figure \ref{fig:HI_deficient}. 

Figure \ref{fig:Choir_sample_context} also shows the running median line and its 16th and 84th percentile of the isolated sample of galaxies (Analysis of the Interstellar Medium of Isolated GAlaxies, AMIGA, \citealt{VerdesMontenegro2005Amiga, Fernandez2013Amiga, Jones2018Amiga}). There is no large difference between the AMIGA and the \HIPASS\ sample as seen their median gas-mass fractions. As noted in \citet{Dzudzar2019}, we find a number of galaxies with an elevated \HI\ mass fraction with respect to galaxies in isolation. The existence of central group galaxies that are more \HI-rich than the isolated central galaxies was shown by \citet{Janowiecki2017}. They suggested that the group central galaxies have large \HI\ reservoirs either due to gas accretion from cosmic web filaments or due to gas-rich mergers. We consider both scenarios: \textbf{i)} With respect to the global environment, we find Choirs both near and far away from the cosmic web filaments (see Section \ref{sec:filaments}). \textbf{ii)} With respect to the local environment, several Choir central galaxies are experiencing interactions and exhibit signs of a merger events. Since our galaxy group sample is small we can not make a definite conclusion which effect makes a larger contribution to their \HI\ gas content. Future large-scale observations of the \HI, with WALLABY survey \citep{Koribalski2012W}, and surveys with MeerKAT \citep{Camilo2018} and Apertif \citep{Osterloo2010} will be help to discern between these scenarios. 

\subsection{{\sc H\thinspace{i}} distribution in galaxies}
\label{sec:extremes}

\begin{figure}
    \centering
        \centering
        \includegraphics[width=\columnwidth]{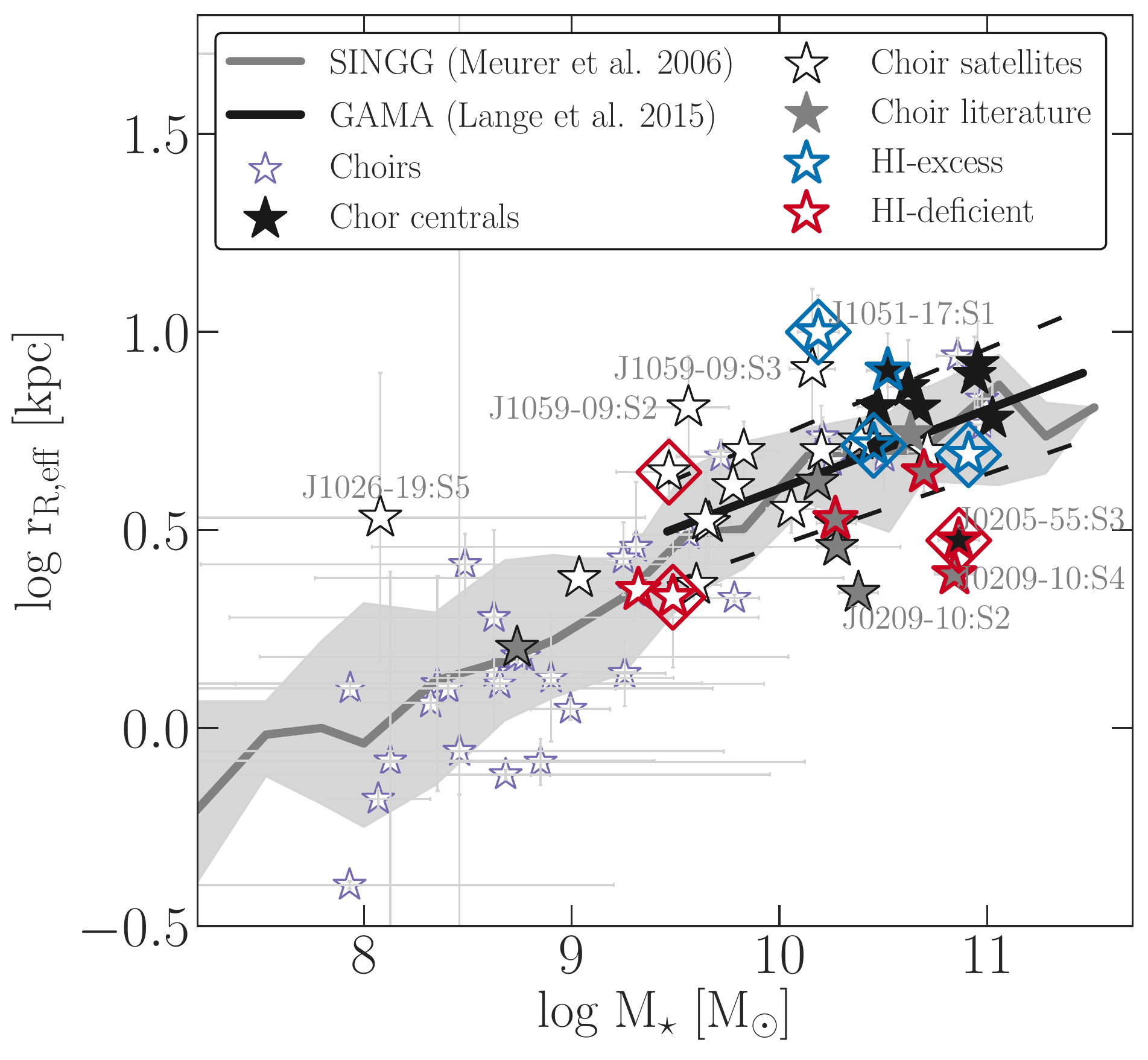} 
        \caption{Relationship between galaxy R-band effective radii and the stellar mass. The solid markers show Choir central galaxies and the open markers show Choir satellite galaxies. The purple open stars show Choir galaxies without known \HI\ content. The blue and red outlined stars show the Choir \HI-excess and \HI-deficient galaxies as discussed in Section \ref{sec:extremes}, while diamonds denote extreme outliers from Figure \ref{fig:HI_deficient} and \ref{fig:HI_excess}. The solid grey line is the running median of the SINGG parent sample and the shaded regions show 16th and 84th percentile. The solid black line shows mass-size relation of late-type galaxies from the GAMA survey, while dashed lines denote 1$\sigma$ of the relation \citep{Lange2015}.}
\label{fig:Reff_Mstar}
\end{figure}

\begin{figure}
        \centering
         \includegraphics[width=1\columnwidth]{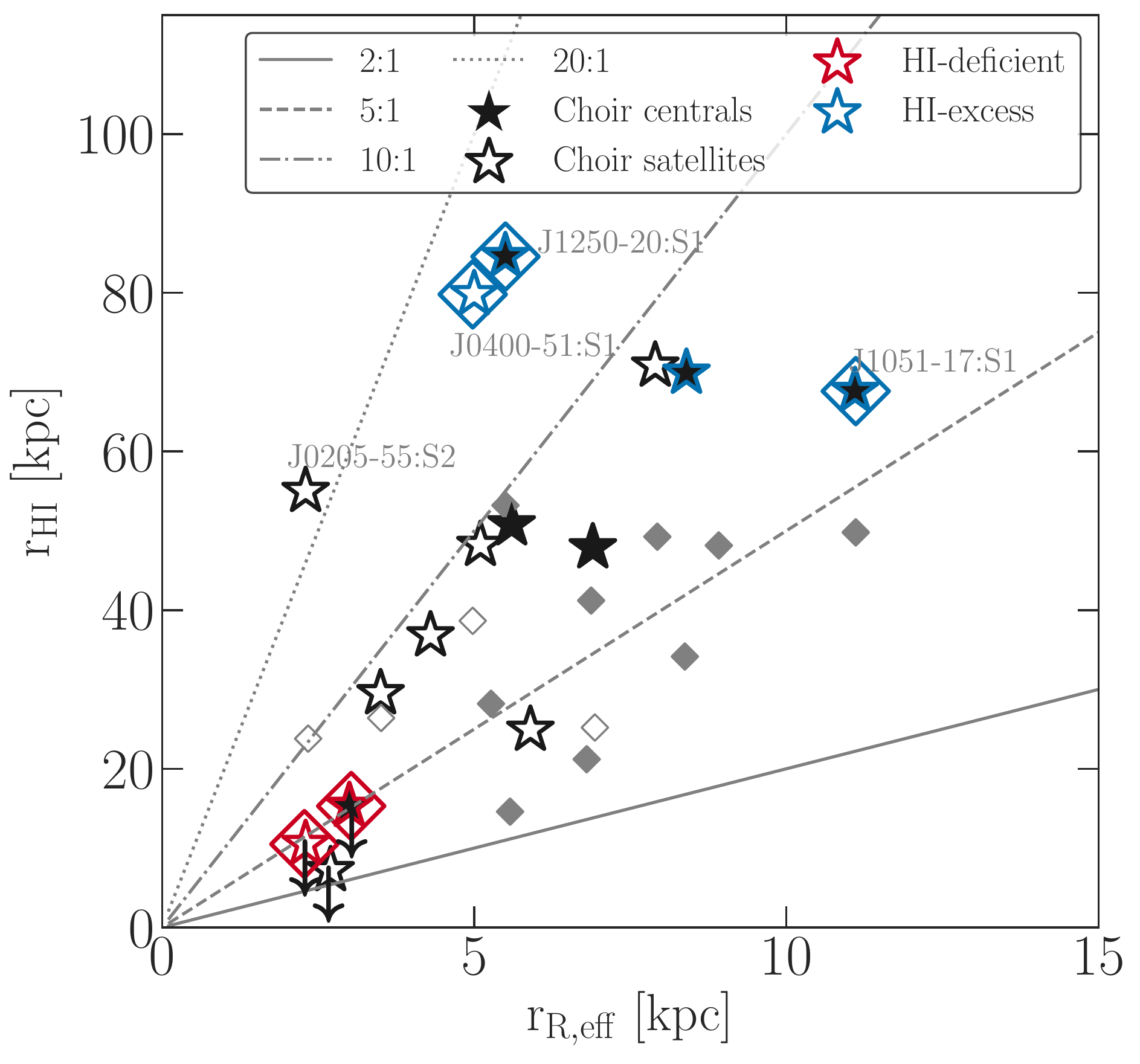}
        \caption{Relationship between galaxy \HI\ radii measured at 3$\times$10$^{19}$cm$^{-2}$ and R$_{\textrm{eff}}$ radii. The solid markers show Choir central galaxies and the open markers show Choir satellite galaxies. The blue and red outlined stars show the Choir \HI-excess and \HI-deficient galaxies as discussed in Section \ref{sec:extremes}, while diamonds denote extreme outliers from Figure \ref{fig:HI_deficient} and \ref{fig:HI_excess}. The grey diamonds show Choir galaxies with the \HI-radius measured at 1M$\odot$ pc$^{-2}$ \citep{Dzudzar2019}.}
\label{fig:Radii}
\end{figure}

In this section we present and discuss the \HI\ distribution in Choir group galaxies in order to determine whether environmental processing has an impact on it. We highlight six Choir galaxies which are respectively at the largest distance below (three most \HI-deficient galaxies) and above (three most \HI-excess galaxies) the running median line of the \HIPASS\ galaxies, representing most extreme outliers (see Figure \ref{fig:Choir_sample_context}). We track these galaxies in the following analysis.

In \citet{Dzudzar2019} we have shown that Choir galaxies follow the established \HI\ size-mass relation (e.g. \citealt{Wang2016}). We present the stellar size-mass relation in Figure \ref{fig:Reff_Mstar}. We show the median of the stellar mass-size relation of the Choir parent SINGG sample and its 16th and 84th percentile. Moreover, we compare SINGG size-mass relation to the relation for the late-type galaxies from the Galaxy And Mass Assembly (GAMA) survey, where they are overlapping at stellar masses above 10$^{9.5}$ M$_{\odot}$ (\citealt{Lange2015}, relation for \textit{r}-band). We find six (out of 69) Choir galaxies lie outside the SINGG percentiles and GAMA scatter. This infers a potential underlying morphological difference resulting in these outliers.

We compare the ratio of the \HI\ to optical radii of Choir galaxies to see if we can quantify the difference between the \HI-excess and \HI-deficient galaxies (see Figure \ref{fig:Radii}). We use the effective R-band radii obtained from the SINGG survey \citep{Meurer2006}. The optical disc sizes are often compared to the \HI\ disc sizes measured at 1 M$_{\odot}$pc$^{-2}$ \citep{Wang2016}, however, in our most \HI-deficient galaxies the \HI\ intensity does not reach this density, thus we use a column density of 3$\times$10$^{19}$cm$^{-2}$ and measure along the minor axis of the synthesised beam to avoid beam smearing effects (some of these measurements are upper limits, as indicated in Table \ref{tab:Choirs_extremes}, and they are in agreement with analytically derived \HI\ sizes from the \HI\ size-mass relation by \citealt{Wang2016}). With this approach, we are comparing the extent of the low \HI\ column density in galaxies. 

\begin{figure*}
    \centering
        \centering
        \includegraphics[width=0.68\columnwidth]{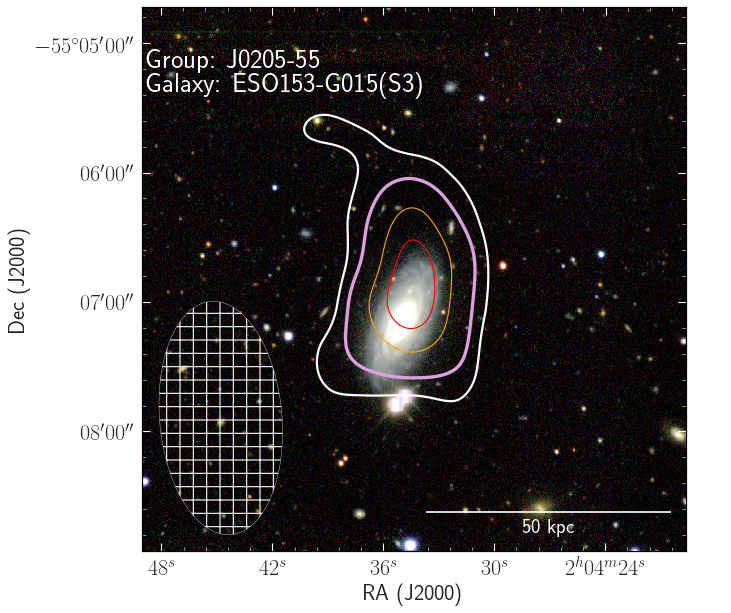}
        \includegraphics[width=0.68\columnwidth]{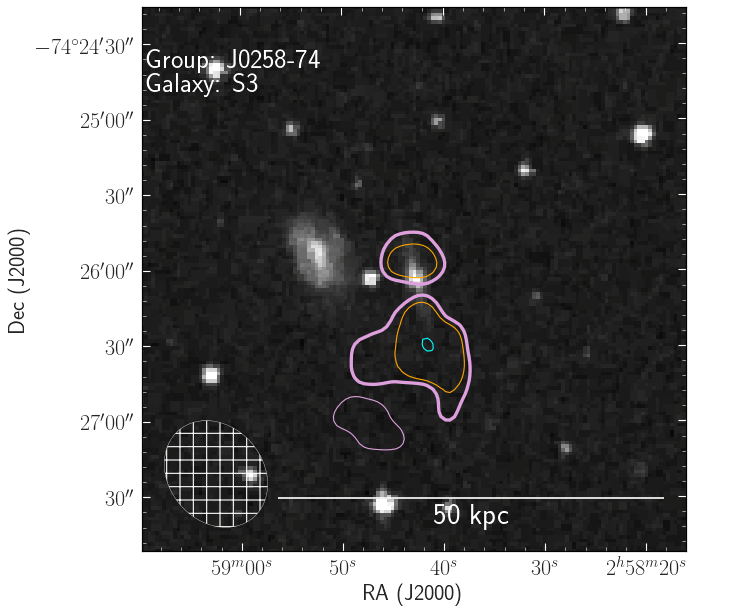}
        \includegraphics[width=0.7\columnwidth]{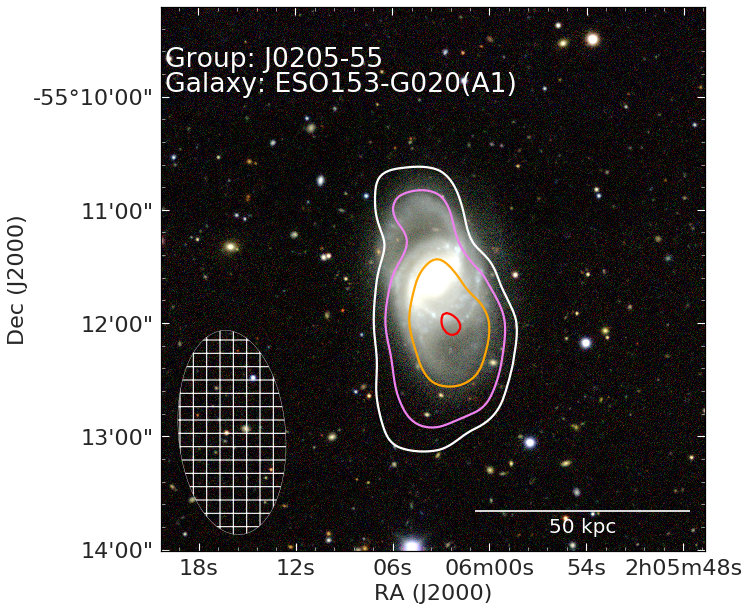}
        \caption{The \HI\ intensity distribution overlaid on the DECam \textit{g}, \textit{r}, \textit{i} colour composite image (J0258-74:S3 is on the DSS2B-band image). Three \HI-deficient galaxies are selected based on their distance below the $\Delta$f$_{\textrm{\HI}}$ of the \HIPASS\ median line. The \HI-density contours are as follows: 1.5, 3, 6 and 9 $\times$10$^{19}$cm$^{-2}$, respectively white, plum, orange and red.
        Galaxy name and its group are noted in the upper left corner in each panel. The synthesized beam is shown in the bottom left corner, and the scale bar in the bottom right corner shows 50 kpc at the group distance.}
\label{fig:HI_deficient}
\end{figure*}

\begin{figure*}
    \centering
        \centering
        \includegraphics[width=0.68\columnwidth]{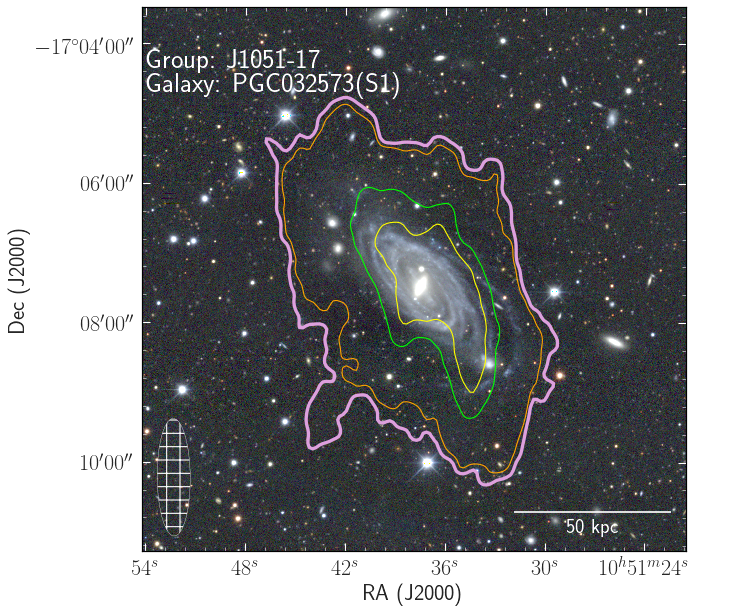}
        \includegraphics[width=0.68\columnwidth]{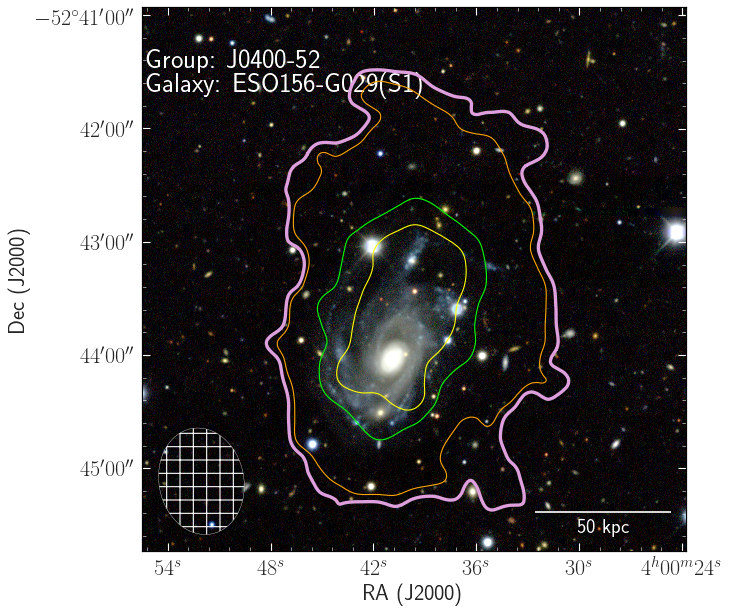}
        \includegraphics[width=0.68\columnwidth]{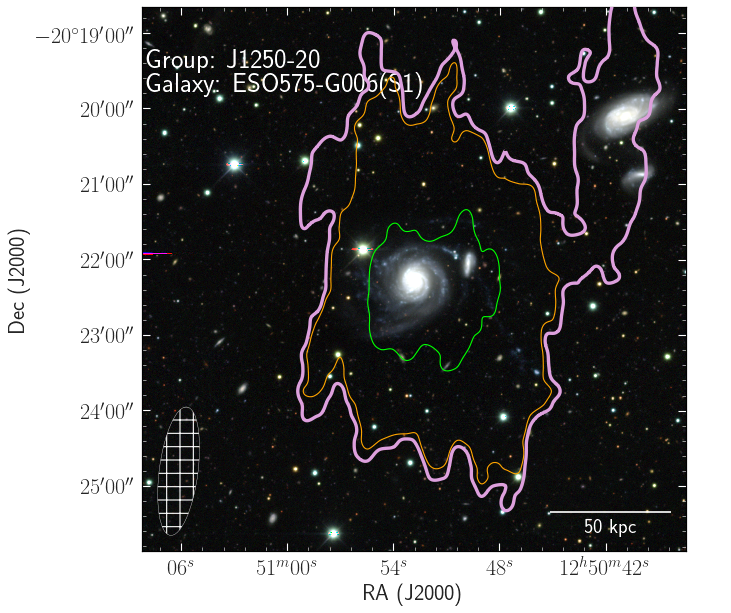}
        \caption{The \HI\ intensity distribution overlaid on the DECam \textit{g}, \textit{r}, \textit{i} colour composite image. Three \HI-excess galaxies are selected based on their distance above the $\Delta$f$_{\textrm{\HI}}$ of the \HIPASS\ median line. The \HI-density contours are as follows: 1.5,
        3, 6, 35 and 50 $\times$10$^{19}$cm$^{-2}$, respectively white, plum, orange, green and yellow contour. Galaxy name and its group are noted in the upper left corner in each panel. The synthesized beam is shown in the bottom left corner, and the scale bar in the bottom right corner shows 50 kpc at the group distance.}
\label{fig:HI_excess}
\end{figure*}

We find that 12 (out of 15) Choir galaxies have 2--10 times larger \HI\ radii when compared to the r$_{\textrm{R,eff}}$ (see Figure \ref{fig:Radii}). J0205-55:S2, J0400-51:S1 and J1250-20:S1 have the largest ratio of \HI\ radius to r$_{\textrm{R,eff}}$, of around 20 times, for \HI\ radii measured 3$\times$10$^{19}$cm$^{-2}$ (corresponding to around 10 times for \HI\ radii measured at 1 M$_{\odot}$pc$^{-2}$ from \citealt{Dzudzar2019}). The \HI\ intensity distribution in these galaxies reaches densities larger than 50$\times$10$^{19}$ cm$^{-2}$ as seen in Figure \ref{fig:HI_excess}. Visually inspecting DECam optical images (Figure \ref{fig:HI_excess}), we see that these galaxies have fainter stellar outer discs. Moreover, each of these galaxies have a peculiar morphology that can be connected with an interaction and possible accretion of the \HI\ gas. J0400-51:S1 has visible faint stellar streams in the North, and we discussed in \citet{Dzudzar2019b} how these may be related to a merger. J1250-20:S1 had clear ongoing multiple interactions and a stellar stream reaching out of a spiral arm \citep{Dzudzar2019} and see Figure \ref{fig:j1250_multiw}. Furthermore, we find that three out of four \HI-excess galaxies are interacting, or show signs of past interactions, thus it is possible that they acquired some of their \HI\ content from mergers. 

The most \HI-deficient galaxies have small \HI\ discs with respect to their optical discs (see Figure \ref{fig:HI_deficient}). Their \HI\ size is comparable to the synthesized beam size, thus they are not fully resolved. The \HI-deficient galaxies that we mapped in \HI\ lack high \HI-column densities, when compared to the \HI-excess galaxies (Figure \ref{fig:HI_excess}); their \HI-column densities on average reach $\sim$10$\times$10$^{19}$ cm$^{-2}$. Six out of eight \HI-deficient galaxies in our sample show irregular structure (optical or \HI) which indicates that they most likely interacted (or currently interacting) with other galaxies within the group. This indicates that the tidal interaction is the primary cause of the \HI-deficiency in Choir galaxies. In the next section we discuss individual \HI-deficient and \HI-excess galaxies.

\subsubsection{Physical processes in the \HI-deficient and \HI-excess galaxies}
    
The two most \HI-deficient galaxies J0205-55:S3 and J0258-74:S3, with stellar masses of 10$^{10.8}$ M$_{\odot}$ and 10$^{9.5}$ M$_{\odot}$, appear to have \HI\ centres shifted from their optical centres. J0205-55:S3 was classified as S0 galaxy \citep{Sweet2013} however, inspecting our deep DECam images we can re-classify it as a spiral galaxy. J0258-74:S3 is edge-on galaxy with a bright bulge, which is seen on the top right panel in Figure \ref{fig:HI_deficient}. J0258-74:S3 is in projection close to the larger J0258-74:S2 galaxy (to the east). We show J0258-74 group in Figure \ref{fig:j0258_group}, and it is seen that the \HI\ contours of J0258-74:S2 and J0258-74:S3 galaxy overlap. Being very close, it is likely that J0258-74:S2 and J0258-74:S3 are interacting. J0205-55:A1 (ESO153-G020) is an SBab galaxy with a stellar mass of 10$^{10.59}$ M$_{\odot}$. \citet{Schmitt2003} classifies ESO153-G020 as Seyfert 2 galaxy, and \citet{Soto2019} finds indications for gas outflows at $\sim$400 km s$^{-2}$.

Three most \HI-excess galaxies in our sample are shown in the upper panel: J1051-17:S1 is a polar disc galaxy with extended low surface brightness disc, we discuss it in details in Kilborn et al. (in prep). J0400-50:S1 has highly asymmetric \HI\ intensity distribution and disturbed gas kinematics. We presented this galaxy in \citet{Dzudzar2019b} as an example of being shaped by the group environment as it is falling into a cluster. J1250-20:S1 is interacting with other galaxies in its group (with S2, S3) and it is connected to a stellar stream that contains S6 and S7. We presented this group in \citet{Dzudzar2019} and in Section \ref{sec:1250}, we also show the \HI\ stream between S1 and S2 galaxy in Figure \ref{fig:j1250_multiw}. 

The presence of the interactions in \HI-deficient and \HI-excess galaxies indicates that the galaxy environmental processing happens through two channels: i) tidal stripping, which contributes to the loss of \HI\ in galaxies and ii) mergers, which contributes to the enhancement of the \HI\ in galaxies.

\subsection{Choirs in mid-infrared}
\label{sec:WISE}

\begin{figure}
    \centering
        \centering
        \includegraphics[width=\columnwidth]{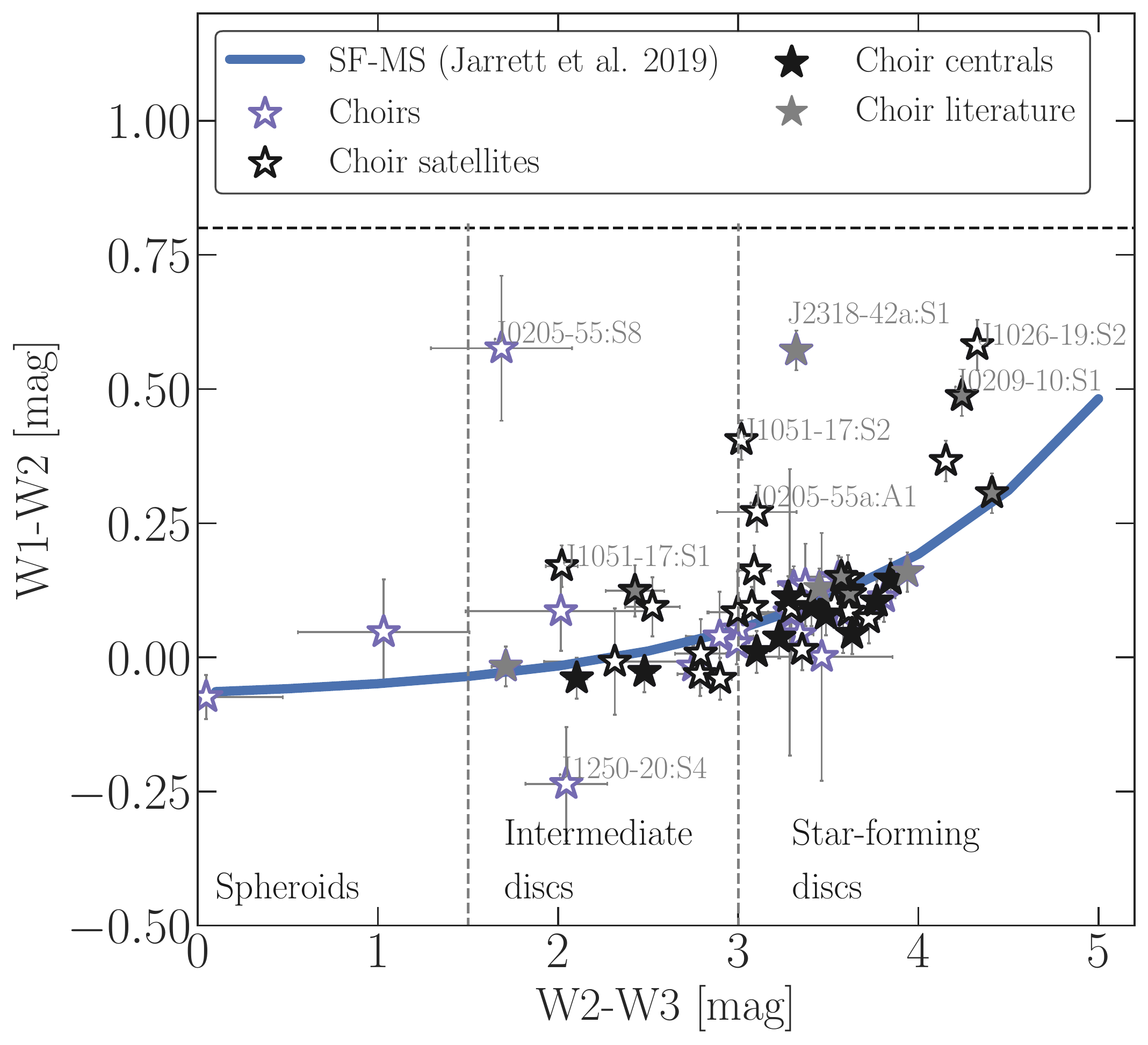} 
        \caption{\textit{WISE} color-color diagram of the Choir galaxies. The solid blue line shows the \textit{WISE} colour-colour sequence \citep{Jarrett2019}. The horizontal dashed line denotes the border above which the AGN emission dominates over the host galaxy, whereas vertical dashed lines separate mid-infrared  morphological classification based on colour. The solid black stars are Choir central galaxies and the open stars with the black outline are satellite galaxies for which we mapped \HI\ content. The solid grey stars are Choir galaxies that we have \HI\ mass from the literature. The purple open stars are all other Choir galaxies for which \textit{WISE} bands could be measured.}
\label{fig:Choir_WISE_COLOR}
\end{figure}

We analyse the mid-infrared properties of Choir galaxies using ``drizzled'' \textit{WISE} imaging \citet{Jarrett2012}, in order to determine whether their star-formation activity is impacted by the group environment. The photometry was obtained from reconstructed images from \textit{WISE} using the ICORE co-addition software \citet{Masci2013}. We show the \textit{WISE} measurements for Choir galaxies in Table \ref{tab:Choirs_WISE}. For details on the \textit{WISE} measurements see \citet{Jarrett2012} and \citet{Jarrett2013}.

\begin{figure}
    \centering
        \centering
        \includegraphics[width=\columnwidth]{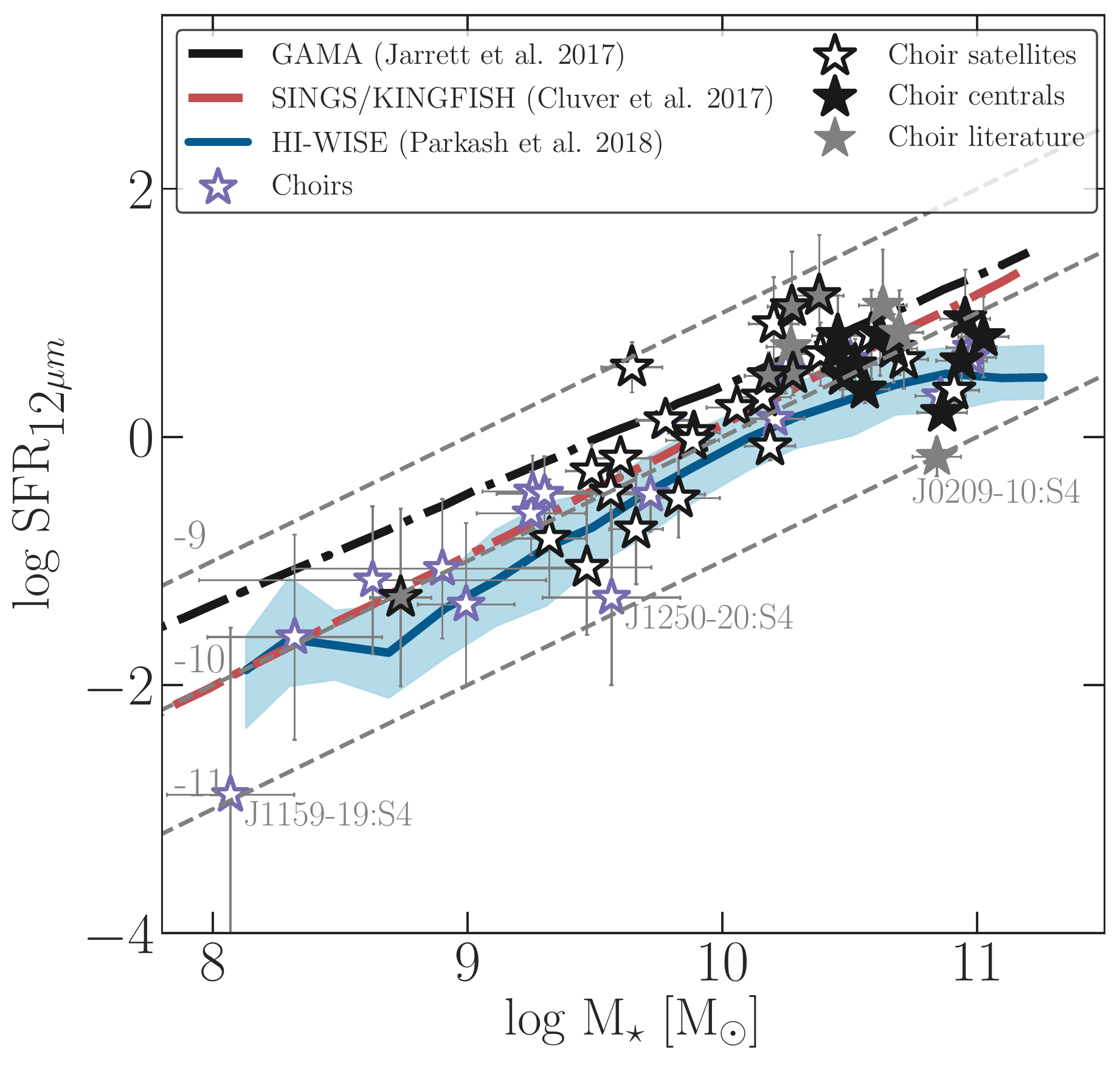}
        \caption{The \textit{WISE} star formation versus stellar mass. The star symbol denote Choir galaxies as described in Figure \ref{fig:Choir_WISE_COLOR}. The horizontal dashed lines are denoting constant specific star-formation rates. The solid blue line is the running median of the star formation rate in bins of the stellar mass, obtained using galaxies from the HI-WISE data of \citet{Parkash2018}. The dot-dashed red line shows the best fit line to the SINGS/KINGFISH sample, used for SFR calibration \citep{Cluver2017}. The dot-dashed black line shows the average `sequence' of the equatorial GAMA G12 field \citep{Jarrett2017}. }
\label{fig:Choir_WISE_SFR_MS}
\end{figure}

The flux of the W1 band traces the stellar light dominated by old stars, thus is it a good tracer of the galaxy's stellar mass \citep{Cluver2014}. We obtain the W1 `in-band' (relative to the Sun) luminosity from the W1-W2 colour, and derive the stellar mass based on the prescription of \citet{Cluver2014}. The \textit{WISE} W3 mid-infrared luminosity traces the dust-obscured star-formation activity \citep{Cluver2017, Jarrett2019}. We determine the star formation rate from the W3 band flux after removing the contribution from evolved stars, based on the prescription of \citet{Cluver2017}. 

We show the mid-infrared colour-colour (W1-W2 versus W2-W3) diagram of the Choir galaxies in Figure \ref{fig:Choir_WISE_COLOR}. It has been shown that the galaxy activity (e.g. star-forming or AGN) can be roughly determined based on galaxy's mid-infrared colours (e.g. \citealt{Jarrett2019}). We see that the majority of Choir galaxies are scattered around the \textit{WISE} colour-colour sequence (star-forming sequence) which was determined by \citet{Jarrett2019}, who also show that W1-W2 colours well above the sequence have excess infrared emission due to host-dust-enshrouded AGN. The increase of the W1-W2 colour towards the right indicates the elevation of the W2 band emission by the hot dust from star-formation \citep{Jarrett2019}. We find several Choir galaxies with elevated W1-W2 colours with respect to the star-forming sequence which may be Seyfert galaxies with excess hot dust emission associated with the AGN torus (see annotations in Figure \ref{fig:Choir_WISE_COLOR}). However, none of them are within the AGN-dominated region (above the horizontal dashed line at W1-W2$\sim$0.8). The most extreme outlier is J0205-55:S8, it is a small compact dwarf galaxy. J0209-10:S1 and J0205-55:A1 are classified as Seyfert 2 galaxies \citep{Carvalho1999, Schmitt2003}, and J2318-42a:S1 as Seyfert 1 \citep{Veron2006}. J1026-19:S2 is an irregular galaxy which is tidally interacting with J1026-19:S1 (see Figure \ref{fig:j1026_multiw}), and J1051-17:S2 is a high surface brightness spiral galaxy \citep{Sweet2013}.  

Denotation of infrared morphological galaxy classification: spheroids, intermediate discs and star-forming discs are assumed based on the colour, which was suggested in \citet{Jarrett2017}. In the `Spheroids' region are expected to reside galaxies with the prominent bulges, such as E or S0 galaxies, most often considered gas-depleted high mass galaxies or low burner dwarf spheroidal galaxies. Choir galaxies in this region are J1059-09:S10 (W2-W3$\sim$0 mag) and J1159-19:S4 (W2-W3$\sim$1 mag) and these are respectively classified as dwarf spheroidal [dS] and dwarf [D] galaxy by \citet{Sweet2013}. The `Intermediate discs' are described as Milky Way type galaxies, with prominent bulges and semi-quiescent discs (1.5$<$W2-W3$<$3). Galaxies in this region are possibly undergoing quenching of their star formation \citep{Jarrett2019}. The `Star-forming discs' are galaxies with W2-W3$>$3, and these galaxies have ongoing active star-formation. 

In Figure \ref{fig:Choir_WISE_SFR_MS} we show the star-formation properties of the Choir galaxies. The dashed lines represent constant specific star-formation rates, and we see that the Choir galaxies are spread between 10$^{-9}$ and 10$^{-11}$ yr$^{-1}$, thus we do not find them to be significantly quenched. We use the HI-WISE data from \citet{Parkash2018} and show the running median of the star formation rate in bins of the stellar mass. We show the running median line in Figure \ref{fig:Choir_WISE_SFR_MS} with the blue solid line, while the shaded region shows the 16th and 84th percentile. For comparison we show the best fit line to the SINGS/KINGFISH sample, used for the calibration of the star formation rate from 12$\mu$m \citep{Cluver2017}, and we also show the star-formation biased `sequence' of the the equatorial GAMA G12 field \citep{Jarrett2017}.

Three galaxies with a specific star formation rate of $\sim$10$^{-11}$ yr$^{-1}$ are: J1159-19:S4 a dwarf galaxy; J1250-20:S4 an Irr galaxy, and J0209-10:S4 an SApec galaxy. All three of these galaxies are experiencing tidal interactions, J1159-19:S4 -- as seen in \HI, while J1250-20:S4 and J0209-10:S4 as seen in \HI\ and optical. 

\subsection{Choirs in context of Hickson compact groups and their evolution}

\begin{figure}
    \centering
        \centering
        \includegraphics[width=1\columnwidth]{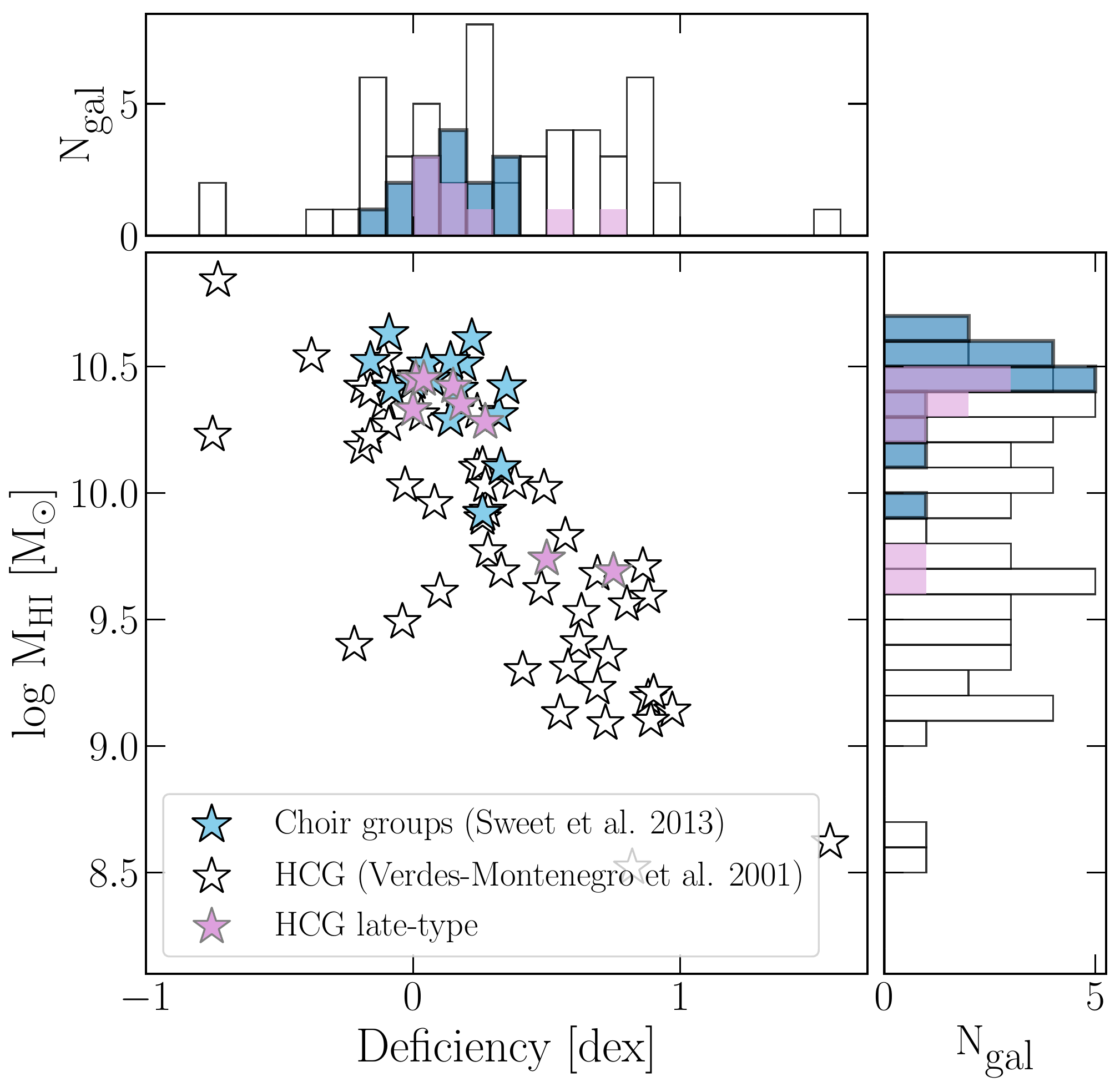} 
        \caption{The \HI\ mass versus deficiency for Choir groups with comparison to the Hickson compact groups. The blue stars show Choir groups. The white stars show HCGs from \citet{VerdesMontenegro2001}. The pink stars show the the late-type dominated HCGs, based on the `MType' from \citet{Hickson1982} and \HI\ properties from \citet{VerdesMontenegro2001}. }
\label{fig:ChoirHCGMHIDEF}
\end{figure}

Based on the group compactness, Choir groups are most similar to HCGs, as shown by the projected separation between the two most luminous galaxy members \citep{Sweet2013}. Therefore, in order to place the Choirs into the broader context of groups, we compare them to Hickson Compact Groups (HCG; \citealt{Hickson1982, VerdesMontenegro2001}). In Figure \ref{fig:ChoirHCGMHIDEF} we show group \HI\ mass versus group deficiency of Choir groups and HCGs. For each Choir, the total group \HI\ mass is taken from \HIPASS, while the group \HI\ deficiency is taken from \citet{Sweet2013}. We find that Choir groups have a high mass and a low deficiency values, while HCGs are predominantly tailing towards lower group \HI\ masses and a higher deficiency values. It is known that HCGs are comprised of galaxies with varying morphologies, and many of these galaxies are early-type galaxies. Based on the morphological type noted in \citep{Hickson1982} and available \HI\ data from \citet{VerdesMontenegro2001} we select HCGs that are comprised of only late-type galaxies: HCG2, HCG16, HCG31, HCG38, HCG80, HCG88, HCG89, HCG100 for comparison. We find that these late-type dominated HCGs are more similar to the Choir groups (see Figure \ref{fig:ChoirHCGMHIDEF}). 

In \citet{Dzudzar2019}, we suggested that the Choir groups would fit into phase 1 and phase 2 of the proposed \HI\ evolutionary scenario of HCGs \citep{VerdesMontenegro2001}. We place the phase classification for each Choir group in Table \ref{tab:phase_classification}. In summary the \citet{VerdesMontenegro2001} evolutionary scenario of the gas content in HCGs is as follows: i) Groups in the phase 1 have an average gas content and a low level of galaxy interaction. ii) Groups in phase 2 contain \HI\ tidal features. iii$_{\textrm{a}}$) Groups in the phase 3a do not have \HI\ within galaxies (it can be in the intragroup medium). iii$_{\textrm{b}}$) Groups in the phase 3b contain the \HI\ gas in the form of a cloud that envelopes entire group. Over time, it is likely that majority of galaxies in Choir groups will continue to deplete their gas content through star formation and interactions due to low velocity dispersion (tidal interactions, mergers), and they will become progressively more \HI\ deficient. 

\begin{figure*}
    \centering
        \centering
        \includegraphics[width=2\columnwidth]{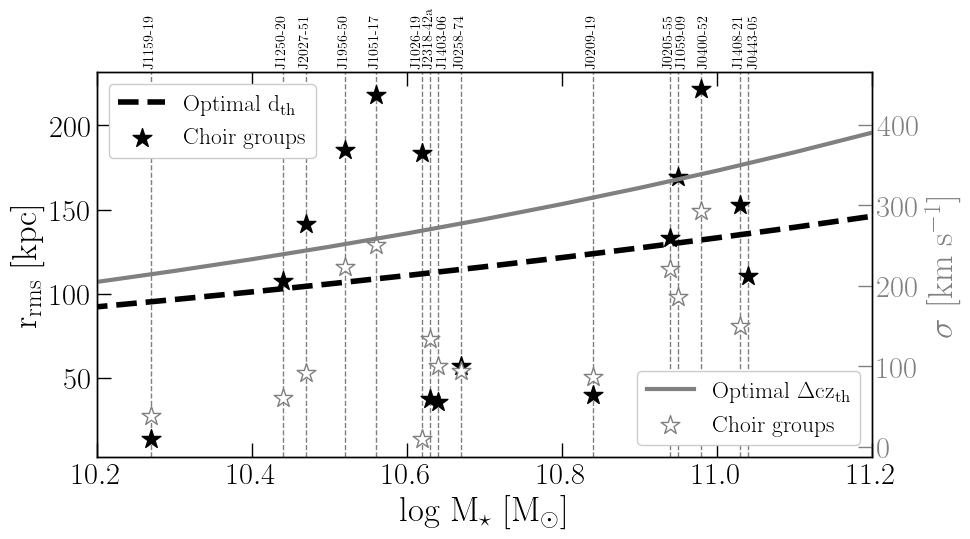} 
        \caption{Evolution of the optimal velocity threshold $\Delta$cz$_{\textrm{th}}$ (grey solid line) and projected distance d$_{\textrm{th}}$ threshold (black dashed line) with the stellar mass of the central galaxy within given Choir group (theoretical lines are obtained from \citet{Pfister2020}). Each Choir group is represented with the solid black star in relation to the optimal d$_{\textrm{th}}$ criteria, and with the open grey star in relation to the optimal $\Delta$cz$_{\textrm{th}}$ criteria (respectively Choir groups are presented with their projected radius and radial velocity dispersion). The vertical dashed lines connect the same Choir groups in both criteria, and they are annotated at the top of the Figure.}
\label{fig:Choir_sample_interactions}
\end{figure*}

Comparing gas depletion timescales from \citet{Dzudzar2019} and the crossing time for Choir groups (see Table \ref{tab:Choirs}), we find that the crossing time is on average shorter than the \HI\ depletion time. Therefore, galaxy mergers that will happen in the future will most likely be gas-rich mergers\footnote{Under assumptions that galaxies will go thought the gravitational potential of the group.}. To investigate which Choir group has conditions for galaxy mergers, we use approximated conditions from \citet{Pfister2020} for the galaxy pair mergers. Utilizing the Horizon-AGN simulation, \citet{Pfister2020} formulated a threshold on the projected distance d$_{\textrm{th}}$ and redshift difference $\Delta$z$_{\textrm{th}}$ below which galaxy pairs are likely to merge. To adapt the \citet{Pfister2020} relations for Choir groups, we use the group projected radius and radial velocity dispersion. Comparing the Choirs group velocity dispersion and the \citet{Pfister2020} criteria for optimal $\Delta$z$_{\textrm{th}}$, all Choir groups have optimal condition for galaxy mergers. Whereas, looking at the group projected radii, we find that seven groups do not satisfy imposed criteria, two are on the threshold and six satisfy imposed criteria. Even though this analysis is rough, since groups have more massive gravitational potential than galaxy pairs, with such approach we obtained lower limits on the given conditions for galaxy mergers. Choir groups J1026-19, J1059-09 and J1408-21 are above optimal d$_{\textrm{th}}$ threshold, however, they contain galaxies that are already interacting, while J0400-52 is infalling into a cluster. To better constrain minimal galaxy separation for mergers, this work requires further studies of Choir group analogues in simulations.

Based on the mentioned parameters, we suggest that some of the Choir groups will go through a phase of compact groups or they will end up as an optical fossil group, depending on the global environment around them. A fossil group is usually defined as a galaxy group with a single dominant elliptical galaxy with an X-ray luminosity of L$_{\textrm{X,bol}} \geq 10^{42}$ h$_{50}^{-2}$ erg s$^{-1}$ and with at least 2.0 magnitude difference from the second brightest group galaxy (e.g. \citealt{Jones2003}). We define an optical fossil group, where the magnitude difference between the brightest galaxy and second brightest galaxy is larger than 2 magnitudes, but we do not impose that they are X-ray bright as in the definition of fossil groups by \citet{Ponman1994}.

Nine of the Choir groups have galaxies in the neighbourhood and thus they have a possibility of forming a HCG like group in the future, assuming that these galaxies will `fall' towards the same potential well (see Section \ref{sec:global_environment} and Appendix \ref{sec:appendix_images}). These groups are: J0205-55, J0209-10, J0443-05, J1051-17, 1059-09, J1159-19, J1403-06, J2027-51 and J2318-42a.

Choir J1159-19 is a compact group which contains a one armed spiral galaxy and three dwarf galaxies. J1159-19 meets the criteria for an optical fossil group. All galaxies within this group are interacting, as seen in their \HI\ distribution (see Figure \ref{fig:j1159_group}) and they will most likely merge in the future. A further five Choirs lie in a fairly isolated environment are more likely to create an optical fossil group, similar to the current state of J1159-19. In these cases, there are fewer galaxies that can merge and thus they will not be able to satisfy criteria to be classified as HCGs. These groups are: J0258-74, J1026-19, J1250-20, J1408-21 and J1956-50. 

One Choir group, J0400-51, is infalling into a cluster. \citet{Dzudzar2019b} proposed that the gas-rich galaxy within this group (ESO156-G029) will be under the effects of ram-pressure once the galaxy approaches closer to the cluster centre. Moreover, we used this galaxy as an example to show that group pre-processing is important in its evolution.

\section{Global environment around the Choir groups}
\label{sec:global_environment}

In this section we explore the wider environment of the Choir groups and its impact on their \HI\ content. We present results from the local and global environment around Choir groups, the former corresponds to the neighbouring galaxies and the latter corresponds to the position of Choirs within cosmic web filaments. 

\subsection{Wider environment of the Choir groups}
\label{sec:obtaining_environment}
We analyse the wider environment of the Choir groups to determine the environment outside the 15\arcmin\ of the SINGG field of view. We made a script in \texttt{python} to query for sources around the \HIPASS\ detection in NASA/IPAC Extragalactic Database (NED). We use RA, Dec and the recessional velocity V$_{\HI}$ of Choir groups from \citet{Sweet2013}, and make a query typically $\sim$4 degrees around each Choir group. The initial result is sub-sampled for only `G' objects i.e. only galaxies, as there are sources which mark position of groups, triplets etc. The obtained catalogue is cross-matched with the 2MASS Extended Source Image Server to obtain the Ks-band total magnitude in order to make a magnitude limited sample. 

We use galaxies with known Ks-magnitude, recessional velocity between $\pm$1000 km s$^{-1}$ from the Choir group velocity, and $<$0.5 degree projected separation from the HIPASS centre, to compute the Ks weighted mean position of the group centroid. Using this sample of galaxies we re-compute the projected distances for each galaxy from the group centroid. We use this data to map the galaxies with their velocity difference (with respect to the Choir group) around all Choir groups (see panel c in Appendix \ref{sec:appendix_images}). We also use the projected angular separation versus the velocity difference of each galaxy from the group mean velocity to show Choir caustics curves in order to map the group potential and find potential galaxies that will merge with Choir groups (see panel d in Appendix \ref{sec:appendix_images}).

We find Choirs in low density environments (J0258-74, J1026-19, J1250-20,J1408-21 and J1956-50), Choirs with nearby, possible infalling galaxies (J0205-55, J1051-17) and Choirs which are potentially part of a larger structure (J0400-52, J1159-19 and J2027-51). We present results for each Choir group individually in Appendix \ref{sec:note_on_groups} and Appendix \ref{sec:appendix_images}: panel \textbf{c} and \textbf{d} in Figures. 

\subsection{Choirs position with respect to the filaments}
\label{sec:filaments}

We explore whether the Choir groups have a `special' position in the cosmic web. \citet{Janowiecki2017} suggested that the \HI-rich central galaxies in the group environment are \HI-rich due to gas accretion from cosmic web filaments, as groups can be positioned in the intersection of those filaments. \citet{Kleiner2017} used the 6 degree Field Galaxy Survey (6dFGS) and the Discrete Persistent Structures Extractor (DisPerSe) to map the filaments in the local Universe. Furthermore, combining filaments and data from \HIPASS, they found that galaxies near filaments, with M$\star \geq$10$^{11}$ M$\odot$, have higher \HI\ mass fractions than the control sample. Utilizing catalogue of the filaments by \citet{Kleiner2017}, we derive the distances to the nearest filament for Choir galaxy groups, see Table \ref{tab:Filaments}. 

Exploring 14 Choir galaxy groups (J0400-52 is excluded as it is near  galaxy cluster), we find that seven are within $\leq$1 Mpc, one is at $\sim$2 Mpc and six are at $\geq$4 Mpc from the nearest filament. We find only two Choir groups near the intersection of filaments: J1408-21 and J1059-09. Based on the \HI-mass fraction, central galaxies in J1408-21 and J1059-09 have average gas content. Furthermore, we find two Choir groups that are more than 10 Mpc away from the nearest filament: J0258-74 and J1956-50. These groups can be classified as being in a void, as distances of $\sim$10 Mpc are considered as voids \citep{Kuutma2017}. The central galaxy in J0258-74 has an average \HI-mass fraction, while central in J1956-50 has a higher than average \HI-mass fraction. We conclude that Choir groups span a range of global environments, from being in a void to being near an intersection of filaments, thus they are not within a unique position in the cosmic web. While gas accretion from the cosmic web may be occurring in some Choir groups, it is not shown to drastically enhance the \HI\ content as the Choir groups with similar \HI\ properties span all global environments.

\begin{table}
\caption{Distances to the nearest filaments}
\label{tab:Filaments}
\begin{threeparttable}
\begin{tabular}{lrrcr}
\toprule
ID &  RA [deg] &  DEC [deg] &   z &  D [Mpc] \\ 
(1) & (2) & (3) & (4) & (5) \\ \hline \hline
Choirs & & & & \\
J0205-55 &   31.272800 & -55.111801 &  0.02172 &                            4.29 \\
J0209-10 &   32.427898 & -10.183700 &  0.01261 &                            1.06 \\
J0258-74 &   44.527000 & -74.456299 &  0.01634 &                           11.90 \\
J0400-52\tnote{a} &   60.170101 & -52.734100 &  0.03526 &                           \dots \\
J0443-05 &   70.932899 &  -5.319420 &  0.01611 &                            0.22 \\
J1026-19 &  156.669998 & -19.051100 &  0.03152 &                            1.67 \\
J1051-17 &  162.906006 & -17.124800 &  0.01938 &                            0.11 \\
J1059-09 &  164.817993 &  -9.793930 &  0.02849 &                            0.16 \\
J1159-19 &  179.876007 & -19.265800 &  0.00584 &                            0.62 \\
J1250-20 &  192.720001 & -20.371000 &  0.02662 &                            4.18 \\
J1403-06 &  210.854004 &  -6.069210 &  0.00957 &                            6.61 \\
J1408-21 &  212.175003 & -21.597200 &  0.02989 &                            0.87 \\
J1956-50 &  299.190002 & -50.055599 &  0.02568 &                           13.70 \\
J2027-51 &  307.027008 & -51.691601 &  0.02031 &                            3.86 \\
2318-42a &  349.597992 & -42.369999 &  0.00537 &                            0.61 \\ 
\bottomrule
\end{tabular}
\begin{tablenotes}
\item[a] Excluded, because it is a part of a galaxy cluster.
\item (1) Group name: HIPASS+ID; (2) Right Ascension (J2000) from HIPASS; (3) Declination (J2000) from HIPASS; (4) Redshift, computed from the distances from \citet{Dzudzar2019} as z$=$ (d$\times$H$_{0}$)/c, where d is the distance to the source in Mpc, H$_{0}$ = 70 km s$^{-1}$ Mpc$^{-1}$, c = 3$\times$10$^{5}$ km s$^{-1}$; (5) Distance to the nearest filament in Mpc, for details see \citet{Kleiner2017}.

\end{tablenotes}
\end{threeparttable}
\end{table}

\subsection{Global classification of Choir groups}
\label{sec:global_classification}
In summary, the Choir groups whose \HI\ content we mapped can be classify into several broad categories:

\begin{itemize}
    \item \textbf{Isolated groups}, embedded in the low density global environment:  J0258-74, J1026-19,  J1250-20, J1408-21, J1956-50.
    \item \textbf{Cluster infalling group}: J0400-52.
    \item \textbf{Groups with tidal interactions:} J1159-19, J1059-09, J1026-19, J1408-21, J1250-20, J0400-52. 
    \item \textbf{Merging subgroups:} J0205-55 (composed of J0205-55a and J0205-55b), J1051-17
    \item \textbf{Potentially part of the larger structure:} J2027-51, J1159-19.
\end{itemize}

\section{Summary and conclusions}
\label{sec:conclusion}

In this paper we present and explore the resolved \HI\ content of galaxies in \HI-rich and late-type dominated groups named `Choirs'. We present the Choir galaxy sample in context by comparing it to its parent sample Survey of Ionization in Neutral Gas Galaxies, SINGG \citep{Meurer2006} and grandparent sample \HI\ Parkes All Sky Survey, \HIPASS \citep{Barnes2001} as well as to several galaxy samples from the literature: Analysis of the Interstellar Medium od Isolated GAlaxies, AMIGA \citep{VerdesMontenegro2005Amiga}, The extended GALEX Arecibo SDSS Survey, xGASS \citep{Catinella2018}, and Hickson Compact Groups, HCG (\citealt{Hickson1982, VerdesMontenegro2001}).

Within 13 Choir groups discussed in this paper, 44 galaxies, out of 78, are detected in \HI\ emission, see Figure \ref{fig:Choir_detections}. Deriving the \HI\ detection limits and the angular sizes of the galaxies (see Figure \ref{fig:hi-non-detections}), we find that eight galaxies, out of 34, are not detected due to a potential \HI\ deficiency, and the expected \HI\ content of remaining galaxies lies below our detection limit. 

We quantify the dispersion in the \HI\ mass fraction (f$_{\textrm{\HI}}$) of the Choir galaxies with respect to their \HIPASS\ grandparent sample, using \HIPASS\ running median of the \HI\ mass fraction in the stellar mass bins of 0.3 dex (see Figure \ref{fig:Choir_sample_context}). We find that the \HI\ mass fraction of the Choir galaxies is dispersed around the median, $\Delta$f$_{\textrm{\HI}}$, in the range -1.4$\leq \Delta$f$_{\textrm{\HI}}\textrm{[dex]}\leq$0.7 (see Figure \ref{fig:Delta_fraction}). Using the standard deviation from the \HIPASS\ running median of the \HI\ mass fraction, we find that 4 galaxies are \HI-excess ($\Delta$f$_{\textrm{\HI}}>$0.4 dex), 32 galaxies have an average \HI\ content ($-$0.4$<\Delta$f$_{\textrm{\HI}}\textrm{ [dex]}<$0.4) and 8 galaxies are \HI-deficient ($\Delta$f$_{\textrm{\HI}}<$($-$0.4 dex)). The existence of the \HI-excess and \HI-deficient galaxies in Choir groups indicates the presence of the environmental processing (e.g. \citealt{Chung2009, Denes2016, Elison2018}), which we further explore.

We map the \HI\ distribution in galaxies within the Choir groups, and highlight six with an extreme \HI\ content, the three most \HI-excess and the three most \HI-deficient, in order to determine whether their \HI\ content is a result of galaxy environmental processing. The most \HI-excess galaxy is J1051-17:S1, with $log$M$_{\star}$=10.19$\pm$0.10 and $log$M$_{\textrm{\HI}}$=10.43$\pm$0.02. J1051-17:S1 has $\sim$4.7 times more \HI\ than expected of a galaxy with a similar stellar mass. The most \HI-deficient galaxy is J0205-55:S3, with $log$M$_{\star}$=10.86$\pm$0.10 and $log$M$_{\textrm{\HI}}$=8.65$\pm$0.13. J0205-55:S3 has $\sim$23 times less \HI\ than expected of a galaxy with a similar stellar mass. We show that the \HI-excess galaxies have a large \HI--to--R-band radii ratio and exhibit signs of past or current galaxy-galaxy interactions (see Figure \ref{fig:HI_excess}). On the other hand, the \HI-deficient galaxies have a small \HI--to--R-band radii ratio (see Figure \ref{fig:HI_deficient}) and their \HI\ column densities do not reach high values (peak at $\sim$10$\times$10$^{19}$ cm$^{-2}$). Six out of eight \HI\ deficient galaxies have irregular structure (optical or \HI) which indicates that they most likely experienced \HI\ stripping due to tidal interactions with the other galaxies within the group. The \HI-excess galaxies show a range of peculiarities in their optical morphology, e.g. galaxy with a polar ring (J1051-17:S1), galaxy with asymmetric \HI\ distribution (J0400-52:S1), and galaxy with stellar streams (J1250-20:S1). These peculiarities indicate past and/or current galaxy interactions which can lead to the enhancement of the \HI\ content i.e. an environmental processing `boosts' the \HI\ content (e.g. \citealt{Elison2018}). Another possibility of the acquiring of the \HI\ is through cold gas accretion from cosmic web filaments \citep{Keres2005} however, our observations are not sensitive enough to detect it. The presence of the \HI-deficient and \HI-excess galaxies shows that the galaxy environmental processing happens through two channels: i) tidal stripping, which contributes to the loss of \HI\ in galaxies and ii) mergers, which contributes to the enhancement of the \HI\ in galaxies.

We explore the impact of the environmental processing on the mid-infrared galaxy properties using the data from \textit{WISE}. We show that all but three Choir galaxies have regular star formation rate. Three galaxies have a low specific star formation rate of $\sim$10$^{-11}$ yr$^{-1}$, two are dwarf galaxies with low star formation and one is SApec galaxy which may be on a path to quenching (e.g. \citealt{Jarrett2019}), and all three of them are experiencing tidal interactions. Out of the 56 galaxies with obtained 12$\mu$m we do not find any heavily dust-obscured, nor starburst-like star formation rates in Choir galaxies. We find that seven, out of 48, Choir galaxies have elevated W1-W2 colours with respect to the star-forming sequence thus they are potentially Seyfert galaxies (see Figure \ref{fig:Choir_WISE_COLOR}). This indicates that galaxy environmental processing is more perceptible in the \HI\ content than in the mid-infrared properties of Choir galaxies.

We discuss Choir groups in the context of Hickson compact groups and explore their evolutionary scenarios. We discuss the possible future of Choir groups taking into consideration the \HI\ depletion timescale, crossing time, velocity dispersion and the global environment around these groups. Since the crossing time is shorter than the depletion time, mergers that will happen in the future are likely to be gas-rich. This scenario points out that the \HI\ from galaxies is potentially going to be stripped off during the galaxy interactions, forming tidal bridges and tails which will likely disperse into the intra-group medium. Exploring the global environment around Choir groups (see Appendix \ref{sec:appendix_images}), we find that nine Choir groups might go through a phase of being more compact, similar to HCGs: J0205-55, J0209-10, J0443-05, J1051-17, 1059-09, J1159-19, J1403-06, J2027-51 and J2318-42a. Such phase is possible due to the low velocity dispersion of Choirs and a number of nearby galaxies which are likely to merge with a group. Choir groups that are embedded in isolated environments may merge only with group members, possibly forming a HIPASS J1159-19 type of system (see Figure \ref{fig:j1159_group}). These groups are: J0258-74, J1026-19, J1250-20, J1408-21 and J1956-50. The only exception to these scenarios is the J0400-52 group as it is infalling into a cluster \citet{Dzudzar2019b}. The rapid \HI-removal process, ram-pressure, is expected in the S1 galaxy within J0400-52 once it approaches the cluster centre \citet{Dzudzar2019b}. J0400-52 already contains galaxies that are potentially \HI-deficient (see Appendix \ref{appendix:detection_limit}) which is due to either group pre-processing, or it is possible that they are the `backsplash' galaxies i.e. already fell through the cluster potential well, as discussed in Section \ref{section:j0400}. In our future work we aim to further probe possible evolutionary pathways for Choir groups by tracing their analogues in simulations. 

We present each Choir group individually showing its \HI\ content and global environment. We explore the global environment around the Choir groups and their position in the cosmic web. We find that seven Choir groups are within $\leq$1 Mpc, one is at $\sim$2 Mpc and six are at $\geq$4 Mpc from the nearest cosmic web filament. We conclude that Choir groups do not have a unique position in the cosmic web. Our observations show that environmental processing in the Choir groups, in the form of tidal interaction and mergers, is present regardless of their global environment. We find galaxy interactions in the group that is in isolation (in void, J0258-74) and in the groups that are near the intersection of the cosmic web filaments, J1059-09 and J1408-21. 

All our results indicate that the galaxy environmental processing is already present in the \HI-rich groups that are dominated by the late-type galaxies. With the upcoming \HI\ WALLABY \citep{Koribalski2012W} survey, and surveys with Apertif \citep{Osterloo2010} and MeerKAT \citep{Camilo2018} we will obtain a large statistical sample of such \HI-rich galaxy groups. Using those data we will be able to further probe the initial stages of galaxy environmental processing in groups.

\section*{Data Availability}

The data underlying this article are available in the article. 

\section*{Acknowledgements}

We thank the anonymous referee for their comments and
suggestions which improved this paper.\\
Robert D\v{z}ud\v{z}ar (RD) acknowledges support by a Swinburne University Postgraduate Research Awards (SUPRA) scholarship throughout major part of this work.\\
RD would like to thank Chandrashekar Murugeshan and Michelle Cluver for insightful discussions. RD would also like to thank Fiona Audcent-Ross for help with the SINGG catalogue, thank you extends to the entire SINGG survey team.\\ 
RD acknowledges financial support (Covid-19 fund) from the Astronomical Society of Australia that helped finishing this paper.\\
Parts of this research were conducted by the Australian Research Council Centre of Excellence for All Sky Astrophysics in 3 Dimensions (ASTRO 3D), through project number CE170100013.\\
Dane Kleiner acknowledges funding received from the European Research Council (ERC) under the European Union's Horizon 2020 research and innovation programme (grant agreement no. 679627)\\
The Australia Telescope Compact Array is part of the Australia Telescope National Facility which is funded by the Australian Government for operation as a National Facility managed by CSIRO.\\
The National Radio Astronomy Observatory is a facility of the National Science Foundation operated under cooperative agreement by Associated Universities, Inc.\\
Based on observations at Cerro Tololo Inter-American Observatory, National Optical Astronomy Observatory (NOAO Prop. AAT/13A/02; PI: Sarah M. Sweet), which is operated by the Association of Universities for Research in Astronomy (AURA) under a cooperative agreement with the National Science Foundation. \\
This work was performed on the OzSTAR national facility at Swinburne University of Technology. OzSTAR is funded by Swinburne University of Technology and the National Collaborative Research Infrastructure Strategy (NCRIS).\\
This work was written on the collaborative Overleaf platform \url{https://www.overleaf.com}.\\
This publication makes use of data products from the Wide-field Infrared Survey Explorer, which is a joint project of the University of California, Los Angeles, and the Jet Propulsion Laboratory/California Institute of Technology, funded by the National Aeronautics and Space Administration.\\
This research has made use of the NASA/IPAC Ex- tragalactic Database (NED), which is operated by the Jet Propulsion Laboratory, California Institute of Technology, under contract with the National Aeronautics and Space Administration.\\
This publication makes use of data products from the Two Micron All Sky Survey, which is a joint project of the University of Massachusetts and the Infrared Processing and Analysis Center/California Institute of Technology, funded by the National Aeronautics and Space Administration and the National Science Foundation.\\
This research has made use of the VizieR catalogue access tool, CDS, Strasbourg, France. The original description of the VizieR service was published in A\&AS 143, 23.\\
This project used public archival data from the Dark Energy Survey (DES). Funding for the DES Projects has been provided by the U.S. Department of Energy, the U.S. National Science Foundation, the Ministry of Science and Education of Spain, the Science and Technology FacilitiesCouncil of the United Kingdom, the Higher Education Funding Council for England, the National Center for Supercomputing Applications at the University of Illinois at Urbana-Champaign, the Kavli Institute of Cosmological Physics at the University of Chicago, the Center for Cosmology and Astro-Particle Physics at the Ohio State University, the Mitchell Institute for Fundamental Physics and Astronomy at Texas A\&M University, Financiadora de Estudos e Projetos, Funda{\c c}{\~a}o Carlos Chagas Filho de Amparo {\`a} Pesquisa do Estado do Rio de Janeiro, Conselho Nacional de Desenvolvimento Cient{\'i}fico e Tecnol{\'o}gico and the Minist{\'e}rio da Ci{\^e}ncia, Tecnologia e Inova{\c c}{\~a}o, the Deutsche Forschungsgemeinschaft, and the Collaborating Institutions in the Dark Energy Survey.
The Collaborating Institutions are Argonne National Laboratory, the University of California at Santa Cruz, the University of Cambridge, Centro de Investigaciones Energ{\'e}ticas, Medioambientales y Tecnol{\'o}gicas-Madrid, the University of Chicago, University College London, the DES-Brazil Consortium, the University of Edinburgh, the Eidgen{\"o}ssische Technische Hochschule (ETH) Z{\"u}rich,  Fermi National Accelerator Laboratory, the University of Illinois at Urbana-Champaign, the Institut de Ci{\`e}ncies de l'Espai (IEEC/CSIC), the Institut de F{\'i}sica d'Altes Energies, Lawrence Berkeley National Laboratory, the Ludwig-Maximilians Universit{\"a}t M{\"u}nchen and the associated Excellence Cluster Universe, the University of Michigan, the National Optical Astronomy Observatory, the University of Nottingham, The Ohio State University, the OzDES Membership Consortium, the University of Pennsylvania, the University of Portsmouth, SLAC National Accelerator Laboratory, Stanford University, the University of Sussex, and Texas A\&M University.

This research has made use of \texttt{python} \url{https://www.python.org} and python packages: \texttt{astropy} \citep{Astropy2013, Astropy2018}, \texttt{matplotlib} \url{http://matplotlib.org/} \citep{Hunter2007}, \texttt{APLpy} \url{https://aplpy.github.io/}, \texttt{pandas} \citep{Pandas}, \texttt{Jupyter notebook} \url{https://github.com/jupyter/notebook}, \texttt{NumPy} \url{http://www.numpy.org/} \citep{VanDerWalt2011}, \texttt{SciPy} \url{https://www.scipy.org/} \citep{Jones2001} and \texttt{CMasher} \url{https://github.com/1313e/CMasher} \citep{CMasher}.

\bibliographystyle{mnras}
\bibliography{Choir_groups}

\appendix
\section{Properties of Choir galaxies and $\Delta \lowercase{\textrm{f}}_\textrm{\HI}$ }

Our analysis was carried out by making distinction between central and satellite galaxies, marking galaxies that are obtained from the literature, as well as those that are \HI-excess and \HI-deficient. Therefore, in this appendix we show Choir galaxy properties and the $\Delta{\textrm{f}}_\textrm{\HI}$ (the distance of the \HI\ mass fraction from the \HIPASS\ running median). 

\begin{figure}
    \centering
        \centering
        \includegraphics[width=\columnwidth]{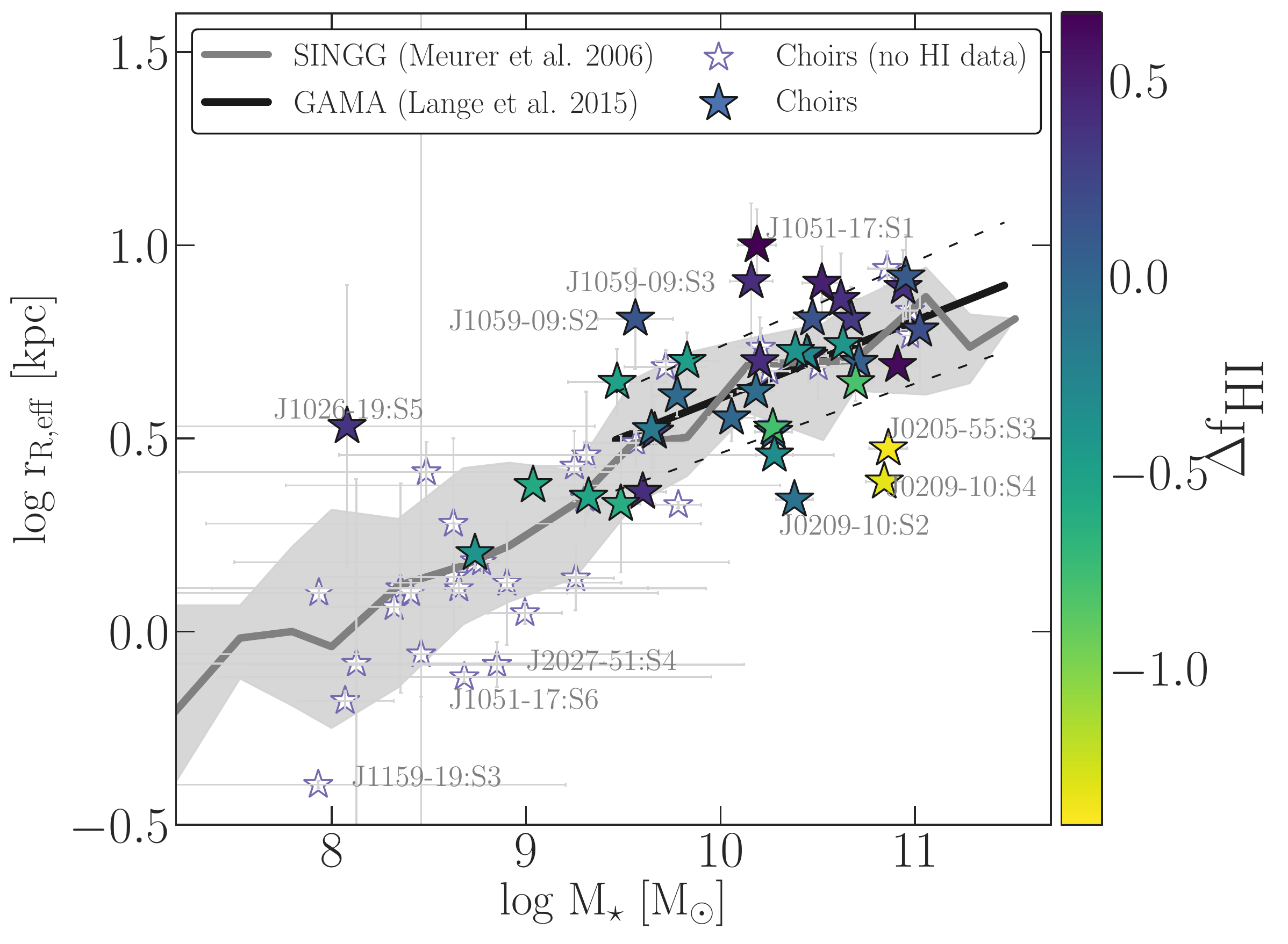}
        \caption{Reproduced Figure \ref{fig:Reff_Mstar}: Relationship between galaxy R-band effective radii and the stellar mass. Here we show $\Delta{\textrm{f}}_\textrm{\HI}$ on the colorbar.
        }
\label{fig:Reff_Mstar_HI}
\end{figure}

In Figure \ref{fig:Reff_Mstar_HI}, the relationship between galaxy R-band effective radii and the stellar mass, we find that at fixed stellar mass range 10$^{10}$ $\leq$ M$_{\star}$ M$_{\odot}$ $\leq$10$^{11}$, with the increase of the r$_\textrm{R,\textrm{eff}}$ the galaxy is more \HI-rich.

\begin{figure}
        \centering
         \includegraphics[width=1\columnwidth]{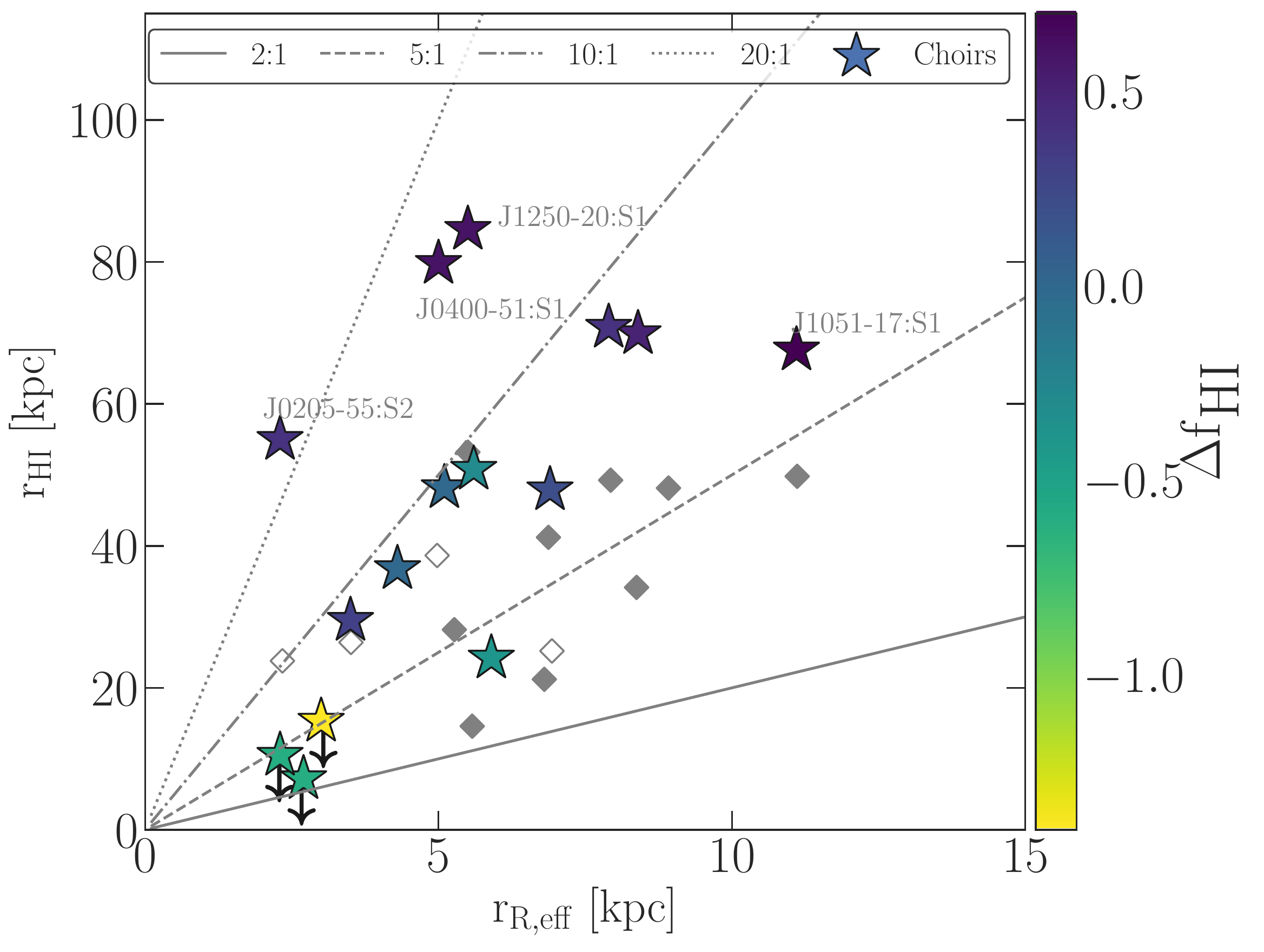}
        \caption{Reproduced Figure \ref{fig:Radii}: Relationship between galaxy \HI\ radii measured at 3$\times$10$^{19}$cm$^{-2}$ and R$_{\textrm{eff}}$ radii. Here we show $\Delta{\textrm{f}}_\textrm{\HI}$ on the colorbar. The star markers show Choir galaxies. The grey diamonds show Choir galaxies with the \HI-radius measured at 1M$\odot$ pc$^{-2}$ \citep{Dzudzar2019}.}
\label{fig:Radii_HI}
\end{figure}

In Figure \ref{fig:Radii_HI}, the relationship between galaxy \HI\ radii measured at 3$\times$10$^{19}$cm$^{-2}$ and R$_{\textrm{eff}}$ radii, we find that the larger the $\Delta{\textrm{f}}_\textrm{\HI}$, galaxy is more likely to have larger \HI\ radius.

\begin{figure}
    \centering
        \centering
        \includegraphics[width=\columnwidth]{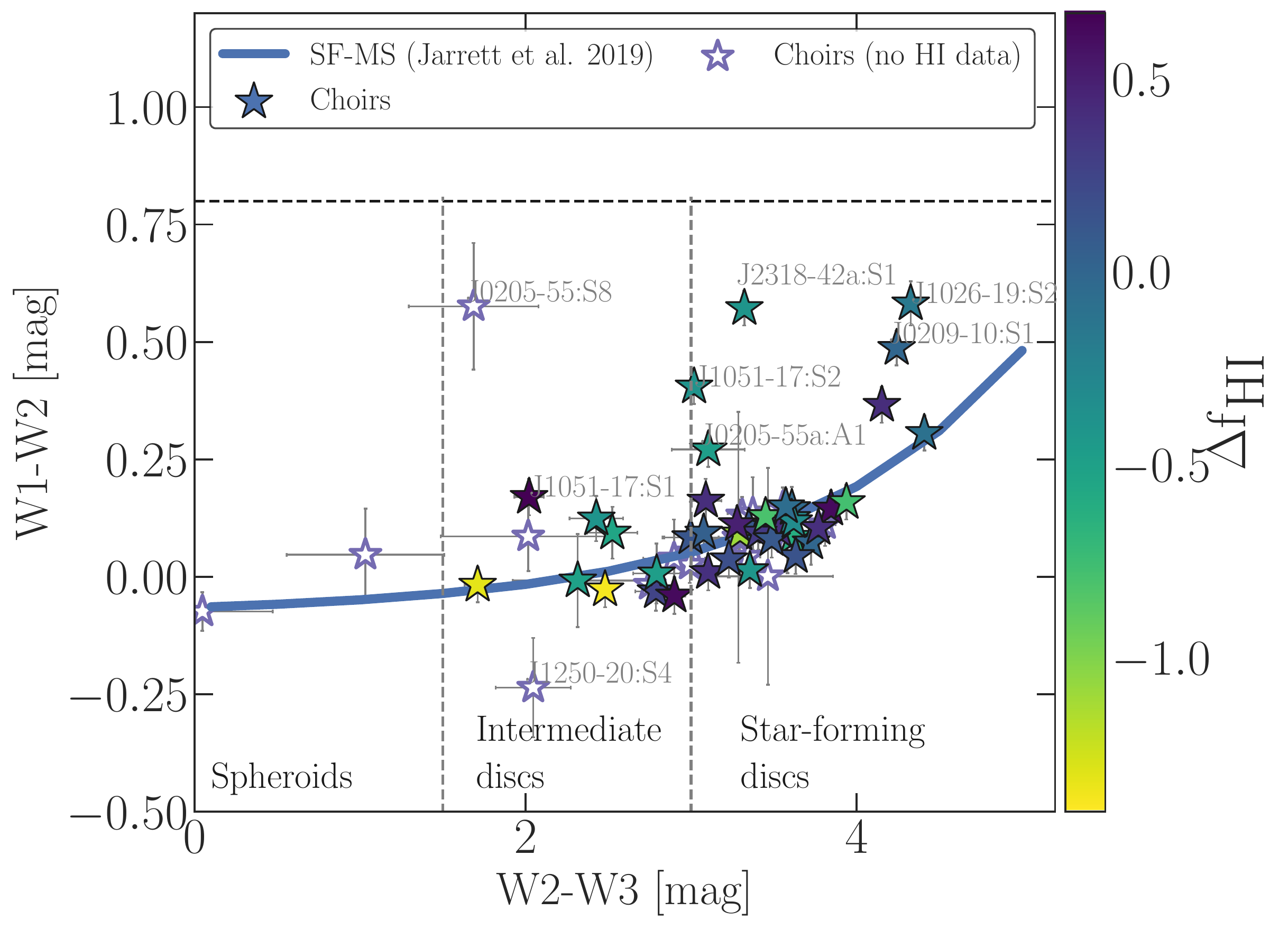} 
        \caption{Reproduced Figure \ref{fig:Choir_WISE_COLOR}: \textit{WISE} color-color diagram of the Choir galaxies. Here we show $\Delta{\textrm{f}}_\textrm{\HI}$ on the colorbar. The solid blue line shows the \textit{WISE} colour-colour sequence \citep{Jarrett2019}. 
        }
\label{fig:Choir_WISE_COLOR_HI}
\end{figure}

In Figure \ref{fig:Choir_WISE_COLOR_HI}, \textit{WISE} color-color diagram of the Choir galaxies, we do not find a clear relationship between \textit{WISE} colours and $\Delta{\textrm{f}}_\textrm{\HI}$. However, we find that galaxies with elevated W1-W2 colours with respect to the star-forming sequence have $\Delta{\textrm{f}}_\textrm{\HI}$ $<$ 0 dex. 

\begin{figure}
    \centering
        \centering
        \includegraphics[width=\columnwidth]{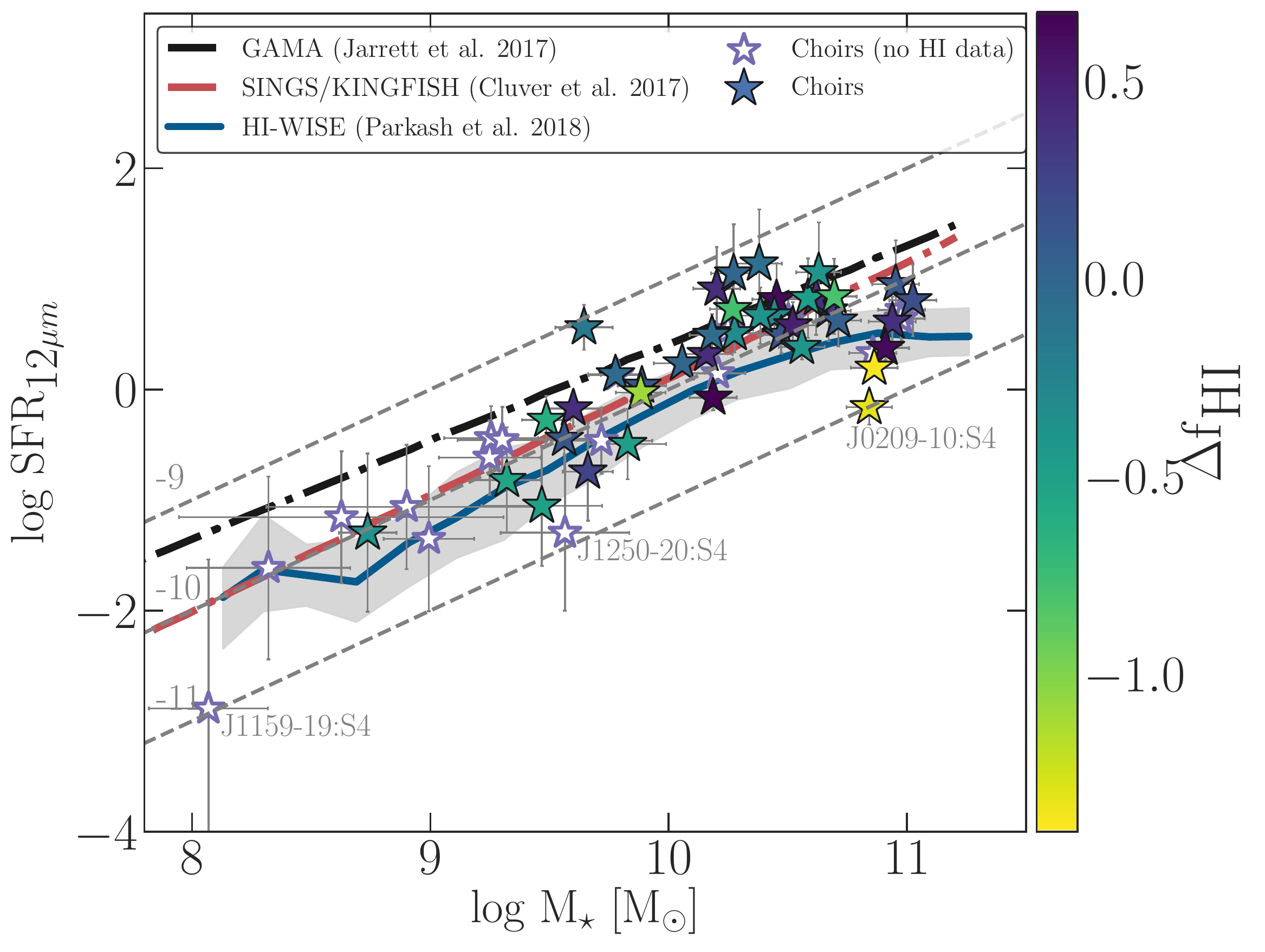} 
        \caption{Reproduced Figure \ref{fig:Choir_WISE_COLOR}: The \textit{WISE} star formation versus stellar mass. Here we show $\Delta{\textrm{f}}_\textrm{\HI}$ on the colorbar. The star markers show Choir galaxies. The horizontal dashed lines are denoting constant specific star-formation rates.
        }
\label{fig:Choir_WISE_SFR_MS_HI}
\end{figure}

In Figure \ref{fig:Choir_WISE_SFR_MS_HI}, the \textit{WISE} star formation versus stellar mass, we do not find a clear relationship with $\Delta{\textrm{f}}_\textrm{\HI}$. J0209-10:S4 has lower specific star formation rate and its an \HI-deficient, and interacting galaxy as seen in the \HI\ and optical imaging (e.g. \citealt{Jones2019}).

\section{Table of Group properties}
\label{sec:table_choir_properties}
We summarise properties of the Choir groups in Table \ref{tab:Choirs} and here we describe how they were computed. The projected velocity dispersion of a group is derived from:
\begin{equation}
    \sigma_{ \textup{group}} = \sqrt{ \frac{1}{N} \sum_{i}^{N} \left ( v_{i}-\left \langle v \right \rangle \right )^{2} }.
\end{equation}
where $N$ is number of galaxies in a group, $v_{i}$ is a galaxy recessional velocity and $<v_{i}>$ is mean group velocity. Values of r$_{200}$ and M$_{200}$ are derived using equations from \citet{Poggianti2010}:

\begin{equation}
    r_{200} = 1.73 \frac{\sigma_{\textup{group}} }{1000 \textup{km} \textup{s}^{-1} \sqrt{{\Omega_{\Lambda}} + {\Omega_{0}} \left ( 1 + z \right )^{3} } } h_{100}^{-1} [\textup{Mpc}]
\end{equation}
and
\begin{equation}
    M_{200} = 1.2 \left (\frac{\sigma_{\textup{group}} }{1000 \textup{km} \textup{s}^{-1}}  \right )^{3}\frac{1}{ \sqrt{\Omega_{\Lambda} + \Omega_{0} \left ( 1 + z \right )^{3} } } h_{100}^{-1} [ 10^{15} \textup{M}\odot ],
\end{equation}
where $\Omega_{\Lambda}$ = 0.7; $\Omega_{0}$ = 0.3; $h$ = 0.7. 

Crossing time is derived using equation from \citet{Konstantopoulos2010}:

\begin{equation}
    t_{c} = \frac{R_{\textup{rms}}}{\sigma _{\textup{group}}},
\end{equation}
where R$_{\textrm{rms}}$ is the projected group radius which is derived using equation from \citet{Berlind2006}:

\begin{equation}
    R_{\textup{rms}} = \sqrt{\frac{1}{N} \sum_{i}^{N} r_{i}^{2}},
\end{equation}
where r$_{i}$ is a galaxy's projected distance from the group centroid. Due to the projected velocity dispersion (and not the three dimensional one) and the small number of galaxies, the crossing time could be a factor of two larger \citep{Konstantopoulos2010}.

Note of caution: these equations are under assumption of the virial equilibrium, which may not be established in Choir groups, thus the derived values are only approximation of these properties. 

\begin{landscape}
\begin{table}
\caption{Summary of the Choir groups}
\label{tab:Choirs}
\begin{threeparttable}
\begin{tabular}{lrrrrrrrrrl}
\toprule
Group ID &  RA$_{\textrm{cen}}$  &  DEC$_{\textrm{cen}}$ &  $\sigma_{\mathrm{group}}$ &  r$_{200}$ &  M$_{200}$ &  M$_{\HI}$ &  Scale &    R$_{\textrm{rms}}$ &  t$_{\textrm{cross}}$ & Comment\\
 &  J2000 [deg]  &  J2000 [deg] &  [km s$^{-1}$] &  [Mpc]  & [M$\odot$] &  [dex] &  [kpc/$\arcsec$] &    [kpc] &  [Gyr] & \\
(1)      &  (2) & (3)   &  (4)  & (5)   & (6)   &    (7)   &    (8)     &   (9)   &  (10)    &     (11) \\ \hline \hline
J0205-55  &       31.24005 &       -55.18582 &  221 &          0.54 &  1.83 &  10.51 &         0.439 &  133.2 & 0.60  & Two subgroups merging \\
J0209-10  &       32.38124 &       -10.14940 &   86 &          0.21 &  0.11 &  10.31 &         0.258 &  39.7 &  0.46 &  \HI\ mapped in \citet{Jones2019} \\  
J0258-74  &       44.58151 &       -74.42768 &   94 &          0.23 &  0.14 &  10.41 &         0.333 &  56.7 &  0.60 &  Isolated group  \\
J0400-52  &       60.22478 &       -52.75107 &  293 &          0.71 &  4.24 &  10.61 &         0.702 &  221.4 & 0.76  & Cluster infalling group \citep{Dzudzar2019b} \\ 
J0443-05  &       70.99437 &        -5.35875 &   75\tnote{*} & 0.18 &  0.07 &  10.41 &         0.328 &  110.5 & 1.47  & Not mapped in \HI\ \\
J1026-19  &      156.6397 & -19.07035 &  9\tnote{*} &   0.02  & 1.23$\times$10$^{9}$  & 10.63 & 0.630 & 183.5 & 20.4  & Tidal interactions  \\
J1051-17  &      162.84646 &       -17.09431 &  250 &          0.61 &  2.66 &  10.45 &         0.393 &  218.3 & 0.87  & Kilborn et al. in prep \\
J1059-09  &      164.78250 &        -9.81185 &  186 &          0.45 &  1.09 &  10.42 &         0.571 &  169.2 & 0.91  & Tidal interactions  \\
J1159-19  &      179.88779 &-19.30826 & 38\tnote{*} & 0.09 &  9.38$\times$10$^{10}$    & 9.92 & 0.120 &  13.2 & 0.35 &  Tidal interactions; \HI\ mapped in \citet{Phookun1992}  \\ 
J1250-20  &      192.70101 &       -20.37571 &   60\tnote{*} & 0.15 &  0.04 &  10.51 &         0.535 &  107.4 & 1.79  & Isolated group; \HI\ mapped in \citet{Dzudzar2019}  \\
J1403-06  &      210.85138 &        -6.07759 &  100 &          0.25 &  0.17 &  10.29 &         0.196 &  35.6 &  0.36 &  \HI\ mapped in Clemens M.S. 1998 Ph.D. Thesis  \\ 
J1408-21  &      212.18354 &       -21.58828 &  150 &          0.37 &  0.57 &  10.52 &         0.599 &  152.7 & 1.02  & Tidal interactions  \\
J1956-50  &      299.02451 &       -50.00326 &  223 &          0.54 &  1.88 &  10.52 &         0.517 &  185.6 & 0.83  & Isolated group  \\
J2027-51  &      306.97329 &       -51.64226 &   92 &          0.23 &  0.13 &  10.44 &         0.411 &  141.4 & 1.54  & Local group analogous  \\ 
J2318-42a &      349.71742 &       -42.30871 &  134 &          0.33 &  0.41 &  10.10 &         0.111 &  37.1 &  0.28 &  \HI\ mapped in \citet{Dahlem2005}  \\ \hline
\end{tabular}
\begin{tablenotes}
\item (1) Group name: HIPASS+Group ID; (2) Right Ascension (J2000) - for group centroid, derived as mass weighted average; (3) Declination (J2000) - for group centroid, derived as mass weighted average; (4) $\sigma_{\mathrm{group}}$ - Radial velocity dispersion of the group [km s$^{-1}$]; (5) r$_{200}$ [Mpc] of the group; (6) M$_{200}$ $\times$10$^{13}$  [M$_{\odot}$], if not otherwise specified, of the group. The r$_{200}$ and M$_{200}$ are based on equations from \citet{Poggianti2010}, assuming virialized structures; (7) M$_{\HI}$ - Logarithm of the group \HI\ mass from \citet{Sweet2013}; (8) Scale factor; (9) R$_{\textrm{rms}}$ - Projected group radius, based on Eq.8 from \citet{Berlind2006}; (10) t$_{\textrm{cross}}$ - Group crossing time, based on \citet{Konstantopoulos2010}; (11) A comment about group. Used equations are summarised in this Section \ref{sec:table_choir_properties}.
\item[*] For J0443-05 and J1026-19 velocity dispersion is computed with three galaxies; J1159-19 and J1250-20 with two, thus they are highly uncertain. 
\end{tablenotes}
\end{threeparttable}
\end{table}
\end{landscape}

\section{Galaxy properties}
\label{app:galaxy_properties}
In Table \ref{tab:Choirs_literature} we show Choirs which data we utilise from the literature, comparing literature values with ours. The largest difference is the distance to J1159-29 group, in which \citet{Phookun1992} adopted 10 Mpc, while our adopted distance is 25 Mpc (similar to Hubble flow).

\begin{table*}
\caption{Summary of the Choir galaxies within groups that are obtained from the literature}
\label{tab:Choirs_literature}
\begin{threeparttable}
\begin{tabular}{lcccccccc}
\toprule
  ID &    RA &      DEC & D &M$_{\HI}$ & D$_{\textrm{our}}$ & M$_{\HI,\textrm{our}}$ & $\Delta$f$_{\textrm{\HI}}$ $\pm$ $\sigma$ & Reference \\
&  [deg] & [deg] & [Mpc] & [dex] & [Mpc] & [dex] & [dex] & \\
(1) &  (2) & (3) & (4) & (5) & (6) & (7) & (8) & (9) \\ \hline \hline
 J0209-10:S1 &   32.4279 & -10.1837 & 55.2  &  9.79  & 54   & 9.71 & 0.0$\pm$0.1 & \citep{Jones2019} \\
 J0209-10:S2 &   32.4103 & -10.1461 & 55.2  &  9.73  & 54   & 9.55 & -0.1$\pm$0.1 &\ditto \\
 J0209-10:S3 &   32.3518 & -10.1363 & 55.2  &  9.15  & 54   & 9.13 & -0.8$\pm$0.1 &\ditto \\
 J0209-10:S4 &   32.3362 & -10.1329 & 55.2  &  8.70  & 54   & 8.68 & -1.3$\pm$0.1 &\ditto \\
 J1159-19:S1 &  179.876 & -19.2658 & 10  &  9.01  &  25  & 9.76 & -0.8$\pm$0.1 & \citep{Phookun1992}  \\
 J1159-19:S2 &  179.873 & -19.3332 & 10  &  7.93  &  25  & 8.68  & -0.4$\pm$0.1 &\ditto  \\
 J1159-19:S3 &  179.899 & -19.3174 & 10  &  \dots  &  25  & \dots & \dots &\dots  \\
 J1159-19:S4 &  179.908 & -19.3292 & 10  &  \dots  &  25  & \dots  & \dots &\dots\\
J2318-42a.S1 &  349.598 & -42.3700 & 22  &  9.52  &  23  & 9.56  & -0.4$\pm$0.1 & \citep{Dahlem2005}  \\
J2318-42a.S2 &  349.728 & -42.2386 & 22  &  9.47  &  23  & 9.51  & -0.3$\pm$0.1 &\ditto  \\
J2318-42a.S3 &  349.823 & -42.2620 & 22  &  9.70  &  23  & 9.74 & -0.1$\pm$0.1  &\ditto  \\
J2318-42a:S4 &  349.7098 & -42.3968 & 22  &  \dots  &  23  & \dots & \dots &\dots \\
\bottomrule
\end{tabular}
\begin{tablenotes}
\item (1) ID: Galaxy ID;  (2) Right Ascension (J2000); (3) Declination (J2000); (4) Distance used in the literature; (5) Mass used in the literature; (6) Distance used in this work; (7) Mass used in this work; (8) Distance of the \HI\ mass fraction from the \HIPASS\ running median line; (9) Reference.
\end{tablenotes}
\end{threeparttable}
\end{table*}

\section{New observed Choirs and detections}
\label{app:new_observed_choirs}
Here we present the \HI\ and stellar masses of the galaxies from new observed groups, in addition to those from our previous work in \citet{Dzudzar2019}. We did not resolve individual galaxies in J1159-19, thus we are placing it in Table \ref{tab:Choirs_literature}. J0205-55 in \citet{Dzudzar2019} has pre-CABB data, and here we present data with new correlator, we also have more \HI\ detections in this group than in our previous work. J025-74:S3 galaxy is very near to S2 galaxy in this group, and in this work we have a better extraction of the \HI\ content within this group. J2027-51:S3 and J1051-17:A3 are galaxies which \HI\ was extracted below 3$\sigma$ level, thus we consider them only as tentative detections.

\begin{table*}
\caption{Summary of the Choir galaxies from new observations and new detections}
\label{tab:Choirs_newobs}
\begin{threeparttable}
\begin{tabular}{lllcccc}
\toprule
ID &  RA &  DEC &  D & M$_{\textrm{\HI}}$ &  M$_{\star}$ & Comment \\
 &  [deg] &  [deg] &  [Mpc] & [dex] &  [dex]  & \\
(1) &  (2) & (3) & (4) & (5) & (6)  & (7)\\ \hline \hline
J0205-55:S1 &   31.27286 & -55.11182 &        93 &  10.43 $\pm$   0.10 &      10.94 $\pm$  0.10 & Data from new CABB observations\\
J0205-55:S2 &   31.21211 & -55.217130 &        93 &     9.93 $\pm$     0.10 &       9.60 $\pm$ 0.12    & \ditto \\
J0205-55:S3 &   31.14558 & -55.119310 &        93 &     8.65 $\pm$      0.13 &      10.86 $\pm$  0.10  & \ditto \\
J0205-55:S4 &   31.08276 & -55.230710 &        93 &    10.0 $\pm$      0.10 &      10.71 $\pm$   0.10  & \ditto \\
J0205-55:A1 &   31.51224 & -55.194627 &        93 &     9.43 $\pm$      0.10 &      10.59 $\pm$ 0.10 & \ditto  \\

J1956-50:S1 &  299.19006 & -50.05563 &       110 &  10.38 $\pm$    0.10 &      10.52 $\pm$ 0.10 & \ditto\\
J1956-50:S2 &  298.97214 & -50.036330 &       110 &     9.56 $\pm$    0.10 &       9.78 $\pm$  0.12 & \ditto \\
J1956-50:A1 &  298.88959 & -49.882370 &       110 &     9.75 $\pm$     0.13 &       9.89 $\pm$  0.10 & \ditto \\

J0258-74:S3 &   44.677720 & -74.434310 &        70 &     8.87 $\pm$       0.20 &       9.49$\pm$  0.10 & New extraction of the \HI\ \\
J2027-51:S3 &  306.952240 & -51.738960 &        87 &     8.88 $\pm$      0.13 &       9.32 $\pm$  0.14 & Tentative detection, below 3$\sigma$ \\
J1051-17:A3 &  162.632092 & -17.064126 &        83 &     8.54 $\pm$      0.10 &       9.88$\pm$  0.10   & \ditto \\
\bottomrule
\end{tabular}
\begin{tablenotes}
\item (1) ID: Galaxy ID;  (2) Right Ascension (J2000); (3) Declination (J2000); (4) Group distance; (5) \HI\ mass and its uncertainty; (6) Stellar mass from the \textit{WISE}; (7) Comment about the galaxy.   
\end{tablenotes}
\end{threeparttable}
\end{table*}

\section{\HI\ detection limit}
\label{appendix:detection_limit}

\begin{figure}
    \centering
    \includegraphics[width=\columnwidth]{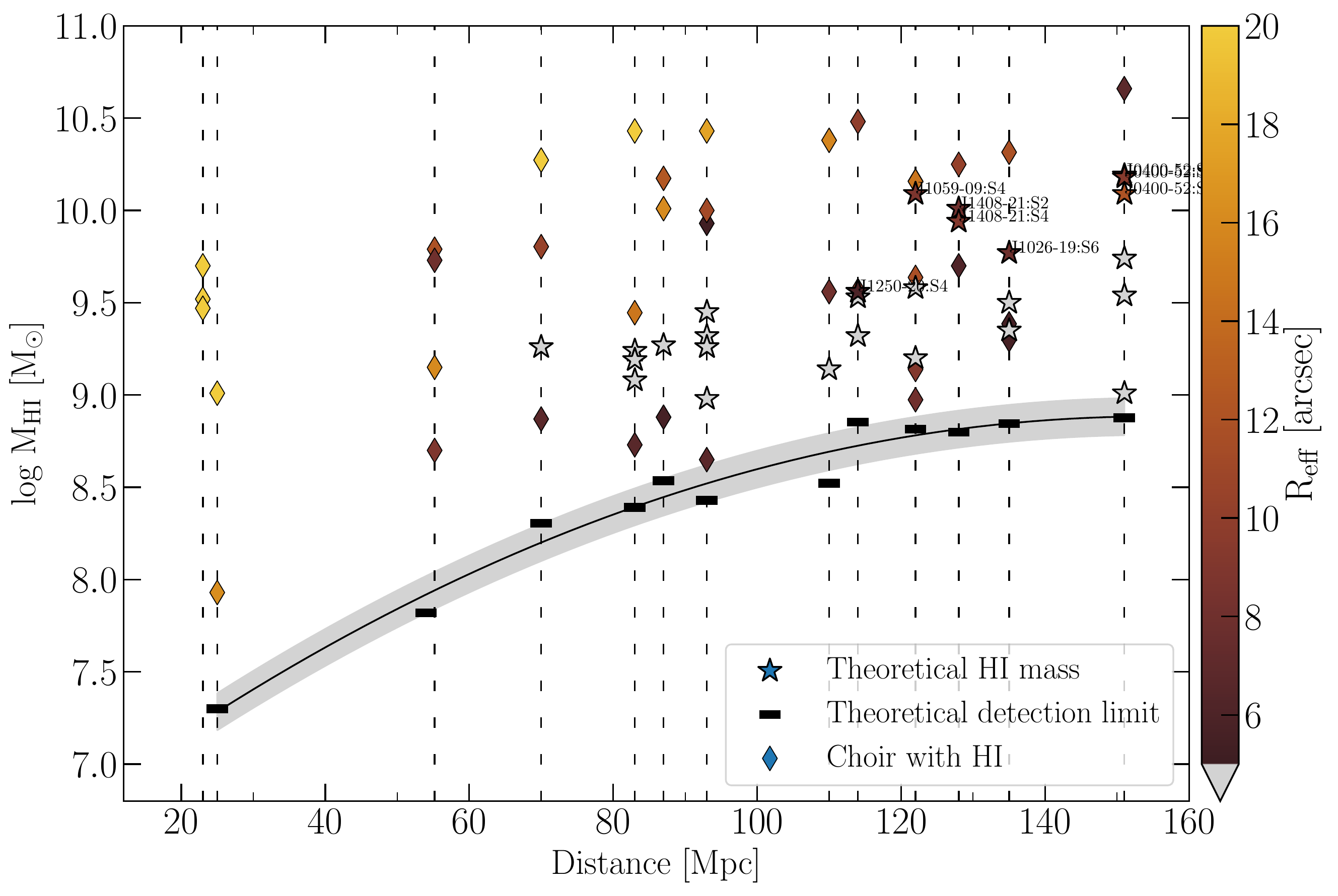}
    \caption{The \HI\ mass versus the distance for Choir galaxies. The vertical dashed lines are connecting galaxies within the same Choir group. The diamonds are Choir galaxies detected in \HI, and the colour shows their R$_{\textrm{eff}}$ radius. The star symbols show the expected \HI\ mass for Choir galaxies, based on the M$_{\textrm{R}}$ to M$_{\HI}$ scaling relation \citep{Sweet2013, Denes2014}. The grey stars are those whose R$_{\textrm{eff}}$ radius is below 5 arcsec (as the smallest detected galaxy has 5.2 arcsec), and thus we consider them to below our detection limit. The coloured stars are those that have high enough radius (and stellar mass from, see Figure \ref{fig:Choir_detections}) to be detected, thus we consider them to be \HI-deficient galaxies. The black rectangles show the theoretical mass limit for each Choir group, while the curved line shows the polyfit line to those limits, with a standard deviation shown as grey shaded region. }
\label{fig:hi-non-detections}
\end{figure}

The \HI\ mass limit was determined using standard relation for mass (see Equation \ref{eq:mass}) where the flux is based on the observed limit. The flux upper limit is based on the relation from \citet{Sardone2019}: F$_{\HI,\textrm{lim}}$ $=$ 3 $\sigma_{\textrm{rms}}$ $\sqrt{\textrm{W} dv}$, where the rms is the measured noise in the data cube and W is the expected \HI\ emission line width, we adopt it to be 300 km s$^{-1}$, and the $dv$ is the velocity resolution, see Figure  \ref{fig:hi-non-detections} and Figure \ref{fig:Choir_detections} in Section \ref{sec:non_detections}. Using these data, we find the probable \HI-deficient galaxy and show we them in Table \ref{tab:deficient}. 

\begin{table}
\caption{\HI-deficient galaxies determined from our \HI\ non-detection analysis}
\label{tab:deficient}
\begin{threeparttable}
\begin{tabular}{lrrccc}
\toprule
ID & RA  &  DEC & M$_{\HI,\textrm{exp}}$  & M$_{\textrm{lim}}$ & Def \\ 
 & [deg] &  [deg] & [dex]  & [dex] & [dex] \\ 
(1) & (2) & (3) & (4) & (5) & (6) \\ \hline \hline

J1026-19:S6  & 156.605 &	-19.17649 & 9.8 & 8.8 & 1 \\
J1059-09:S4  & 164.69545 &	-9.84515  & 10.2 & 8.8 & 1.3 \\
J1250-20:S4  & 192.66643 & -20.34798  & 9.6 & 8.9 & 0.7 \\
J1408-21:S2  &  212.24107 &	-21.64791 & 10 & 8.7 & 1.3 \\
J1408-21:S4  & 212.17101 &	-21.628 & 9.9 & 8.7 & 1.2 \\
J0400-52:S4  &  60.32599 & -52.70749  & 10.2 & 8.9 & 1.3 \\
J0400-52:S5  & 60.22048 & -52.82712 & 10.2 & 8.9 & 1.3 \\
J0400-52:S6  & 60.32479 & -52.80079  & 10.1 & 8.9 & 1.2 \\

\bottomrule
\end{tabular}
\begin{tablenotes}
\item (1) ID: Galaxy ID; (2) Right Ascension (J2000); (3) Declination (J2000); (4) M$_{\HI,\textrm{exp}}$ - The logarithm of galaxy's expected \HI\ mass, based on the M$_{\textrm{R}}$ to M$_{\HI}$ scaling relation \citep{Sweet2013, Denes2014}; (5) M$_{\HI,\textrm{lim}}$ - The logarithm of the theoretical limit for the observed \HI\ mass; (6) Def - The \HI\ deficiency, obtained as the difference between the expected \HI\ mass and the limit \HI\ mass. 
\end{tablenotes}
\end{threeparttable}
\end{table}

\section{Galaxy radii}
\label{appendix:galaxy_radii}

In Table \ref{tab:Choirs_extremes} we summarise the measurements of the galaxy radii from Section \ref{sec:extremes}. 

\begin{table*}
\caption{Measurements of galaxy radii}
\begin{threeparttable}
\begin{tabular}{llccccc}
\toprule
                  Optical ID &  Our ID  &  $\Delta$f$_{\textrm{\HI}}$ $\pm$ $\sigma$ &  r$_{\textrm{\HI}}$ &  D$_{\textrm{\HI}}$ &  r$_{\textrm{R},\textrm{eff}}$\tnote{b} &  \HI\ content\\
                  & & [dex] & [kpc] & [kpc] & [kpc] & \\
                  (1)      &  (2) & (3)   &  (4)  & (5)   & (7)  &  (9)\\ \hline \hline
                    PGC032573  &  J1051-17:S1 &       0.7$\pm$0.1 &       67.6 &        99.6 &       11.1    & \HI-excess \\
                 ESO156-G029 &   J0400-52:S1 &       0.6$\pm$0.1 &       79.8 &        77.3 &        5.0     & \HI-excess \\
                 ESO575-G006 &   J1250-20:S1 &       0.6$\pm$0.1 &       84.6 &       106.4 &        5.5 &  \HI-excess\\
                      IC4909 &   J1956-50:S1 &       0.5$\pm$0.1 &       69.9 &        68.3\tnote{a} &        8.4 & \HI-excess \\
 GALEXASC J202808.43-512830.5&   J2027-51:A1\tnote{d} &       0.4$\pm$1.3  &       \dots &       25.9     &     \dots  & \HI-average \\
                MCG-01-28-012 &   J1059-09:S3 &       0.4$\pm$0.1 &          \dots &        \dots &        8.7 & \ditto \\
                 ESO234-G032 &   J2027-51:S1 &       0.4$\pm$0.1 &          \dots &        56.4 &        5.3 &     \ditto \\
              ESO 153-IG 016 &  J0205-55:S2 &       0.4$\pm$0.1 &       55.0 &        47.6 &        2.3 &        \ditto \\ 
                 ESO568-G011 &   J1026-19:S1 &       0.4$\pm$0.1 &          \dots &        \dots &        7.8 &        \ditto \\
                 ESO153-G017 &  J0205-55:S1 &       0.4$\pm$0.1 &       70.8 &        98.5 &        7.9 &       \ditto \\
 WISEA J102641.63-190742.0   &    J1026-19:S5\tnote{d} &      0.4$\pm$1.3  &        \dots &      \dots &      3.7  &   \ditto \\
                 ESO031-G005 &   J0258-74:S1 &       0.3$\pm$0.1 &       47.9 &        82.4 &        6.9 &       \ditto \\
 GALEXASCJ025852.12-742554.8 &   J0258-74:S2 &       0.3$\pm$0.1 &       29.5 &        52.8 &        3.5 &       \ditto \\
 GALEXASCJ140846.58-212708.8 &   J1408-21:A1\tnote{d} &      0.2$\pm$1.3 &       \dots &       \dots &       \dots &    \ditto \\
                 ESO578-G026 &   J1408-21:S1 &       0.2$\pm$0.1 &          \dots &        \dots &        6.4 &    \ditto \\
                 ESO234-G028 &   J2027-51:S2 &       0.1$\pm$0.1 &          \dots &        42.4 &        6.8 &     \ditto \\
GALEXASC J105906.72-094506.7 &   J1059-09:S2 &       0.1$\pm$0.1 &          \dots &        50.4 &        6.9 &    \ditto \\
                 ESO233-G007 &   J1956-50:A1 &      0.1$\pm$0.1 &          42.7  &        \dots &         \dots &  \ditto \\
               MCG-01-28-013 &   J1059-09:S1 &       0.1$\pm$0.1 &          \dots &        96.3 &        8.9 &     \ditto \\
                 ESO153-G013 &  J0205-55:S4 &       0.0$\pm$0.1 &       48.2 &        \dots &        5.1 &         \ditto \\
 GALEXASCJ140841.07-213741.5 &   J1408-21:S3 &      0.0$\pm$0.1 &          \dots &        \dots &        3.8 &      \ditto \\
    2MASSX J19555339-5002124 &   J1956-50:S2 &      0.0$\pm$0.1 &       36.8 &        \dots &        4.3 &         \ditto \\
     2MASX J10265008-1904310 &   J1026-19:S2 &      -0.2$\pm$0.1 &          \dots &        \dots &        3.6 &      \ditto \\
        WISEA J105049.71-171411.2 &  J1051-17:A2\tnote{d}&      -0.3$\pm$1.3 &       9.3\tnote{c} &  \dots & \dots &  \ditto \\
                 ESO575-G004 &   J1250-20:S2 &      -0.3$\pm$0.1 &       50.8 &        29.2 &        5.6 &         \ditto \\
                     NGC3431 &   J1051-17:S2 &      -0.4$\pm$0.1 &       24.9 &        \dots &        5.9 &       \ditto \\
                MGC-03-28-016&  J1051-17:A1   &   -0.4$\pm$0.1 &      \dots&  \dots&  \dots&  \ditto\\
GALEXASC J105930.89-094425.7 &   J1059-09:S5 &      -0.5$\pm$0.2 &          \dots &        \dots &        5.4 &      \ditto \\ 
GALEXASC J105908.60-094312.8 &   J1059-09:S6 &      -0.5$\pm$0.3 &          \dots &        \dots &        4.8 &\ditto \\
   WISEA J105135.87-165919.3 &   J1051-17:S3\tnote{d} &      -0.6$\pm$1.3 &       7.2\tnote{c} &         \dots &        2.7 &     \ditto \\
                ESO153-G020  &   J0205-55:A1  &    -0.5$\pm$0.1 &      29.3 & \dots &  \dots  &  \HI-deficient \\
     2MASS J20274857-5144205 &   J2027-51:S3\tnote{e} &      -0.6$\pm$0.1 &          \dots &        \dots &        2.3 &     \HI-deficient \\
   GALEXASCJ02584292-7426028 & J0258-74:S3 &      -0.6$\pm$0.1  &       10.5\tnote{c} &        \dots &        2.3 &    \HI-deficient \\
             MCG -03-28-013  &  J1051-17:A3\tnote{e} &      -1.1$\pm$0.1 &       \dots &  \dots & \dots & \HI-deficient \\
                 ESO153-G015 &  J0205-55:S3 &      -1.4$\pm$0.1 &       15.3\tnote{c} &        \dots &        3.0 &       \HI-deficient \\
\bottomrule

\end{tabular}
\begin{tablenotes}
\item (1) Optical ID: Galaxy ID from NED; (2) Our ID: HIPASS+Group+Galaxy identification; (3)  $\Delta$f$_{\textrm{\HI}}$: Distance of the \HI\ mass fraction from the \HIPASS\ running median line; (4) r$_{\textrm{\HI}}$: \HI\ radius measured at the \HI\ column density of 3$\times$10$^{19}$ cm$^{-2}$; (5) D$_{\textrm{\HI}}$: \HI\ diameter, measured at at 1 M$\odot$ pc$^{-2}$ \citep{Dzudzar2019}; (6) r$_{\textrm{R},\textrm{eff}}$: R-band effective radius from SINGG; (7) \HI\ content: classification based on $\Delta$log(f$_{\textrm{\HI}}$) into \HI-excess, \HI-average or \HI-deficient galaxy. 
\item[a] Measured in this work, following \citet{Dzudzar2019} procedure.
\item[b] From the SINGG survey \citep{Meurer2006}.
\item[c] Upper limits on the \HI-radius, as described in Section \ref{sec:extremes}.
\item[d] Has highly uncertain stellar mass ($\sim$1.3 dex).
\item[e] The HI emission in galaxies is detected below 3$\sigma$.
\item \textbf{Comments:} \\
\textbf{i)} The first three galaxies are highlighted in Figure \ref{fig:HI_excess} as most \HI-excess, while the bottom three are highlighted in Figure \ref{fig:HI_deficient} as most \HI-deficient; excluding J1051-17:A3 and J2027-51:S3 since their \HI\ mass is measured at 2$\sigma$. \textbf{ii)} We do not include r$_{\textrm{\HI}}$ for galaxies in the following groups because the lowest \HI-column density is higher than the 3$\times$10$^{19}$ cm$^{-2}$: J1026-19, J1059-09, J1408-21 and J2027-51. 

\end{tablenotes}
\end{threeparttable}
\label{tab:Choirs_extremes}
\end{table*}

\section{\textit{WISE} measurements}
\label{app:wise_measurements}
The mid-infrared properties of Choir galaxies are obtained using ``drizzled'' \textit{WISE} imaging \citet{Jarrett2012}. The photometry was obtained from reconstructed images from \textit{WISE} using the ICORE co-addition software \citet{Masci2013}. We show the \textit{WISE} measurements for Choir galaxies in Table \ref{tab:Choirs_WISE}. For details on the \textit{WISE} measurements see \citet{Jarrett2012} and \citet{Jarrett2013}.

The W3pah and W4dust are fluxes that are obtained after removing the expected (model) stellar continuum. If the continuum-subtraction from the W3 flux (measured at the 2$\sigma$ level) is below zero (or zero) we do not have measurements of the star formation rate.

\onecolumn

\begin{landscape}
\begin{table*}
\caption{\textit{WISE} measurements of Choir members and their \HI\ mass}
\begin{threeparttable}
\begin{tabular}{lrrrccccccccc}
\toprule

ID &  RA &  DEC & D & W1 & W2 &  W3pah &   W4dust & (W1-W2)$\pm$ $\sigma$ & (W2-W3)$\pm$ $\sigma$ & M$_{\star}$ $\pm$ $\sigma$ & SFR$_{12}$ $\pm$ $\sigma$ & M$_{\HI}$ $\pm$ $\sigma$\\

 & [deg] &  [deg] & [Mpc] & [mJy] & [mJy] &      [mJy] &   [mJy] & [mag] & [mag] & [dex] & [M$_{\odot}$ $yr^{-1}$] & [dex] \\

(1) & (2) &  (3) & (4) & (5) & (6) &  (7) & (8) &    (9) &   (10) &  (11) & (12) & (13) \\ \hline \hline

J0205-55:S1  & 31.27286  & -55.11182 & 93  & 24.722  & 13.752  & 36.888  & 58.294     &  0.01	$\pm$	0.039  &  3.103	$\pm$	0.055   & 10.94 $\pm$ 0.10  & 4.10  $\pm$ 1.43 & 10.43 $\pm$ 0.10\tnote{a} \\
J0205-55:S2  & 31.21211  & -55.21713 & 93  & 2.753 & 1.761 & 4.723 & 11.720           &  0.162	$\pm$	0.047  &  3.089	$\pm$	0.094   & 9.60  $\pm$ 0.12  & 0.67  $\pm$ 0.24 & 9.93 $\pm$ 0.12\tnote{a} \\
J0205-55:S3  & 31.14558  & -55.11931 & 93  & 16.712  & 8.987 & 12.383  & 11.422       &  -0.027	$\pm$	0.038  &  2.48	$\pm$	0.055    & 10.86 $\pm$ 0.10  & 1.57  $\pm$ 0.55 & 8.65 $\pm$ 0.13\tnote{a} \\
J0205-55:S4  & 31.08276  & -55.23071 & 93  & 23.964  & 14.403  & 37.868  & 53.625     &  0.094	$\pm$	0.037  &  3.076	$\pm$	0.04   & 10.71 $\pm$ 0.10  & 4.20  $\pm$ 1.45 & 10.0 $\pm$ 0.1\tnote{a} \\
J0205-55:S5  & 31.22966  & -55.14355 & 93  & 0.093 & 0.057 & \dots  & \dots           &  0.11	$\pm$	0.226  &  \dots   & 8.49  $\pm$ 1.27  & \dots  & \dots \\
J0205-55:S6  & 31.23776  & -55.22611 & 93  & 0.438 & 0.261 & 0.216 & \dots            &  0.086	$\pm$	0.074  &  2.016	$\pm$	0.529   & 8.99  $\pm$ 0.19  & 0.04  $\pm$ 0.02 & \dots \\
J0205-55:S7  & 31.34851  & -55.23729 & 93  & 0.041 & 0.034 & \dots  & \dots           &  0.47	$\pm$	0.228  &  \dots    & 8.13  $\pm$ 1.27  &  \dots & \dots \\
J0205-55:S8  & 31.12363  & -55.2156  & 93  & 0.181 & 0.170 & 0.108 & \dots            &  0.576	$\pm$	0.135  &  1.686	$\pm$	0.392    & 8.32  $\pm$ 0.34  & 0.02  $\pm$ 0.01 & \dots \\
J0205-55:A1  & 31.51664  & -55.19627 & 93  & 33.180  & 23.475  & 64.485  & 159.368    &  0.271	$\pm$	0.037  &  3.104	$\pm$	0.22  &  10.59 $\pm$ 0.10  & 6.70  $\pm$ 2.60 & 9.43 $\pm$ 0.10\tnote{a} \\
J0209-10:S1 & 32.42817  & -10.18376 & 54  & 47.941  & 41.366  & 342.754 & 1805.609    &  0.487	$\pm$	0.037  &  4.242	$\pm$	0.038  &  10.27 $\pm$ 0.10  & 11.19 $\pm$ 3.88 &  9.71\tnote{b} \\
J0209-10:S2 & 32.41058  & -10.14634 & 54  & 61.337  & 44.820  & 432.838 & 1418.895    &  0.306	$\pm$	0.037  &  4.409	$\pm$	0.038  &  10.38 $\pm$ 0.10  & 13.74 $\pm$ 4.76  & 9.55\tnote{b} \\
J0209-10:S3 & 32.35258  & -10.13578 & 54  & 83.744  & 51.983  & 199.119 & 403.120     &  0.129	$\pm$	0.037  &  3.45	$\pm$	0.038  &  10.70 $\pm$ 0.09  & 6.94  $\pm$ 2.40  & 9.13\tnote{b} \\
J0209-10:S4 & 32.33679  & -10.13312 & 54  & 50.069  & 27.167  & 14.437  & 24.159      &  -0.017	$\pm$	0.037  &  1.71	$\pm$	0.079  &  10.84 $\pm$ 0.10  & 0.69  $\pm$ 0.24 & 8.68\tnote{b} \\
J0258-74:S1 & 44.52795  & -74.45632 & 70  & 38.541  & 23.165  & 84.705  & 122.573     &  0.094	$\pm$	0.036  &  3.406	$\pm$	0.038  &  10.67 $\pm$ 0.09  & 5.17  $\pm$ 1.79 & 10.27 $\pm$ 0.03\tnote{c} \\
J0258-74:S2 & 44.71847  & -74.43176 & 70  & 1.809 & 0.969 & 1.864 & 3.128             &  -0.031	$\pm$	0.041  &  2.788	$\pm$	0.125   & 9.66  $\pm$ 0.11  & 0.18  $\pm$ 0.06 & 9.8 $\pm$ 0.1\tnote{c} \\
J0258-74:S3 & 44.67772  & -74.43431 & 70  & 2.379 & 1.417 & 6.357 & 9.928             &  0.085	$\pm$	0.04  &  3.614	$\pm$	0.051  &  9.49  $\pm$ 0.10  & 0.53  $\pm$ 0.18 & 8.87 $\pm$ 0.2\tnote{a} \\
J0258-74:S4 & 44.37181  & -74.37624 & 70  & 0.121 & 0.097 & \dots  & \dots            &  0.406	$\pm$	0.142  &  \dots  &  8.35  $\pm$ 1.27  &  \dots  \\
J0400-52:S1 & 60.17053  & -52.73405 & 151 & 6.546 & 3.478 & 7.514 & 8.362             &  -0.04	$\pm$	0.039  &  2.899	$\pm$	0.065  &   10.91 $\pm$ 0.10  & 2.37  $\pm$ 0.83 & 10.66 $\pm$0.09\tnote{c} \\
J0400-52:S2 & 60.20074  & -52.68383 & 151 & 0.010 & 0.009 & \dots  & \dots            &  0.485	$\pm$	0.884  &  \dots  &  7.93  $\pm$ 1.28  & \dots & \dots \\
J0400-52:S3 & 60.02511  & -52.65909 & 151 & 1.188 & 0.726 & 3.937 & 4.324             &  0.112	$\pm$	0.046  &  3.809	$\pm$	0.064  &  9.78  $\pm$ 0.12  & 1.35  $\pm$ 0.47 & \dots \\
J0400-52:S4 & 60.32599  & -52.70749 & 151 & 13.128  & 7.591 & 18.515  & 25.185        &  0.052	$\pm$	0.038  &  3.006	$\pm$	0.045  &  10.98 $\pm$ 0.10  & 5.25  $\pm$ 1.82 & \dots \\
J0400-52:S5 & 60.22048  & -52.82712 & 151 & 11.142  & 6.288 & 15.075  & 23.084        &  0.026	$\pm$	0.039  &  2.992	$\pm$	0.042  &  10.98 $\pm$ 0.10  & 4.38  $\pm$ 1.52 & \dots \\
J0400-52:S6 & 60.32479  & -52.80079 & 151 & 6.622 & 3.592 & 6.654 & 9.859             &  -0.017	$\pm$	0.039  &  2.75	$\pm$	0.058  &  10.86 $\pm$ 0.10  & 2.13  $\pm$ 0.74 & \dots \\
J0400-52:S7 & 60.28706  & -52.82552 & 151 & 0.288 & 0.171 & 0.560 & \dots             &  0.08	$\pm$	0.084  &  3.296	$\pm$	0.339  &  9.25  $\pm$ 0.21  & 0.24  $\pm$ 0.11 & \dots \\
J1026-19:S1 & 156.67012 & -19.05137 & 135 & 9.704 & 5.883 & 30.689  & 49.012          &  0.104	$\pm$	0.039  &  3.769	$\pm$	0.043  &  10.62 $\pm$ 0.10  & 6.72  $\pm$ 2.33 & 10.3 $\pm$ 0.1\tnote{c} \\
J1026-19:S2 & 156.70862 & -19.07548 & 135 & 1.807 & 1.702 & 15.305  & 60.496          &  0.582	$\pm$	0.047  &  4.327	$\pm$	0.057  &  9.65  $\pm$ 0.12  & 3.64  $\pm$ 1.27 & 9.4 $\pm$ 0.1\tnote{c} \\
J1026-19:S3 & 156.57893 & -18.96424 & 135 & 0.292 & 0.160 & \dots  & \dots            &  -0.007	$\pm$	0.127  &  \dots  &  9.31  $\pm$ 1.27  &  \dots & \dots \\
J1026-19:S4 & 156.6017  & -19.03421 & 135 & 0.102 & 0.056 & 0.217 & \dots             &  0.001	$\pm$	0.231  &  3.464	$\pm$	0.393  &  8.90  $\pm$ 0.59  & 0.09  $\pm$ 0.04 & \dots \\
J1026-19:S5 & 156.67381 & -19.12782 & 135 & 0.017 & 0.014 & \dots  & \dots            &  0.399	$\pm$	0.72  &  \dots  &  8.08  $\pm$ 1.28  &  \dots & 9.3 $\pm$ 0.1\tnote{c} \\
J1026-19:S6 & 156.605 & -19.17649 & 135 & 0.836 & 0.478 & 1.040 & 3.031               &  0.039	$\pm$	0.083  &  2.897	$\pm$	0.187  &  9.72  $\pm$ 0.21  & 0.34  $\pm$ 0.13 & \dots \\
J1051-17:S1 & 162.90611 & -17.12477 & 83  & 13.983  & 9.013 & 7.660 & 13.094          &  0.17	$\pm$	0.039  &  2.021	$\pm$	0.089  &  10.19 $\pm$ 0.10  & 0.84  $\pm$ 0.30 & 10.43 $\pm$ 0.02\tnote{c} \\
J1051-17:S2 & 162.81264 & -17.00815 & 83  & 26.310  & 21.056  & 53.591  & 121.628     &  0.405	$\pm$	0.037  &  3.018	$\pm$	0.04  &  10.39 $\pm$ 0.10  & 4.66  $\pm$ 1.61 & 9.45 $\pm$ 0.05\tnote{c} \\
J1051-17:S3 & 162.89915 & -16.98857 & 83  & 0.412 & 0.244 & \dots  & \dots            &  0.08	$\pm$	0.12  &  \dots  &  9.04  $\pm$ 1.27  &  \dots & 8.73 $\pm$ 0.07\tnote{c} \\
J1051-17:S4 & 162.85841 & -17.08458 & 83  & 0.097 & 0.032 & \dots  & \dots            &  -0.554	$\pm$	0.655  &  \dots  &  8.41  $\pm$ 1.27  &  \dots & \dots \\
J1051-17:S5 & 162.9621  & -16.97549 & 83  & 0.171 & 0.101 & \dots  & \dots            &  0.078	$\pm$	0.188  &  \dots  &  8.65  $\pm$ 1.27  &  \dots & \dots \\
J1051-17:S6 & 162.9283  & -17.1096  & 83  & 0.183 & 0.093 & \dots  & \dots            &  -0.089	$\pm$	0.196  &  \dots  &  8.68  $\pm$ 1.27  & \dots & \dots \\
J1051-17:S7 & 162.88902 & -17.14346 & 83  & 0.224 & 0.124 & \dots  & \dots            &  0.005	$\pm$	0.152  &  \dots  &  8.77  $\pm$ 1.27  &  \dots & \dots \\
J1051-17:S8 & 162.85745 & -17.13794 & 83  & 0.725 & 0.394 & \dots  & \dots            &  -0.016	$\pm$	0.064  &  \dots  &  9.28  $\pm$ 1.27  &  \dots & \dots \\
J1051-17:A1 & 163.06651 & -17.13014 & 83  & 13.189  & 7.360 & 25.448  & 32.342        &  0.014	$\pm$	0.038  &  3.355	$\pm$	0.058  &  10.56 $\pm$ 0.10  & 2.42  $\pm$ 0.84 & 9.48 $\pm$ 0.03\tnote{c} \\
J1051-17:A2 & 162.70667 & -17.23528 & 83  & 0.317  & 0.176  & \dots  & \dots      &  0.012	$\pm$ 0.13  &  \dots  & 8.92 $\pm$ 1.27  & \dots& 9.01 $\pm$ 0.03\tnote{c} \\
J1051-17:A3 & 162.6321  & -17.06358 & 83  & 4.392 & 2.633 & 8.635 & 19.437            &  0.092	$\pm$	0.041  &  3.296	$\pm$	0.059  &  9.88  $\pm$ 0.10  & 0.94  $\pm$ 0.33 &  8.54 $\pm$ 0.10\tnote{a} \\

\bottomrule
\end{tabular}
\begin{tablenotes}
\item (1)  ID: HIPASS+Group+Galaxy identification, S -- within SINGG, A -- additional galaxy, discovered in \HI-emission; (2) Right Ascension (J2000); (3) Declination (J2000); (4) Distance; (5) and (6) are W1 and W2 fluxes; (7) and (8) W3pah and W4dust are fluxes of the ISM emission after removing the expected (model) stellar continuum; (9) The W1-W2 colour; (10) The W2-W3 colour; (11) Logarithm of the stellar mass; (12) 12$\mu$m star formation rate; (13) Logarithm of the \HI\ mass.
\item[a] \HI\ data from this work.
\item[b] \HI\ data obtained from the literature: J2318-42a \citep{Dahlem2005}, J1159-19 \citep{Phookun1992} and J0209-10 \citep{Jones2019}.
\item[c] \HI\ data obtained from our previous work, from  \citet{Dzudzar2019}.
\end{tablenotes}
\end{threeparttable}
\label{tab:Choirs_WISE}
\end{table*}
\end{landscape}

\begin{landscape}

\begin{table*}
\contcaption{\textit{WISE} measurements of Choir members}
\begin{threeparttable}
\begin{tabular}{lrrrccccccccc}
\toprule

ID &  RA &  DEC & D & W1 & W2 &  W3pah &   W4dust & (W1-W2)$\pm$ $\sigma$ & (W2-W3)$\pm$ $\sigma$ & M$_{\star}$ $\pm$ $\sigma$ & SFR$_{12}$ $\pm$ $\sigma$ & M$_{\HI}$ $\pm$ $\sigma$\\

 & [deg] &  [deg] & [Mpc] & [mJy] & [mJy] &      [mJy] &   [mJy] & [mag] & [mag] & [dex] & [M$_{\odot}$ $yr^{-1}$] & [dex] \\

(1) & (2) &  (3) & (4) & (5) & (6) &  (7) & (8) &    (9) &   (10) &  (11) & (12) & (13) \\ \hline \hline

J1059-09:S1 & 164.81769 & -9.79417  & 122 & 22.226  & 13.180  & 52.149  & 77.593     &  0.08	$\pm$	0.039  &  3.487	$\pm$	0.044  &  10.95 $\pm$ 0.10  & 8.96  $\pm$ 3.11 & 10.16 $\pm$ 0.10\tnote{c} \\
J1059-09:S2 & 164.77763 & -9.75148  & 122 & 0.926 & 0.551 & 1.337 & 2.290            &  0.084	$\pm$	0.076  &  2.997	$\pm$	0.169  &  9.56  $\pm$ 0.19  & 0.36  $\pm$ 0.13 & 9.6 $\pm$ 0.1\tnote{c} \\
J1059-09:S3 & 164.81587 & -9.81705  & 122 & 3.973 & 2.397 & 9.955 & 17.688           &  0.098	$\pm$	0.043  &  3.535	$\pm$	0.065  &  10.16 $\pm$ 0.11  & 2.09  $\pm$ 0.73 & 10.2 $\pm$ 0.1\tnote{c} \\
J1059-09:S4 & 164.69545 & -9.84515  & 122 & 10.664  & 6.633 & 22.095  & 50.761       &  0.131	$\pm$	0.039  &  3.309	$\pm$	0.053  &  10.50 $\pm$ 0.10  & 4.21  $\pm$ 1.46 & \dots \\
J1059-09:S5 & 164.87964 & -9.74003  & 122 & 1.095 & 0.608 & 1.181 & 2.013            &  0.007	$\pm$	0.064  &  2.792	$\pm$	0.142  &  9.83  $\pm$ 0.16  & 0.32  $\pm$ 0.12 & 9.1 $\pm$ 0.1\tnote{c} \\
J1059-09:S6 & 164.78612 & -9.71987  & 122 & 0.435 & 0.238 & 0.273 & -1.985           &  -0.008	$\pm$	0.099  &  2.315	$\pm$	0.392  &  9.47  $\pm$ 0.25  & 0.09  $\pm$ 0.04 & 8.9 $\pm$ 0.1\tnote{c} \\
J1059-09:S7 & 164.83879 & -9.79738  & 122 & 0.615 & 0.384 & 1.362 & 3.027            &  0.135	$\pm$	0.077  &  3.373	$\pm$	0.152  &  9.26  $\pm$ 0.20  & 0.36  $\pm$ 0.13 & \dots \\
J1059-09:S8 & 164.7572  & -9.87943  & 122 & 0.074 & 0.041 & \dots  & \dots           &  0.008	$\pm$	0.437  &  \dots  &  8.63  $\pm$ 1.27  &  \dots & \dots \\
J1059-09:S9 & 164.6862  & -9.89128  & 122 & 0.047 & 0.022 & \dots  & \dots           &  -0.18	$\pm$	0.509  &  \dots  &  8.43  $\pm$ 1.27  &  \dots & \dots \\
J1059-09:S10  & 164.76097 & -9.88889  & 122 & 3.606 & 1.856 & \dots  & \dots         &  -0.074	$\pm$	0.041  &  0.047	$\pm$	0.425  &  10.49 $\pm$ 0.11  &  \dots & \dots \\
J1159-19:S1 & 179.87596 & -19.26519 & 25  & 174.993 & 111.692 & 687.798 & 1177.216   &  0.159	$\pm$	0.037  &  3.939	$\pm$	0.04  &  10.27 $\pm$ 0.10  & 5.33  $\pm$ 1.85  & 9.76\tnote{b} \\
J1159-19:S2 & 179.87216 & -19.331 & 25  & 4.178 & 2.580 & 3.445 & 5.929              &  0.124	$\pm$	0.048  &  2.427	$\pm$	0.162  &  8.74  $\pm$ 0.12  & 0.05  $\pm$ 0.02 & 8.68\tnote{b} \\
J1159-19:S3 & 179.89937 & -19.31749 & 25  & 0.357 & 0.171 & \dots  &\dots            &  -0.152	$\pm$	0.174  &  \dots  &  7.93  $\pm$ 1.27  &  \dots & \dots \\
J1159-19:S4 & 179.90811 & -19.32914 & 25  & 0.573 & 0.330 & 0.055 & \dots            &  0.047	$\pm$	0.098  &  1.032	$\pm$	0.476  &  8.07  $\pm$ 0.25  & 0.0013  $\pm$ 0.001 & \dots\\
J1250-20:S1 & 192.72018 & -20.37128 & 114 & 11.791  & 7.423 & 41.787  & 64.010       &  0.145	$\pm$	0.039  &  3.846	$\pm$	0.046  &  10.45 $\pm$ 0.10  & 6.54  $\pm$ 2.27 & 10.48 $\pm$ 0.09\tnote{c} \\
J1250-20:S2 & 192.67047 & -20.33507 & 114 & 10.310  & 6.375 & 28.482  & 63.978       &  0.125	$\pm$	0.039  &  3.608	$\pm$	0.042  &  10.44 $\pm$ 0.10  & 4.67  $\pm$ 1.62 & 9.6 $\pm$ 0.1\tnote{c} \\
J1250-20:S3 & 192.70738 & -20.36734 & 114 & 0.577 & 0.343 & 1.489 & 5.659            &  0.082	$\pm$	0.074  &  3.582	$\pm$	0.14  &  9.30  $\pm$ 0.19  & 0.35  $\pm$ 0.13 & \dots \\
J1250-20:S4 & 192.66643 & -20.34798 & 114 & 0.489 & 0.217 & 0.165 & \dots            &  -0.236	$\pm$	0.106  &  2.046	$\pm$	0.227  &  9.57  $\pm$ 0.27  & 0.05  $\pm$ 0.02  & \dots \\
J1250-20:S5 & 192.74641 & -20.47055 & 114 & 0.124 & 0.074 & 0.239 & \dots            &  0.084	$\pm$	0.267  &  3.286	$\pm$	0.449  &  8.63  $\pm$ 0.68  & 0.07  $\pm$ 0.03 & \dots \\
J1408-21:S1 & 212.17491 & -21.59719 & 128 & 18.599  & 10.605  & 32.353  & 46.150     &  0.037	$\pm$	0.039  &  3.228	$\pm$	0.052  &  11.03 $\pm$ 0.10  & 6.41  $\pm$ 2.23 & 10.25 $\pm$ 0.22\tnote{c} \\
J1408-21:S2 & 212.24107 & -21.64791 & 128 & 5.974 & 3.773 & 16.022  & 25.289         &  0.148	$\pm$	0.042  &  3.555	$\pm$	0.048  &  10.25 $\pm$ 0.11  & 3.45  $\pm$ 1.20 & 9.7 $\pm$ 0.1\tnote{c} \\
J1408-21:S3 & 212.17101 & -21.628 & 128 & 2.467 & 1.454 & 7.254 & 5.565              &  0.073	$\pm$	0.047  &  3.725	$\pm$	0.06  &  10.06 $\pm$ 0.12  & 1.72  $\pm$ 0.60 & \dots \\
J1408-21:S4 & 212.13867 & -21.60202 & 128 & 2.925 & 1.677 & 5.741 & 6.087            &  0.043	$\pm$	0.045  &  3.343	$\pm$	0.075  &  10.21 $\pm$ 0.12  & 1.40  $\pm$ 0.49 & \dots \\
J1408-21:A1 & 212.19417 & -21.45222 & 128 & 0.191 & 0.104 & \dots  & \dots           &  -0.012	$\pm$	0.157  &  \dots  &  9.08  $\pm$ 1.27  &  \dots & 9.5 $\pm$ 0.2\tnote{c} \\
J1956-50:S1 & 299.19006 & -50.05563 & 110 & 12.247  & 7.484 & 24.142  & 44.917       &  0.112	$\pm$	0.039  &  3.278	$\pm$	0.05  &  10.52 $\pm$ 0.10  & 3.80  $\pm$ 1.32 & 10.38 $\pm$ 0.10\tnote{a}  \\
J1956-50:S2 & 298.97214 & -50.03633 & 110 & 2.685 & 1.692 & 7.575 & 14.167           &  0.146	$\pm$	0.045  &  3.609	$\pm$	0.057  &  9.78  $\pm$ 0.12  & 1.37  $\pm$ 0.48 & 9.56 $\pm$ 0.10\tnote{a} \\
J1956-50:S3 & 299.03409 & -50.03929 & 110 & 0.062 & 0.033 & \dots  & 2.498           &  -0.044	$\pm$	0.303  &  \dots  &  8.46  $\pm$ 1.27  &  \dots & \dots \\
J1956-50:A1 & 298.88959 & -49.88237 & 110 & 2.751 & 1.672 & 5.784 & 9.758            &  0.106	$\pm$	0.04  &  3.349	$\pm$	0.064  &  9.89  $\pm$ 0.10  & 1.08  $\pm$ 0.38 & 9.75 $\pm$ 0.13\tnote{a} \\
J2027-51:S1 & 307.02701 & -51.6916  & 87  & 15.698  & 12.118  & 92.091  & 374.980    &  0.366	$\pm$	0.038  &  4.153	$\pm$	0.039  &  10.20 $\pm$ 0.10  & 8.15  $\pm$ 2.82 & 10.2 $\pm$ 0.1\tnote{c} \\
J2027-51:S2 & 306.88406 & -51.65551 & 87  & 11.789  & 6.766 & 30.820  & 53.893       &  0.044	$\pm$	0.038  &  3.632	$\pm$	0.044  &  10.47 $\pm$ 0.10  & 3.11  $\pm$ 1.08 & 10.0 $\pm$ 0.1\tnote{c} \\
J2027-51:S3 & 306.95224 & -51.73896 & 87  & 1.115 & 0.671 & 0.991 & \dots            &  0.094	$\pm$	0.055  &  2.524	$\pm$	0.152  &  9.32  $\pm$ 0.14  & 0.15  $\pm$ 0.06 & 8.88 $\pm$ 0.13\tnote{a} \\
J2027-51:S4 & 306.97769 & -51.6349  & 87  & 0.244 & 0.145 & \dots  & \dots           &  0.079	$\pm$	0.108  &  \dots  &  8.85  $\pm$ 1.27  &  \dots & \dots \\
J2027-51:A1 & 307.03687 & -51.47511 & 87  & 0.087 & 0.054 & \dots  & \dots           &  0.129	$\pm$	0.118  &  \dots  &  8.40  $\pm$ 1.27  &  \dots & 9.49 $\pm$ 0.13\tnote{c} \\
J2318-42a.S1  & 349.59833 & -42.3704  & 23  & 600.930 & 561.039 & 1941.192 &6602.433 &  0.572	$\pm$	0.037  &  3.322	$\pm$	0.037  &  10.63 $\pm$ 0.09  & 11.47 $\pm$ 3.97  & 9.56\tnote{b} \\
J2318-42a.S2  & 349.72839 & -42.23901 & 23  & 164.991 & 101.376 & 458.249 & 725.787  &  0.118	$\pm$	0.037  &  3.62	$\pm$	0.038  &  10.28 $\pm$ 0.10  & 3.23  $\pm$ 1.12  & 9.51\tnote{b} \\
J2318-42a.S3  & 349.83566 & -42.25764 & 23  & 160.482 & 101.586 & 438.978 & 640.655  &  0.15	$\pm$	0.037  &  3.572	$\pm$	0.217  &  10.18 $\pm$ 0.09  & 3.11  $\pm$ 1.20  & 9.74\tnote{b} \\
J2318-42a:S4  & 349.70981 & -42.39675 & 23  & 0.027 & 0.016 & \dots  & \dots         &  0.08	$\pm$	0.308  &  \dots &  6.73  $\pm$ 1.27  &  \dots  & \dots\\

\bottomrule
\end{tabular}
\begin{tablenotes}
\item (1)  ID: HIPASS+Group+Galaxy identification, S -- within SINGG, A -- additional galaxy, discovered in \HI-emission; (2) Right Ascension (J2000); (3) Declination (J2000); (4) Distance; (5) and (6) are W1 and W2 fluxes; (7) and (8) W3pah and W4dust are fluxes of the ISM emission after removing the expected (model) stellar continuum; (9) The W1-W2 colour; (10) The W2-W3 colour; (11) Logarithm of the stellar mass; (12) 12$\mu$m star formation rate; (13) Logarithm of the \HI\ mass.
\item[a] \HI\ data from this work.
\item[b] \HI\ data obtained from the literature: J2318-42a \citep{Dahlem2005}, J1159-19 \citep{Phookun1992} and J0209-10 \citep{Jones2019}.
\item[c] \HI\ data obtained from our previous work, from  \citet{Dzudzar2019}.
\end{tablenotes}
\end{threeparttable}
\end{table*}
\end{landscape}

\twocolumn
\section{Note on individual Choir groups}
\label{sec:note_on_groups}

This section and the next one are presented as supplementary online material in the accepted paper version. Here we present results of each Choir group individually whose \HI\ content we mapped. Based on \citet{VerdesMontenegro2001} classification, we place the \HI\ phase class for each Choir group in Table \ref{tab:phase_classification}. Moreover, we present our Figures of each group with their \HI\ content and their environment.

\subsection{HIPASS J0205-55}
\label{sec:j0205-55}
Group HIPASS J0205-55 is composed of two subgroups (HIPASS J0205-55a and HIPASS J0205-55b) and it is thought that they are merging based on their recessional velocities \citep{Sweet2013}. The SINGG survey covered nine galaxies within this field. These galaxies are in the velocity range between 5758 \kms\ and 6490 \kms and R-band absolute magnitudes between --15.37 mag and --22.57 mag \citep{Sweet2013, Sweet2014}. With the ATCA we resolved the \HI\ content of five galaxies: S1, S2, S3, S4 and one new galaxy (A1) which is not covered in the SINGG field of view. The \HI\ emission, velocity field and the \HI\ spectra are shown in Figure \ref{fig:j0205_group}. S5, a dwarf galaxy is embedded within the S1 \HI\ envelope and it is not resolved individually in our observations. S6, a dwarf galaxy, is within the S2 \HI\ envelope. Galaxies: S7, S8 and S9 are small compact dwarf galaxies are not detected in the \HI\ emission. 

We examined the local environment of J0205-55 field within the radius of 100\arcmin\ centered on the S1 galaxy and in the velocity range between 4600 and 7600 \kms. J0205-55 appears to be within 1--1.5 Mpc in the projected distance from the Low-density-contrast group LDCE122 \citep{Crook2007} and galaxy pair ESO153-IG004 \citep{Tully2015a} towards south-west (see panel c and d in Figure \ref{fig:j0205_group}). The nearby group LDCE122 has the same recessional velocity as the J0205-55b group and it is positioned on the J0205-55 caustics curves (see Appendix \ref{sec:appendix_images} description). These lines are computed using M$_{200}$, where M$_{200}$ is determined based on the radial velocity dispersion using equation of \citet{Poggianti2010} and it is 1.8$\times$10$^{13}$ M$_{\odot}$ (see Appendix \ref{sec:table_choir_properties}).

\subsection{HIPASS J0258-74}

Group HIPASS J0258-74 (hereafter, J0258-74) contains four known galaxy members, three of which we detected in \HI\ (see Figure \ref{fig:j0258_group}). The galaxy without an \HI\ detection is S4 and it is a small dwarf irregular galaxy. S3 galaxy is an edge-on galaxy and it is one of the most deficient galaxies in our sample. We have shown its \HI\ distribution in Figure \ref{fig:HI_deficient}. The \HI\ is offset from its optical centre and due to proximity to S2 galaxy, it is possible that they are interacting. 

J0258-74 is the most isolated group of galaxies in our sample, it is considered being in a void (see Section \ref{sec:filaments}). The closest known neighbouring galaxy to J0258-74 group is a small, edge-on, late-type galaxy 2MASX J02463321-7341405 with a projected separation of around 66 arcmin which corresponds to $\sim$1.3 Mpc at the group distance of 70 Mpc. The next closest galaxy in projected separation from S1 is ESO031-G013 ($\sim$80 arcmin) however, this galaxy is at a much larger systemic velocity of $\sim$6400 km s$^{-1}$ with respect to the groups systemic velocity of 4805 km s$^{-1}$.

\subsection{HIPASS J0400-52}
\label{section:j0400}
\begin{figure}
    \centering
        \centering
        \includegraphics[width=\columnwidth]{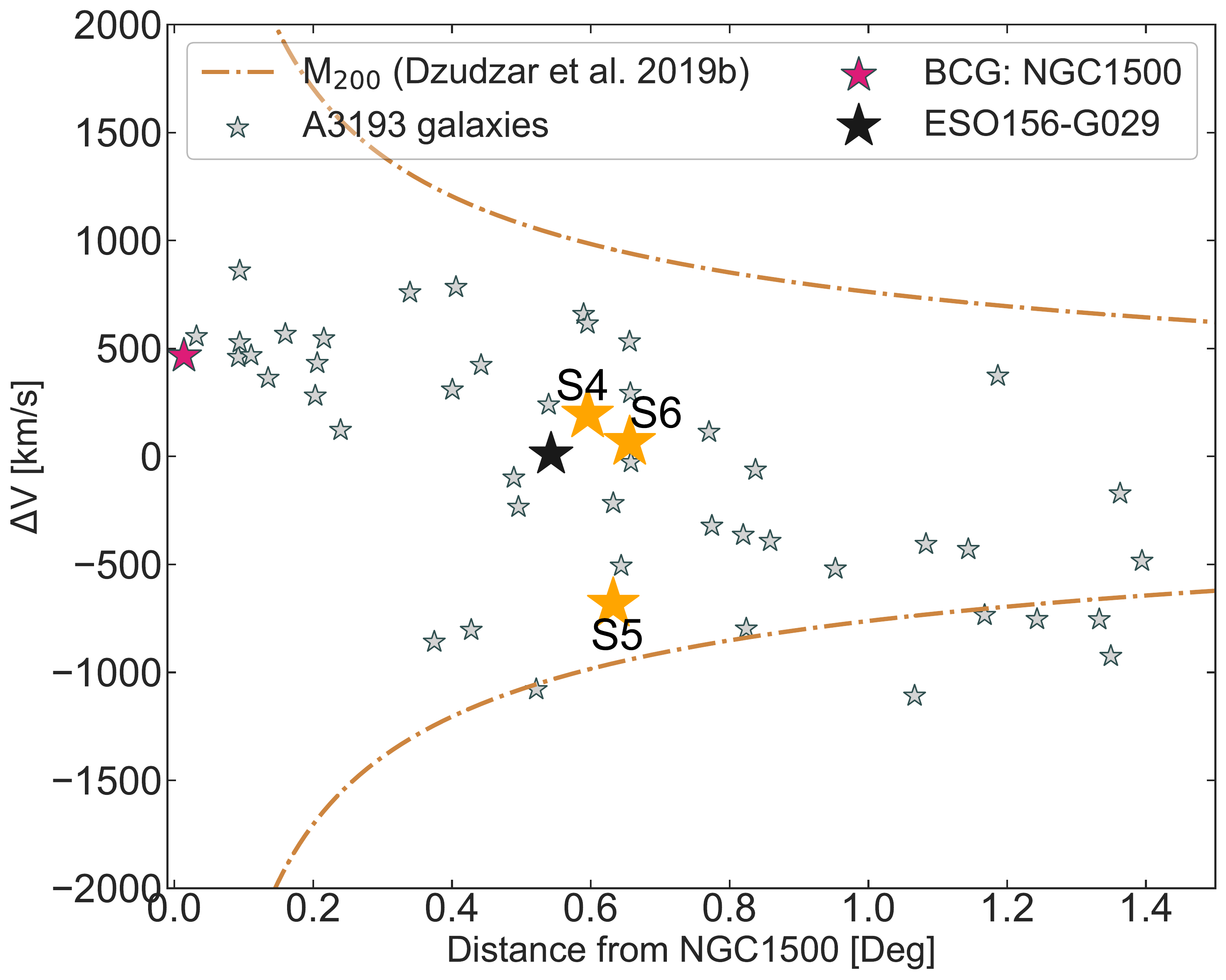}\quad
        \caption{Projected angular separation of the galaxies within cluster A3193 versus recessional velocity difference between the cluster mean velocity and other galaxies. The caustics line show a potential of M$=2.7\times10^{14}$M$_{\odot}$ \citep{Dzudzar2019b}. We highlight the position of S1, S4, S5 and S6 galaxies in J0400-52. }
\label{fig:caustics_j0400}
\end{figure}

Group HIPASS J0400-52 contains nine galaxies \citep{Sweet2013} and it is embedded in a larger structure: it is at the virial radius of cluster Abell 3193. We detected the \HI\ emission in only one galaxy (ESO156-G029). This galaxy was presented the detail in \citet{Dzudzar2019b} as an example of galaxy being ``pre-processed" - that is being shaped by the group environment as it falls into a cluster. 
In Section \ref{sec:non_detections}, we show that non-detected galaxies are typically dwarfs, particularly small compact dwarfs. However, there are three galaxies that are large enough, have a high stellar mass and large angular size, to be detected in this group: J0400-52:S4 [Scd], J0400-52:S5 [SB pec] and J0400-52:S6 [SB pec]. Our analysis show that these galaxies are \HI\ deficient. The stellar masses of these galaxies are $\sim$10$^{11}$ M$_{\odot}$, and using M$_{\textrm{R}}$ to M$_{\HI}$ scaling relation we were expecting M$_{\HI}$ $\sim$ 10$^{10.2}$ M$_{\odot}$ \citep{Sweet2013, Denes2014}. The theoretical limit of our observations in this group is M$_{\HI,\textrm{lim}}$ = 10$^{8.9}$ M$_{\odot}$, using this information we derive the lower limit of \HI\ deficiency to be 1.3 dex. Such large deficiencies are typically found in a cluster environment e.g. \citet{Chung2009}. The \HI\ deficiency in these galaxies is due to either group pre-processing, or it is possible that they are the `backsplash' galaxies i.e. already fell through the cluster potential well (e.g. \citealt{Yoon2017}). The backsplash scenario is most likely for S4 and S6, as the difference in their recessional velocity from the cluster mean velocity ($\Delta$V) is around 0 km s$^{-1}$ (see Figure \ref{fig:caustics_j0400}). 

\subsection{HIPASS J1026-19}

Group HIPASS J1026-19 contains six known member galaxies, three of which we detected in \HI\ (S1, S2 and S5). The non-detected galaxies S3 and S4 are small dwarf galaxies, below our detection limit, while S6 is \HI-deficient. 

\begin{figure}
    \centering
        \centering
        \includegraphics[width=\columnwidth]{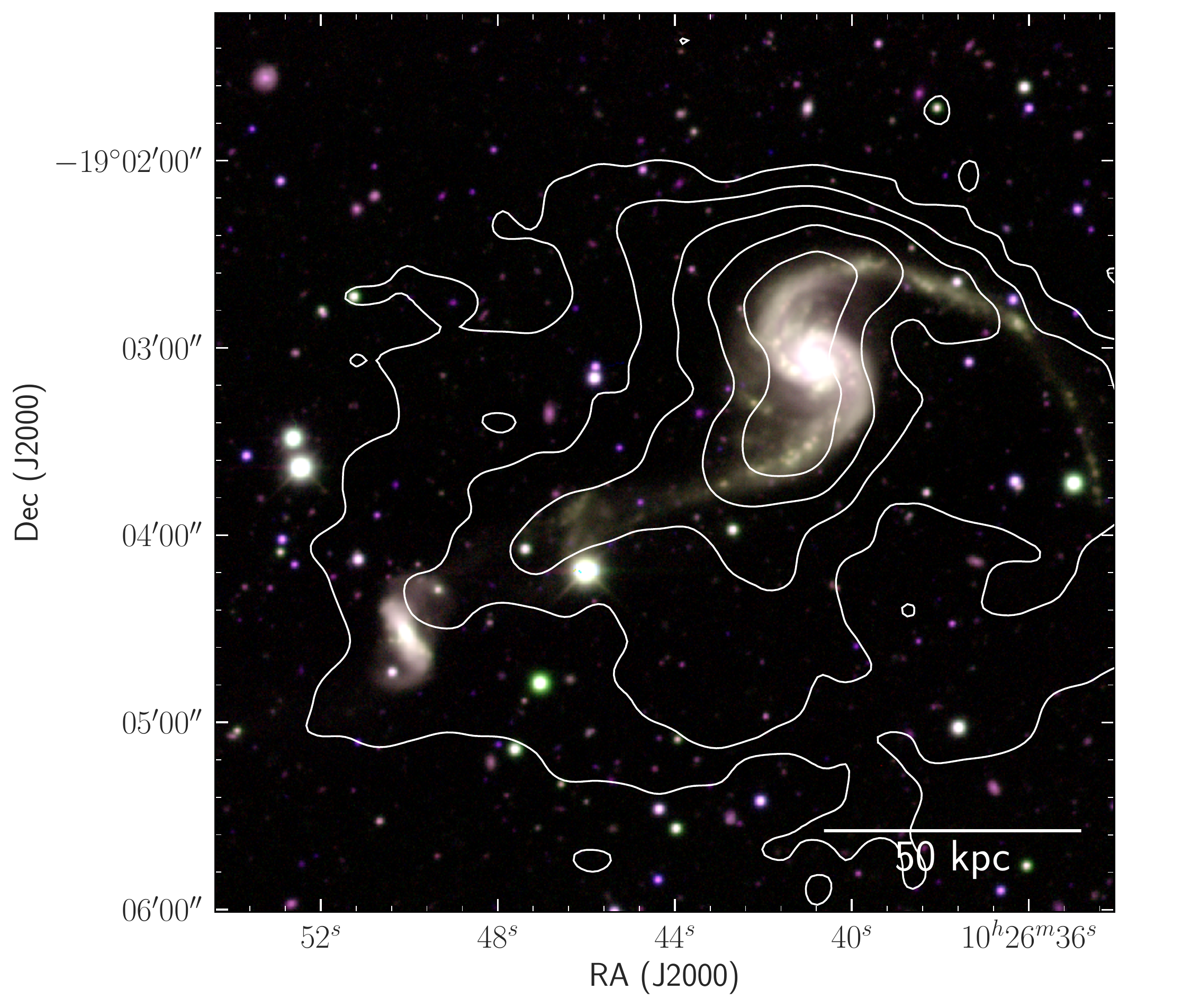}\quad
        \includegraphics[width=\columnwidth]{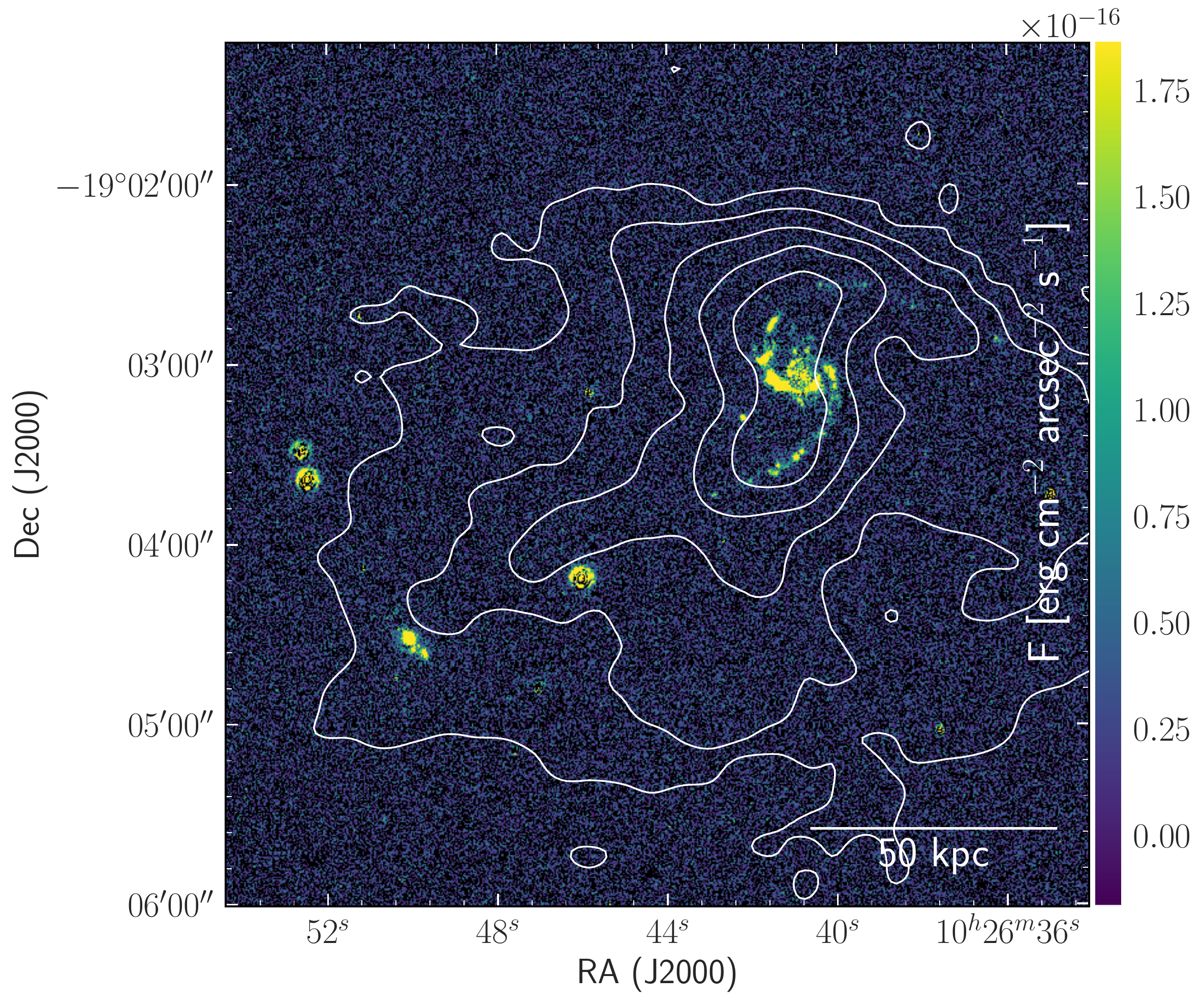}\quad
        \includegraphics[width=\columnwidth]{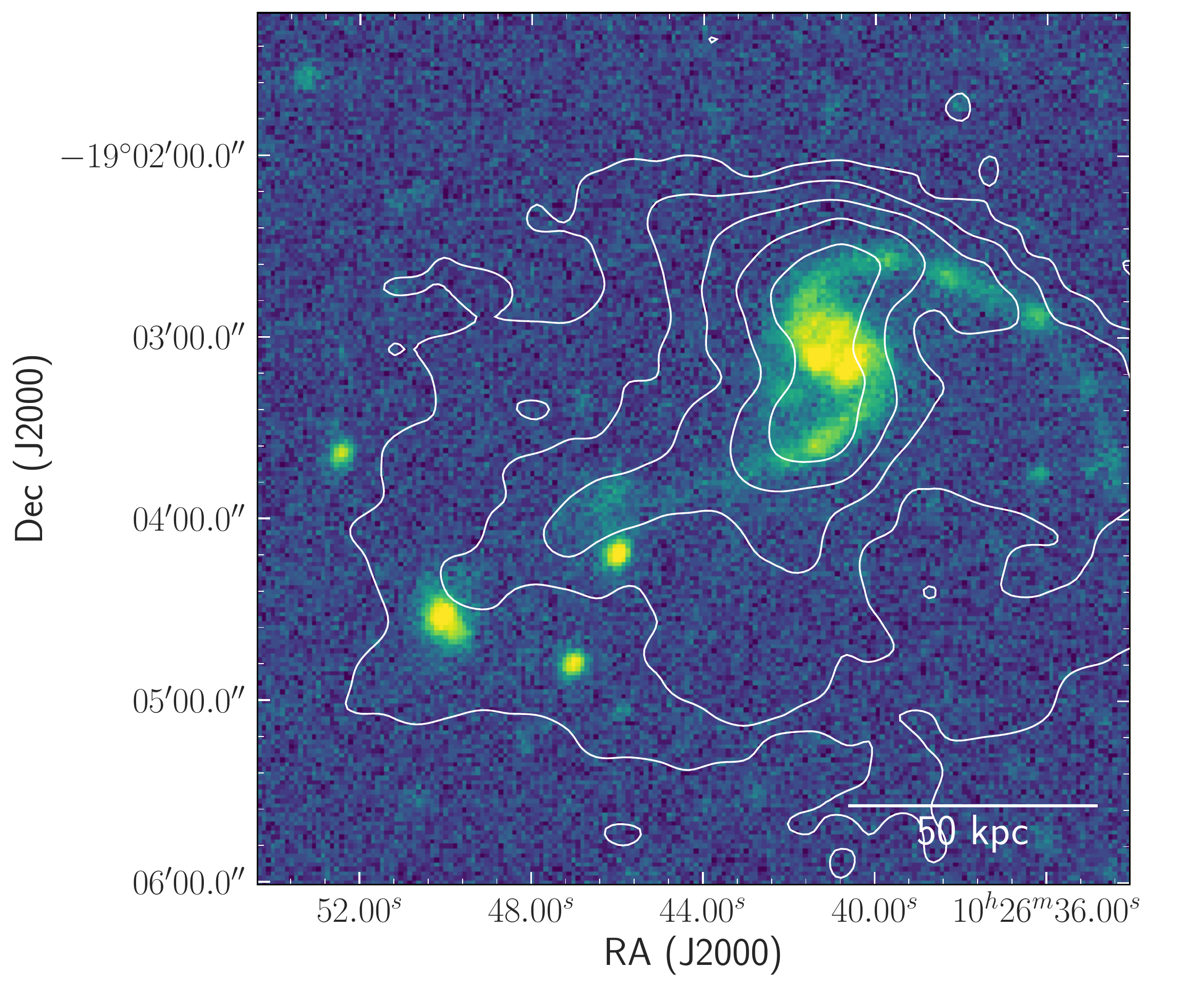}
        \caption{Multi-wavelength composites of the J1026-19 group. Panels with \HI\ intensity distribution of J1026-19 overlaid on the multi-wavelength images. The \HI\ column density contours: 5, 15, 30, 50 and 70$\times10^{19}$. Upper: optical DECam gri composition. Middle: H$\alpha$ imaging from SINGG narrow band image in the background, the colorbar shows surface brightness. Lower: GALEX near-uv image in the background. }
\label{fig:j1026_multiw}
\end{figure}

The largest member of J1026, S1, is connected to its smaller companion S2 (which is either a small edge-on disk or a strongly barred-dwarf galaxy) by a tidal tail which is visible in optical, UV and \HI\ imaging (see Figure \ref{fig:j1026_multiw}). Tracing emission (optical and UV) from the S1 galaxy, we can see the faint stellar stream as a continuation of the spiral arm. This stellar stream broadens roughly at the midpoint between the S1 and S2 (looking at the projected distance between the S1 and S2) and the faint irregular structure resembles a tidal dwarf galaxy candidate. We examine the \HI\ intensity map in the J1026-19 and see that low-density contours overlap S1 and S2 galaxies (see Figure \ref{fig:j1026_group}). The \HI\ contours in the north-west part of the J1026-19 group (S1 galaxy) are compressed with respect to those in the south-east. Moreover, the velocity map of the J1026-19 is being skewed towards the south-east at the roughly recessional velocity of the group. It is possible that a fly-by can explain such \HI\ and stellar features (e.g. \citealt{Kim2017}). In a fly-by scenario, S2 galaxy passed near S1 galaxy (from the west to the east) which skew the gas in the direction of the motion of the S2 galaxy.

Examining the global environment we find that the J1026-19 is fairly isolated, having two nearest neighbouring galaxies at $\sim$1.5 Mpc in projected distance and velocity offset of $\sim$250 km s$^{-1}$. 

\subsection{HIPASS J1051-17}

Group HIPASS J1051-17 contains 20 known member galaxies, majority of them are small dwarf galaxies, within a velocity range between 5220 and 6243 km s$^{-1}$. We detect six galaxies in \HI\ emission within this group, which are potentially part of two sub-groups based on their recessional velocities. Out of six galaxies detected in \HI\ emission, one is an \HI-excess, polar ring galaxy J1051-17:S1 (see Figure \ref{fig:HI_excess}), one has an average \HI\ content four are \HI-deficient galaxies which are a result of the environmental processing (see in depth analysis by Kilborn et al. in prep.).

\subsection{HIPASS J1059-09}

Group HIPASS J1059-09 contains 10 member galaxies, five of which we detected in \HI. The non-detected galaxies S7, S8, S9, S10 are small dwarf galaxies, below our detection limit, while S4 is an SB0 type galaxy and based on its expected \HI\ mass it is \HI-deficient. We show our \HI\ imaging of the entire group in Figure \ref{fig:j1059_group}.

\begin{figure}
    \centering
        \centering
        \includegraphics[width=\columnwidth]{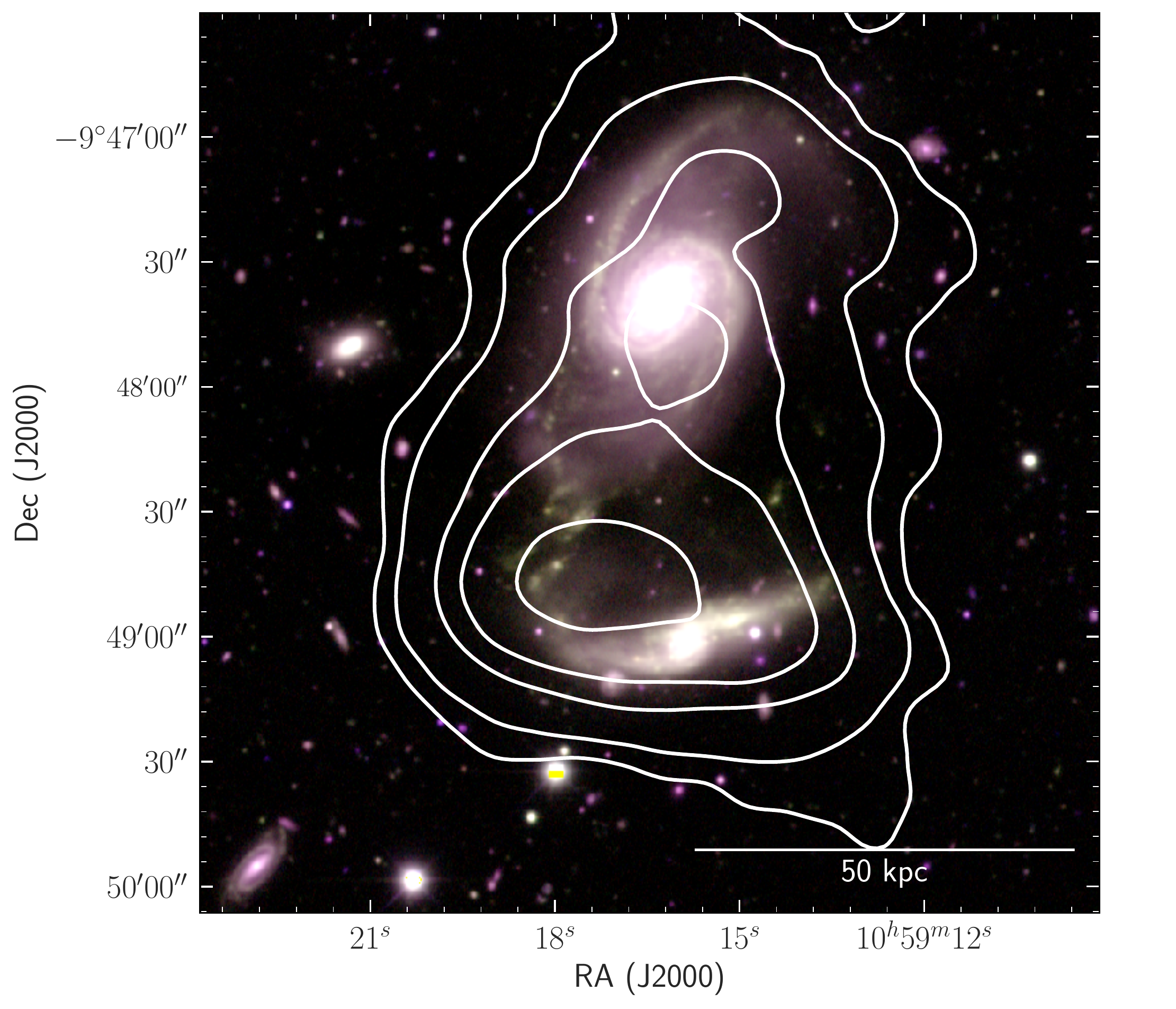}\quad
        \includegraphics[width=\columnwidth]{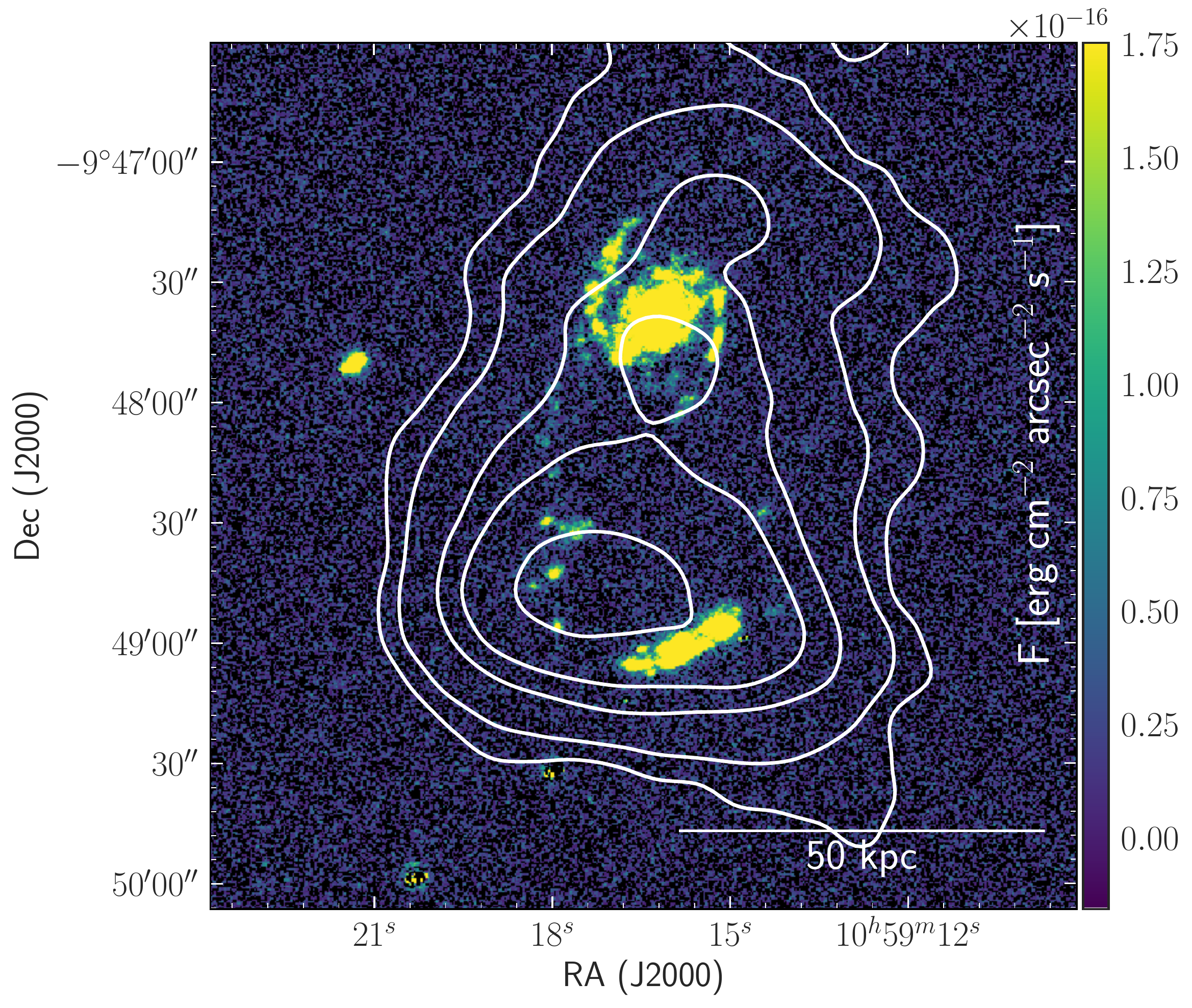}\quad
        \includegraphics[width=\columnwidth]{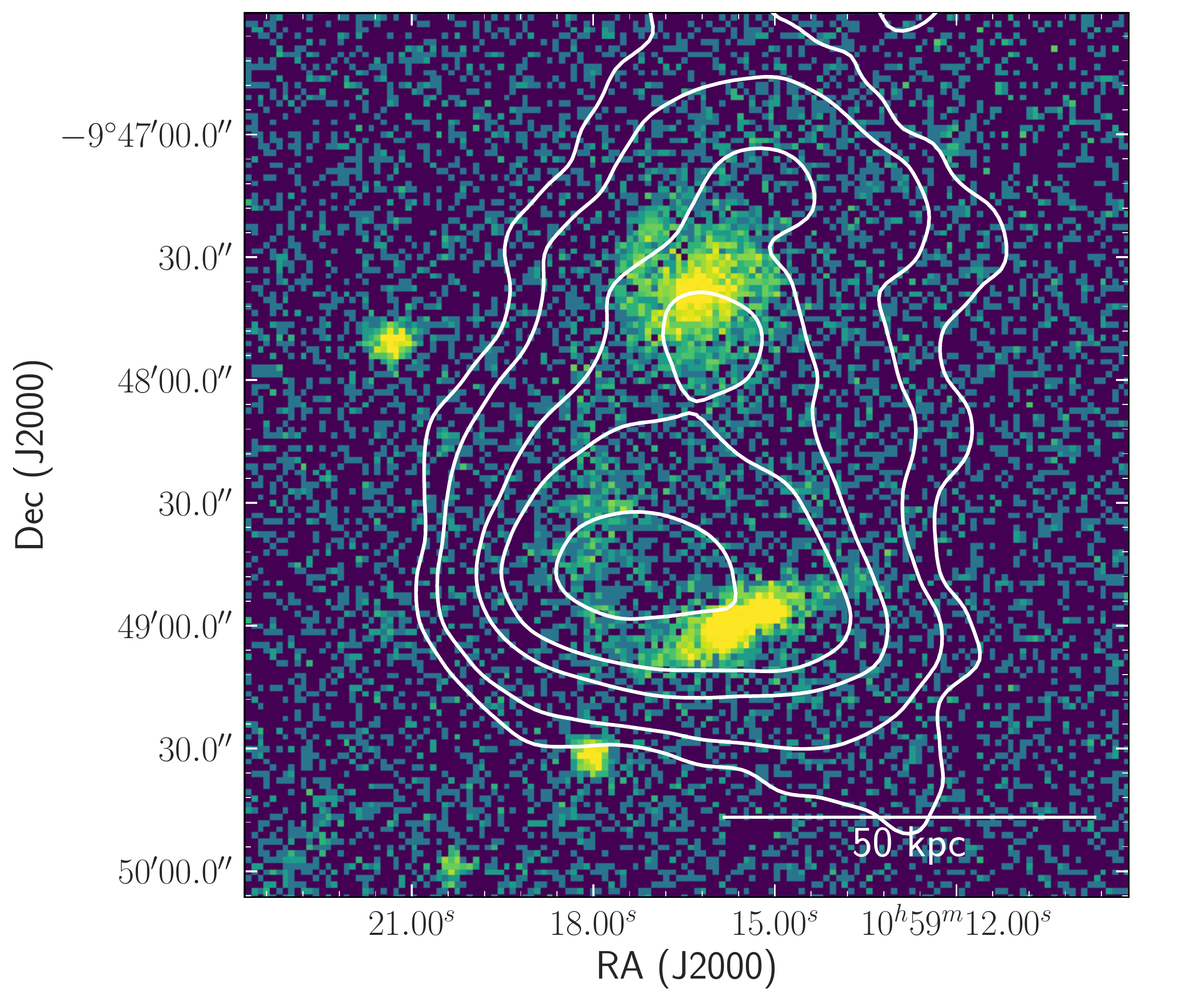}
        \caption{Multi-wavelength composites of the J1059-09 group. Panels with \HI\ intensity distribution of J1059-09 overlaid on the multi-wavelength images. The \HI\ column density contours: 12, 20, 40, 54 and 78$\times10^{19}$. Upper: optical DECam gri composition. Middle: H$\alpha$ imaging from SINGG narrow band image in the background, the colorbar shows surface brightness. Lower: GALEX near-uv image in the background. }
\label{fig:j1059_multiw}
\end{figure}

J1059-19 contains a strongly interacting galaxy pair S1 and S3 (MCG-01-28-013 and MCG-01-28-013, see Figure \ref{fig:j1059_multiw}) which are connected by a tidal tail. The tidal tail between S1 and S3 galaxy is visible in multi-wavelength observations (optical, H$\alpha$, UV, IR and \HI). We note that the peak of \HI\ emission is offset from the stellar centres of both galaxies, both in the direction of interaction. The \HI\ peak of the \HI\ emission that corresponds to the S3 galaxy is in the tidal tail between itself and S1 galaxy (see Figure \ref{fig:j1059_multiw}). The obvious cause of the \HI\ offsets is galaxy interaction, which is most likely triggering the star formation in the tidal tail region (e.g. \citealt{Chromey1998}). 

Examining the global environment we find that the J1059-09 is possibly part of a larger structure, having a number of galaxies that are within a projected separation $<$2 Mpc and velocity offset of $\sim$300 km s$^{-1}$. We also find another possible group to the east of J1059-09 at a velocity offset from the J1059-09 of 400-500 \kms\ and projected distance of $\sim$3 Mpc. J1059-09 is considered to be a group near of the intersection of the cosmic web filaments (see Section \ref{sec:filaments}).  

\begin{figure*}
    \centering
        \centering
        \includegraphics[width=0.9\columnwidth]{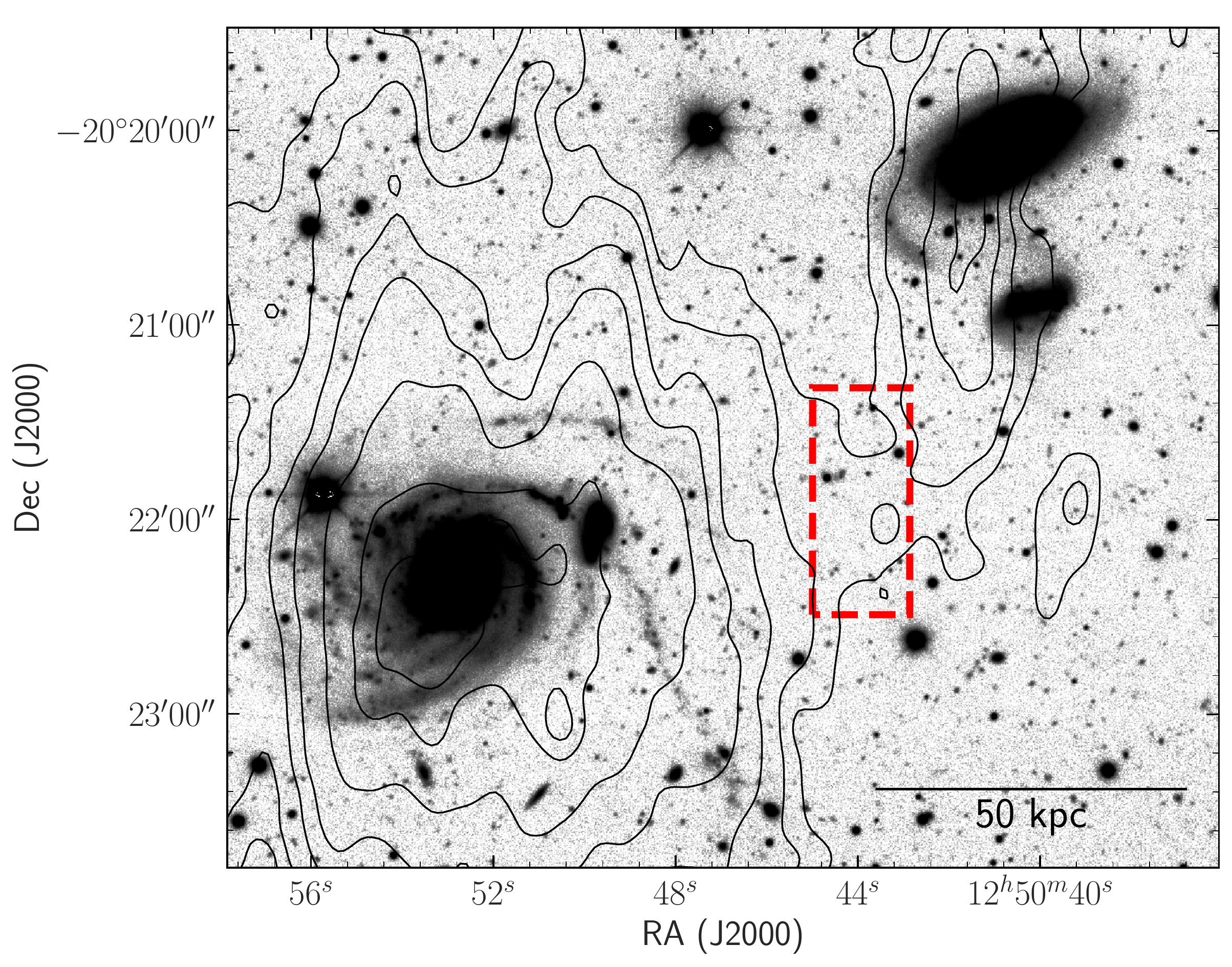}
        \includegraphics[width=0.7\columnwidth]{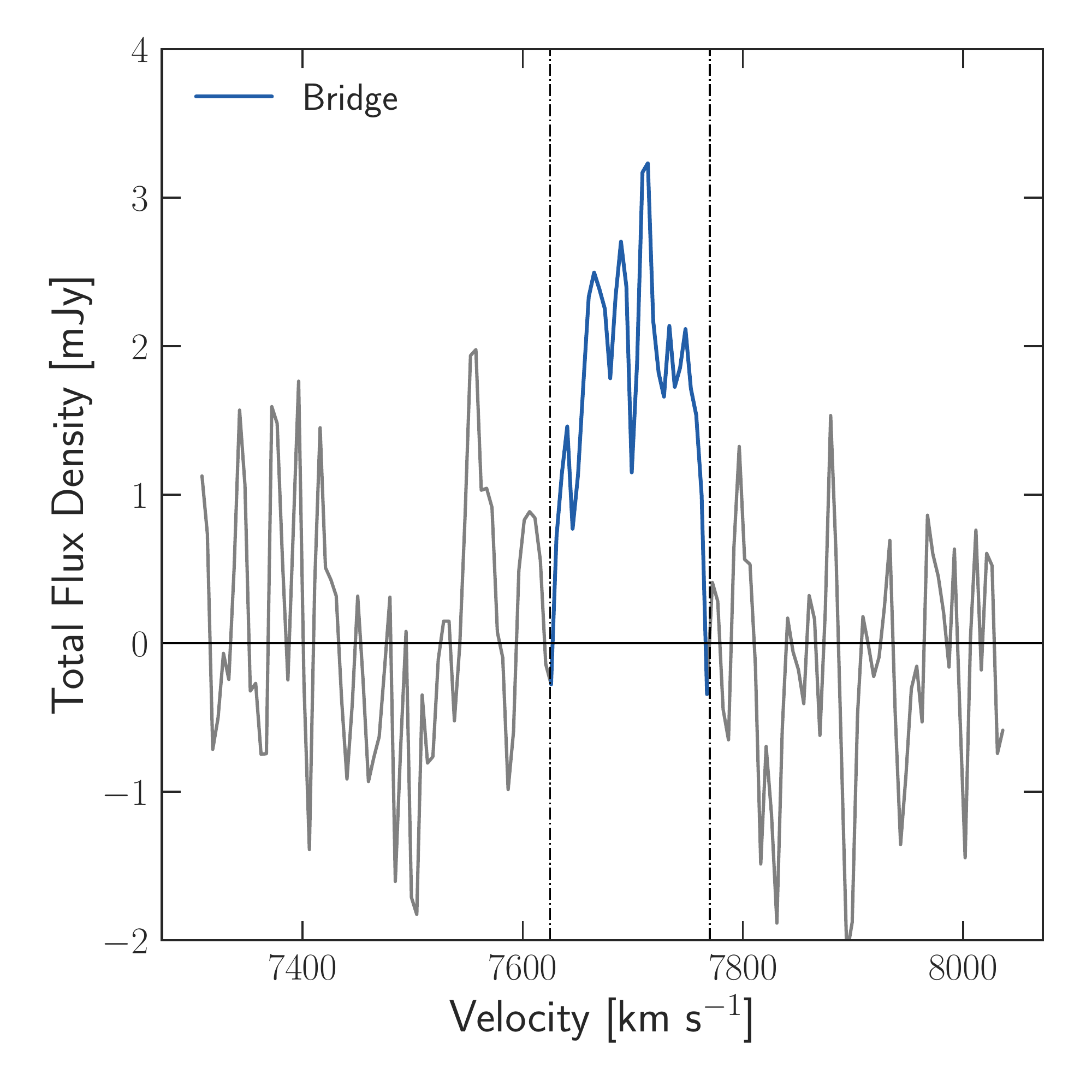}

        \caption{\textbf{Left:} Optical DECam g` band image overlaid with \HI\ density contours and rectangle region showing the \HI\ bridge between galaxies. The \HI\ column density contours: 3.5, 7, 14, 20, 30, 40 and 45$\times10^{19}$. \textbf{Right:} The \HI\ spectrum extracted from the rectangular region on the right. }
\label{fig:j1250_multiw}
\end{figure*}

\subsection{HIPASS J1159-19}

The HIPASS J1159-19 group is a compact galaxy group with four known galaxy members \citep{Sweet2013}. Previous studies analysed resolved \HI\ imaging of this system and found an interaction between NGC4027 (S1) and NGC4027A (S2) \cite{Phookun1992}. Moreover, \citet{Phookun1992} found that NGC4027 has a ring of \HI\ which is most likely thrown out from the NGC4027A galaxy in a flyby. With the SINGG survey two more dwarf galaxies were found in the projected proximity of S2 \citep{Meurer2006}. With our ATCA imaging we can not separate the \HI\ from the individual galaxies (see Figure \ref{fig:j1159_group}). However, we do find that the \HI\ gas in NGC4027 has two components and we can see disturbed kinematics of this group. There is a large magnitude offset between the two brightest galaxies (in R-band: $\sim$3.5 mag), thus we classified this group as an optical fossil group. 

Our analysis shows that group belongs to a larger structure as we find a large number of galaxies within projected separation of 0.5 Mpc around this group (see c and d panels in Figure \ref{fig:j1159_group}). At the north-east from the J1159-19 there is Antennae group (Arp 244) and at the south-west there are several galaxies. We find that there are three HIPASS detection around J1159-19 group: Arp 244, HIPASS J1159-19a and HIPASS J1154-20.

\subsection{HIPASS J1250-20}
\label{sec:1250}
Group HIPASS J1250-20 has seven known members, five of these are spectroscopically confirmed, others are thought to be the group members, as stated in \citep{Sweet2013, Dzudzar2019}. We detect the \HI\ emission which we associated with S1 and S2 galaxies. The non-detected galaxies are small compact dwarfs and we show that one galaxy, S4, is an \HI\ deficient galaxy. The likely reason for \HI-deficiency in S4 is the tidal interaction with S2, see Figure \ref{fig:j1250_multiw}. 

The two most luminous galaxies in J1250-20, S1 and S2 are interacting, see Figure \ref{fig:j1250_multiw}. These galaxies have a comparable stellar masses however, they have vastly different \HI\ mass. The \HI\ contours that we associate with S2 galaxies overlap also on top of its smaller companion with which its interacting, S4. In this work we associate \HI\ emission to S2, as S4 is physically smaller, has stellar mass $\sim$0.9 dex smaller, and has a lower specific star-formation. 

The \HI\ mass extracted from the region between the S1 and S2 galaxies is 10$^{8.9}$M$_\odot$. This is a lower limit of the \HI\ mass between galaxies as there is uncertainty where to place galaxy `borders'. The \HI\ velocity of the extracted region corresponds to the recessional group velocity (in velocity space between S1 and S2), and in this region we do not detect faint stellar streams. 

\subsection{HIPASS J1408-21}

Group HIPASS J1408-21 was reported to have six member galaxies \citep{Sweet2013}, with our ATCA imaging we detect two galaxies and find one new member, denoted as A1. The non-detected galaxies are two small compact dwarfs, which are below our detection limit, while two galaxies are (S2 and S4) are \HI\ deficient. 

\citet{Sweet2013} noted a possible tidal distortion of S1 galaxy towards S3 in H$\alpha$, and now with the \HI\ imaging we can clearly see that they are interacting (see Figure \ref{fig:j1408_group}).
Examining the global environment we find that the J1408-21 near of the intersection of the cosmic web filaments (see Section \ref{sec:filaments}).

\subsection{HIPASS J1956-50}

Group HIPASS J1956-50 was reported to have four possible member galaxies \citep{Sweet2013}, however, the velocity of S3 galaxy is quite offset from the rest of the group members \citep{Sweet2014} thus it is less likely to be a group member. We mapped the \HI\ content with the ATCA and we detected three galaxies in \HI, one of which is a new A1 group member, as previously was outside the angular SINGG field-of-view and thus was not detected (see Figure \ref{fig:j1956_group}). One non-detected galaxy is a compact dwarf, and it is below our detection limit. 

Examining the global environment we find that the J1956-50 is isolated (see panel c and d in Figure \ref{fig:j1956_group}) and it is considered to be a group within a void (see Section \ref{sec:filaments}). One neighbouring galaxy to J1956-50 is a faint compact dwarf galaxy south of S2 \citep{Sweet2013} with $\sim$1000 \kms\ offset from the group velocity. Moreover, there are two galaxies with the velocity offset of $\sim$500 \kms\ albeit at $\sim$2 Mpc in projected distance from the group.

\subsection{HIPASS J2027-51}

Group HIPASS J2027-51 has five known members, four of these were noted by \citet{Sweet2013} and one is a newly discovered member \citep{Dzudzar2019}. S1 and S2 are large spiral galaxies and we resolve the \HI\ within them. S3 is a dwarf irregular galaxy and we only have a tentative \HI\ detection, as it is below 3$\sigma$. We do not detect \HI\ in S4 which is a compact dwarf galaxy and it is below our detection limit. New detected galaxy is an irregular galaxy, marked A1 in Figure \ref{fig:j2027_group}, and it is $\sim$200 kpc north from S4 galaxy.   

Examining the global environment we note that J2027-51 belongs to a larger structure. We find several galaxies west from the S2 galaxy with the similar recessional velocity. ESO234-G024 located $\sim$0.5 Mpc west from S2, \citet{Mathewson1992} observed \HI\ flux of 9.42 Jy km s$^{-1}$, which corresponds to M$_{\HI}$=10$^{10.23}$M$\odot$ adopting the group distance.

\begin{center}
\begin{table}
\caption{The phase classification based on the \HI\ evolutionary scenario by \citet{VerdesMontenegro2001}.}
\label{tab:phase_classification}
\begin{threeparttable}
\begin{tabular}{cc}
\toprule
ID &  Phase class \\ 
(1) & (2) \\ \hline \hline
J0205-55 & Phase 1  \\
J0209-10 & Phase 2 \\
J0258-74 & Phase 1 \\
J0400-52 & Phase 1\\
J0443-05 & \dots \\
J1026-19 & Phase 2 \\
J1051-17 & Phase 1\\
J1059-09 & Phase 2\\
J1159-19 & Phase 2\\
J1250-20 & Phase 2\\
J1403-06 & \dots \\
J1408-21 & Phase 1 \\
J1956-50 & Phase 1\\
J2027-51 & Phase 1\\
2318-42a & Phase 2 \\ 
\bottomrule
\end{tabular}
\begin{tablenotes}
\item (1) Group name: HIPASS+ID; (2) \HI\ Phase classification.
\end{tablenotes}
\end{threeparttable}
\end{table}
\end{center}

\section{Individual groups}
\label{sec:appendix_images}

\begin{figure}

\begin{subfigure}{\columnwidth}
\centering
    \includegraphics[width=0.45\columnwidth]{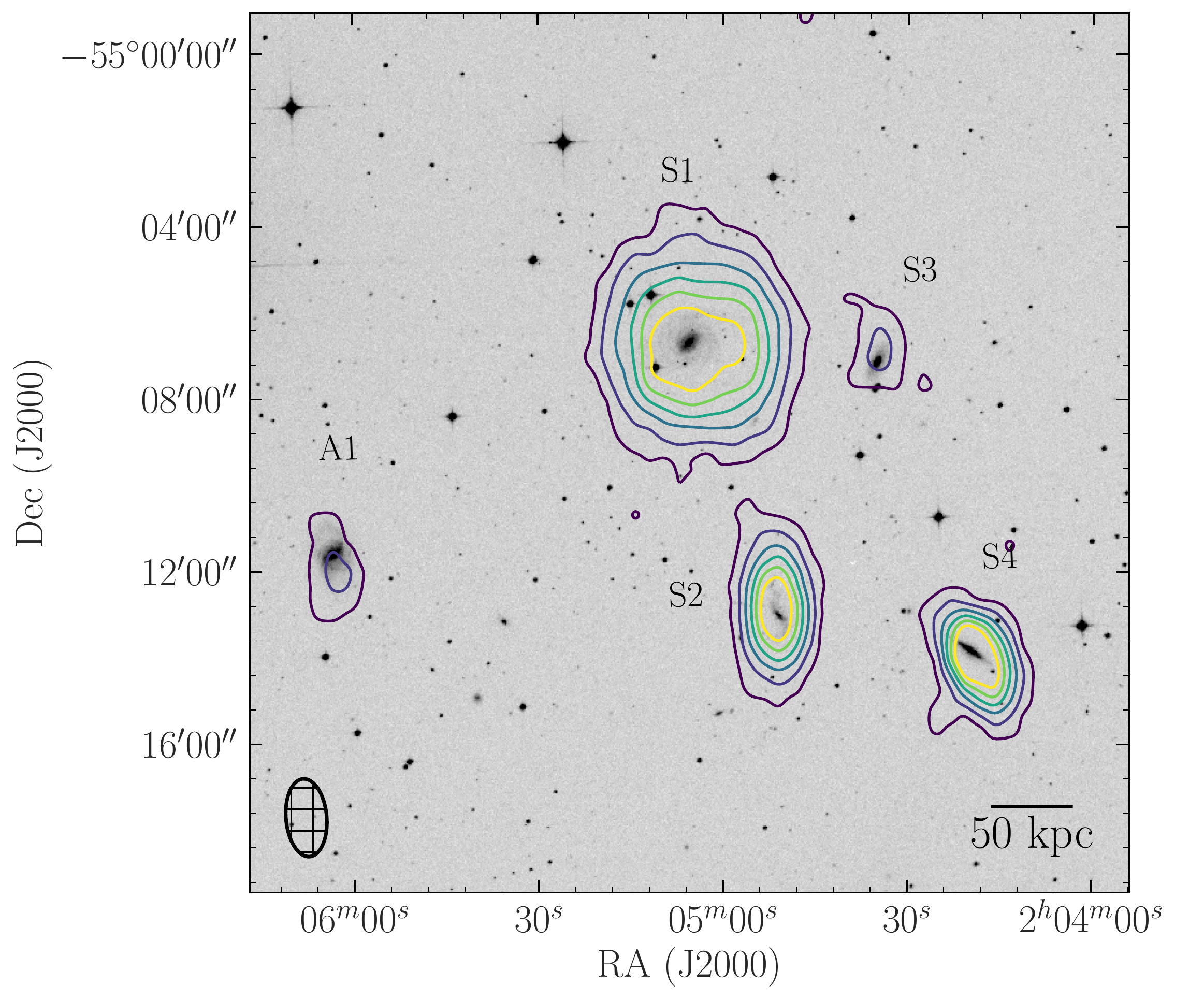}
    \includegraphics[width=0.45\columnwidth]{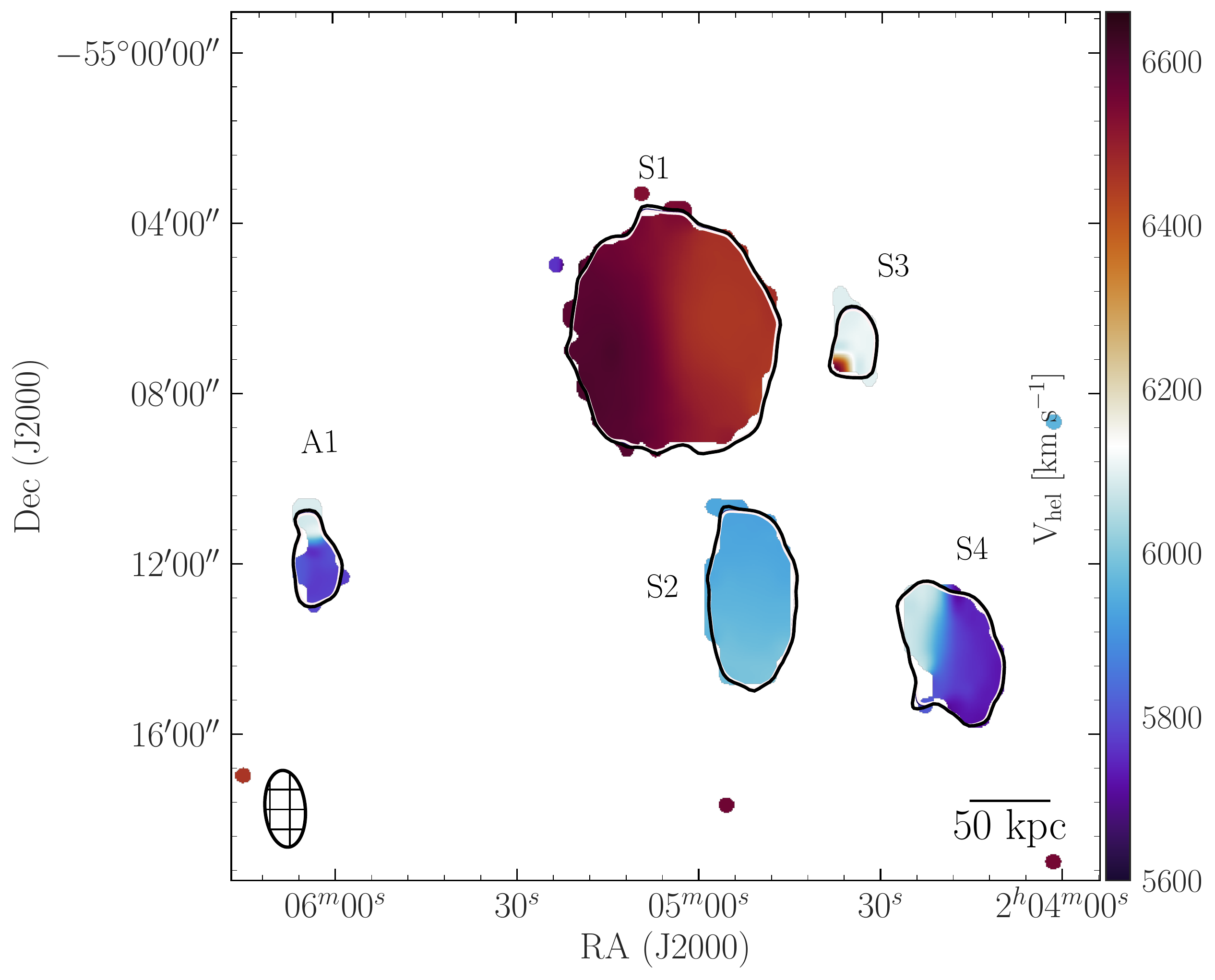}    
    \caption{}
\end{subfigure}

    \centering
    \begin{tabular}[t]{cc}

\begin{subfigure}{0.45\textwidth}
    \centering
    \smallskip
    \includegraphics[width=0.95\linewidth,height=1.4\textwidth]{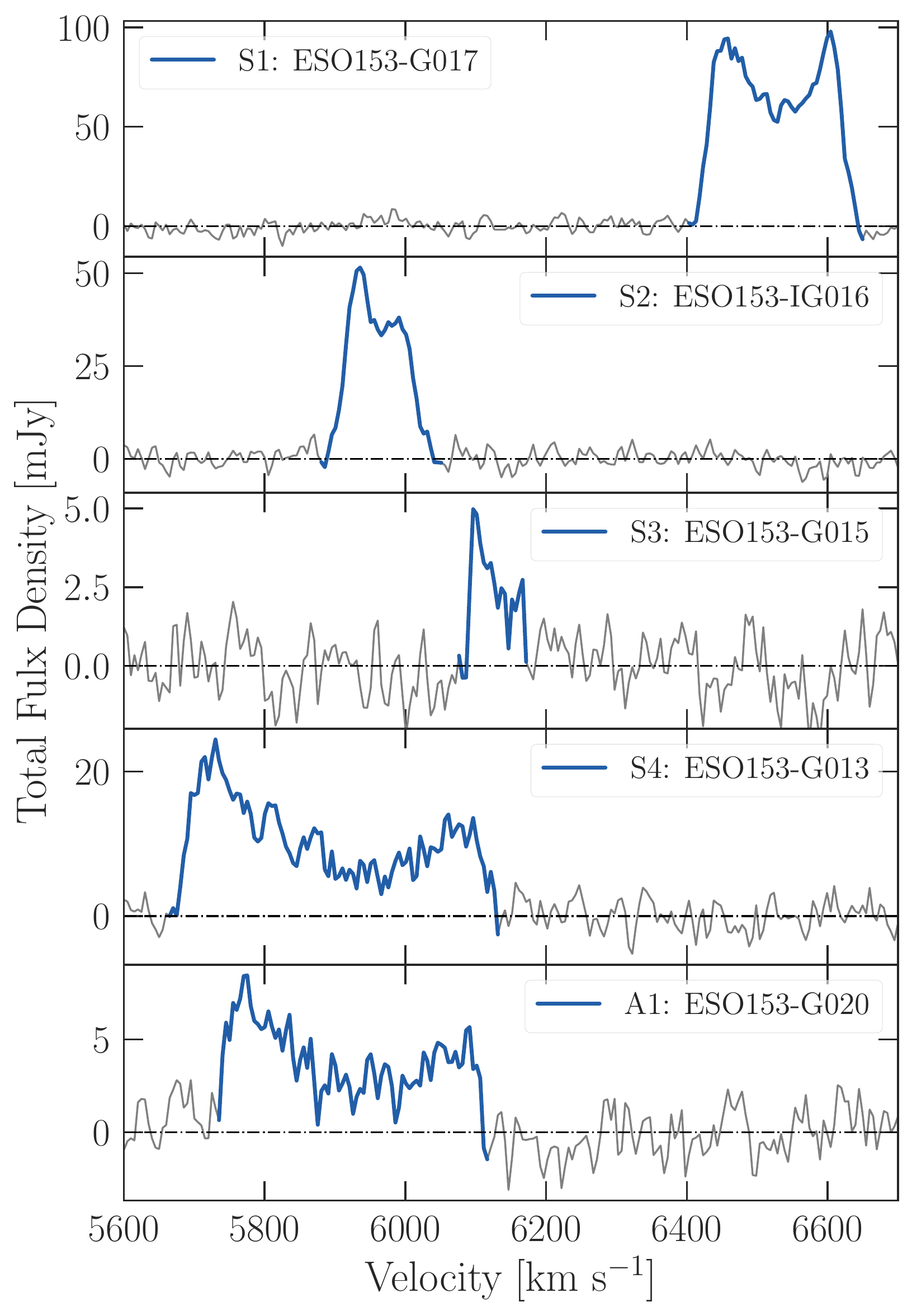}
    \caption{} 
\end{subfigure}
    &
        \begin{tabular}{c}
        \smallskip
            \begin{subfigure}[t]{0.42\textwidth}
                \centering
                \includegraphics[width=0.94\textwidth]{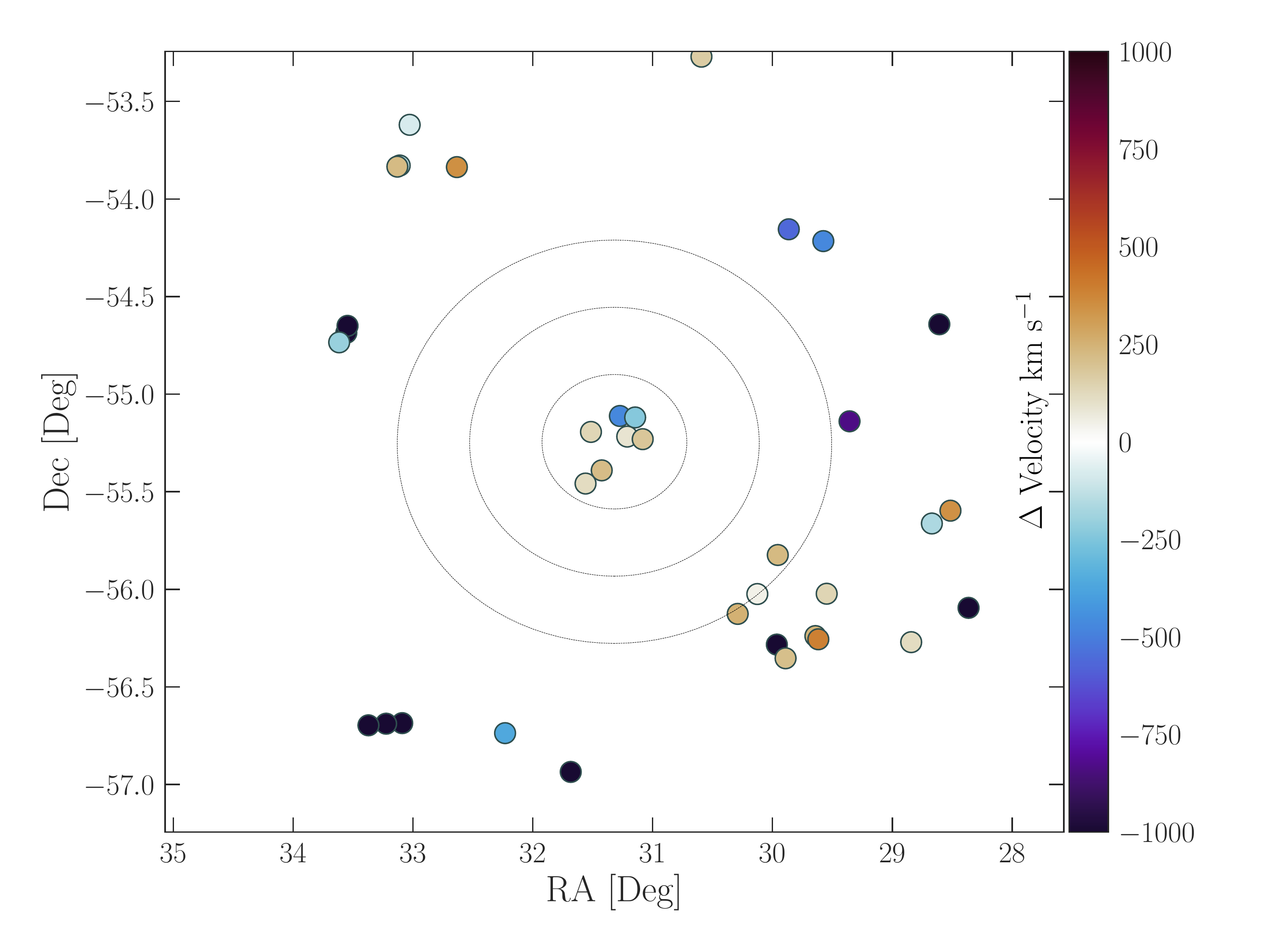}
                \caption{}
            \end{subfigure}\\
            \begin{subfigure}[t]{0.43\textwidth}
                \centering
                \includegraphics[width=0.95\textwidth]{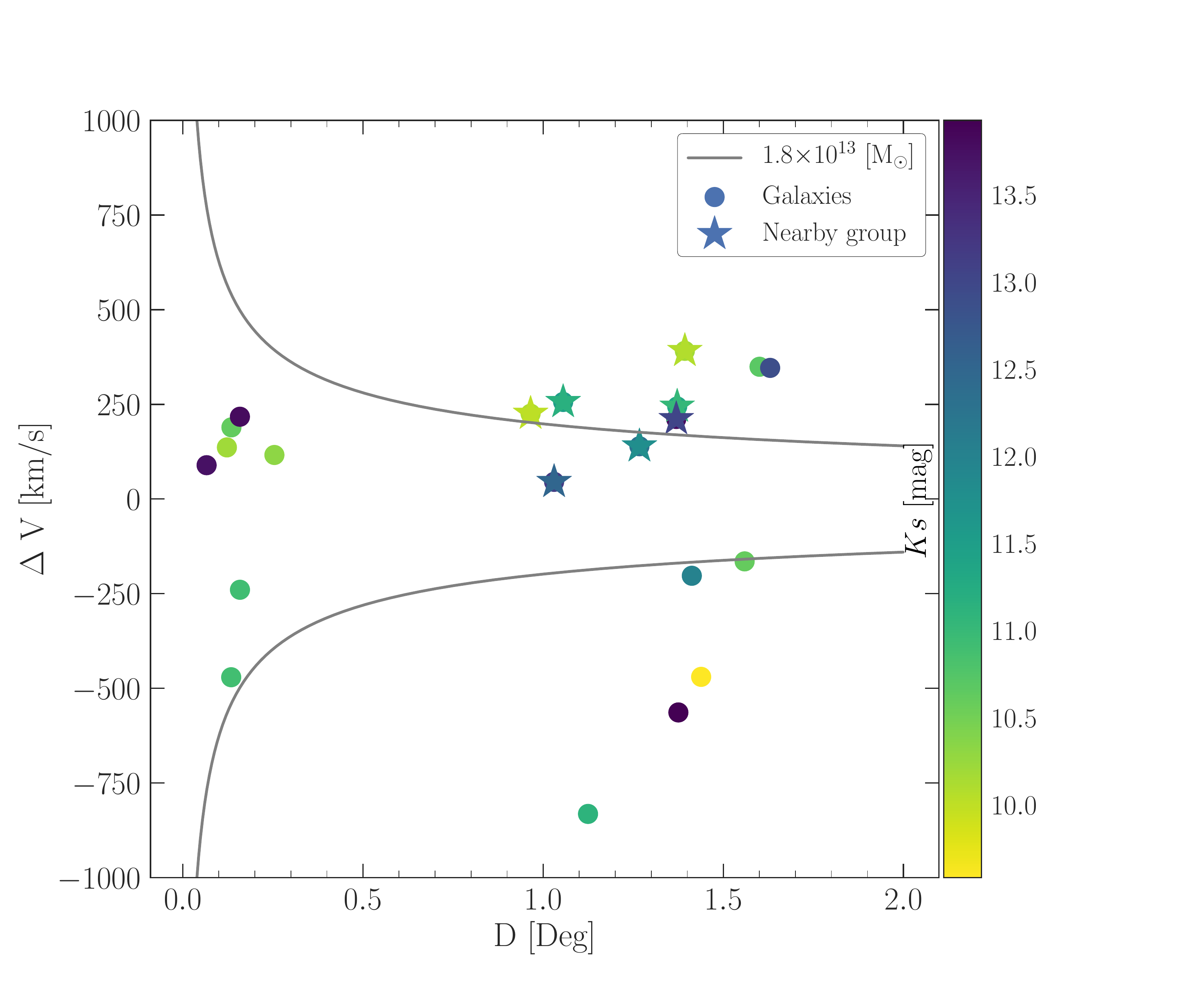}
                \caption{}
            \end{subfigure}
        \end{tabular}\\

    \end{tabular}
\caption{\textbf{a)} \textbf{Left:} The \HI\ emission in the J0205-55 group. The \HI\ emission is shown by the contours overlaid on the optical DSS2Red image. The lowest shown \HI\ column density is a 3$\sigma$ detection over a velocity width of 15 km s$^{-1}$ and corresponds to 1.5$\times$10$^{19}$ cm$^{-2}$. The other \HI\ column density contours are: 7, 14, 21, 28 and 35$\times10^{19}$ cm$^{-2}$. \textbf{Right}: The velocity field of the J0205-55 group in the limit from 5600 to 6660 \kms. In the Figure map are marked group members as well as the lowest \HI\ column density (2$\times$10$^{19}$ cm$^{-2}$). The synthesized beam is shown in the bottom left corner, and the scale bar in the bottom right corner shows 50 kpc at the group distance.
\textbf{b)} Panels of the \HI\ spectra for each detected galaxy in the J0205-55 group. The enhanced colour shows where the spectrum was integrated.
\textbf{c)} Environment around J0205-55, centred on the weighted mean of the group. The black circles are having radii of 0.5, 1, and 1.5 respectively. We show the velocity difference between weighted mean group velocity (6053 km s$^{-1}$) and the other nearby galaxies with magnitudes brighter than 13.5 mag, in the velocity range between 4600 and 7600 \kms and radius of 100\arcmin\ from weighted mean of the group.
\textbf{d)} Projected angular separation of the galaxies within J0205-55 region versus recessional velocity difference between weighted mean group velocity and other nearby galaxies. Galaxies associated with the group are within 0.5 deg, while galaxies associated with the south-west group (in panel c) are shown as stars. Grey solid curved line show simple caustics curves for a potential of M $=$ 1.8$\times$10$^{13}$, assuming that all sources are at the J0205-55 distance of 93 Mpc.}
\label{fig:j0205_group}
\end{figure}

\begin{figure}

\begin{subfigure}{\columnwidth}
\centering
    \includegraphics[width=0.45\columnwidth]{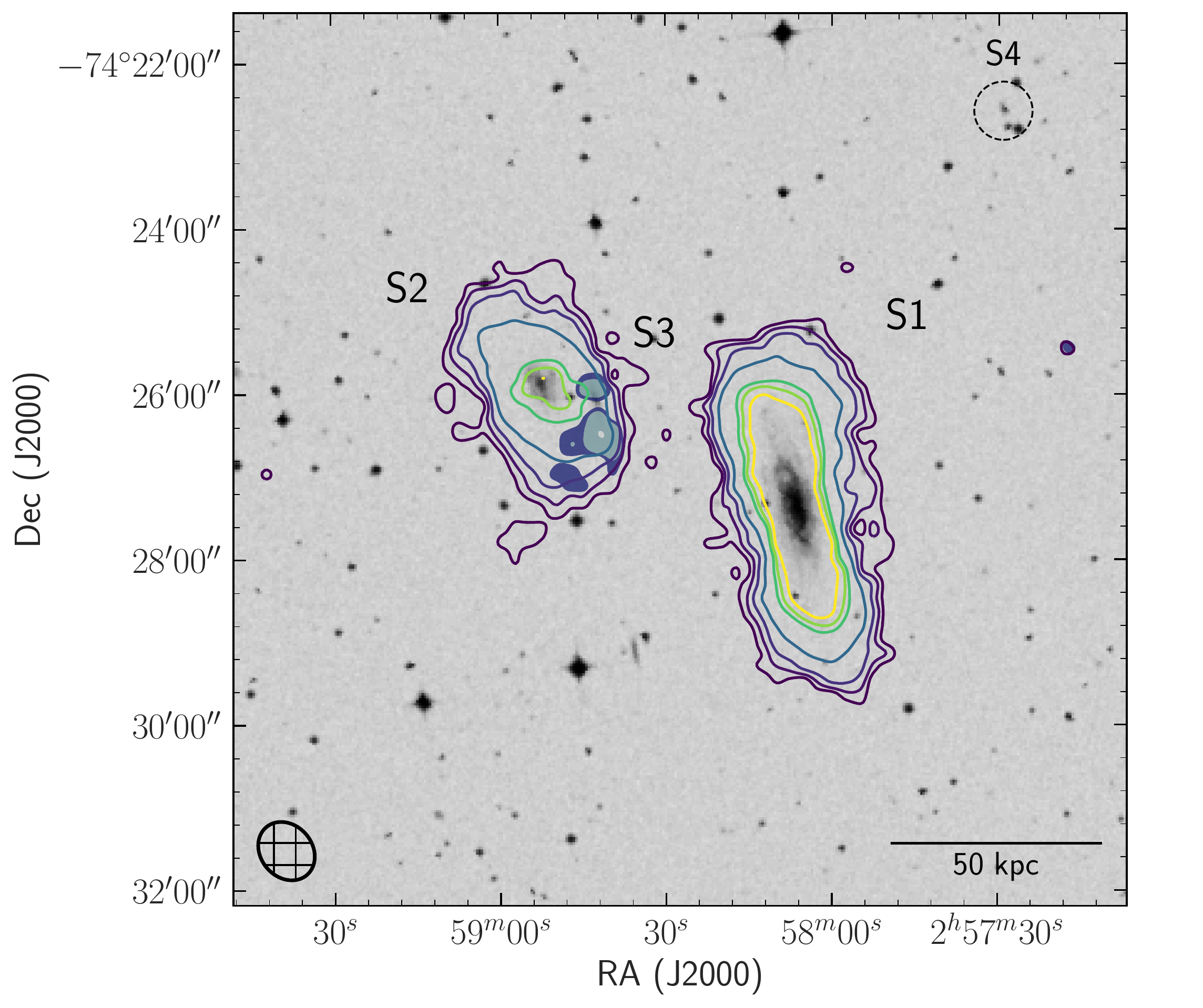}
    \includegraphics[width=0.45\columnwidth]{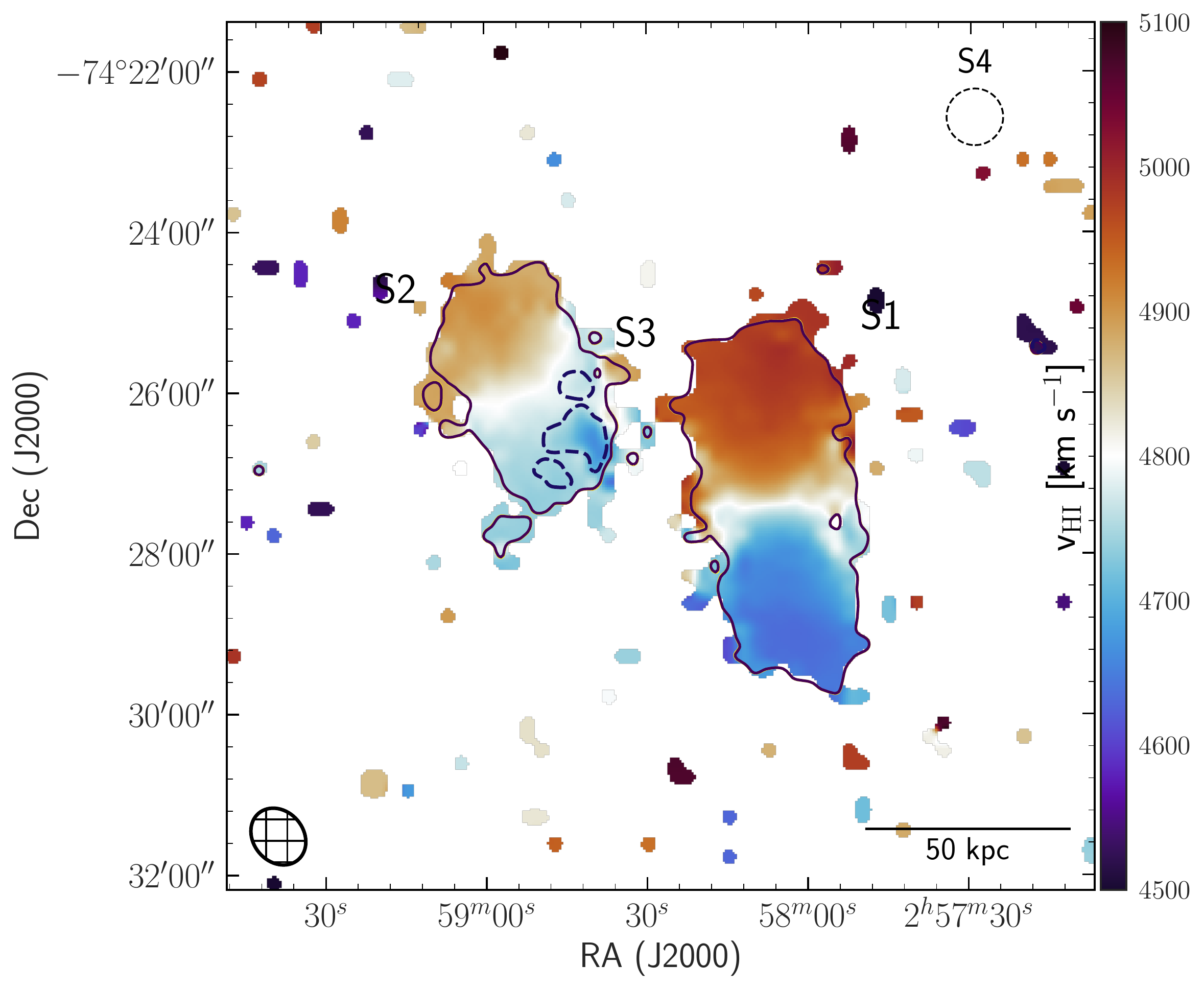}    
    \caption{}
\end{subfigure}

    \centering
    \begin{tabular}[t]{cc}

\begin{subfigure}{0.45\textwidth}
    \centering
    \smallskip
    \includegraphics[width=0.95\linewidth,height=1.4\textwidth]{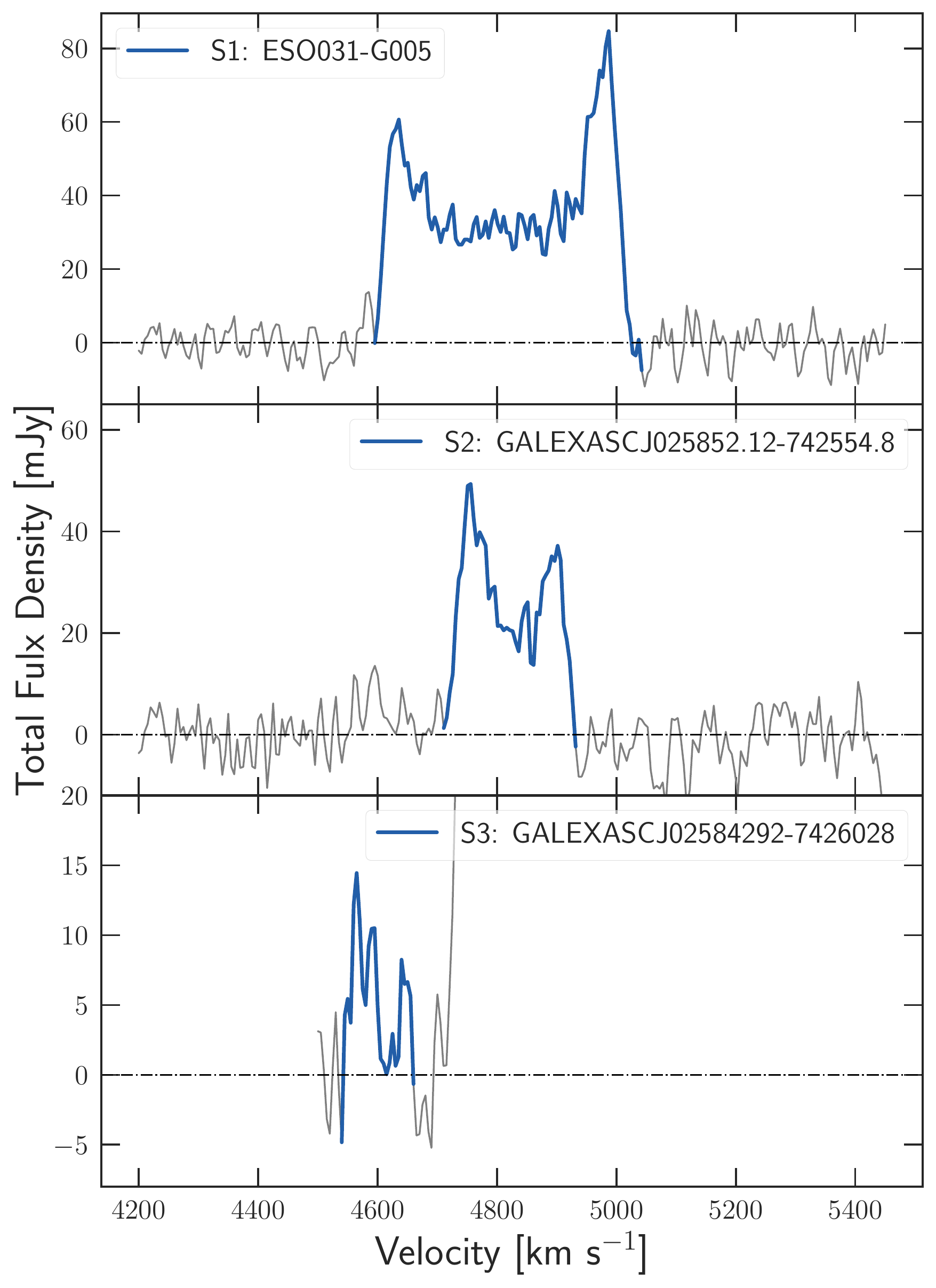}
    \caption{} 
\end{subfigure}
    &
        \begin{tabular}{c}
        \smallskip
            \begin{subfigure}[t]{0.42\textwidth}
                \centering
                \includegraphics[width=0.94\textwidth]{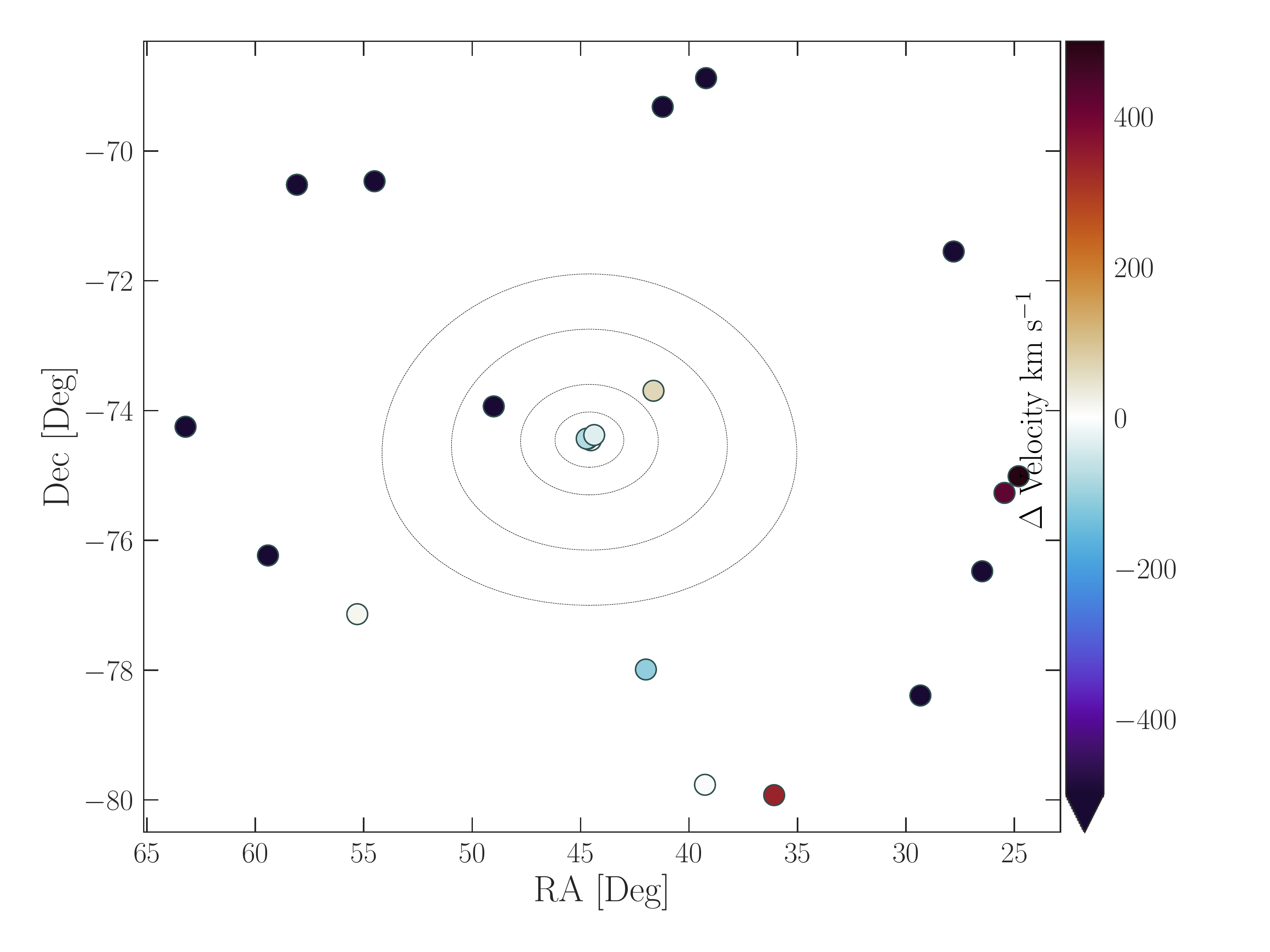}
                \caption{}
            \end{subfigure}\\
            \begin{subfigure}[t]{0.43\textwidth}
                \centering
                \includegraphics[width=0.95\textwidth]{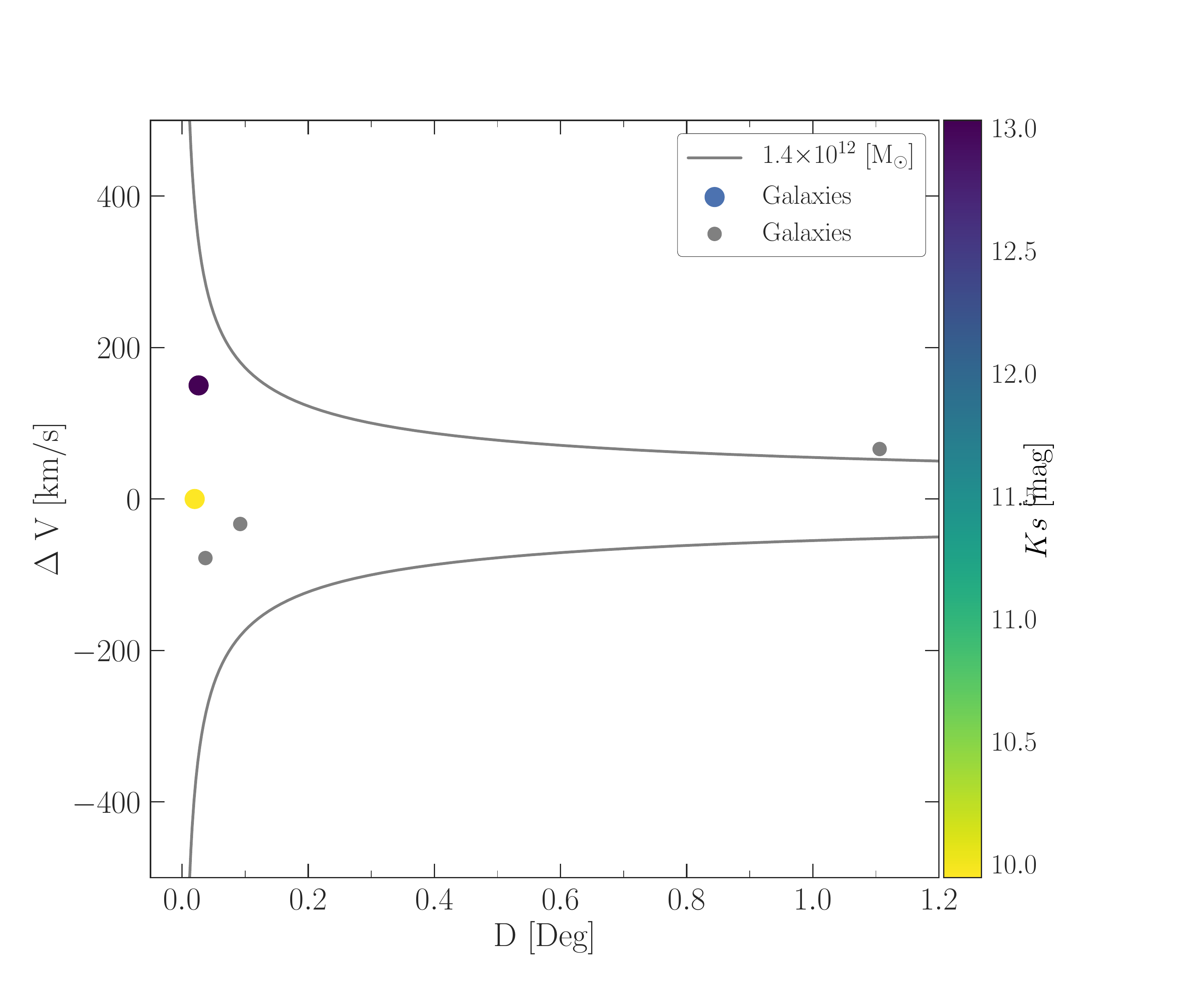}
                \caption{}
            \end{subfigure}
        \end{tabular}\\

    \end{tabular}
\caption{\textbf{a)} \textbf{Left:} The \HI\ emission in the J0258-74 group. The \HI\ emission is shown by the contours overlaid on the optical DSS B-band image. The lowest shown \HI\ column density 3$\times$10$^{19}$ cm$^{-2}$. The other \HI\ column density contours are: 8, 16, 32 and 64, 75 and 90$\times10^{19}$ cm$^{-2}$. The S3 galaxy is shown as shaded contours (between 3 and 6$\times10^{19}$ cm$^{-2}$) to enhance its visibility. The S4 galaxy is not detected in \HI. \textbf{Right:} The velocity field of the J0258-74 group in the limit from 4500 to 5100 \kms. The group members are marked (S1-S4) as well as the lowest \HI\ column density (3$\times$10$^{19}$ cm$^{-2}$). The dashed line show the position of the S3 galaxy to enhance its visibility. The synthesized beam is shown in the bottom left corner, and the scale bar in the bottom right corner shows 50 kpc at the group distance. \textbf{b)} Panels of the \HI\ spectra for each detected galaxy in the J0258-74 group. The enhanced colour shows where the spectrum was integrated. \textbf{c)} The global environment around J0258-74, centred on the weighted mean of the group (all four group galaxies are overlapping in the centre of the Figure). The black circles are denoting radii of 0.5, 1, 2 and 3 Mpc, respectively. We show the velocity difference between the S1 galaxy (\HIPASS\ source) and the other nearby galaxies. \textbf{d)} Projected angular separation of the galaxies within J0258-74 region versus recessional velocity difference between S1 galaxy (\HIPASS\ source) and other nearby galaxies. Galaxies associated with the group are within 0.2 deg. Grey solid curved lines show simple caustics curves for a potential of M $=$ 1.4$\times$10$^{12}$, assuming that all sources are at the J0258-74 distance of 70 Mpc.}
\label{fig:j0258_group}
\end{figure}

\begin{figure}

\begin{subfigure}{\columnwidth}
\centering
    \includegraphics[width=0.45\columnwidth]{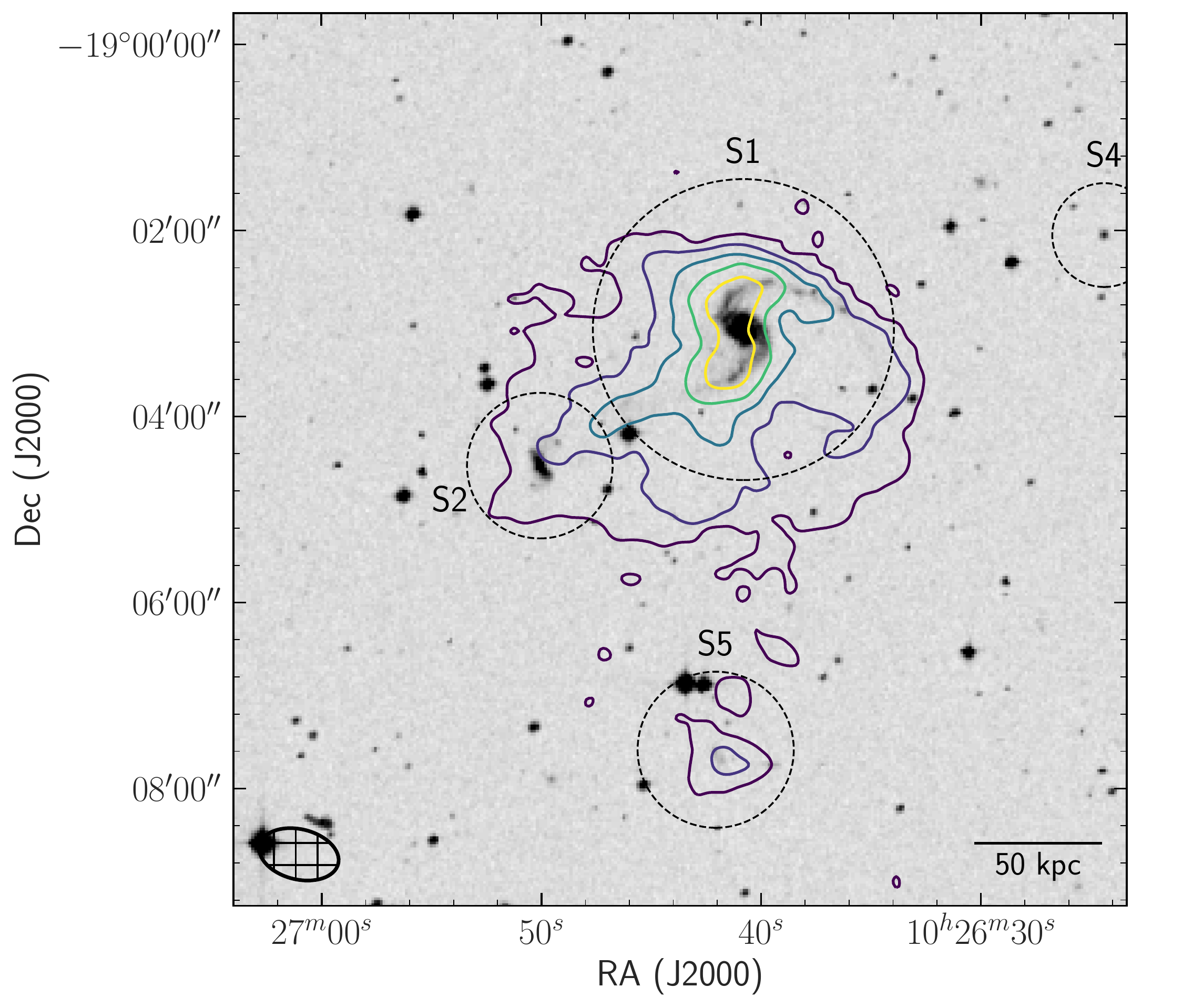}
    \includegraphics[width=0.45\columnwidth]{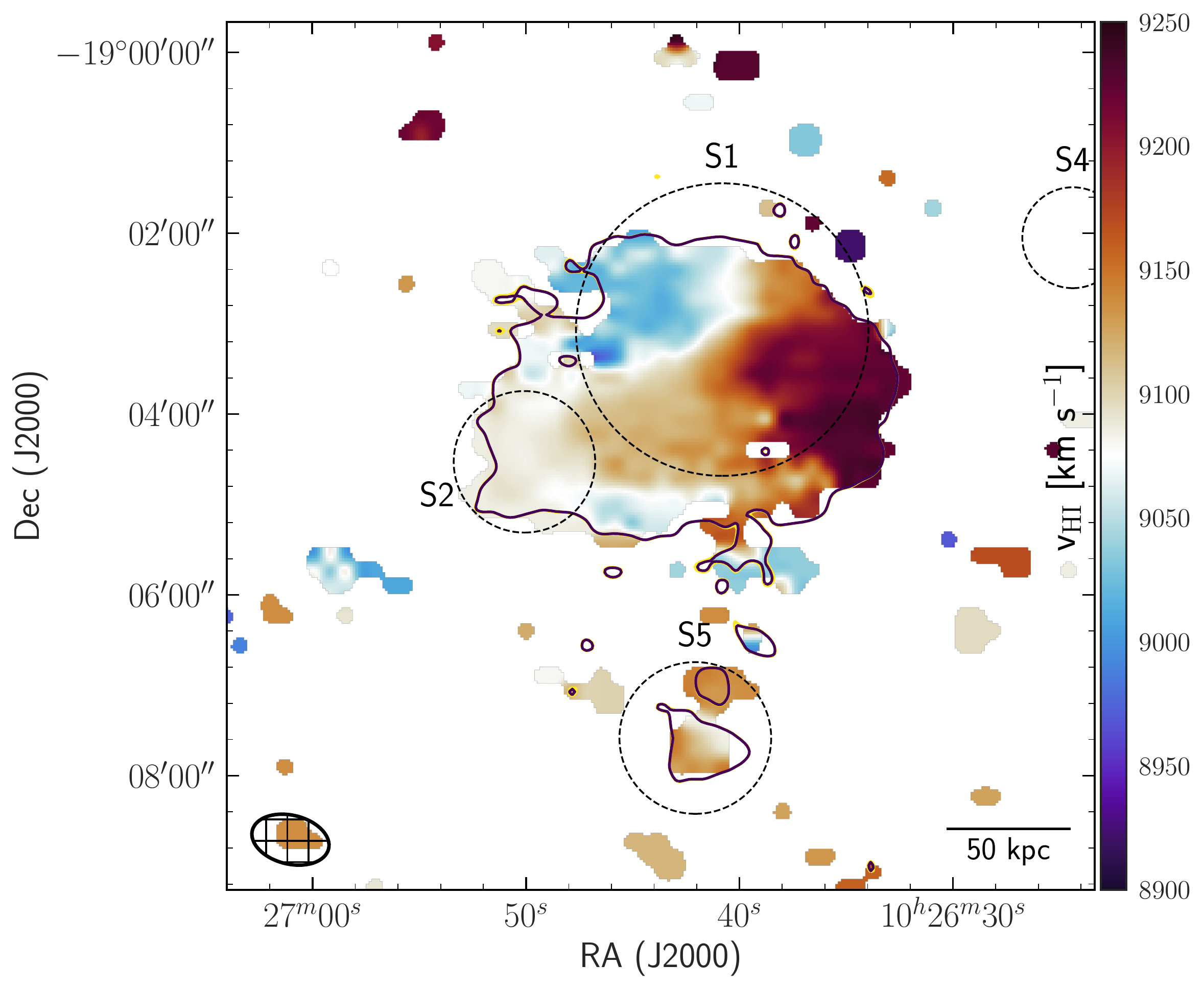}    
    \caption{}
\end{subfigure}

    \centering
    \begin{tabular}[t]{cc}

\begin{subfigure}{0.45\textwidth}
    \centering
    \smallskip
    \includegraphics[width=0.95\linewidth,height=1.4\textwidth]{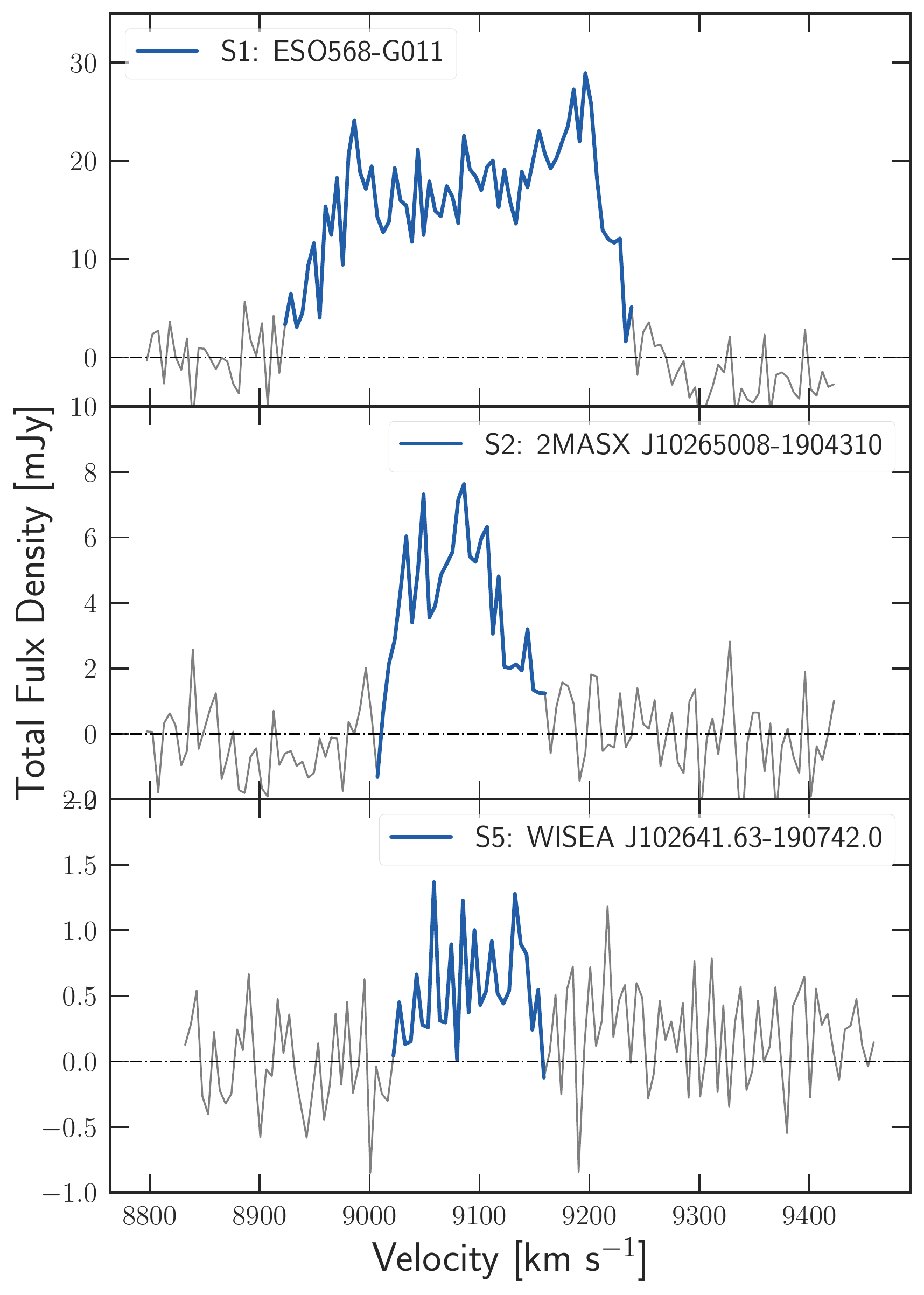}
    \caption{} 
\end{subfigure}
    &
        \begin{tabular}{c}
        \smallskip
            \begin{subfigure}[t]{0.42\textwidth}
                \centering
                \includegraphics[width=0.94\textwidth]{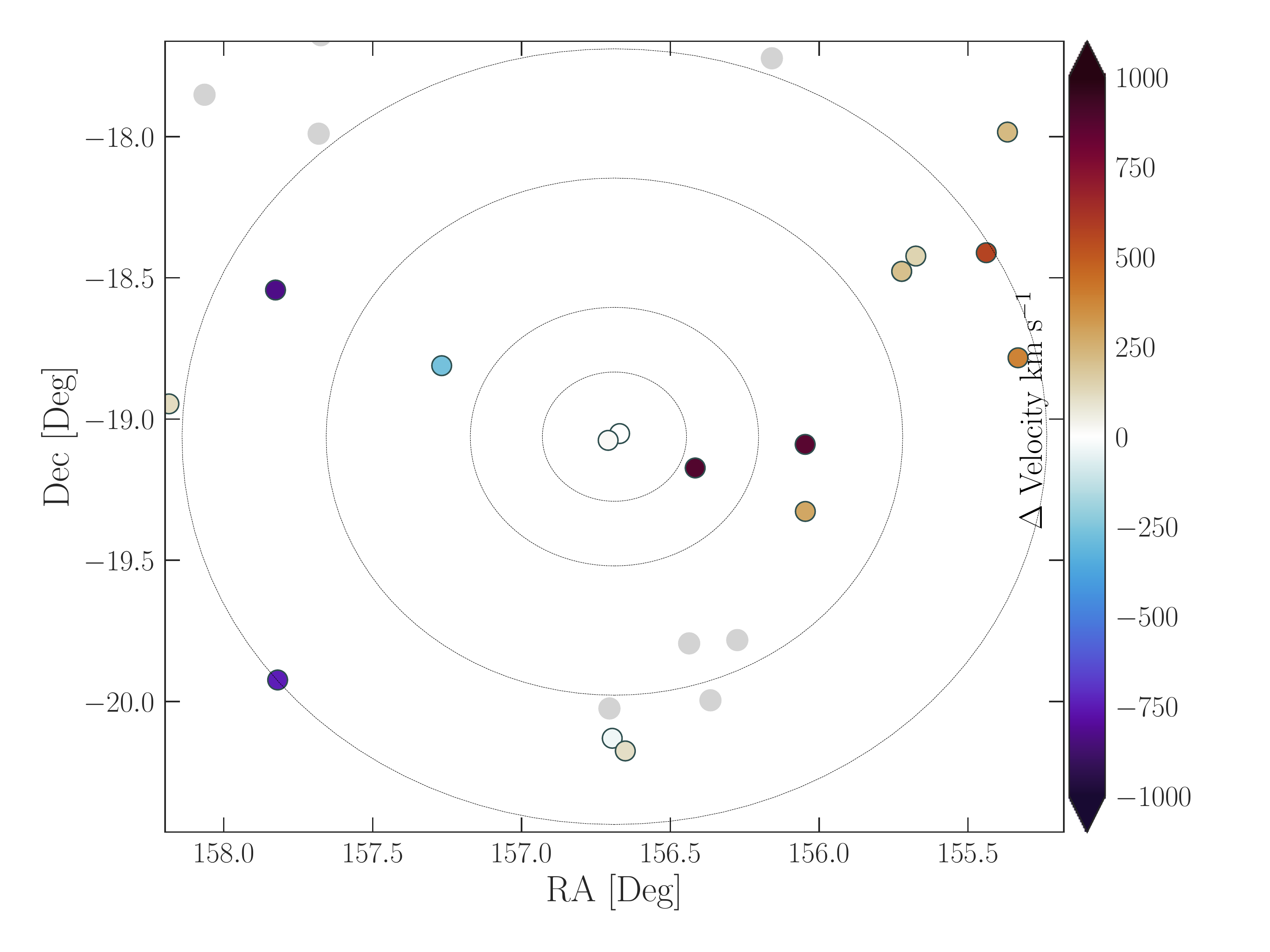}
                \caption{}
            \end{subfigure}\\
            \begin{subfigure}[t]{0.43\textwidth}
                \centering
                \includegraphics[width=0.95\textwidth]{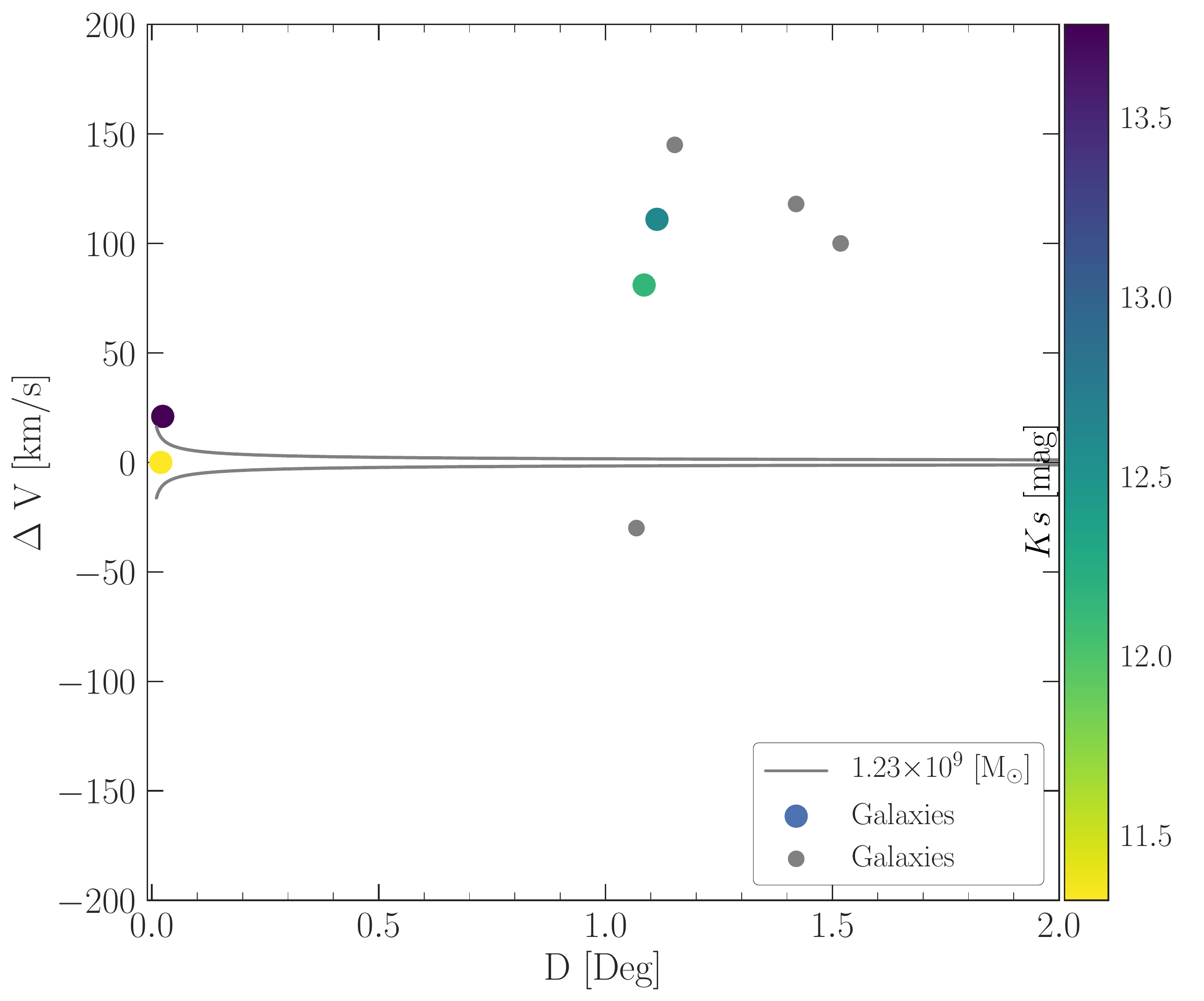}
                \caption{}
            \end{subfigure}
        \end{tabular}\\

    \end{tabular}
\caption{\textbf{a)} \textbf{Left:} The \HI\ emission in the J1026-19 group, zoomed in into the central region. The \HI\ emission is shown by the contours overlaid on the optical DSS B-band image. The lowest shown \HI\ column density 5$\times$10$^{19}$ cm$^{-2}$. The other \HI\ column density contours are: 15, 30, 50, and 70$\times10^{19}$ cm$^{-2}$. The S3, S4 and S6 galaxy are not detected in the \HI. \textbf{Right:} The velocity field of the J1026-19 group in the limit from 8900 to 9250 \kms. The group members that are within the field of view are marked (S1, S2, S4 and S5), and we show the lowest \HI\ column density (5$\times$10$^{19}$ cm$^{-2}$). The synthesized beam is shown in the bottom left corner, and the scale bar in the bottom right corner shows 50 kpc at the group distance. \textbf{b)} Panels of the \HI\ spectra for each detected galaxy in the J1026-19 group. The enhanced colour shows where the spectrum was integrated. \textbf{c)} The global environment around J1026-19, centred on the weighted mean of the group. The black circles are denoting radii of 0.5, 1, 2 and 3 Mpc, respectively. We show the velocity difference between the S1 galaxy (\HIPASS\ source) and the other nearby galaxies. \textbf{d)} Projected angular separation of the galaxies within J1026-19 region versus recessional velocity difference between S1 galaxy (\HIPASS\ source) and other nearby galaxies. Galaxies associated with the group are within 0.1 deg. Grey solid curved lines show simple caustics curves for a potential of 1.23$\times$10$^{9}$ M$\odot$, assuming that all sources are at the J1026-19 distance of 135 Mpc. }
\label{fig:j1026_group}
\end{figure}

\begin{figure}

\begin{subfigure}{\columnwidth}
\centering
    \includegraphics[width=0.45\columnwidth]{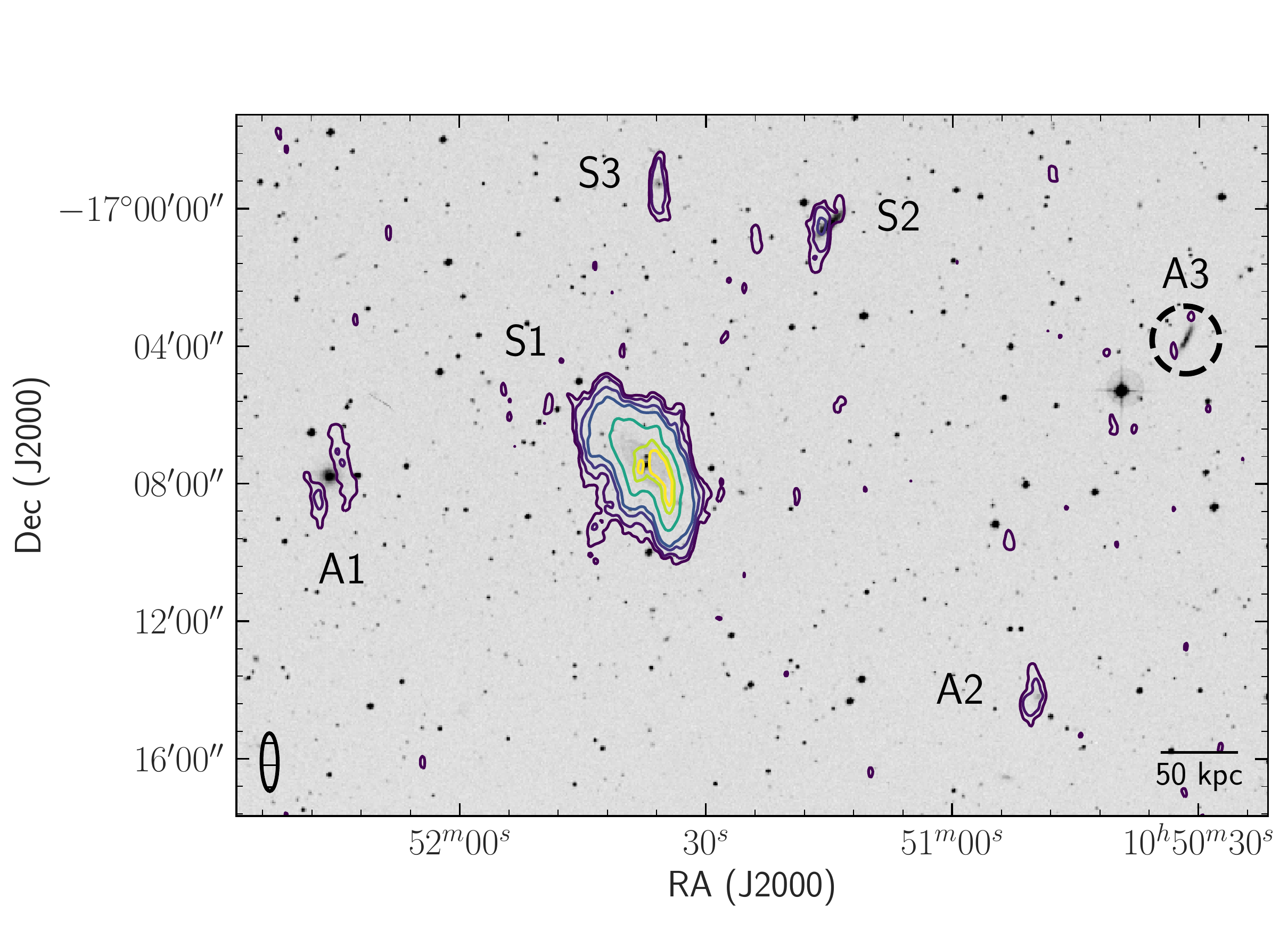}
    \includegraphics[width=0.45\columnwidth]{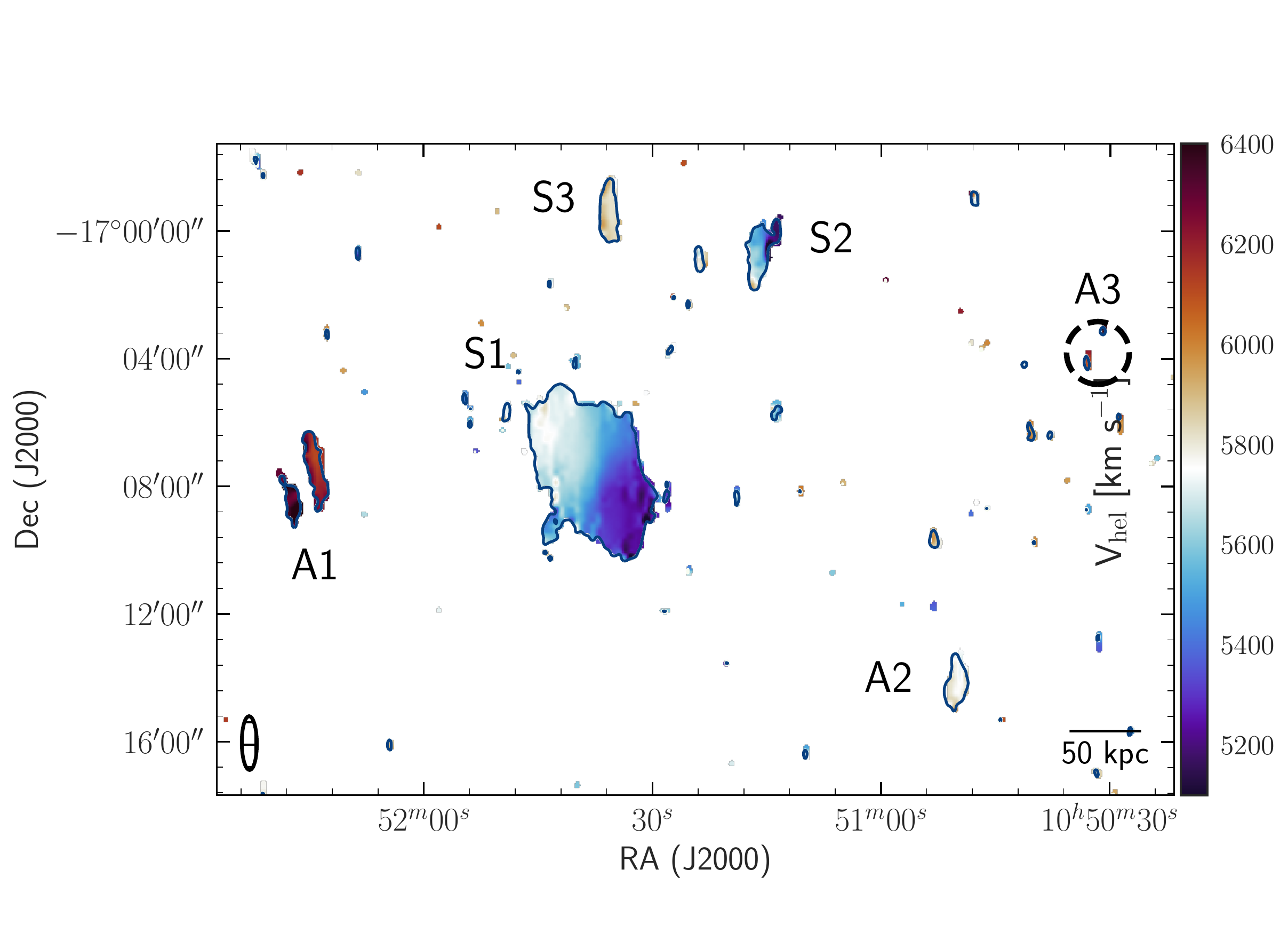}    
    \caption{}
\end{subfigure}

    \centering
    \begin{tabular}[t]{cc}

\begin{subfigure}{0.45\textwidth}
    \centering
    \smallskip
    \includegraphics[width=0.95\linewidth,height=1.6\textwidth]{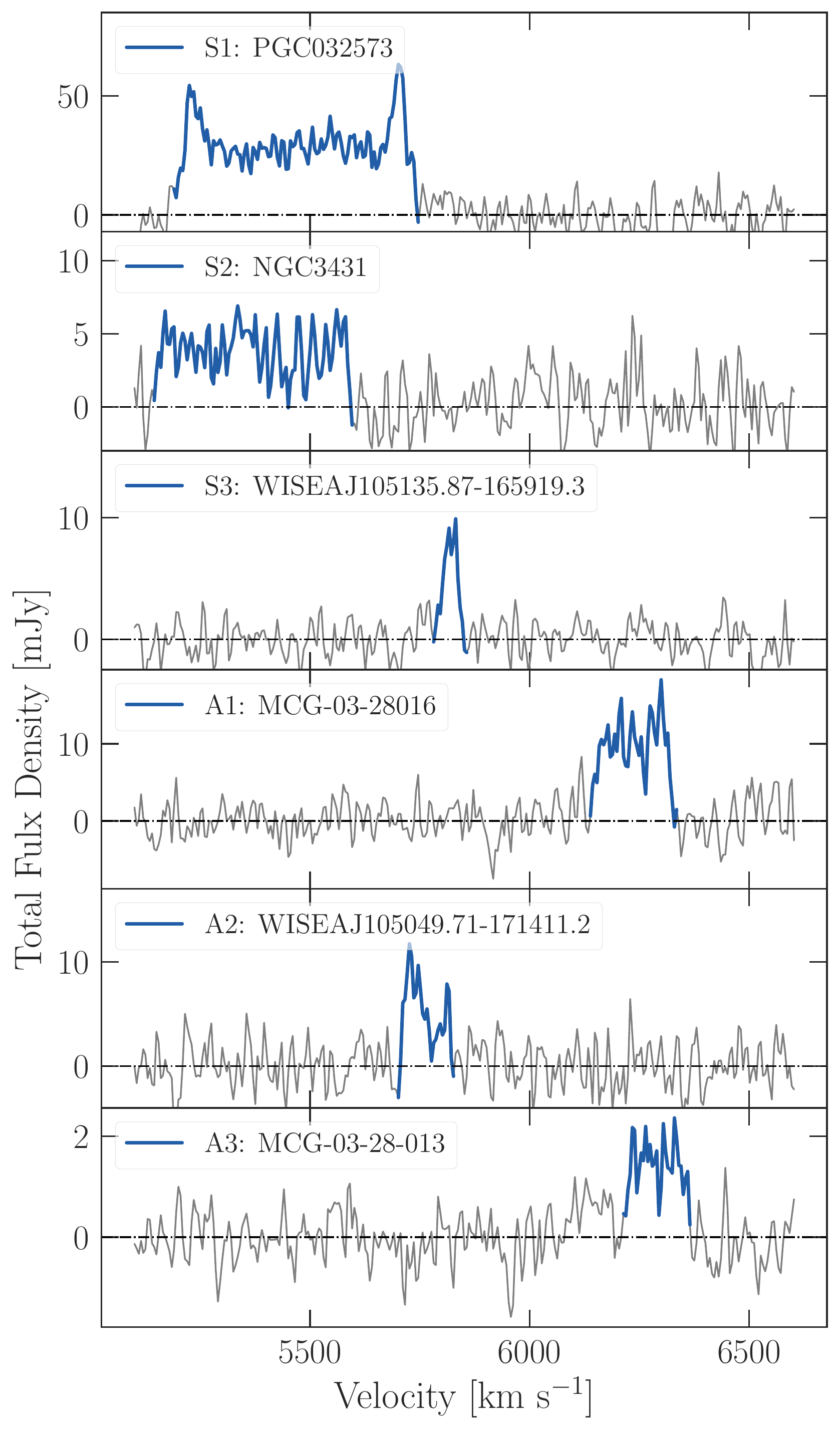}
    \caption{} 
\end{subfigure}
    &
        \begin{tabular}{c}
        \smallskip
            \begin{subfigure}[t]{0.42\textwidth}
                \centering
                \includegraphics[width=0.94\textwidth]{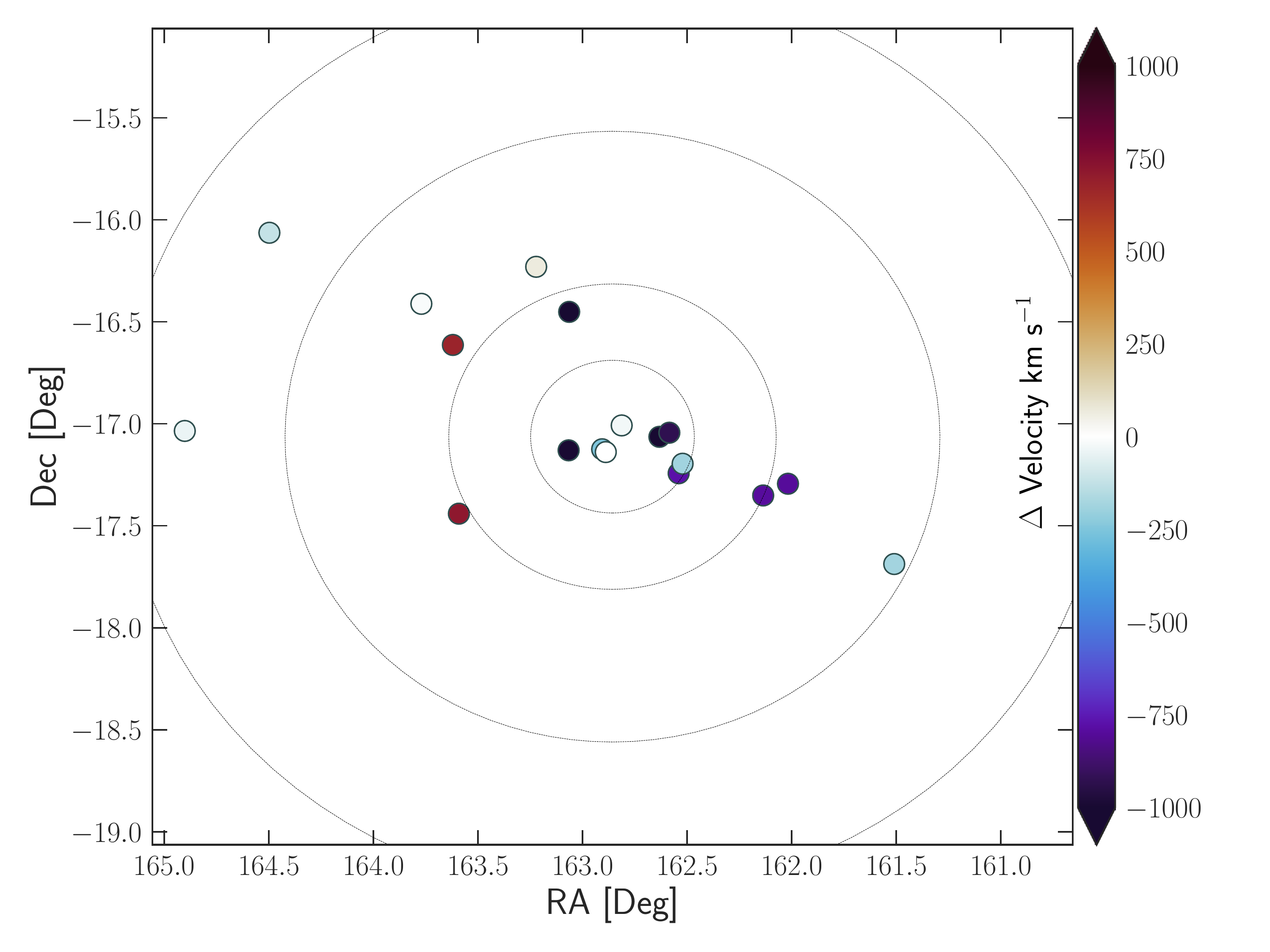}
                \caption{}
            \end{subfigure}\\
            \begin{subfigure}[t]{0.43\textwidth}
                \centering
                \includegraphics[width=0.95\textwidth]{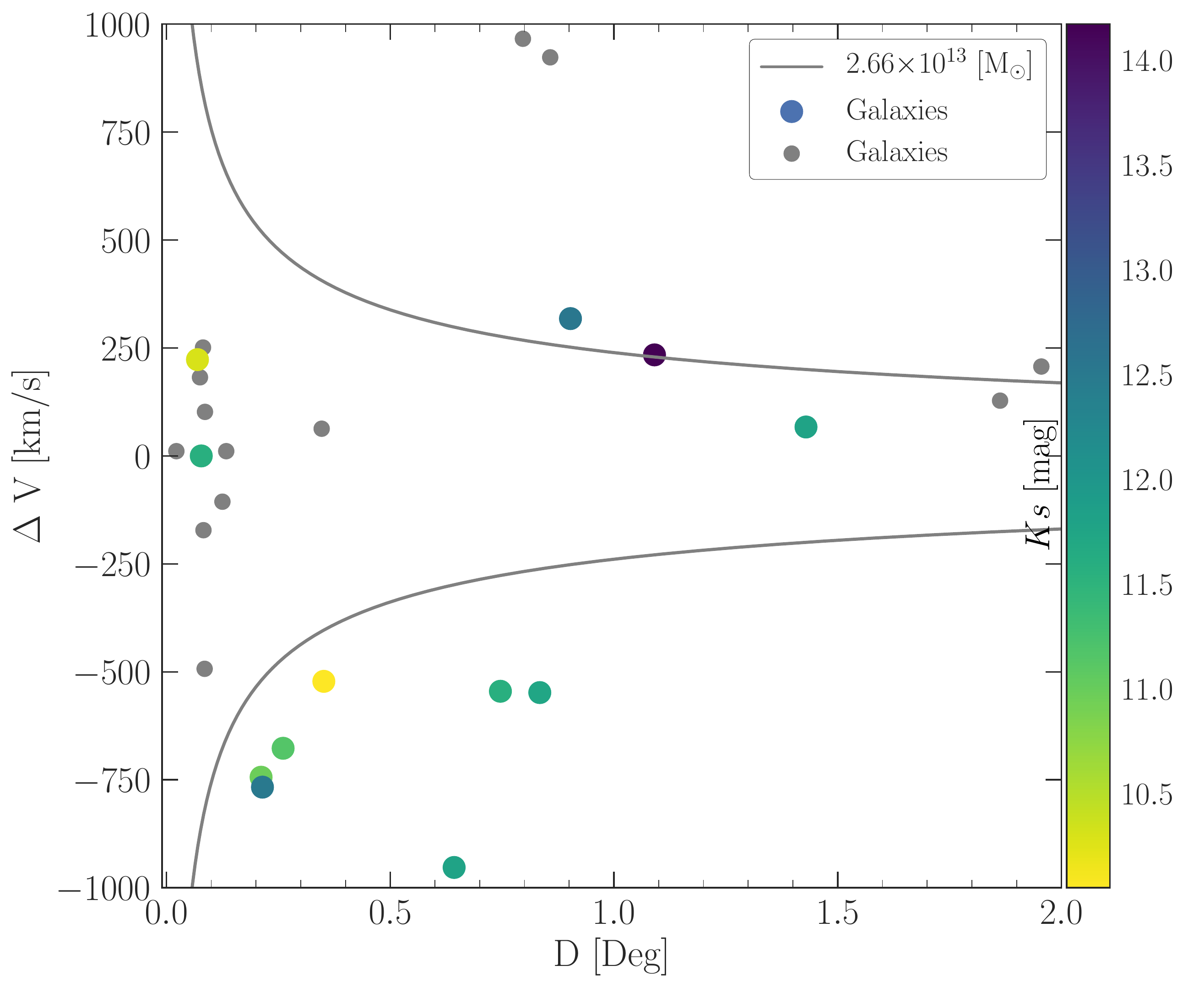}
                \caption{}
            \end{subfigure}
        \end{tabular}\\

    \end{tabular}
\caption{\textbf{a)} \textbf{Left:} The \HI\ emission in the J1051-17 group. The \HI\ emission is shown by the contours overlaid on the optical DSS B-band image. The lowest shown \HI\ column density 3$\times$10$^{19}$ cm$^{-2}$. The other \HI\ column density contours are: 6, 12, 18, 36, 54, and 60$\times10^{19}$ cm$^{-2}$. \textbf{Right:} The velocity field of the J1051-17 group in the limit from 5100 to 6400 \kms. The group members that are detected in \HI\ are marked (S1, S2, S3, A1, A2 and A3), and we show the lowest \HI\ column density (3$\times$10$^{19}$ cm$^{-2}$). The synthesized beam is shown in the bottom left corner, and the scale bar in the bottom right corner shows 50 kpc at the group distance. \textbf{b)} Panels of the \HI\ spectra for each detected galaxy in the J1051-17 group. The enhanced colour shows where the spectrum was integrated. \textbf{c)} The global environment around J1051-17, centred on the weighted mean of the group. The black circles are denoting radii of 0.5, 1, 2 and 3 Mpc, respectively. We show the velocity difference between the S1 galaxy (\HIPASS\ source) and the other nearby galaxies. \textbf{d)} Projected angular separation of the galaxies within J1051-17 region versus recessional velocity difference between S1 galaxy (\HIPASS\ source) and other nearby galaxies. Galaxies associated with the group are within 0.2 deg. Grey solid curved lines show simple caustics curves for a potential of 2.66$\times$10$^{13}$ M$\odot$, assuming that all sources are at the J1051-17 distance of 83 Mpc. }
\label{fig:j1051_group}
\end{figure}

\begin{figure}

\begin{subfigure}{\columnwidth}
\centering
    \includegraphics[width=0.45\columnwidth]{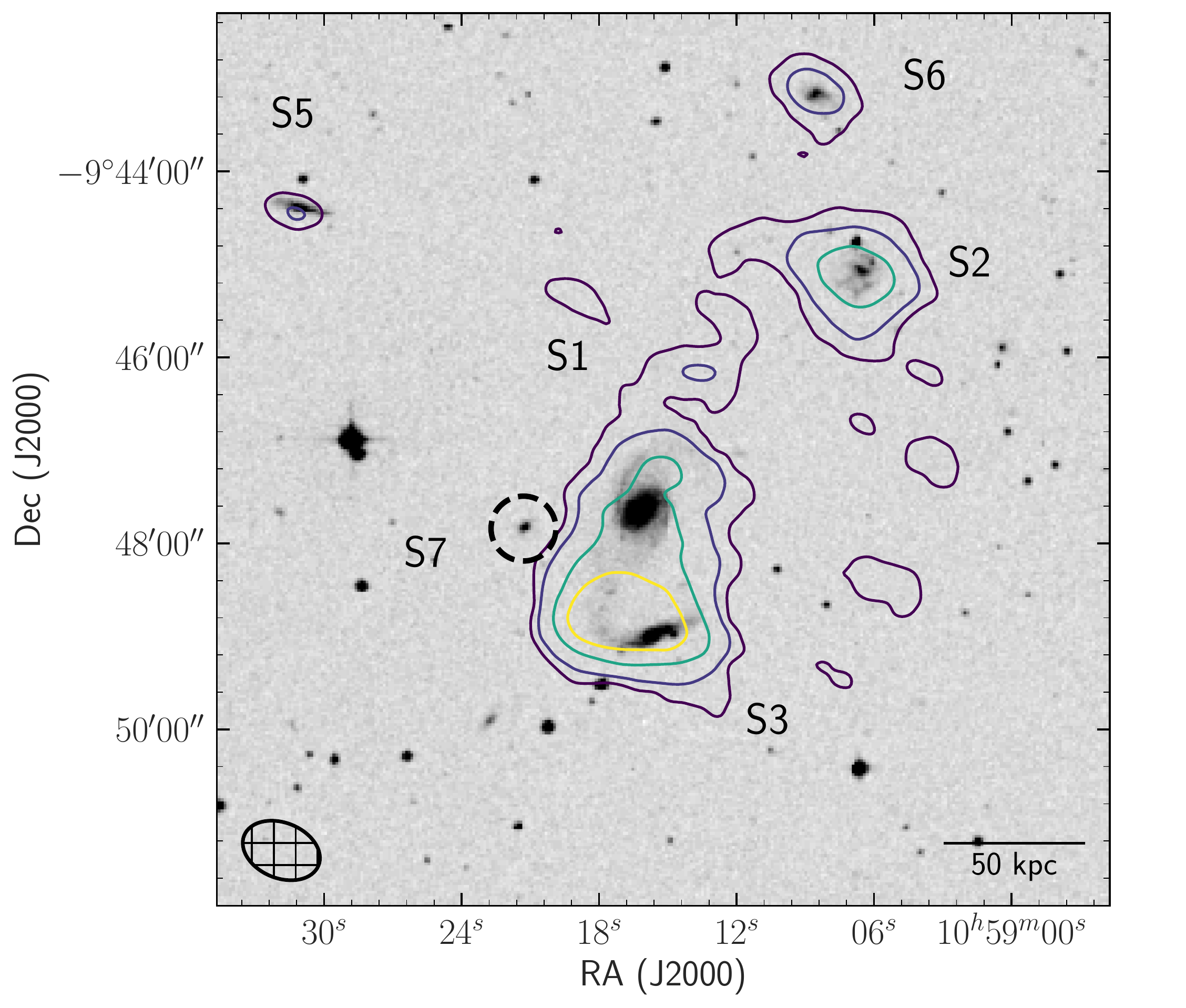}
    \includegraphics[width=0.45\columnwidth]{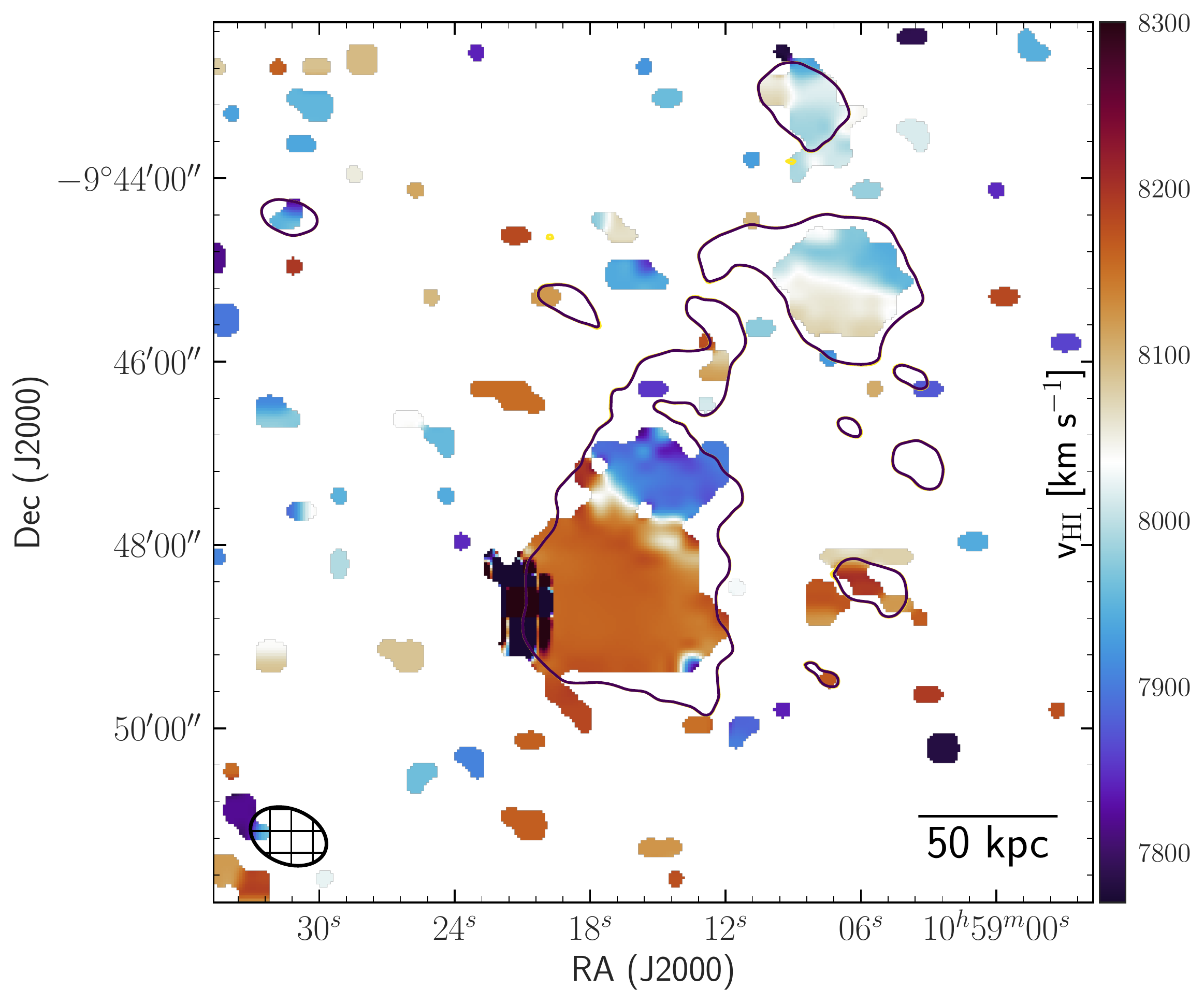}    
    \caption{}
\end{subfigure}

    \centering
    \begin{tabular}[t]{cc}

\begin{subfigure}{0.45\textwidth}
    \centering
    \smallskip
    \includegraphics[width=0.95\linewidth,height=1.4\textwidth]{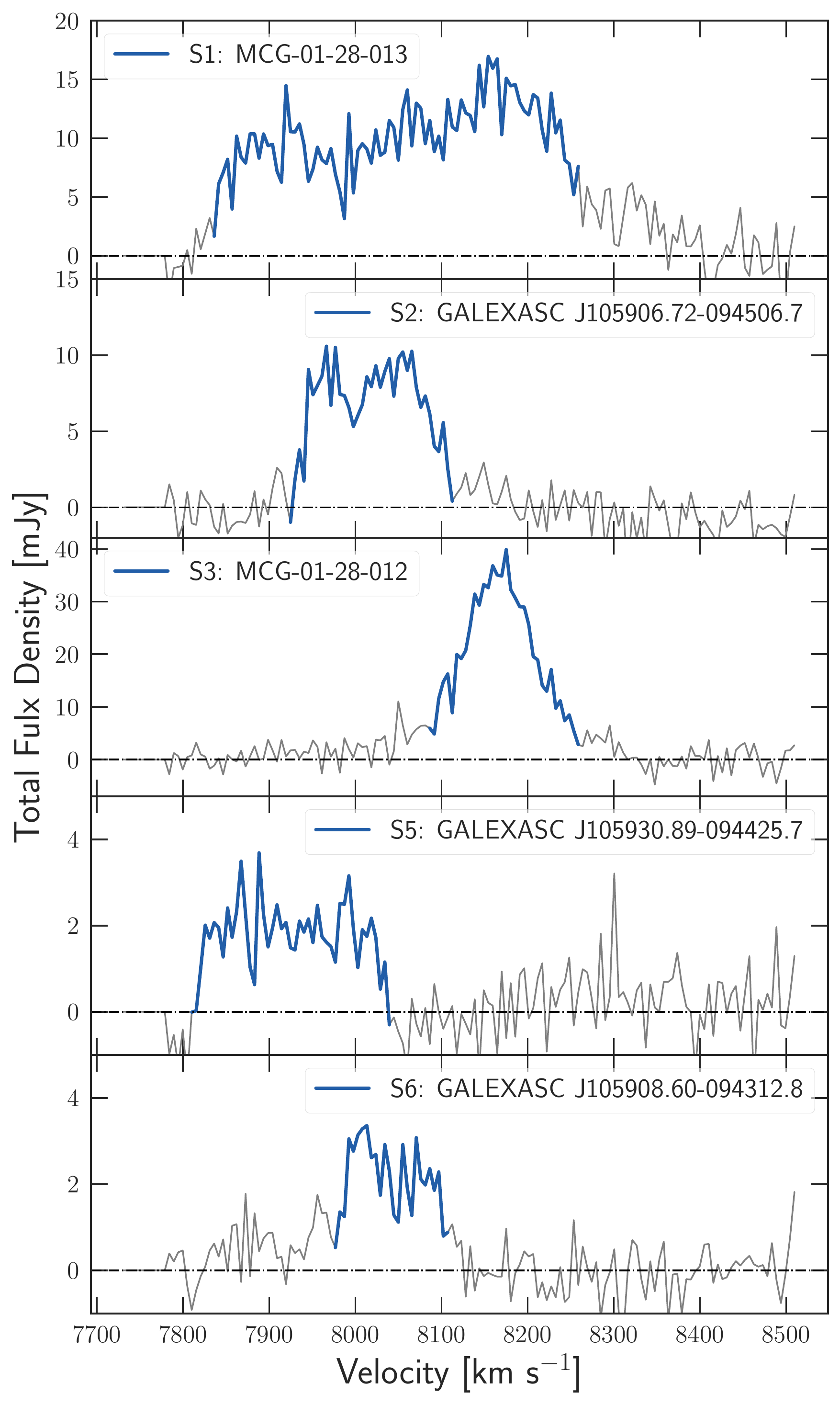}
    \caption{} 
\end{subfigure}
    &
        \begin{tabular}{c}
        \smallskip
            \begin{subfigure}[t]{0.42\textwidth}
                \centering
                \includegraphics[width=0.94\textwidth]{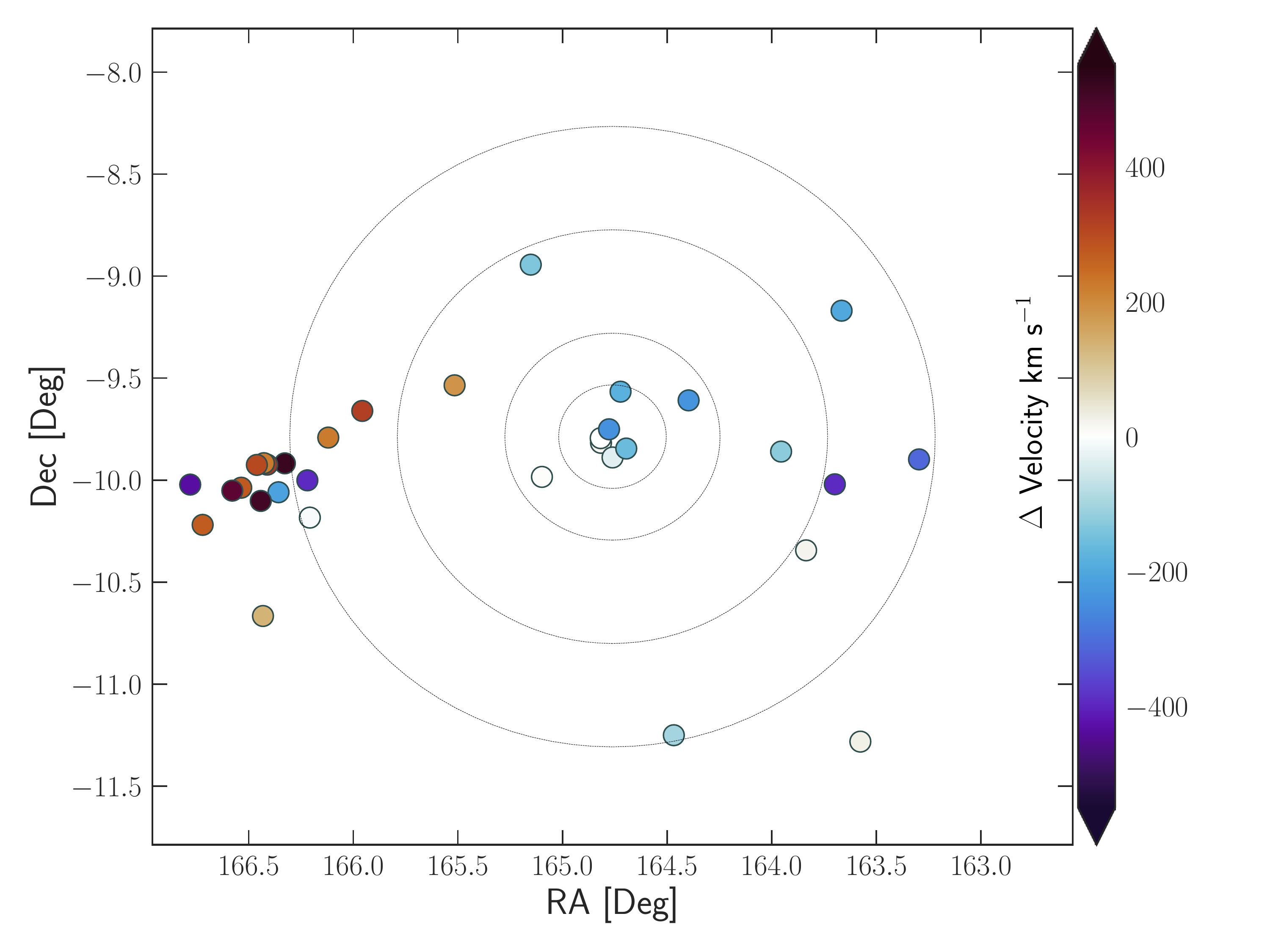}
                \caption{}
            \end{subfigure}\\
            \begin{subfigure}[t]{0.43\textwidth}
                \centering
                \includegraphics[width=0.95\textwidth]{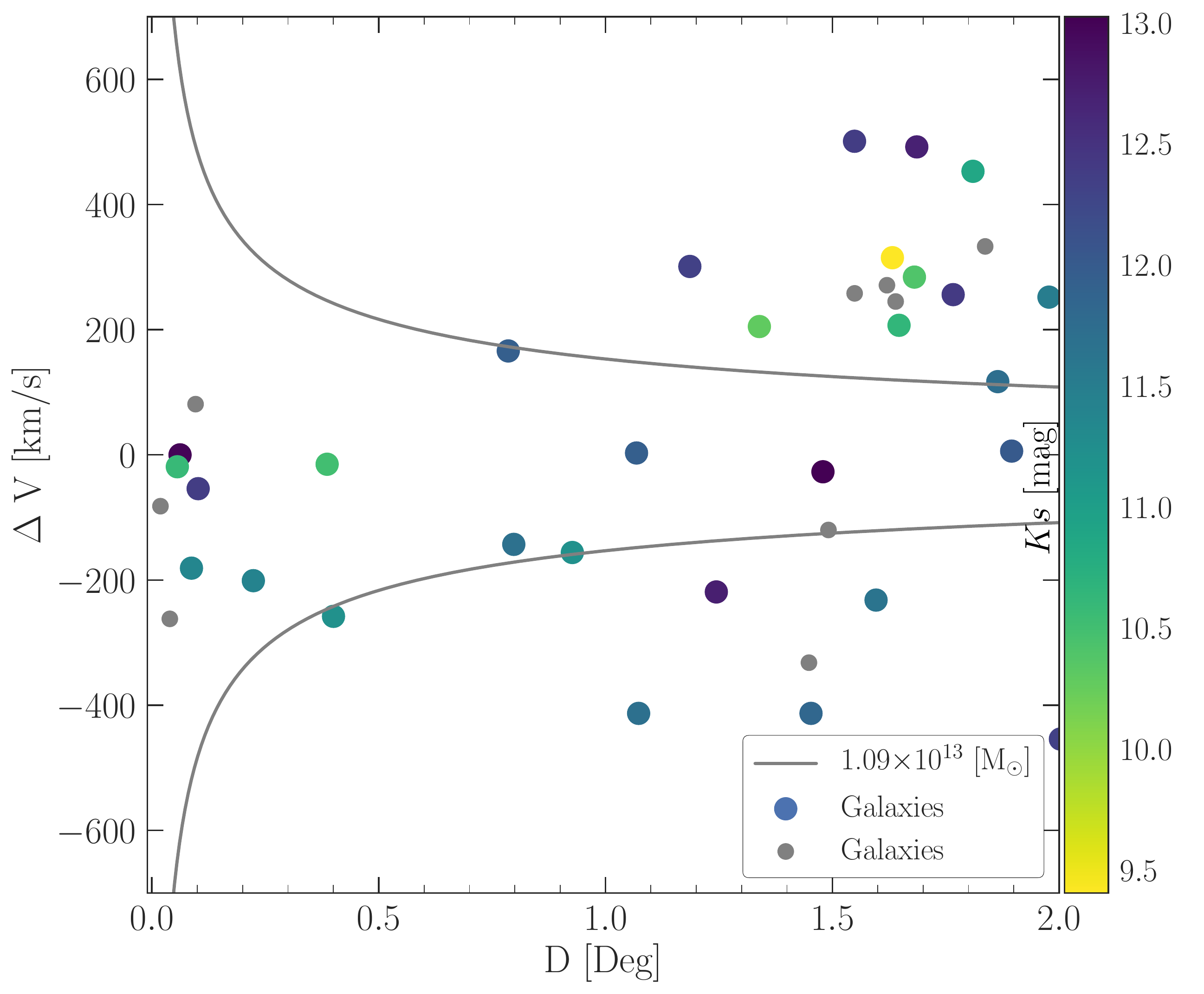}
                \caption{}
            \end{subfigure}
        \end{tabular}\\

    \end{tabular}
\caption{\textbf{a)} \textbf{Left:} The \HI\ emission in the J1059-09 group, zoomed in into the central region. The \HI\ emission is shown by the contours overlaid on the optical DSS B-band image. The lowest shown \HI\ column density 12$\times$10$^{29}$ cm$^{-2}$. The other \HI\ column density contours are: 20, 40, and 60$\times10^{19}$ cm$^{-2}$. \textbf{Right:} The velocity field of the J1059-09 group in the limit from 7770 to 8300 \kms. The group members are marked (S1-S4) as well as the lowest \HI\ column density (12$\times$10$^{19}$ cm$^{-2}$). The synthesized beam is shown in the bottom left corner, and the scale bar in the bottom right corner shows 50 kpc at the group distance. \textbf{b)} The \HI\ spectrum of the J1059-09 group obtained with ATCA. The enhanced colour shows where the spectrum was integrated. \textbf{c)} The global environment around J1059-09, centred on the weighted mean of the group. The black circles are denoting radii of 0.5, 1, 2 and 3 Mpc, respectively. We show the velocity difference between the S1 galaxy (\HIPASS\ source) and the other nearby galaxies. \textbf{d)} Projected angular separation of the galaxies within J1059-09 region versus recessional velocity difference between S1 galaxy (\HIPASS\ source) and other nearby galaxies. Grey solid curved lines show simple caustics curves for a potential of 1.09$\times$10$^{13}$ M$\odot$, assuming that all sources are at the J1059-09 distance of 122 Mpc.
}
\label{fig:j1059_group}
\end{figure}

\begin{figure}

\begin{subfigure}{\columnwidth}
\centering
    \includegraphics[width=0.45\columnwidth]{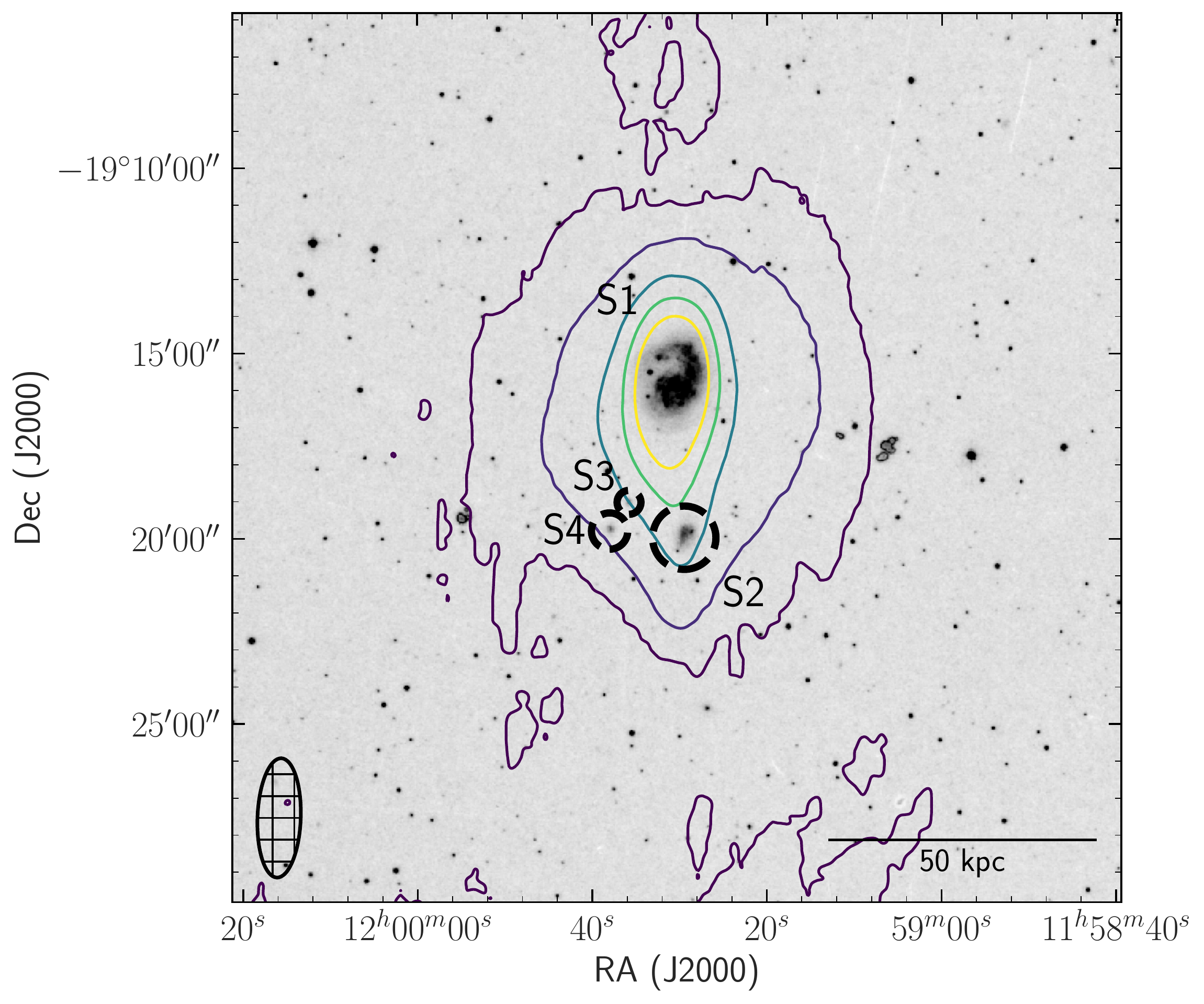}
    \includegraphics[width=0.45\columnwidth]{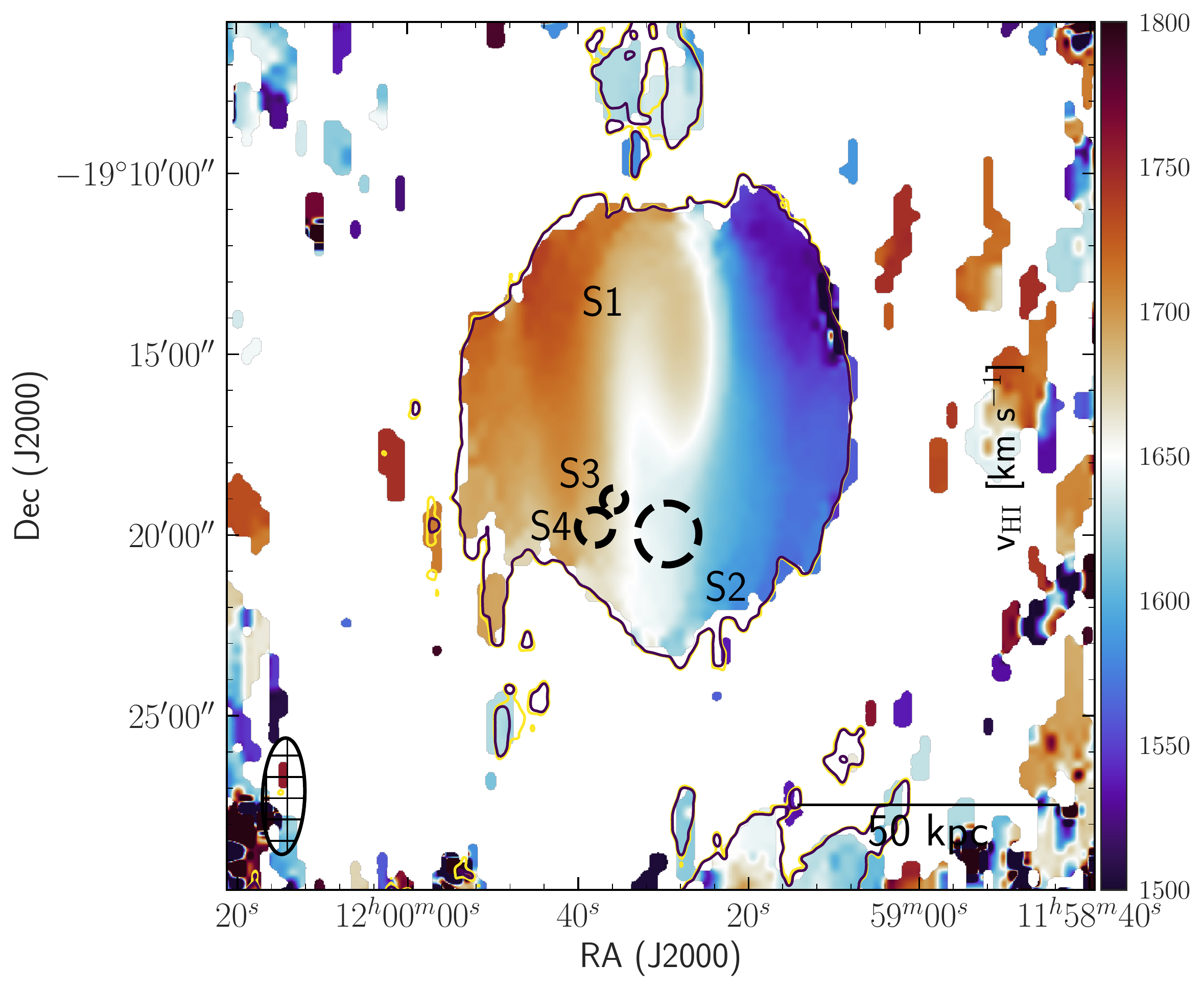}    
    \caption{}
\end{subfigure}

    \centering
    \begin{tabular}[t]{cc}

\begin{subfigure}{0.45\textwidth}
    \centering
    \smallskip
    \includegraphics[width=0.95\linewidth,height=1\textwidth]{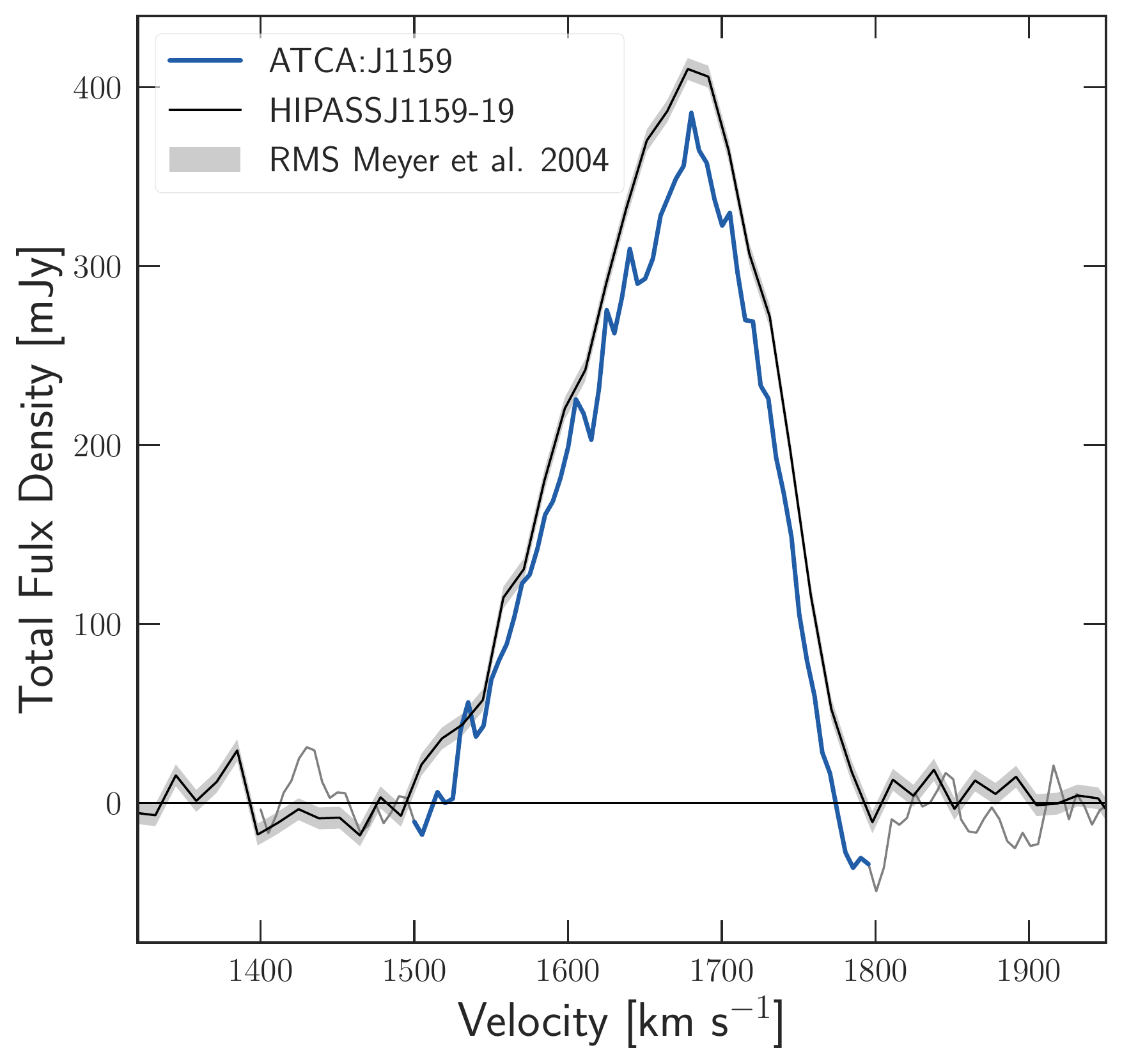}
    \caption{} 
\end{subfigure}
    &
        \begin{tabular}{c}
        \smallskip
            \begin{subfigure}[t]{0.42\textwidth}
                \centering
                \includegraphics[width=0.94\textwidth]{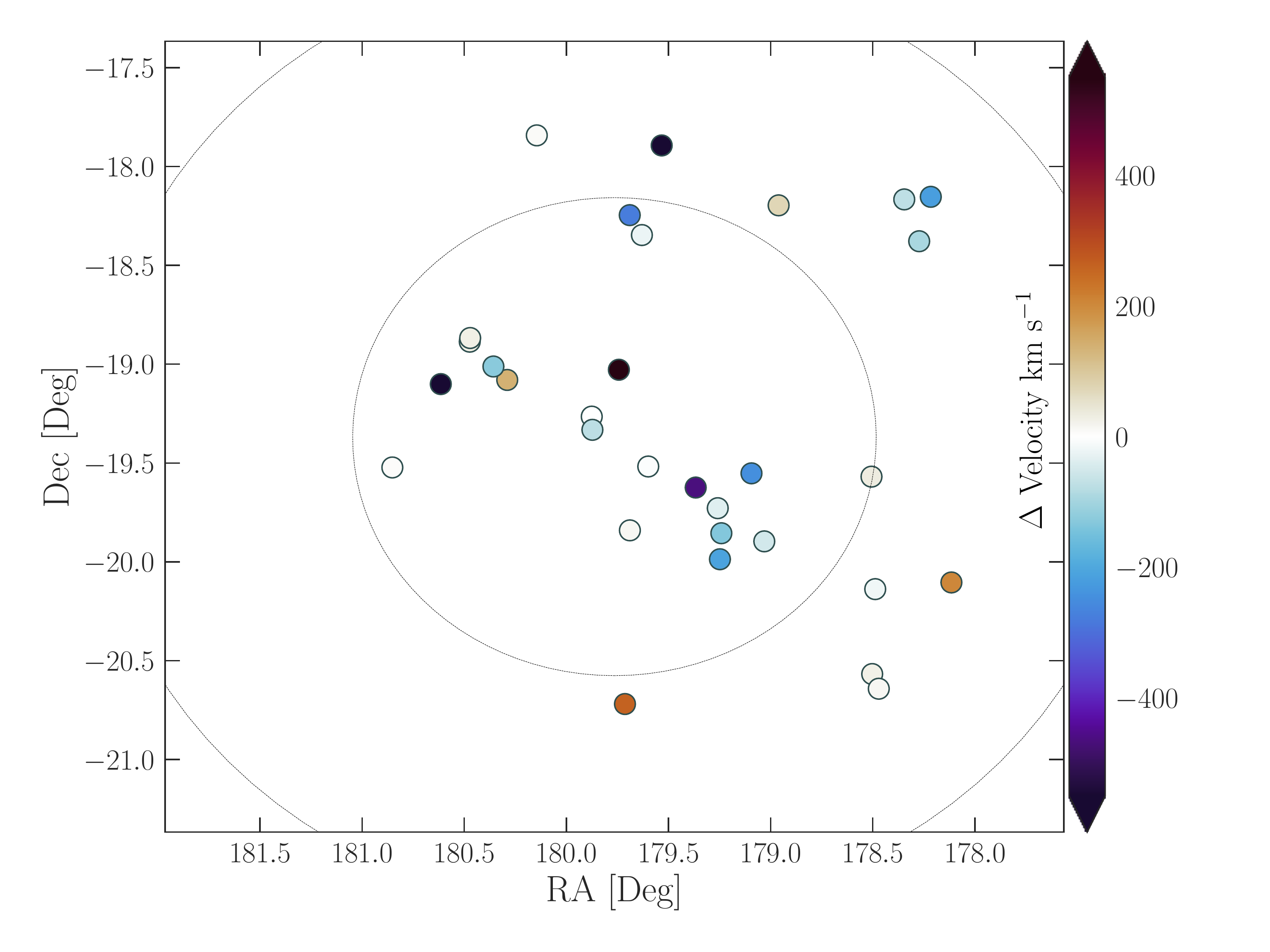}
                \caption{}
            \end{subfigure}\\
            \begin{subfigure}[t]{0.43\textwidth}
                \centering
                \includegraphics[width=0.95\textwidth]{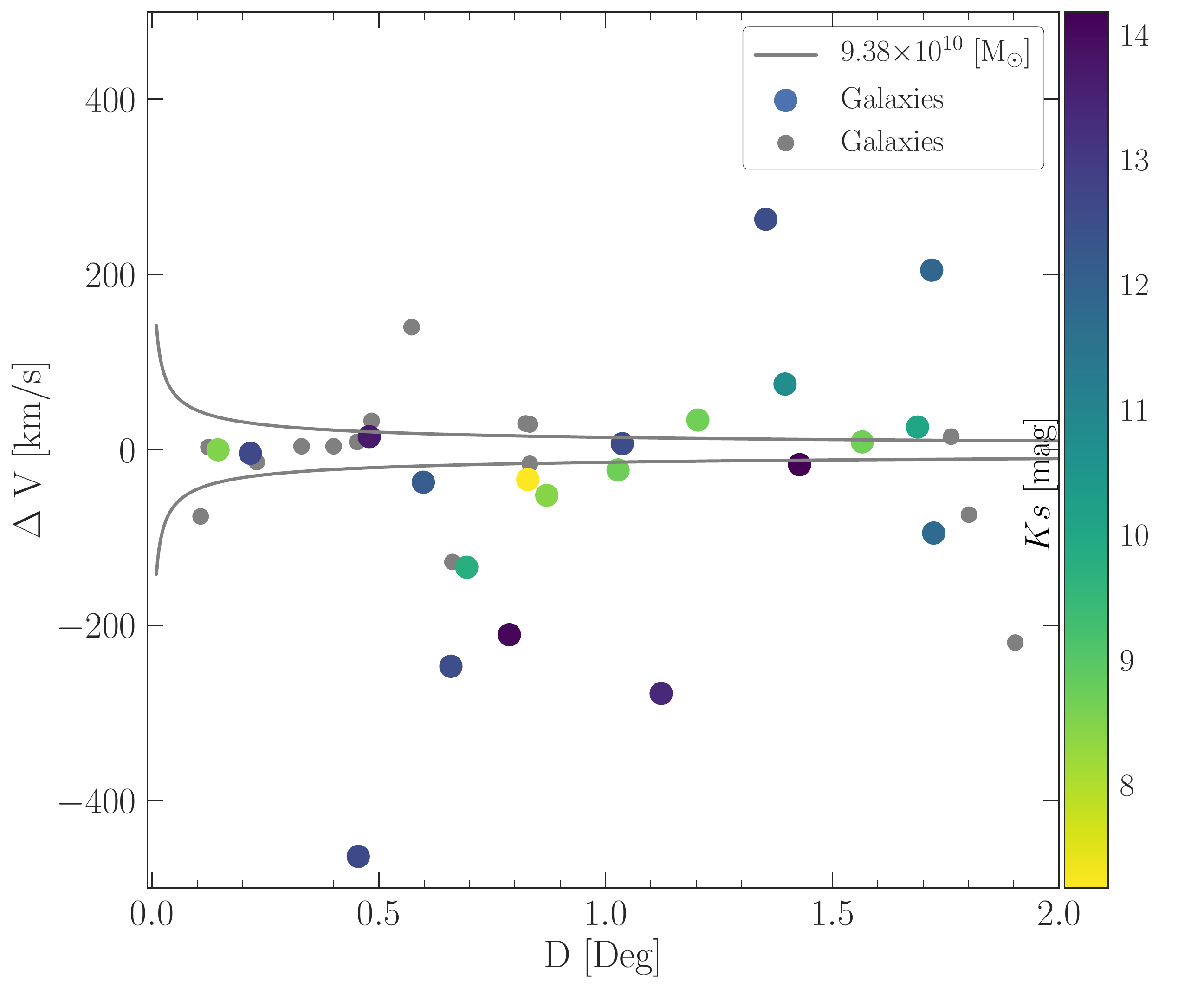}
                \caption{}
            \end{subfigure}
        \end{tabular}\\

    \end{tabular}
\caption{\textbf{a)} \textbf{Left:} The \HI\ emission in the J1159-19 group. The \HI\ emission is shown by the contours overlaid on the optical DSS B-band image. The lowest shown \HI\ column density 1$\times$10$^{19}$ cm$^{-2}$. The other \HI\ column density contours are: 10, 30, 50, and 70$\times10^{19}$ cm$^{-2}$. \textbf{Right:} The velocity field of the J1159-19 group in the limit from 1500 to 1800 \kms. The group members are marked (S1-S4) as well as the lowest \HI\ column density (1$\times$10$^{19}$ cm$^{-2}$). The synthesized beam is shown in the bottom left corner, and the scale bar in the bottom right corner shows 50 kpc at the group distance. \textbf{b)} The \HI\ spectrum of the J1159-19 group obtained with ATCA and HIPASS. The enhanced colour shows where the spectrum was integrated. \textbf{c)} The global environment around J1159-19, centred on the weighted mean of the group. The black circles are denoting radii of 0.5 and 1 Mpc, respectively. We show the velocity difference between the S1 galaxy (\HIPASS\ source) and the other nearby galaxies. \textbf{d)} Projected angular separation of the galaxies within J1159-19 region versus recessional velocity difference between S1 galaxy (\HIPASS\ source) and other nearby galaxies. Grey solid curved lines show simple caustics curves for a potential of M $=$ 9.38$\times$10$^{10}$, assuming that all sources are at the J1159-19 distance of 25 Mpc.}
\label{fig:j1159_group}
\end{figure}

\begin{figure}

\begin{subfigure}{\columnwidth}
\centering
    \includegraphics[width=0.45\columnwidth]{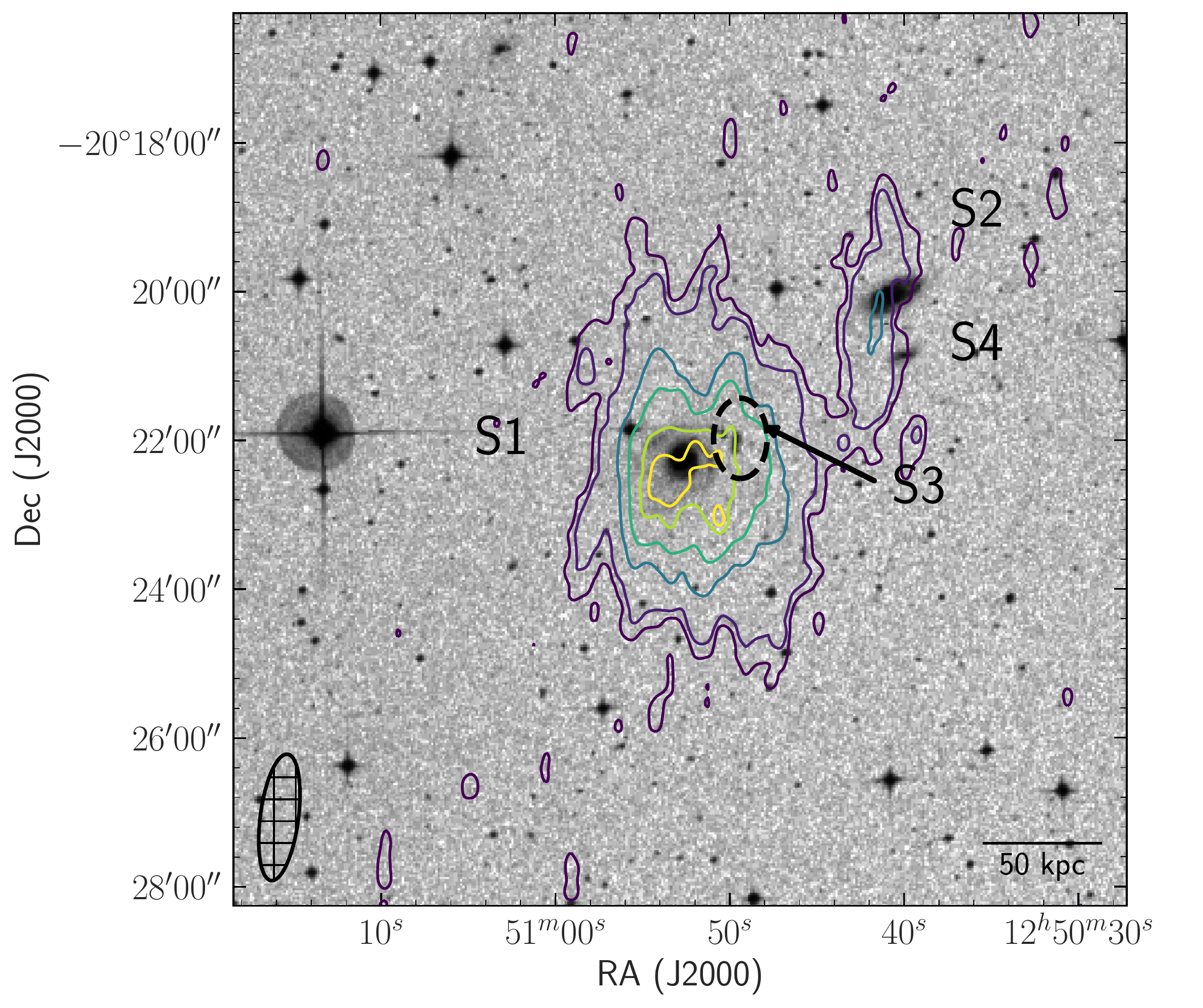}
    \includegraphics[width=0.45\columnwidth]{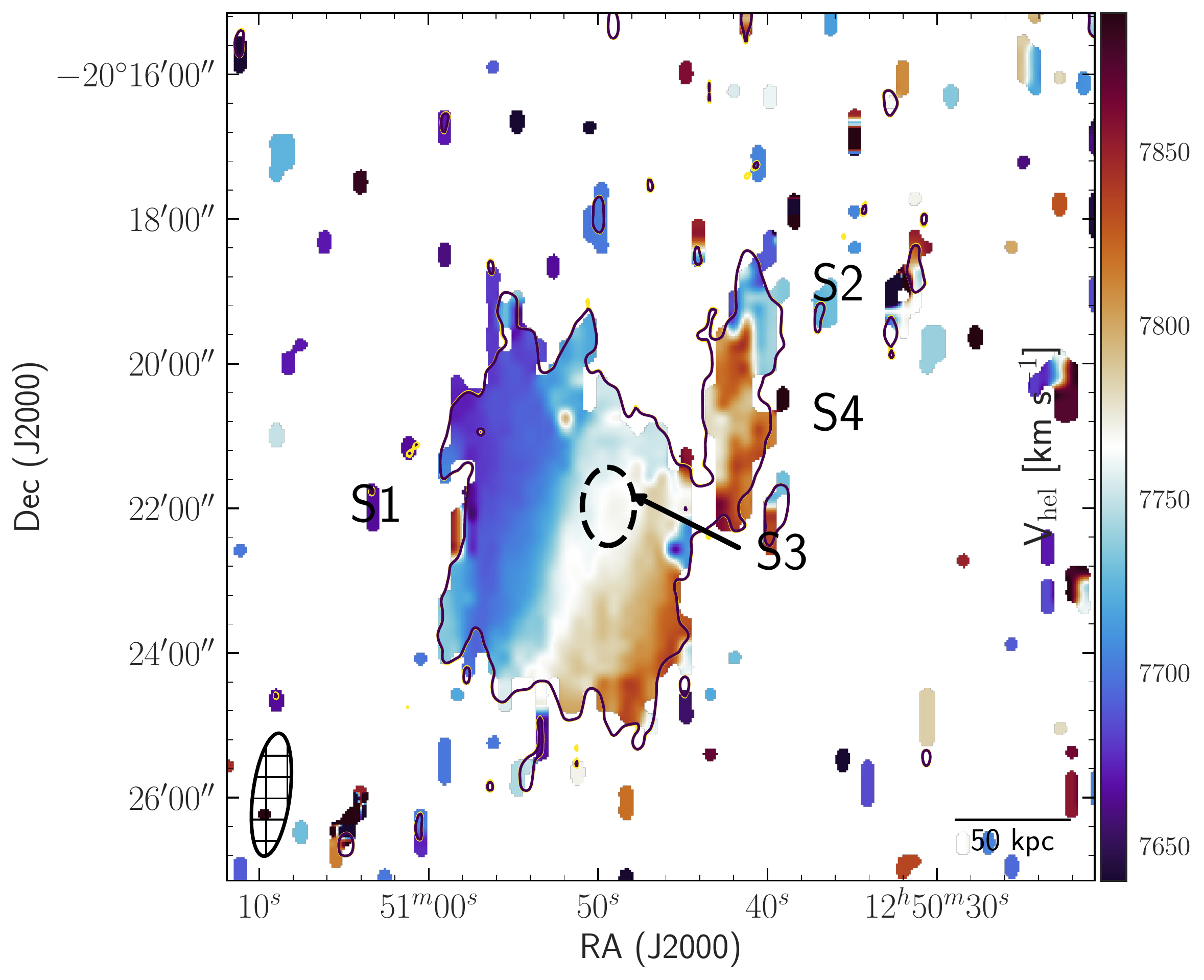}    
    \caption{}
\end{subfigure}

    \centering
    \begin{tabular}[t]{cc}

\begin{subfigure}{0.45\textwidth}
    \centering
    \smallskip
    \includegraphics[width=0.95\linewidth,height=1.2\textwidth]{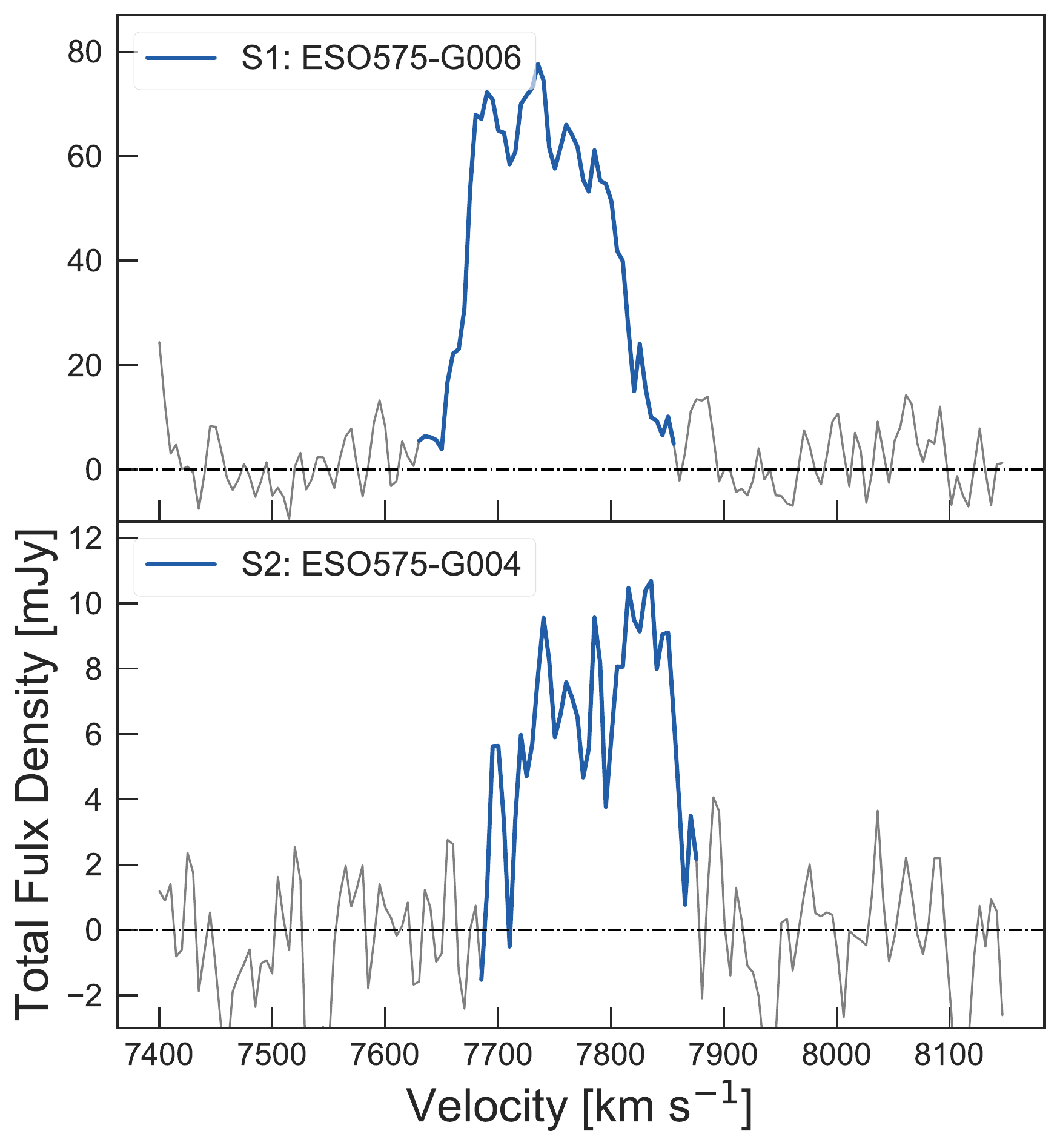}
    \caption{} 
\end{subfigure}
    &
        \begin{tabular}{c}
        \smallskip
            \begin{subfigure}[t]{0.42\textwidth}
                \centering
                \includegraphics[width=0.94\textwidth]{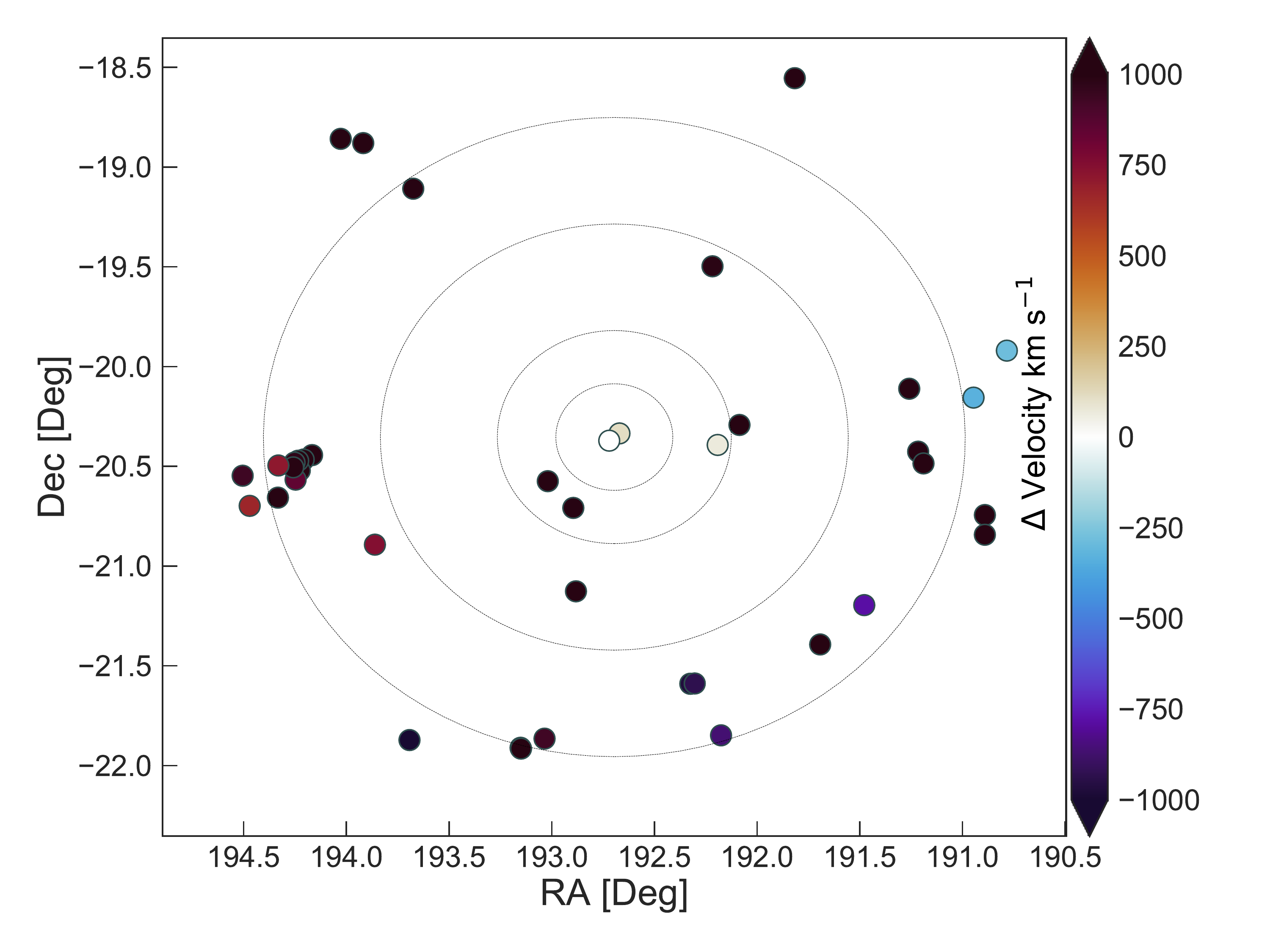}
                \caption{}
            \end{subfigure}\\
            \begin{subfigure}[t]{0.43\textwidth}
                \centering
                \includegraphics[width=0.95\textwidth]{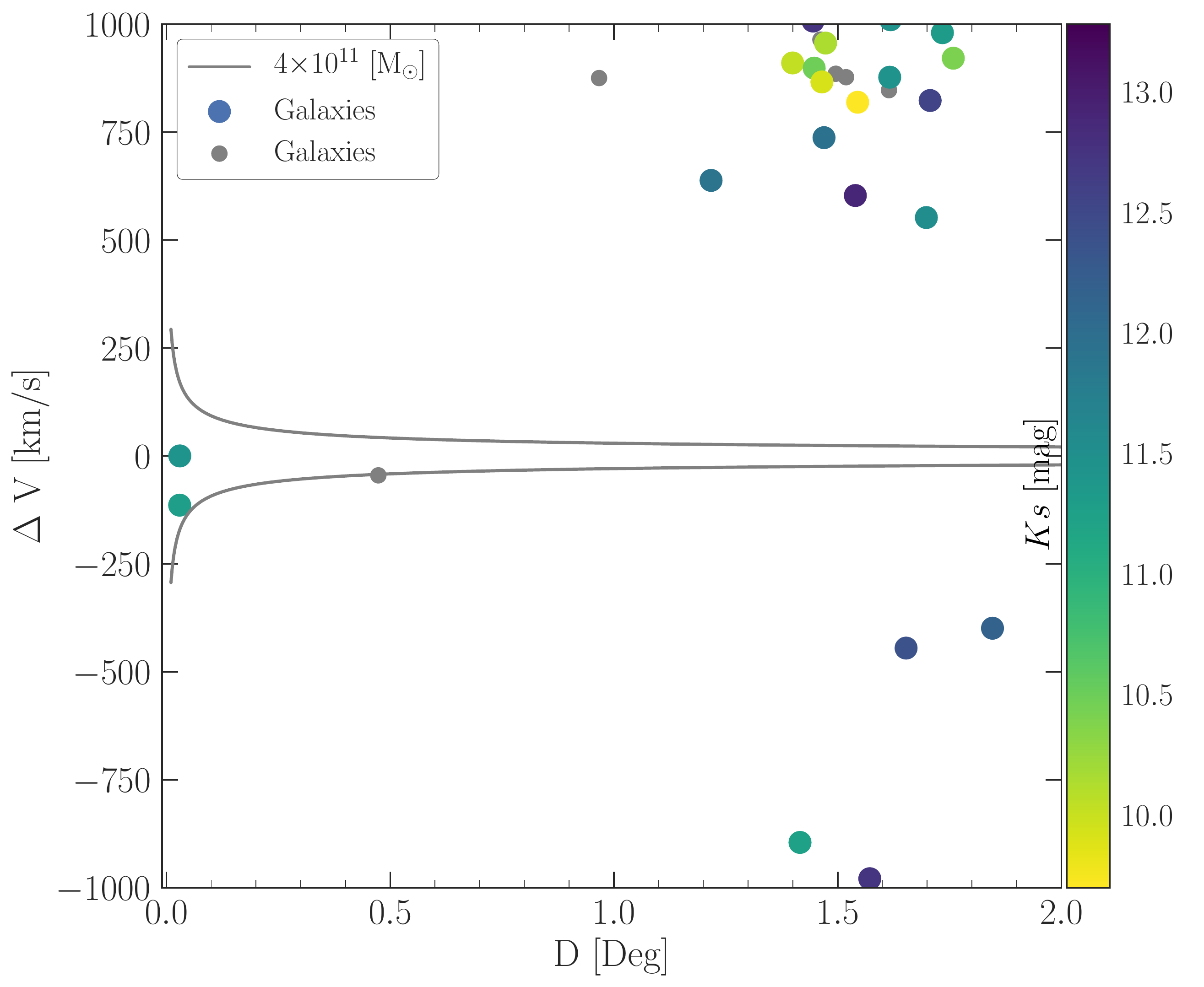}
                \caption{}
            \end{subfigure}
        \end{tabular}\\

    \end{tabular}
\caption{\textbf{a)} \textbf{Left:} The \HI\ emission in the J1250-20 group. The \HI\ emission is shown by the contours overlaid on the optical DSS-B band image. The lowest shown \HI\ column density 3.5$\times$10$^{19}$ cm$^{-2}$. The other \HI\ column density contours are: 7, 20, 30, 40, and 45$\times10^{19}$ cm$^{-2}$. \textbf{Right:} The velocity field of the J1250-20 group in the limit from 7640 to 7890 \kms. The group members that are within the field of view and detected in \HI\ are marked as S1, S2, S3 and S4, and we show the lowest \HI\ column density (3.5$\times$10$^{19}$ cm$^{-2}$). \textbf{b)} Panels of the \HI\ spectra for each detected galaxy in the J1250-20 group. The enhanced colour shows where the spectrum was integrated. \textbf{c)} The global environment around J1250-20, centred on the weighted mean of the group. The black circles are denoting radii of 0.5, 1, 2 and 3 Mpc, respectively. We show the velocity difference between the S1 galaxy (\HIPASS\ source) and the other nearby galaxies. \textbf{d)} Projected angular separation of the galaxies within J1250-20 region versus recessional velocity difference between S1 galaxy (\HIPASS\ source) and other nearby galaxies. Grey solid curved lines show simple caustics curves for a potential of 4$\times$10$^{11}$ M$\odot$, assuming that all sources are at the J1250-20 distance of 114 Mpc.}
\label{fig:j1250_group}
\end{figure}

\begin{figure}

\begin{subfigure}{\columnwidth}
\centering
    \includegraphics[width=0.45\columnwidth]{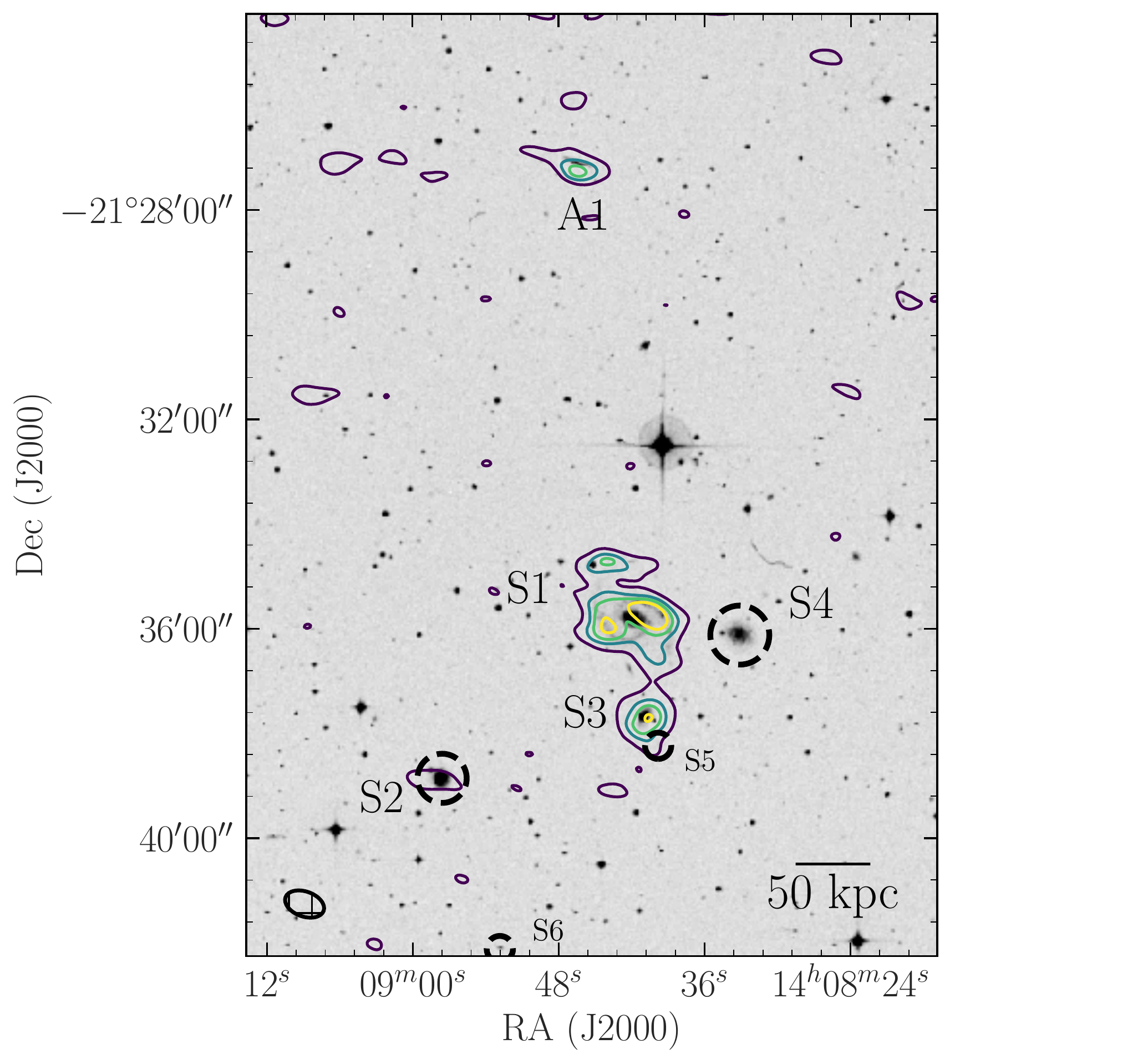}
    \includegraphics[width=0.45\columnwidth]{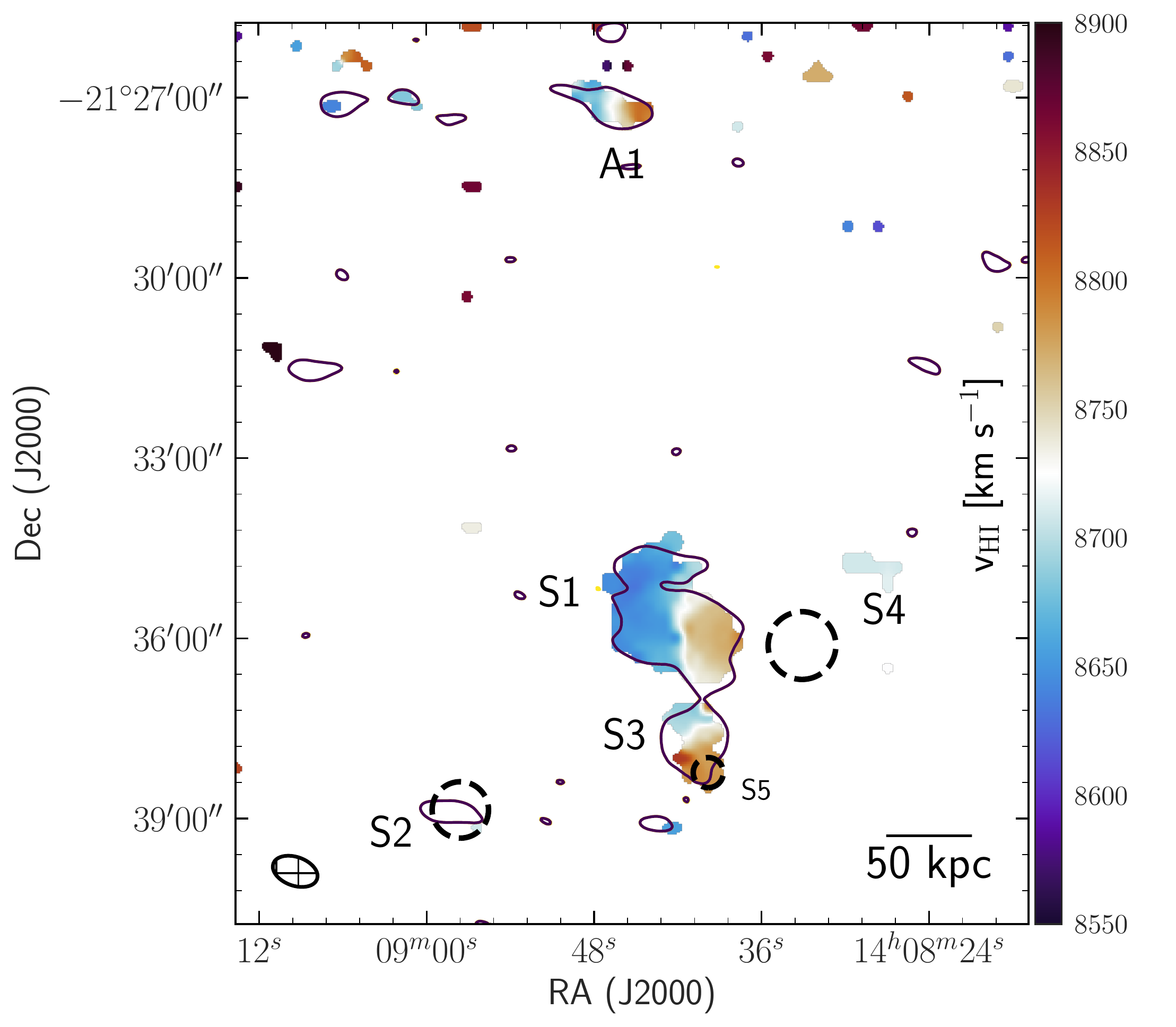}    
    \caption{}
\end{subfigure}

    \centering
    \begin{tabular}[t]{cc}

\begin{subfigure}{0.45\textwidth}
    \centering
    \smallskip
    \includegraphics[width=0.95\linewidth,height=1.4\textwidth]{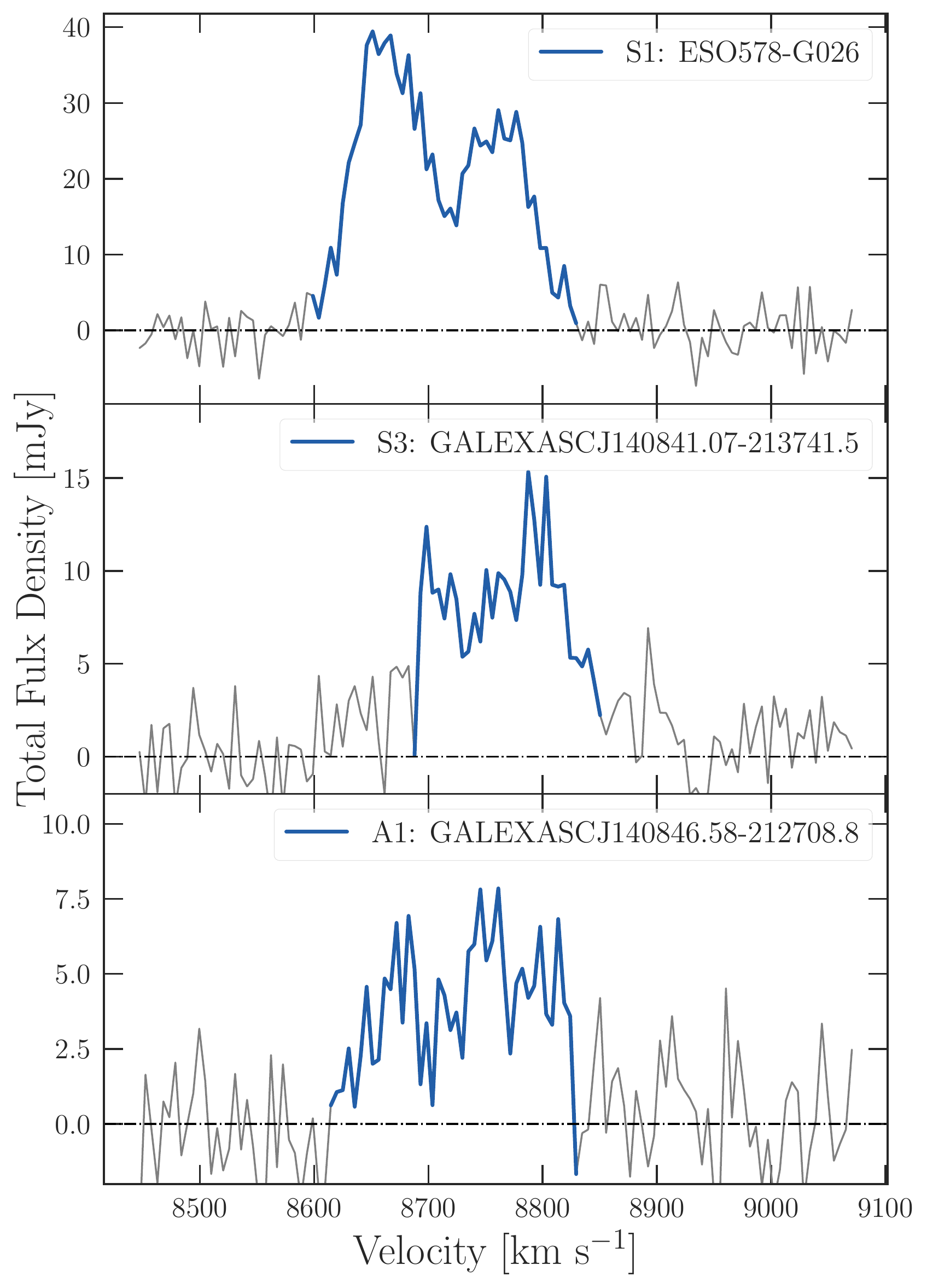}
    \caption{} 
\end{subfigure}
    &
        \begin{tabular}{c}
        \smallskip
            \begin{subfigure}[t]{0.42\textwidth}
                \centering
                \includegraphics[width=0.94\textwidth]{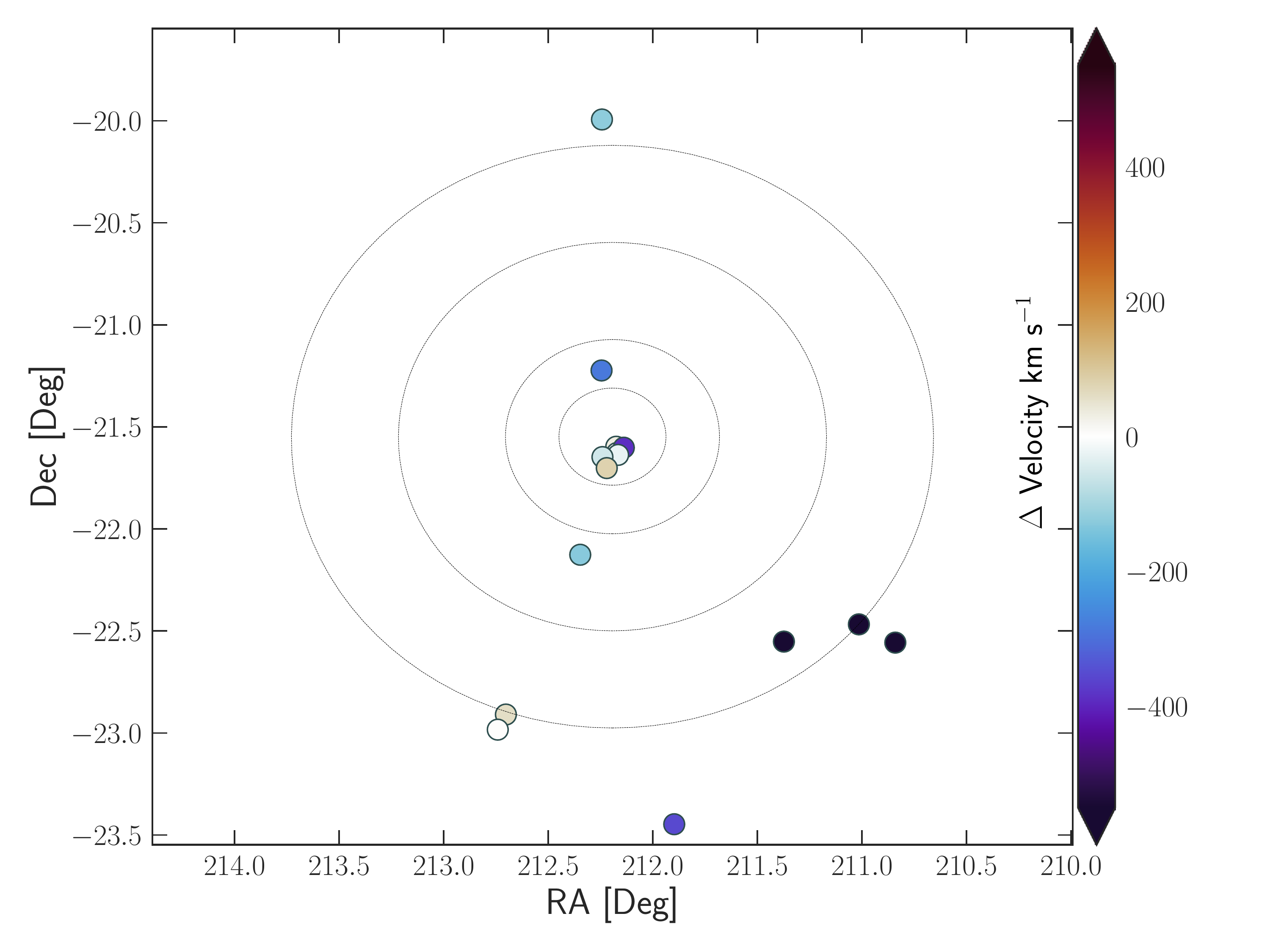}
                \caption{}
            \end{subfigure}\\
            \begin{subfigure}[t]{0.43\textwidth}
                \centering
                \includegraphics[width=0.95\textwidth]{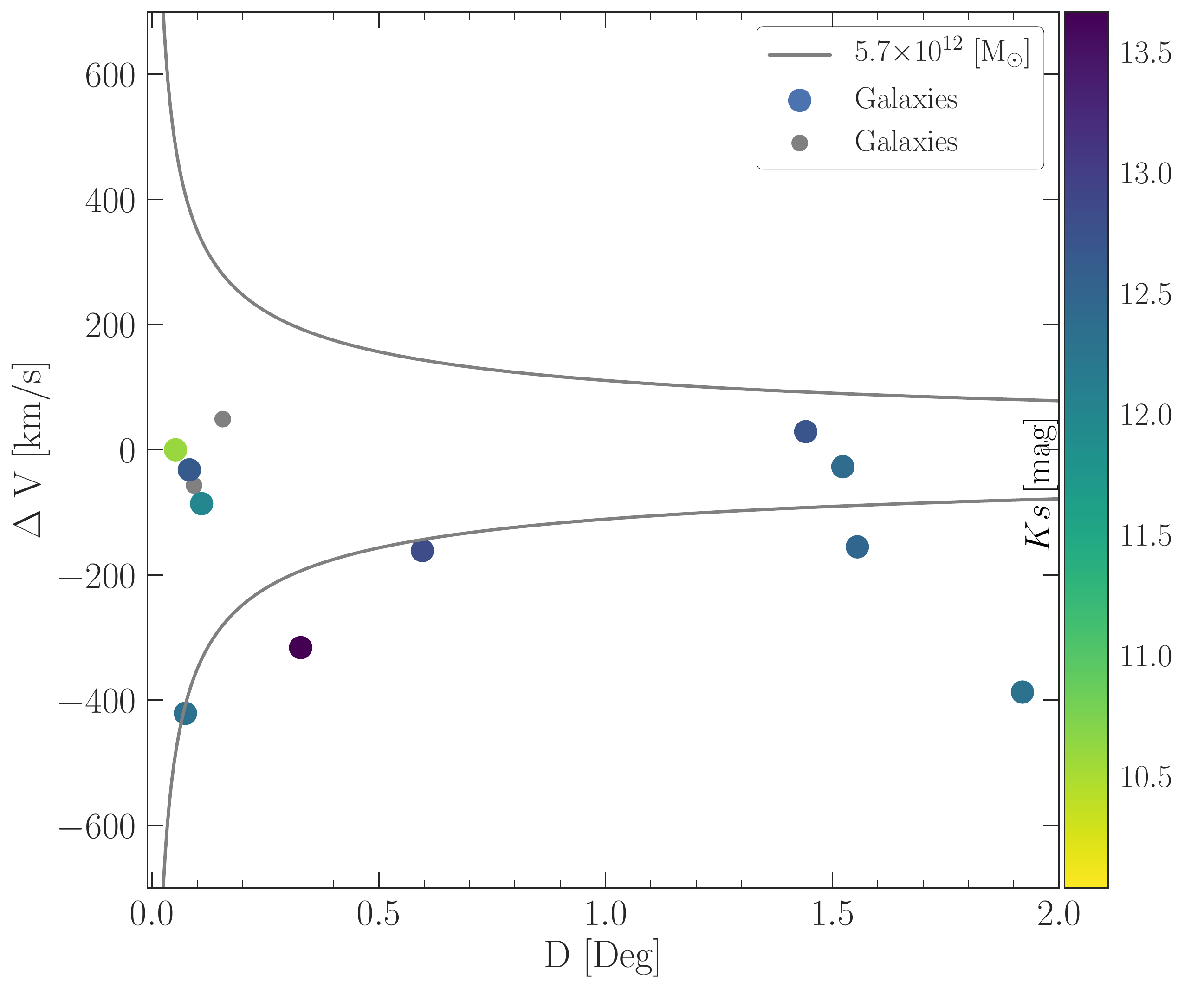}
                \caption{}
            \end{subfigure}
        \end{tabular}\\

    \end{tabular}
\caption{\textbf{a)} \textbf{Left:} The \HI\ emission in the J408-21 group. The \HI\ emission is shown by the contours overlaid on the optical DSS image. The lowest shown \HI\ column density 24$\times$10$^{19}$ cm$^{-2}$. The other \HI\ column density contours are: 40, 50, and 60$\times10^{19}$ cm$^{-2}$. \textbf{Right:} The velocity field of the J408-21 group in the limit from 8550 to 8900 \kms. The group members that are within the field of view and detected in \HI\ are marked as S1, S3 and A1, and we show the lowest \HI\ column density (24$\times$10$^{19}$ cm$^{-2}$). \textbf{b)} Panels of the \HI\ spectra for each detected galaxy in the J1408-21 group. The enhanced colour shows where the spectrum was integrated. \textbf{c)} The global environment around J1408-21, centred on the weighted mean of the group. The black circles are denoting radii of 0.5, 1, 2 and 3 Mpc, respectively. We show the velocity difference between the S1 galaxy (\HIPASS\ source) and the other nearby galaxies. \textbf{d)} Projected angular separation of the galaxies within J1408-21 region versus recessional velocity difference between S1 galaxy (\HIPASS\ source) and other nearby galaxies. Grey solid curved lines show simple caustics curves for a potential of 5.7$\times$10$^{12}$ M$\odot$, assuming that all sources are at the J1408-21 distance of 128 Mpc.}
\label{fig:j1408_group}
\end{figure}

\begin{figure}

\begin{subfigure}{\columnwidth}
\centering
    \includegraphics[width=0.45\columnwidth]{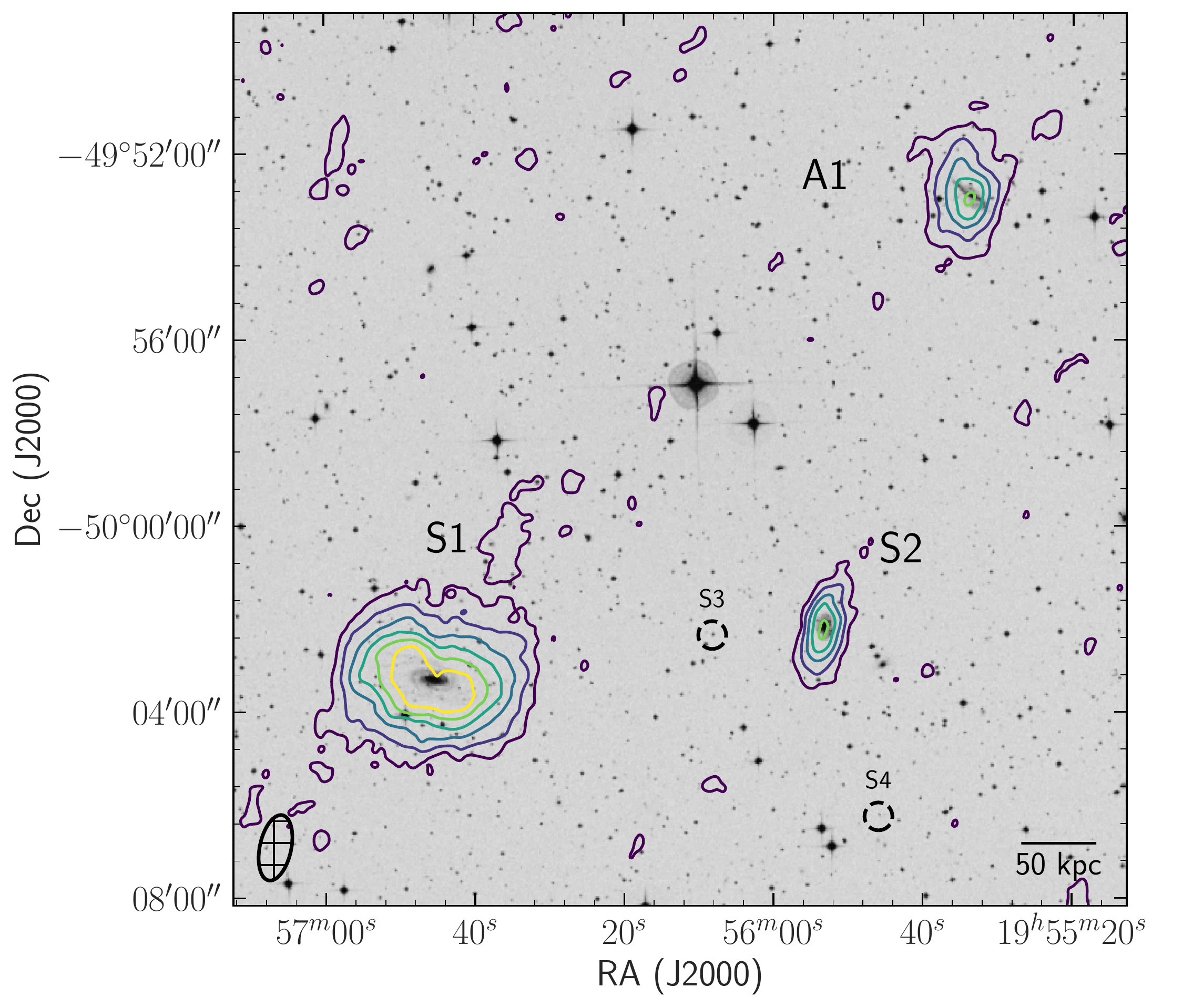}
    \includegraphics[width=0.45\columnwidth]{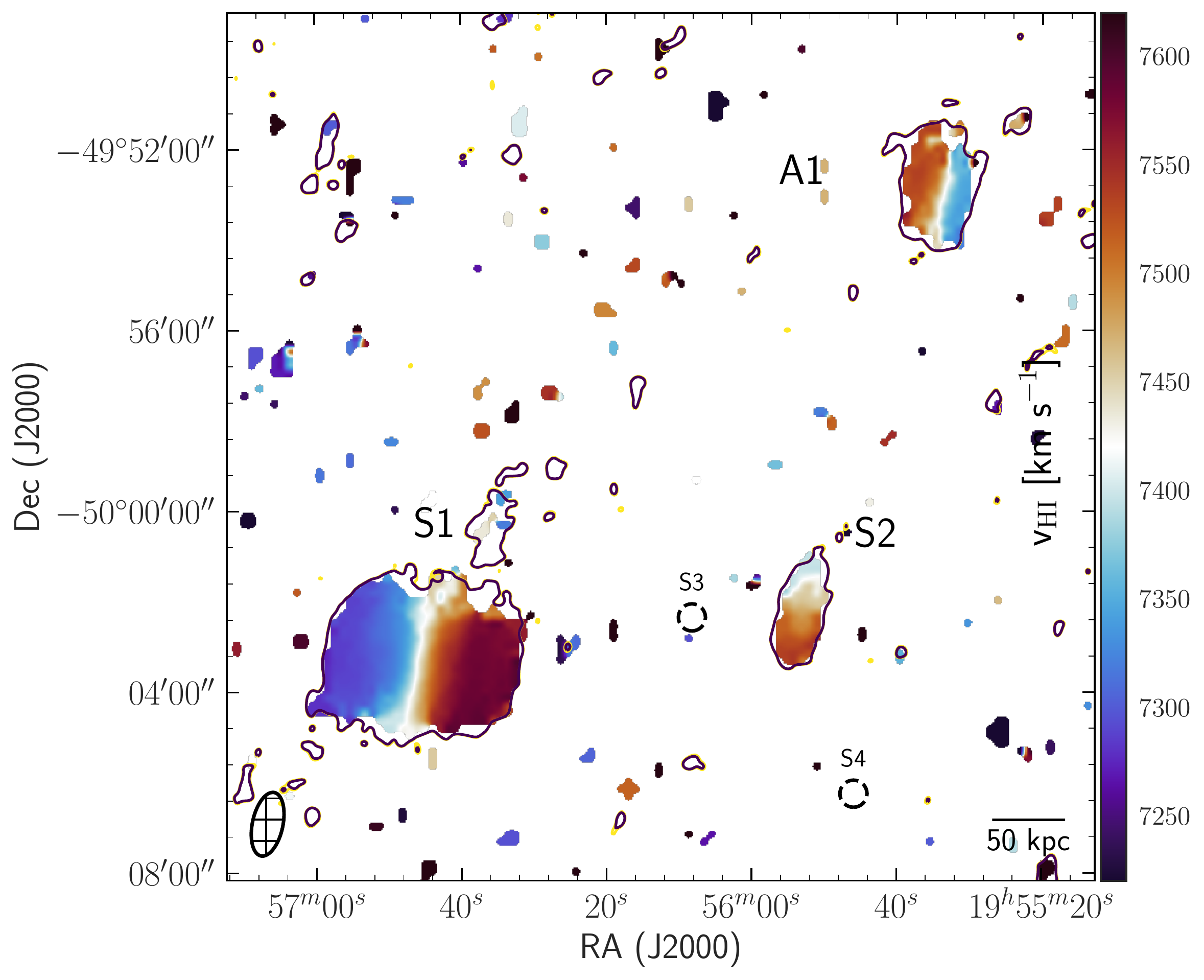}    
    \caption{}
\end{subfigure}

    \centering
    \begin{tabular}[t]{cc}

\begin{subfigure}{0.45\textwidth}
    \centering
    \smallskip
    \includegraphics[width=0.95\linewidth,height=1.4\textwidth]{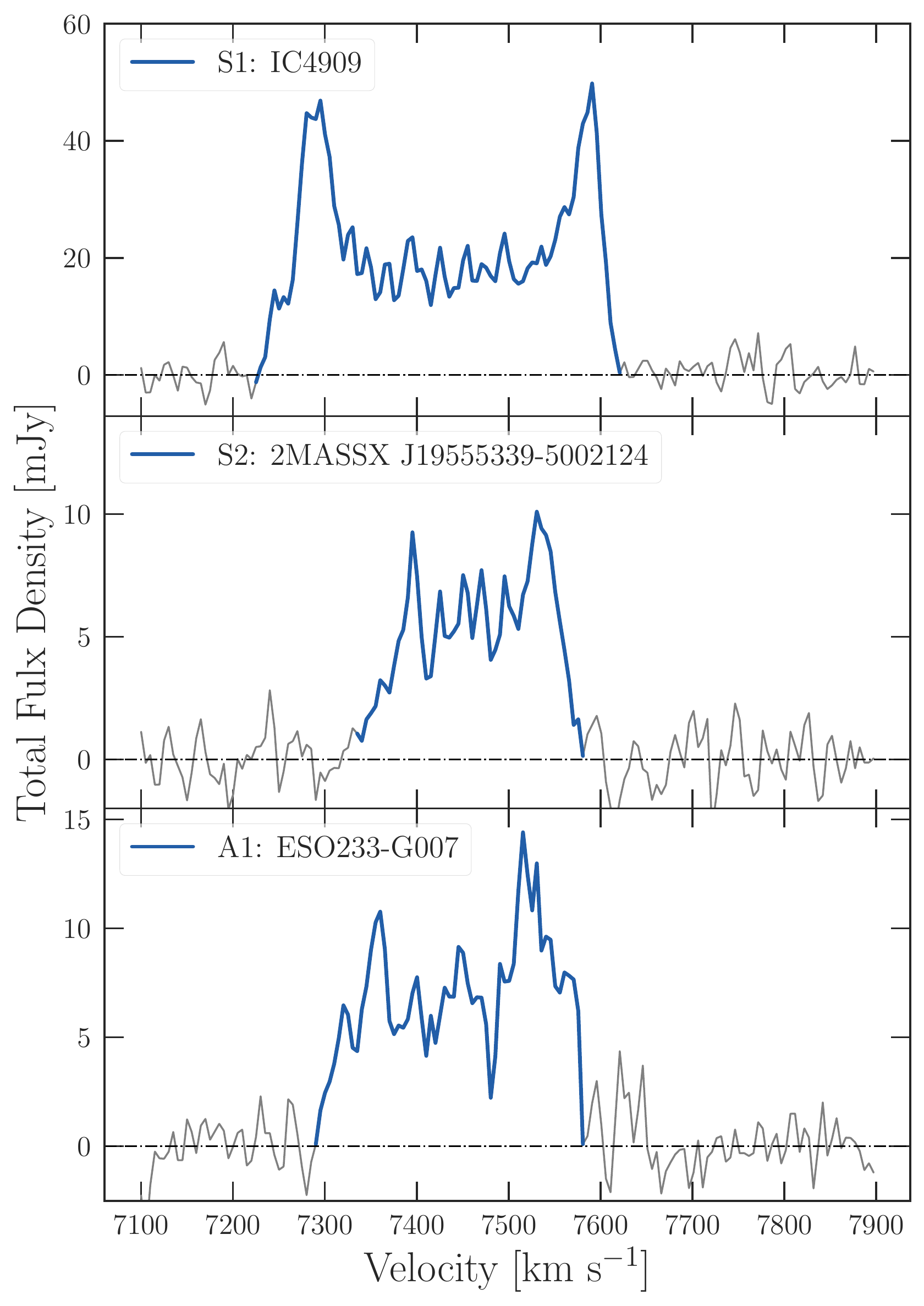}
    \caption{} 
\end{subfigure}
    &
        \begin{tabular}{c}
        \smallskip
            \begin{subfigure}[t]{0.42\textwidth}
                \centering
                \includegraphics[width=0.94\textwidth]{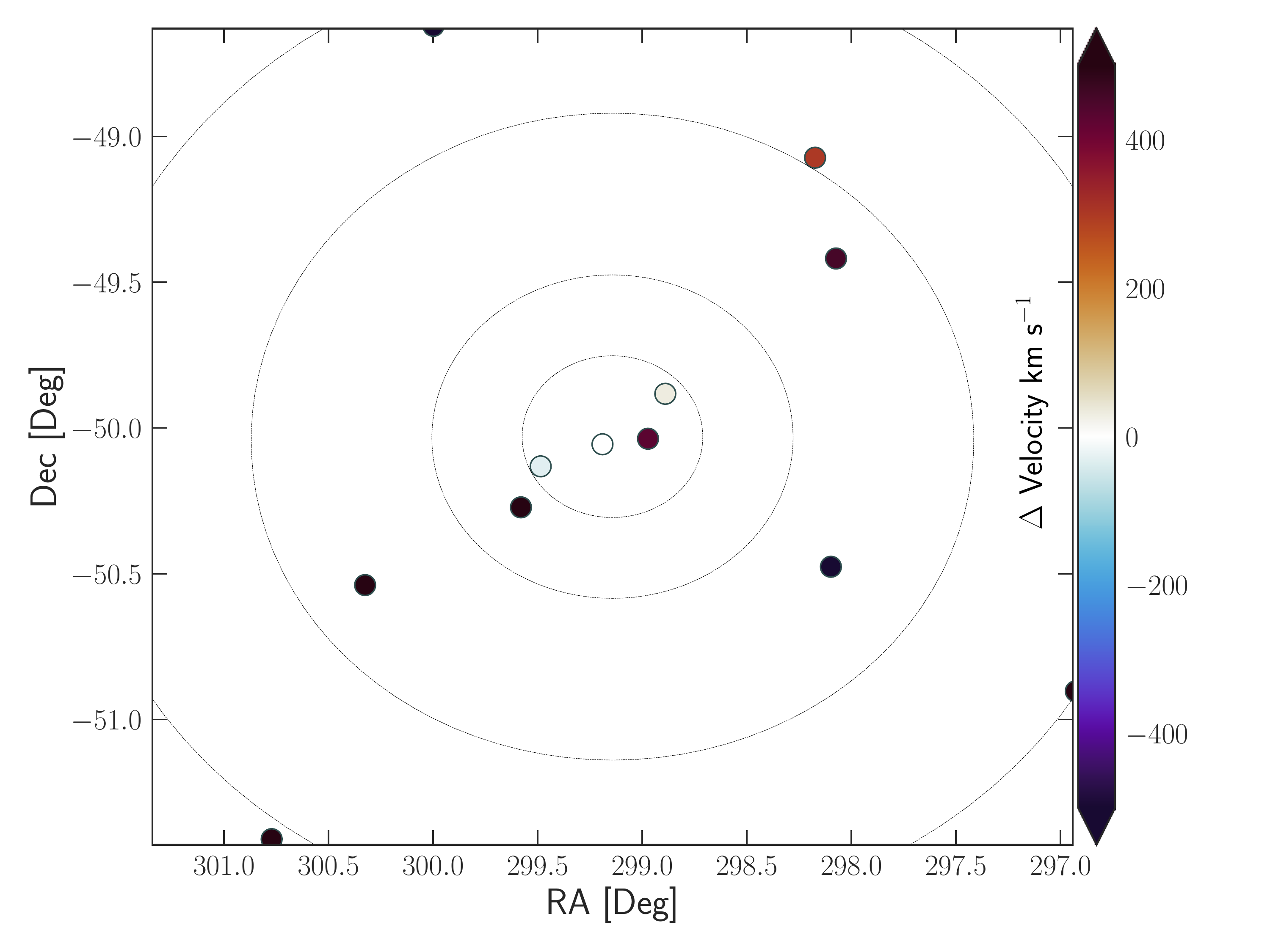}
                \caption{}
            \end{subfigure}\\
            \begin{subfigure}[t]{0.43\textwidth}
                \centering
                \includegraphics[width=0.95\textwidth]{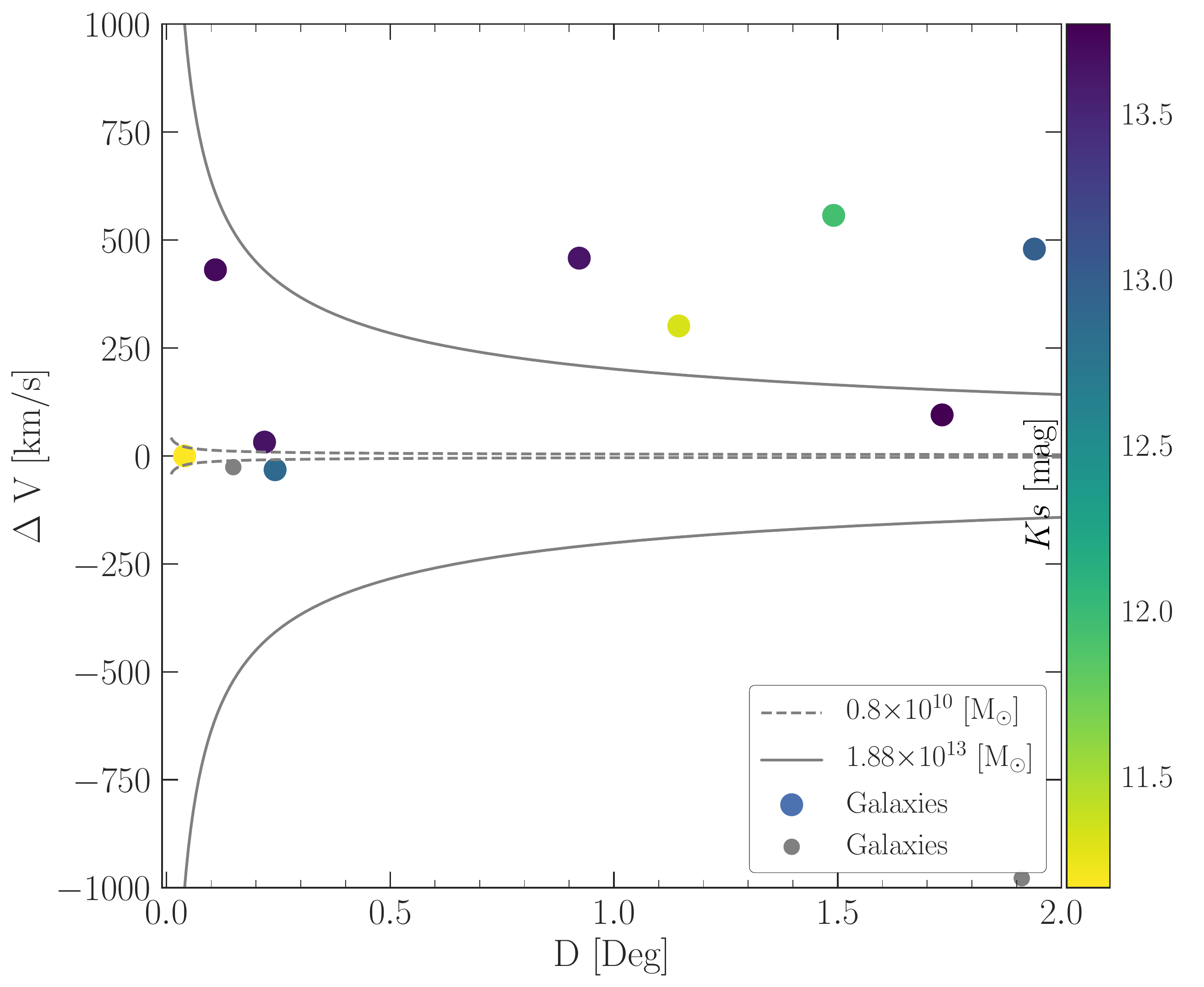}
                \caption{}
            \end{subfigure}
        \end{tabular}\\

    \end{tabular}
\caption{\textbf{a)} \textbf{Left:} The \HI\ emission in the J1956-50 group. The \HI\ emission is shown by the contours overlaid on the optical DSS image. The lowest shown \HI\ column density 1.8$\times$10$^{19}$ cm$^{-2}$. The other \HI\ column density contours are: 7, 14, 21, 28 and 35$\times10^{19}$ cm$^{-2}$. \textbf{Right:} The velocity field of the J1956-50 group in the limit from 7220 to 7620 \kms. The group members that are within the field of view and detected in \HI\ are marked as S1, S2 and A1, and we show the lowest \HI\ column density (1.8$\times$10$^{19}$ cm$^{-2}$). The synthesized beam is shown in the bottom left corner, and the scale bar in the bottom right corner shows 50 kpc at the group distance. \textbf{b)} Panels of the \HI\ spectra for each detected galaxy in the J1956-50 group. The enhanced colour shows where the spectrum was integrated. \textbf{c)} The global environment around J1956-50, centred on the weighted mean of the group. The black circles are denoting radii of 0.5, 1, 2 and 3 Mpc, respectively. We show the velocity difference between the S1 galaxy (\HIPASS\ source) and the other nearby galaxies. \textbf{d)} Projected angular separation of the galaxies around J1956-50 region versus the recessional velocity difference between S1 galaxy (\HIPASS\ source) and other nearby galaxies. Grey solid (including galaxy at $\Delta$v$\sim$500 \kms) and dashed (excluding galaxy at $\Delta$v$\sim$500 \kms) curved lines show simple caustics curves for a potential of M $=$ 1.89$\times$10$^{13}$ and 0.8$\times$10$^{10}$ M$\odot$ respectively, assuming that all sources are at the J1956-50 distance of 110 Mpc.}
\label{fig:j1956_group}
\end{figure}

\begin{figure}

\begin{subfigure}{\columnwidth}
\centering
    \includegraphics[width=0.45\columnwidth]{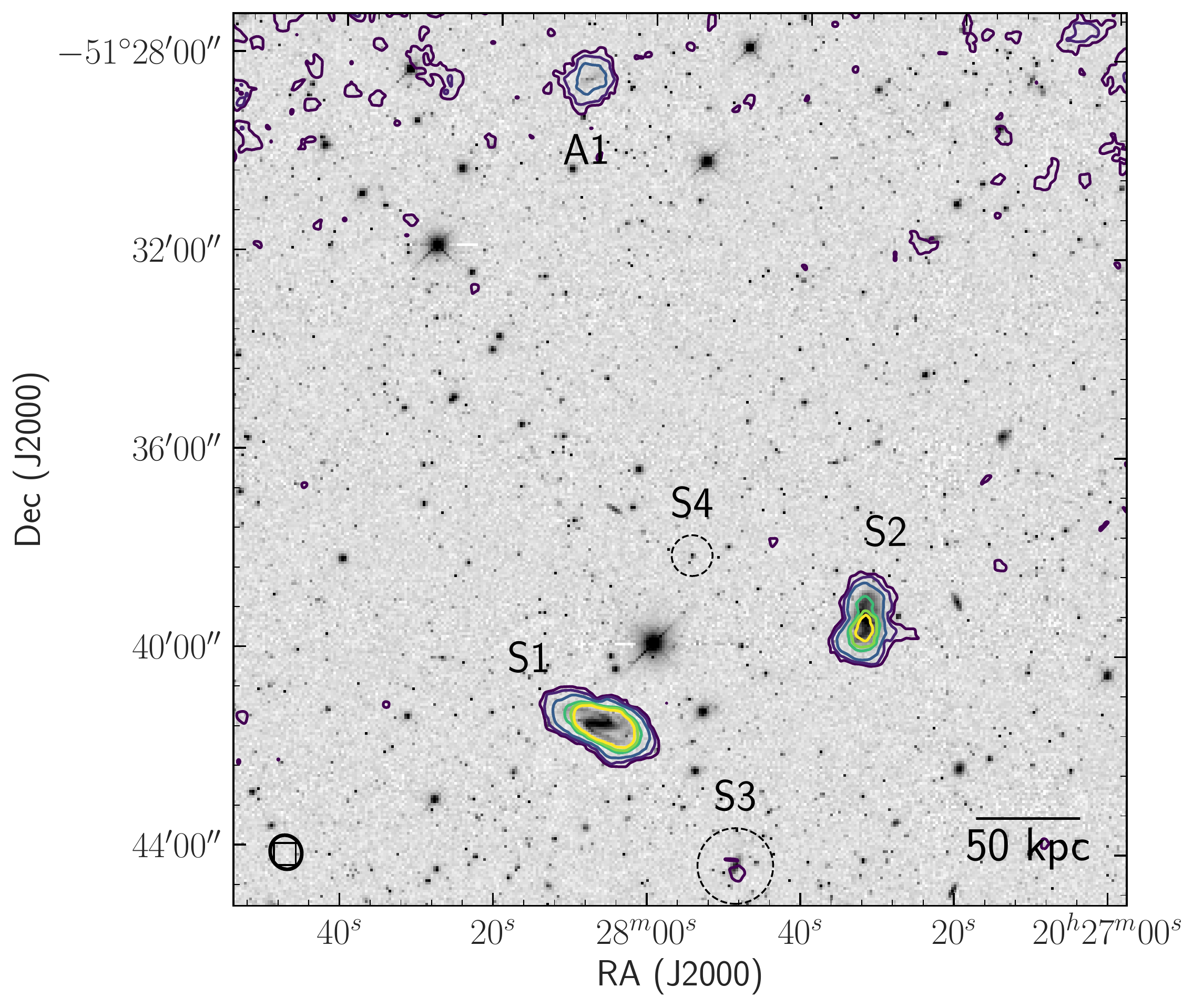}
    \includegraphics[width=0.45\columnwidth]{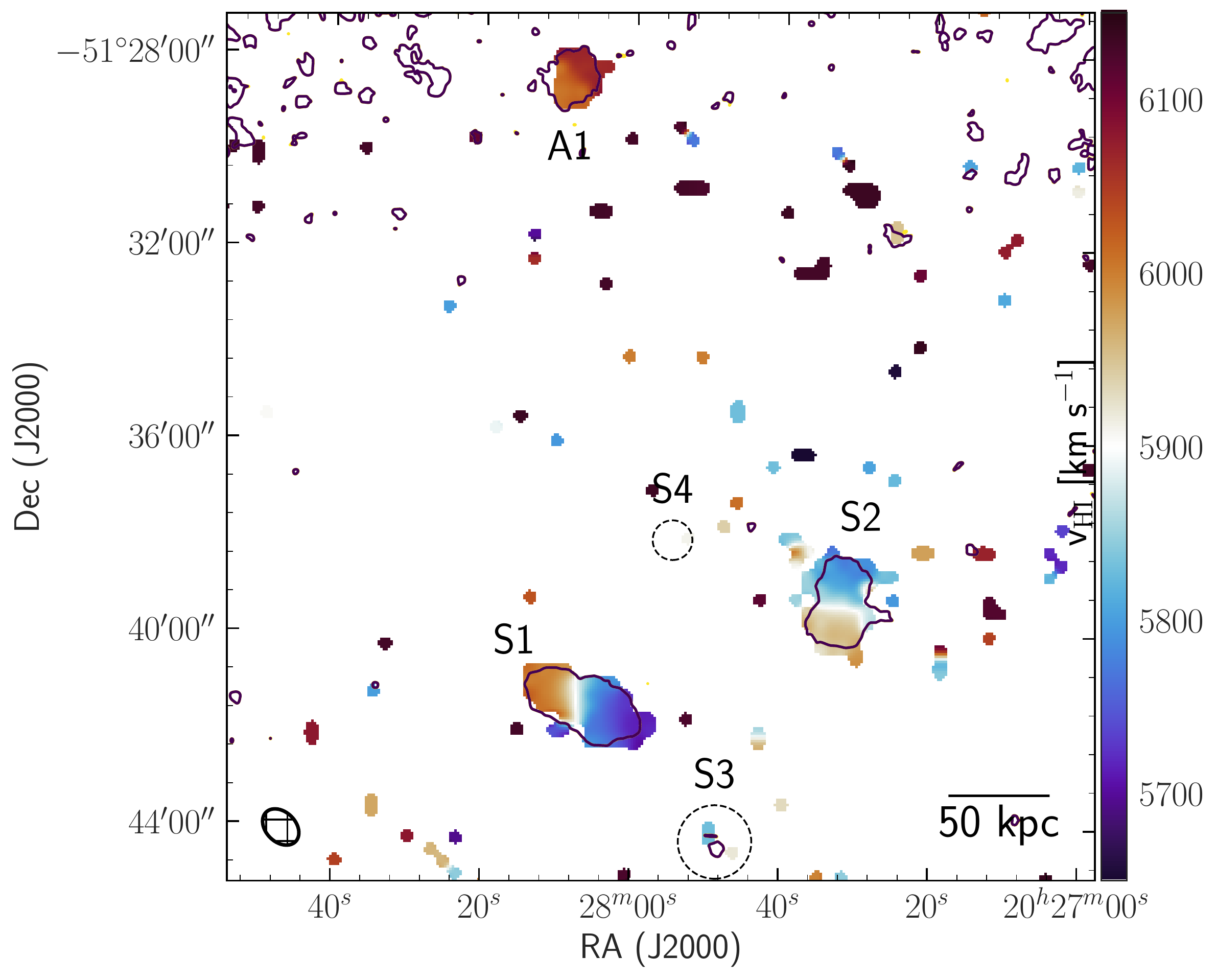}    
    \caption{}
\end{subfigure}

    \centering
    \begin{tabular}[t]{cc}

\begin{subfigure}{0.45\textwidth}
    \centering
    \smallskip
    \includegraphics[width=0.95\linewidth,height=1.4\textwidth]{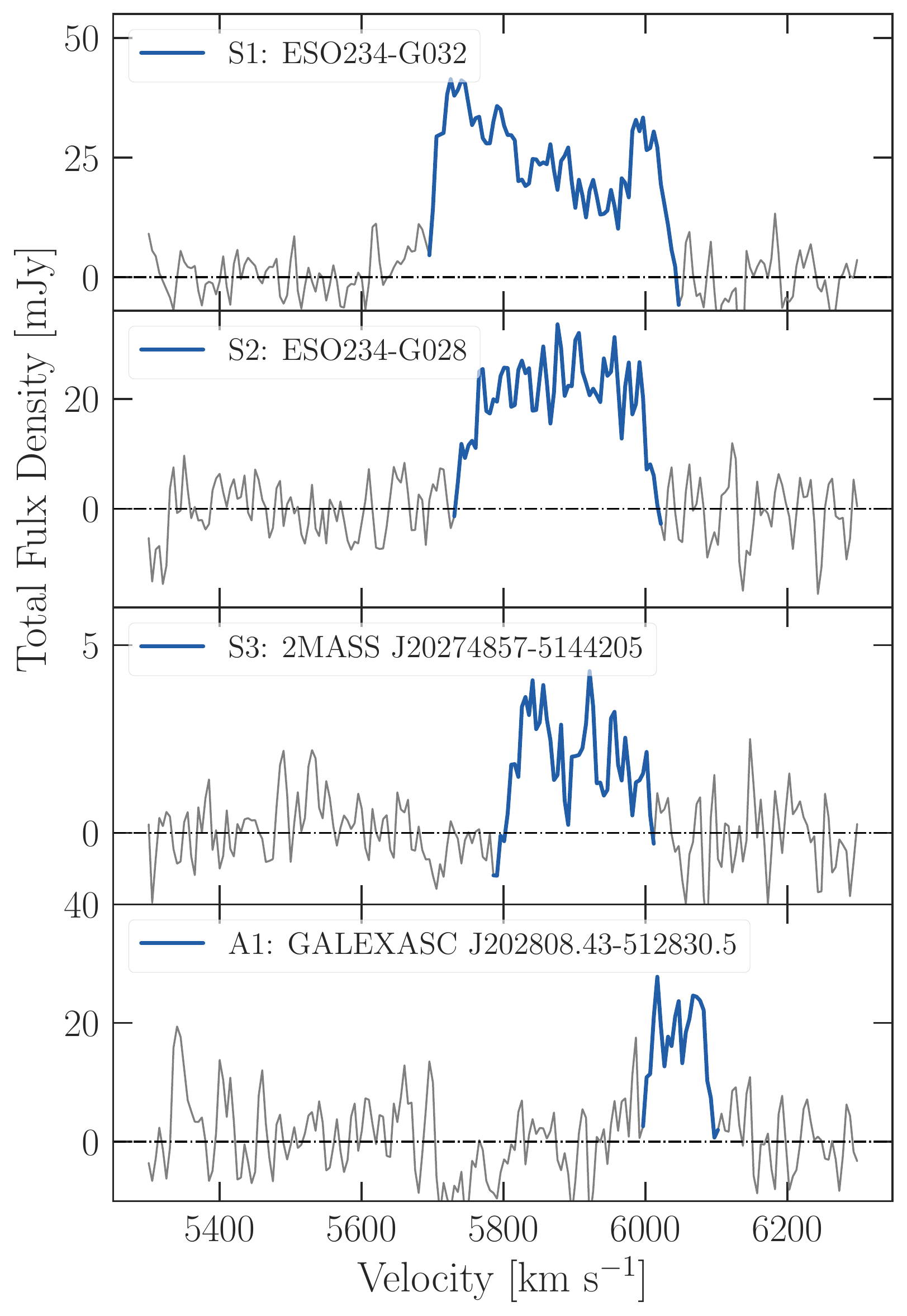}
    \caption{} 
\end{subfigure}
    &
        \begin{tabular}{c}
        \smallskip
            \begin{subfigure}[t]{0.42\textwidth}
                \centering
                \includegraphics[width=0.94\textwidth]{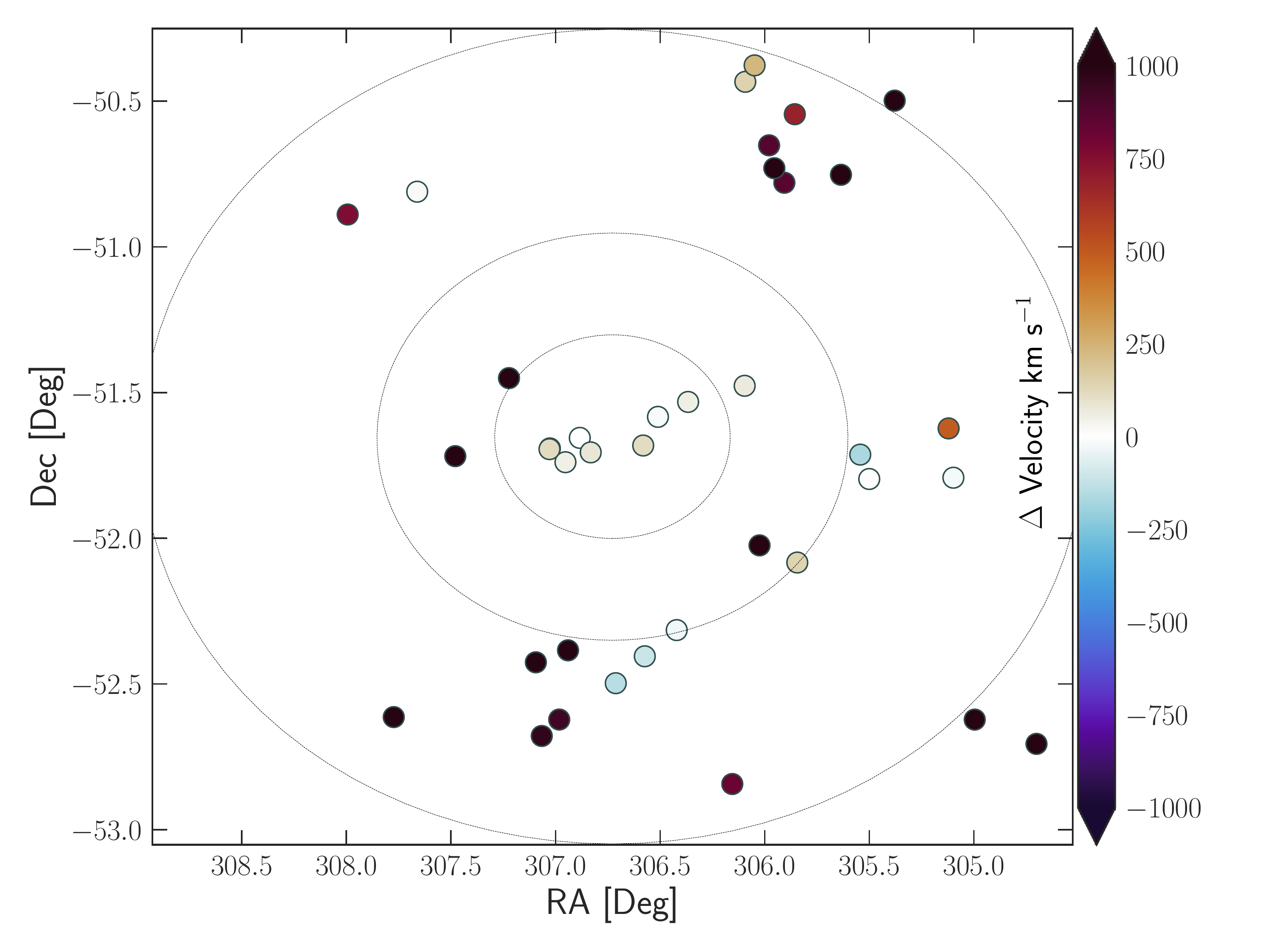}
                \caption{}
            \end{subfigure}\\
            \begin{subfigure}[t]{0.43\textwidth}
                \centering
                \includegraphics[width=0.95\textwidth]{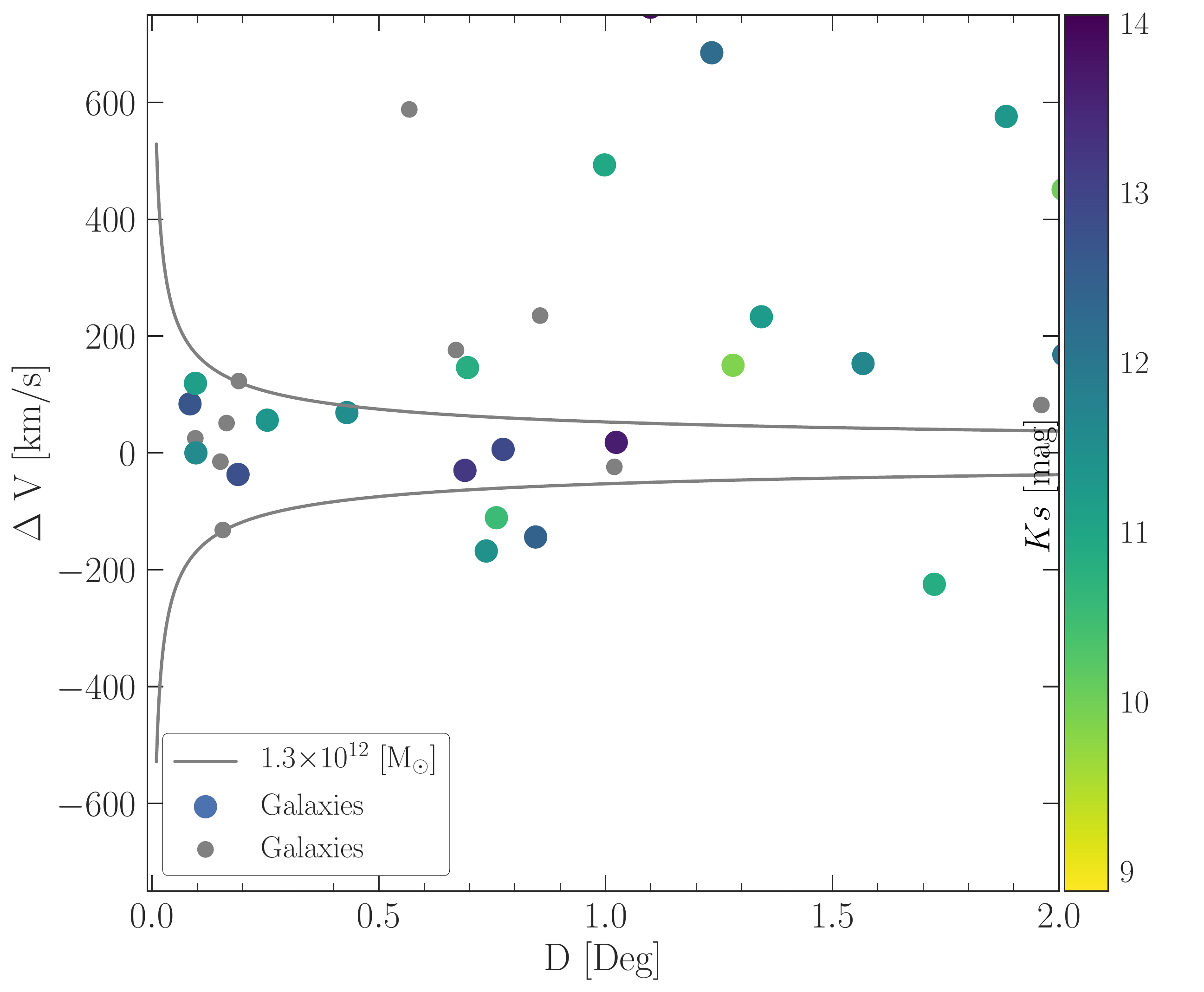}
                \caption{}
            \end{subfigure}
        \end{tabular}\\

    \end{tabular}
\caption{\textbf{a)} \textbf{Left:} The \HI\ emission in the J2027-51 group, zoomed in into the central region. The \HI\ emission in the J2027-51 group. The \HI\ emission is shown by the contours overlaid on the optical DECam \textit{r}-band image. The lowest shown \HI\ column density 8$\times$10$^{19}$ cm$^{-2}$. The other \HI\ column density contours are: 16, 32, 64, 75 and 90$\times10^{19}$ cm$^{-2}$. The S3 galaxy has a weak possible detection (see S3 spectrum in panel b), while the S4 galaxy is not detected in \HI. At the North we detected new galaxy, labelled A1. \textbf{Right:} The velocity field of the J2027-51 group in the limit from 5650 to 6150 \kms. The group members that are within the field of view are marked (S1, S2, S4 and A1), and we show the lowest \HI\ column density (8$\times$10$^{19}$ cm$^{-2}$). The synthesized beam is shown in the bottom left corner, and the scale bar in the bottom right corner shows 50 kpc at the group distance. \textbf{b)} Panels of the \HI\ spectra for each detected galaxy in the J2027-51 group. The enhanced colour shows where the spectrum was integrated. \textbf{c)} The global environment around J2027-51, centred on the weighted mean of the group. The black circles are denoting radii of 0.5, 1, and 2 Mpc, respectively. We show the velocity difference between the S1 galaxy (\HIPASS\ source) and the other nearby galaxies. \textbf{d)} Projected angular separation of the galaxies within J2027-51 region versus recessional velocity difference between S1 galaxy (\HIPASS\ source) and other nearby galaxies. Grey solid curved lines show simple caustics curves for a potential of 1.3$\times$10$^{12}$ M$\odot$, assuming that all sources are at the J2027 distance of 87 Mpc. }
\label{fig:j2027_group}
\end{figure}

In Figures \ref{fig:j0205_group} to \ref{fig:j2027_group} we illustrate \HI\ observations of the groups and their environment for which include the following: 
\begin{itemize}
    \item[\textbf{a)}] The group \HI\ intensity distribution (left) and the \HI\ velocity field of the group (right). We specify in each figure description what is the lowest column density and the \HI\ group velocity range. The detected galaxies in \HI\ are marked with names that are used throughout this work. 
    \item[\textbf{b)}] In each sub-panel we show the \HI\ spectra of the detected galaxies in the group. We show with the highlighted colour range where spectra was integrated to get the \HI\ properties. 
    \item[\textbf{c)}] The position (Right Ascension and Declination) of the galaxies around each group. The colourbar shows the velocity difference of each galaxy from the group mean velocity. In Figure we also show the circles of the constant projected separation from the group centre, which are 0.5, 1 and 2 Mpc, if not otherwise specified in the Figure description.
    \item[\textbf{d)}] The projected angular separation versus the velocity difference of each galaxy from the group mean velocity. The coloured points are showing galaxies above the adopted Ks magnitude limit of $\sim$13.5 mag, while the grey points show galaxies that are below the Ks magnitude limit. The grey lines show the simple caustics curves that are calculated with: $v = \sqrt{\frac{G M_{200}}{2r}}$, assuming that all galaxies are at the group distance. 
    \item[*] Each Figure contains its own description with properties for the shown group.
    \item[*] Process of obtaining environment around Choirs is explained in Section \ref{sec:obtaining_environment}.
\end{itemize}

\clearpage
\onecolumn


\clearpage
\onecolumn


\clearpage
\onecolumn


\clearpage
\onecolumn


\clearpage
\onecolumn


\clearpage
\onecolumn


\clearpage
\onecolumn


\clearpage
\onecolumn


\clearpage
\onecolumn


\clearpage
\onecolumn

\end{document}